\documentclass[review]{elsarticle}
\usepackage[colorlinks,citecolor=blue,linktoc=all,linkcolor=cyan]{hyperref}

\usepackage[utf8]{inputenc}

\usepackage[T1]{fontenc}

\usepackage{dsfont}               
\usepackage{mathrsfs}             
\usepackage{slashed}              
\usepackage{amsmath}
\usepackage{amssymb}
\usepackage{amsbsy}
\usepackage{amsfonts}

\numberwithin{equation}{section}
\numberwithin{table}{section}
\numberwithin{figure}{section}

\journal{Progress in Particle and Nuclear Physics}

\usepackage{titlesec}
\usepackage{sectsty}
\titleformat{\section}{\normalfont\Large\bfseries}{\thesection}{1em}{}
\titleformat{\subsection}{\normalfont\large\bfseries}{\thesubsection}{1em}{}
\titleformat{\subsubsection}{\normalfont\normalsize\bfseries}{\thesubsubsection}{1em}{}

\usepackage{xspace}
\usepackage{graphicx}
\usepackage[svgnames]{xcolor}
\usepackage{rotating}
\usepackage{bm}     
\usepackage{lineno}
\usepackage{listings}
\usepackage[normalem]{ulem}
\usepackage[section]{placeins}
\usepackage{hepunits}
\usepackage{hepparticles}
\usepackage{cancel}
\usepackage{hepnames}
\usepackage{epstopdf}
\usepackage{mathtools}
\usepackage{braket}
\usepackage{multirow}
\usepackage[nowatermark]{fixmetodonotes}

\usepackage{fullpage}
\usepackage{xspace}
\usepackage{xcolor}
\usepackage{subcaption}
\usepackage{xfrac}

\definecolor{NALblue}{rgb}{0,0.2,0.6}
\definecolor{NALgreen}{rgb}{0.317647,0.603922,0.141176}
\definecolor{NALorange}{rgb}{0.858824,0.447059,0.0470588}

\newcommand{\cut}[1]{}                          
\newcommand{\st}[1]{}
\newcommand{\ask}[1]{\textcolor{black}{#1}}

\newcommand{\mm}[1]{\textcolor{NALgreen}{#1}} 

\newcommand{\jgm}[1]{\textcolor{blue}{#1}}

\newlength{\parenbarKernelHeight}
\providecommand{\parenbar}[1]{%
    \settoheight{\parenbarKernelHeight}{\ensuremath{#1}}%
    \addtolength{\parenbarKernelHeight}{0pt}
    \llap{\raisebox{\parenbarKernelHeight}{\scalebox{0.5}[0.5]{(}}}%
    \overline{#1}%
    \rlap{\raisebox{\parenbarKernelHeight}{\scalebox{0.5}[0.5]{)}}}%
}

\newcommand{\numu}{\ensuremath{\nu_\mu}\xspace}
\newcommand{\nue}{\ensuremath{\nu_e}\xspace}
\newcommand{\nutau}{\ensuremath{\nu_{\tau}}\xspace}
\newcommand{\numubar}{\ensuremath{\overline{\nu}_\mu}\xspace}
\newcommand{\nuebar}{\ensuremath{\overline{\nu}_e}\xspace}

\newcommand{\nuany}{\ensuremath{\hspace{0.5mm}\parenbar{\nu}}\xspace}
\newcommand{\numuany}{\ensuremath{\nuany_\mu}\xspace}
\newcommand{\nueany}{\ensuremath{\nuany_{e}}\xspace}

\providecommand{\ssqthonethree}{\ensuremath{\sin^2\theta_{13}}\xspace}
\providecommand{\ssqthtwothree}{\ensuremath{\sin^2\theta_{23}}\xspace}

\providecommand{\deltacp}{\ensuremath{\delta_{\scriptscriptstyle\mathrm{CP}}}\xspace}

\providecommand{\dmsqtwothree}{\ensuremath{\Delta{}m^2_{32}}\xspace}

\newcommand{\dcp}{\ensuremath{\delta_{CP}}\xspace}

\newcommand{\minerva}{MINER$\nu$A\xspace}
\newcommand{\HK}{T2HK\xspace}

\begin{document}

\begin{frontmatter}



\title{
\vspace*{+1.0 cm} NuSTEC Review: Current Experimental and Theoretical
Challenges of \\ Neutrino-Nucleus Scattering}

\date{\today}

\author[UMN]{A.~P.~Furmanski\corref{mycorrespondingauthor}}
\address[UMN]{University of Minnesota, Minneapolis, MN 55414, USA}
\cortext[mycorrespondingauthor]{Corresponding author}
\ead{afurmans@umn.edu}

\author[FNAL]{J.~G.~Morf\'in}
\address[FNAL]{Fermi National Accelerator Laboratory, Batavia, Illinois 60510, USA}

\author[Valencia]{L.~Alvarez-Ruso}
\address[Valencia]{Instituto de F\'{i}sica Corpuscular (IFIC), CSIC-Universitat de Val\'{e}ncia, E-46980 Paterna, Valencia, Spain}

\author[Jagiellonian]{A.~M.~Ankowski}
\address[Jagiellonian]{M. Smoluchowski Institute of Physics, 
Jagiellonian University, 
PL-30348 Krak\'ow, Poland}

\author[TAU]{A.~Ashkenazi}
\address[TAU]{Tel Aviv University, Tel Aviv-Yafo, 6997801, Israel}

\author[Aligarh]{M.~Sajjad Athar}
\address[Aligarh]{Department of Physics, Aligarh Muslim University, Aligarh, 202002, India}


\author[CERN]{S.~Dolan} 
\address[CERN]{CERN, 1211 Geneva 23, Switzerland}


\author[Pitt]{S.~Dytman}
\address[Pitt]{Department of Physics and Astronomy, University of Pittsburgh, Pittsburgh, PA 15260, USA}

\author[FNAL]{S.~Gardiner}

\author[Seville]{R.~Gonz\'alez-Jim\'enez}
\address[Seville]{Departamento de Física Atómica, Molecular y Nuclear, Universidad de Sevilla, 41080 Sevilla, Spain}

\author[Kentucky,FNAL]{R.~J.~Hill}
\address[Kentucky]{University of Kentucky, Department of Physics and Astronomy, Lexington, KY 40506 USA}

\author[Iowa]{M.~Hostert}
\address[Iowa]{Department of Physics and Astronomy, University of Iowa, Iowa City, IA 52242, USA}

\author[Michigan]{J.~Isaacson}
\address[Michigan]{Department of Physics and Astronomy, Michigan State University, East Lansing, MI 48824 USA}

\author[Ghent]{N.~Jachowicz}
\address[Ghent]{Department of Physics and Astronomy, Ghent University, Proeftuinstraat 86, B-9000 Gent, Belgium}

\author[Oxford]{M.~Kabirnezhad}
\address[Oxford]{Department of Physics, University of Oxford, Oxford OX1 3RH, United Kingdom}

\author[Kings]{T.~Katori}
\address[Kings]{King's College London, London WC2R 2LS, UK}

\author[LBNL,Berkeley]{S.~R.~Klein}
\address[LBNL]{Lawrence Berkeley National Laboratory, Berkeley CA 94720 USA}
\address[Berkeley]{Physics Department, University of California, Berkeley, CA 94720}

\author[Bonn]{F.~Kling}
\address[Bonn]{Universit\"at Bonn, Regina-Pacis-Weg 3, 53113 Bonn, Germany}

\author[IPSA,LPHNE]{M.~Martini}
\address[IPSA]{IPSA-DRII,  63 boulevard de Brandebourg, 94200 Ivry-sur-Seine, France}
\address[LPHNE]{Sorbonne Universit\'e, CNRS/IN2P3, Laboratoire
de Physique Nucl\'eaire et de Hautes Energies (LPNHE), Paris, France}

\author[Seville]{G.~D.~Megias}

\author[UMN]{B.~A.~Messerly}

\author[EcolePoly,Ghent]{K.~Niewczas}
\address[EcolePoly]{École polytechnique, IN2P3-CNRS, Laboratoire Leprince-Ringuet, F-91120 Palaiseau, France}

\author[Washington]{A. Nikolakopoulos}
\address[Washington]{University of Washington, Seattle, WA, USA}

\author[Lancs]{J.~Nowak}
\address[Lancs]{Lancaster University, Lancaster LA1 4YW, United Kingdom}

\author[FNAL]{J.~M.~Paley}

\author[FNAL]{V.~Pandey}

\author[LANL]{A.~Papadopoulou}
\address[LANL]{Los Alamos National Laboratory, 
Los Alamos, NM 87545, USA}

\author[RAL]{L.~Pickering}
\address[RAL]{Particle Physics Department, Rutherford Appleton Laboratory, Harwell, OX11 0QX, United Kingdom}

\author[Valencia,FNAL]{N.~Rocco}

\author[Wroclaw]{J.~T.~Sobczyk}
\address[Wroclaw]{Faculty of Physics and Astronomy, University of Wroc\l aw, Plac Maxa Borna 9, 50-204  Wroc\l aw, Poland}

\author[Chalmers]{J.~E.~Sobczyk}
\address[Chalmers]{Department of Physics and Astronomy, Chalmers University of Technology, SE-412 96, Göteborg, Sweden}

\author[Manchester]{C.~Thorpe}
\address[Manchester]{University of Manchester, Manchester M13 9PL, United Kingdom}

\author[Imperial]{C.~Wret}
\address[Imperial]{Imperial College London, London, SW7 2AZ, United Kingdom}

\begin{abstract}
The precise measurement of neutrino properties is among the highest priorities in fundamental particle physics, involving many experimental and theoretical efforts worldwide. Since the experiments rely on the interactions of neutrinos in the nuclear environment, the planned advances in the scope and precision of these experiments require a commensurate effort in understanding and modeling the hadronic and nuclear physics of these interactions, as incorporated in neutrino event generators. This modeling is essential to every phase of the experimental analyses, and the theoretical uncertainties play an important role in interpreting experimental results.

In this Review of current challenges, we summarize the open questions in neutrino-nucleus interactions at GeV-scale energies, and discuss in detail the potential impacts of these unknowns on 
the measurement of neutrino properties, using the determination of oscillation parameters as a central example. 
After an Executive Summary and a concise Overview of the issues, we address how the neutrino event generators treat these challenges, how electron-nucleus interactions can help understanding these challenges and the important unanswered questions for each underlying physics process---from quasi-elastic to deep-inelastic scattering.
Across every topic, we find that current challenges can be met only through active support and collaboration among specialists in strong interactions and electroweak physics, including theorists and experimentalists from both the nuclear and high-energy physics communities.
\end{abstract}


\end{frontmatter}


\newpage
\thispagestyle{empty}
\tableofcontents

 \newpage
 
\section{Executive Summary and Priorities}

\label{exec}
The experimental study of neutrino-nucleus interactions, that began with early bubble chamber experiments, has been essential to the understanding of weak interactions. The earlier studies 
made key contributions to the establishment of the parton structure of the nucleon and the formulation of quantum chromodynamics.
The discovery of non-zero neutrino masses at the end of the twentieth century, which remains one of the very few hints regarding the nature of physics beyond the standard model of particle physics, has demanded more precise knowledge of neutrino-nucleus interactions, particularly in the few-GeV energy regime which hadn't previously been well-explored.

The ability to fully explore this new physics requires high-statistics, high-precision neutrino oscillation experiments. The current global landscape for neutrino oscillation physics, particularly in regards to exploring the potential for CP violation, centers on the construction of two long-baseline experiments: the Deep Underground Neutrino Experiment (DUNE) in the US, and Hyper-Kamiokande in Japan. 
Both have been labeled as among the highest priorities in particle physics for the next decades.
These complementary projects will determine the neutrino mass hierarchy and search for matter-antimatter asymmetry (CP violation), among other goals.
The European community has expressed significant support for both of these efforts through participation, and in the 2020 and 2026 iterations of the European Strategy for Particle Physics.

The exploitation of the physics capabilities of these facilities that include near and far detectors, requires dramatically improving our ability to describe neutrino--nucleus scattering.
The current systematic uncertainties associated with neutrino--nucleus interactions already play a significant role, even after taking into account essential information from the near detectors.
The current state-of-the-art oscillation analyses see an impact from interaction systematic uncertainties of 3--5\% on the number of events at the far detector, compared to statistical uncertainties of 5--10\%.
The next generation experiments will collect orders of magnitude higher statistics, pushing the statistical uncertainties to the 1-2\% level - considerably smaller than current systematic uncertainties.
Indeed, the 2023 P5 report highlighted that in this precision era, \textit{``control of systematic uncertainties (such as those arising from the interaction of neutrinos and nuclei) becomes increasingly crucial''}, and that \textit{``the interaction of neutrinos and nuclei represents a complex many-body quantum problem, and significant theoretical work is required to gain quantitative understanding at a subatomic or nuclear level''}.

The higher precision and the large statistics of these next-generation experiments will also provide essential and complementary information on nuclear and nucleon structure, allowing for the search of new neutrino–matter interactions, as well as providing invaluable information concerning new fundamental particles.
The 2026 European Strategy for Particle Physics states that \textit{``studies of neutrino interactions further expand the opportunities for discovery''},
with the deliberation document detailing that next-generation experiments require \textit{``a significant reduction of systematic uncertainties in neutrino fluxes and cross-sections''}, and highlighting the need for
\textit{``near detectors, hadron production experiments... complemented by advances in nuclear theory to improve the modelling of neutrino–nucleus interactions.''}

A defining challenge for neutrino oscillation experiments is that the oscillations are dependent on the neutrino's energy and the distance traveled. 
In general, however,
the incoming neutrino beam is not mono-energetic, and thus the neutrino's energy must be inferred event-by-event from the outgoing products of the neutrino's interaction in the detector.
In the case of some neutrino sources---such as atmospheric and astrophysical---the incoming neutrino's direction is also not known, and also has to be inferred from the interaction products.
A further complication is that the interactions in question are with nuclei which are complex many-body systems. This leads to an experiment having limited knowledge of both the neutrino and nuclear initial state of any given interaction.
Therefore a measurement of any given final state does not usually uniquely determine the initial state, which is the property of interest.
Experiments must then rely on Monte Carlo simulations (event generators) to connect observations in a detector to distributions of possible true kinematics.
Inaccuracies or biases in these simulations can lead to biases in the inferred oscillation parameters.
To ensure these event generators are as precise and accurate as possible, they must employ state-of-the-art theories and be benchmarked against 
data covering a diverse range of nuclear targets and beam energies, as well as complementary measurements of electron-nucleus and hadron-nucleus scattering.
Continued extensive collaboration between theorists, experimentalists, and Monte Carlo simulation experts is critical.

These Monte Carlo simulations, covering all possible channels of the interactions, are based on a collection of theoretical and phenomenological multi-discipline models, representing the overlapping frontiers of High-Energy Physics (HEP) and Nuclear Physics (NP). While HEP provides the framework to study fundamental particles and the electroweak force, NP provides the complex, many-body environment of the atomic nucleus required to interpret these interactions. The resulting nuclear physics aspect of the model should cover basic components such as the nuclear initial states, nucleon-nucleon correlations and the propagation of interaction products through the nuclear medium. 
It also has to describe modifications of these interactions in the nuclear environments, referred to as ``nuclear effects'', over a broad energy range, yielding, for example, nuclear shadowing and nuclear parton distributions. The nuclear response depends on the specific energy transferred; consequently, the relevant energy scales for even a single incoming neutrino energy can span a wide range, with essentially no lower limit and an upper limit dictated by the neutrino energy.  Simulations and predictions must naturally be able to cover these vastly different energy scales at the same time encompassing phenomena that cross into the domain of high-energy physics fundamentally concerned with the intrinsic interactions between neutrinos and the basic constituents of matter (quarks and leptons) via the weak nuclear force governed by the Standard Model. Because of this, close co-operation between nuclear physics (NP) and high energy particle physics (HEP) is essential.

The NuSTEC White Paper\cite{NuSTEC:2017hzk}, from roughly a decade ago, summarized the state of our knowledge of neutrino nucleus interactions and challenges at the time.
Since then, substantial progress has been made both theoretically and experimentally; however, significant challenges remain. 
In this paper, we summarize the progress, current status, and challenges in the field of neutrino-nucleus interactions, and highlight key priorities for the coming years.

 \vspace{\baselineskip}

Considering their impact on the priorities highlighted by the US P5 process and European Strategy, the following have been identified as critical priorities:
\begin{enumerate}
  \item{\textbf{Support for theoretical development of nuclear models.}} This includes improved theoretical handling of nuclear effects in the relativistic regime, hadronization, and final state interactions. Strong emphasis should be placed on theoretical quantification of uncertainties. Producing predictions of fully exclusive final states is critical. 

  \item{\textbf{Support for event generator development and model integration.}} As the link between theory and experiment, it is critical that the community can rapidly integrate new models into event generators.
  Developing more standardized formats and interfaces that can be shared across event generators would be valuable. 
  Generators should also improve the handling of cross sections as they transition from nucleon degrees of freedom to parton degrees of freedom. 

  \item{\textbf{Detailed measurements of hadronic final states across various topologies in expanded kinematic regions}.} 
  Measurements of neutrons (of which there are very few to-date), protons near the detection threshold, and the kinematics of pions in multi-pion final states are identified as critical.  This should be done using a variety of nuclear targets and fluxes to ensure models correctly handle both nuclear-target dependence and neutrino energy dependence.
  Measurements of correlations between the hadronic and leptonic systems, as well as between the hadrons within the hadronic system, have been shown to be very powerful.

  \item{\textbf{An extensive array of electron-nucleus and hadron-nucleus scattering measurements.}} Due to the ability to determine the scattering kinematics, electron- and hadron-nucleus scattering measurements have unique information that neutrino measurements can never provide. Data exist that include the full event kinematics exist, with measurements completed in the QE-like region - extending these measurements into the resonance-SIS/DIS transition region, including complete measurements of the hadronic final states (including neutrons as far as possible), would be highly informative.  Multiple experiments stand to provide measurements in this space.  Additionally hadron-nucleus scattering measurements provide a unique constraint on intranuclear rescattering (known as ``final state interactions'', or FSI).
  While these measurements provide unique information, they do not measure the axial response of the nucleus, which only neutrinos can provide.

  \item{\textbf{Minimization of model dependence in data.}} Experimental measurements should follow best-practices in targeting model-independent results and include complete (or best-estimate) uncertainty correlations between different measurements from the same experiment and beam. Measurements should further investigate whether reported uncertainties and correlations provide proper coverage, for example in verifying the validity of often-used Gaussian approximations.

  \item{\textbf{Improvements to flux uncertainties.}} 
Neutrino flux uncertainties are a dominant and often limiting source of uncertainty in neutrino-nucleus cross section measurements. Although neutrino experiments can provide some in-situ flux constraints using neutrino-electron elastic scattering, coherent, and inverse muon decay events--all of which would benefit from further work to reduce theoretical uncertainties--robust a-priori flux predictions remain critical. Dedicated hadron production measurements may reduce a-priori beam and atmospheric flux uncertainties to the few-percent level.
\end{enumerate}

%

\newpage
\section{Introduction and Overview of the Current Challenges}
\label{Overview}

\subsection{Introduction: General Challenges}

The recent increased interest in neutrino-nucleus interactions is mostly due to its importance in neutrino oscillation studies.
The next generation of oscillation experiments with a goal to measure CP violation phase (DUNE in the United States and Hyper-Kamiokande in Japan) are costly enterprises requiring international level coordination and cooperation.
It \cut{is not always}\ask{must be} recognized that their success may depend on a significant effort in \ask{understanding, quantifying and} reducing the systematic error coming from modeling neutrino-nucleus interactions.
Apart from a critical importance in neutrino oscillation studies, neutrino interaction research \ask{supplements electron- and photon-scattering studies of}\cut{is also relevant to} hadronic physics, \cut{providing supplementary information to that attainable in electron and photon scattering studies through the inclusion of}\ask{by including the} axial-vector \cut{current}\ask{interactions}.
Additionally, there is significant interest in searches for physics beyond the standard model with next-generation neutrino facilities, and such searches need to be able to predict neutrino-induced backgrounds.

\begin{figure}[btp]
\begin{center}
	\includegraphics[width=0.55\textwidth,trim={50pt 90pt 50pt 90pt},clip]{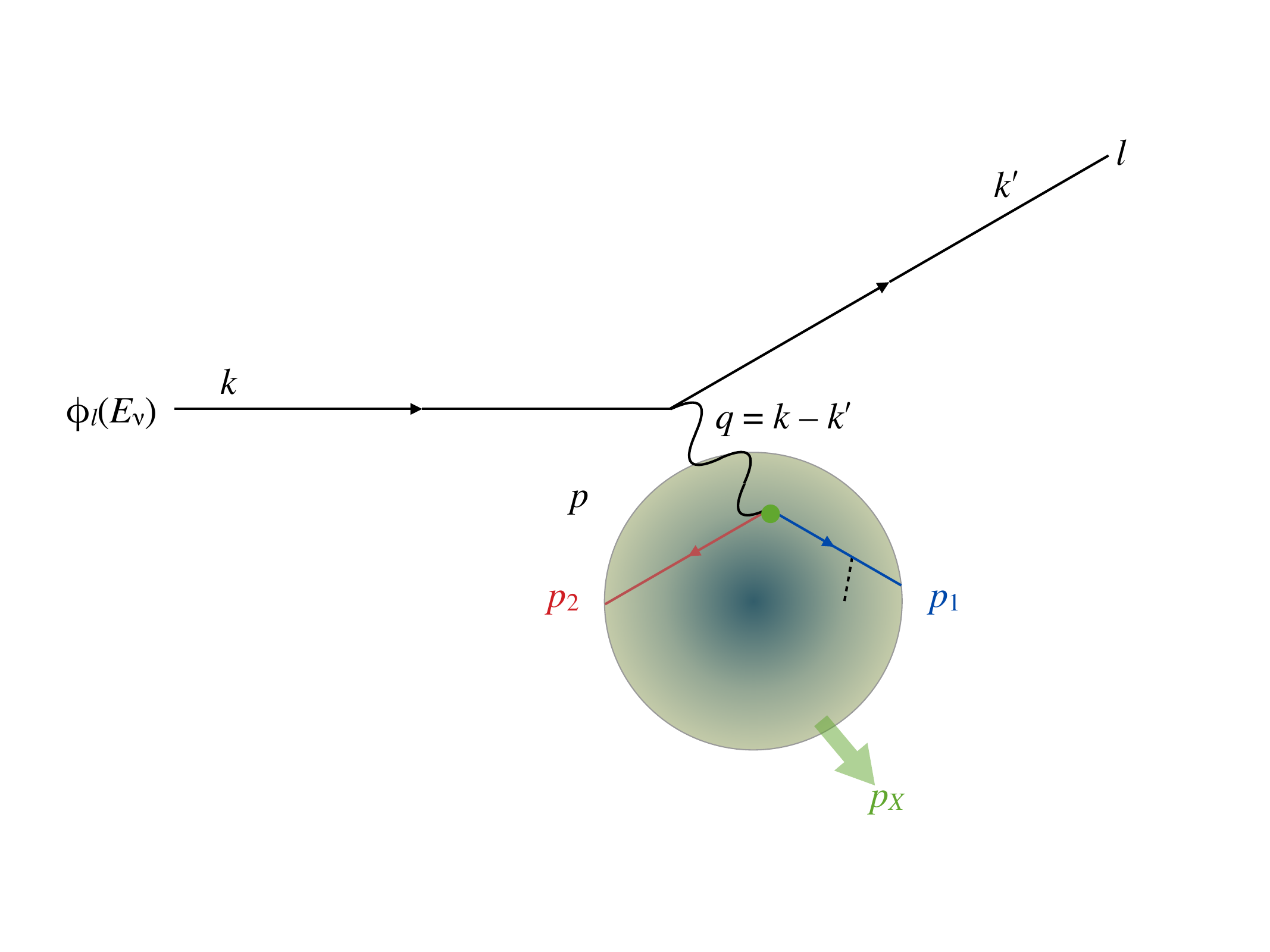}
    \caption{\color{black} In neutrino-nucleus scattering a neutrino of energy $E_\nu$ and flavor $l$ within a beam with energy spectrum~$\phi_l(E_\nu)$, strikes a nucleus of atomic number~$A$. In charged (neutral) current interaction the associated charged lepton~$l$ (neutrino of same flavor) emerges. Hadrons emerge from the initial interaction vertex as well that include one or more nucleons and, typically, pions (black dashed lines).}
    \label{fig:overview:scatter}
\end{center}
\end{figure}
The basic setup of a lepton-nucleus scattering experiment is shown in Fig.~\ref{fig:overview:scatter}. A neutrino of unknown energy enters the detector made of heavier nuclei and interacts.
In charged-current neutrino scattering, the final-state lepton is the charged partner of the incoming flavor while in neutral-current scattering the final state lepton is a neutrino of the same flavor as the incoming neutrino.
Typically, the exchanged $W$ or $Z$ boson interacts with a nucleon, moving with Fermi momentum $p_F$ 
within the nuclear environment, producing an outgoing nucleon of four-momentum~$p_1$ and, if the neutrino energy is high enough, additional hadrons, mostly pions.  The exchanged boson can also interact with a pair of correlated nucleons in which case a second nucleon is released in the initial interaction: these ``two-particle-two-hole'' events are fascinating from the perspective of nuclear physics and, it turns out, of quantitative importance in measuring neutrino-oscillation parameters. These nuclear effects of the initial interaction; including the Fermi momentum of the bound nucleon and the existence of correlated multi-nucleon ensembles, affect the initial kinematic distribution of both the outgoing lepton and hadronic shower.

\begin{figure}
\begin{center}
	\includegraphics[width=0.55\textwidth] {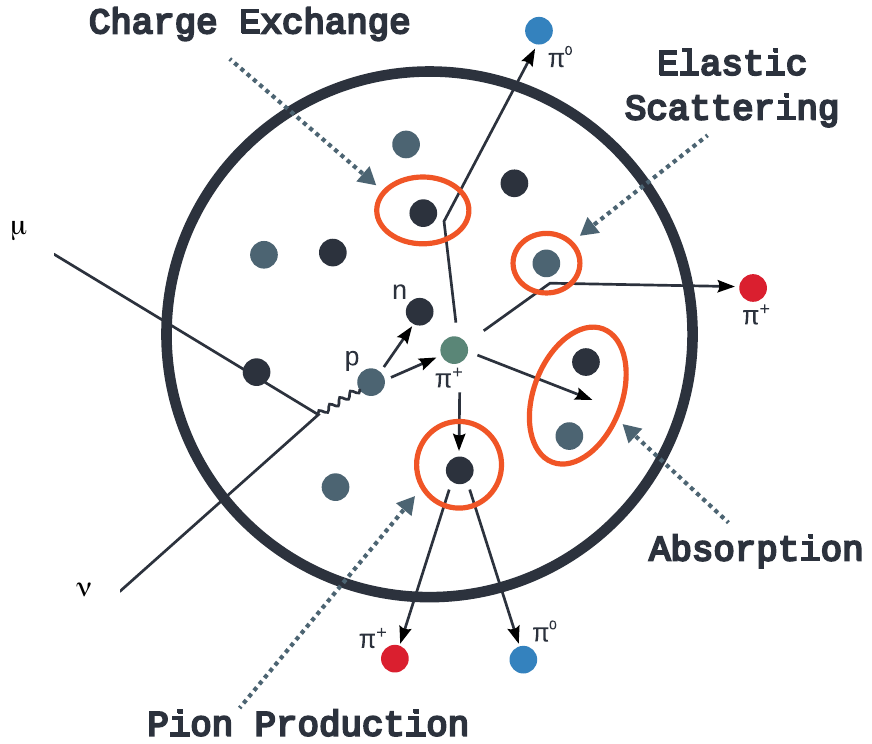}
    \caption{The hadronic shower produced in the initial interaction must still traverse the dense nuclear matter and is then subject to Final State Interactions (FSI) before appearing in the detector.  These FSI include nucleon-nucleon interactions as well as pion-nucleon interactions as illustrated. Figure from Tomasz Golan.}
    \label{fig:overview:FSI}
\end{center}
\end{figure}

The final state lepton invariably escapes the nucleus, however the initially produced hadronic shower undergoes significant further nuclear effects as it proceeds through the dense nuclear matter within the nucleus. As illustrated in Fig.~\ref{fig:overview:FSI}, these final state interactions (FSI) can change the energy and direction and even charge state and multiplicity of the originally produced hadrons.
Hadrons in most nuclei have a reasonable probability of being totally absorbed within the nucleus and not emerging in the detector at all. 
For example a pion in Fig.~\ref{fig:overview:FSI} is absorbed by a pair of nucleons within the nuclear environment to balance energy and momentum.

It cannot be stressed enough that the incident neutrino energy is not a~priori known. This situation differs dramatically from electron or muon scattering studies where the amounts of energy and momentum that are transfered to the nucleus is known precisely on event-by-event basis.
For neutrino nucleus scattering the incoming neutrino energy and initially produced hadronic particles, which have been subject to the above mentioned nuclear effects, can only be estimated from what is observed in the detector.

Since the oscillation paramters impact the neutrino energy spectrum, the strong dependence of the unbiased extraction of neutrino-oscillation parameters on neutrino-interaction physics can best be summarized by noting that the energy and configuration of interactions observed in experimental detectors are, aside from detector effects, the convolution of the energy-dependent neutrino flux, the energy-dependent neutrino-nucleon cross section, and these significant energy-dependent nuclear effects.

Practically, experimenters combine information about the energy dependence of all exclusive cross sections as well as nuclear effects into a \emph{nuclear model}.
This model along with the best estimate of the spectrum of incoming neutrino energies then enters the Monte Carlo predictions of target nucleus response and topology of final states and is a critical component of oscillation analyses.

The above picture of course assumes that processes can be factorized into a two-step process, with intermediate boson interaction on one or two nucleons followed by hadron rescatterings inside nucleus.
While this picture is justified for large enough values of momentum transfer, and provides a reasonable description of many measured interactions, it breaks down at low momentum transfers when the wavelength of the exchanged boson becomes comparable to or larger than typical nucleon separation distances.
Due to the lack of a-priori knowledge of the neutrino energy, or the four-momentum of the exchanged boson, it is impossible to determine even whether the factorisation approximation holds for a given observed neutrino interaction.
Simulations must therefore attempt to capture the impact of long-range correlations and many-body effects within the nucleus as secondary corrections.

To illustrate how oscillation experiments depend on this nuclear model, consider the following illustrative conceptual outline of a two-detector, long-baseline oscillation analysis:
The beam produces a flux of neutrinos, $\phi(E_{\nu})$.
Via the \emph{nuclear model}, the cross section, $\sigma(E_{\nu},\vec{y})$, relates a neutrinos energy to the final state kinematics, $\vec{y}$.
This cross section encodes all nuclear effects, and the quantity $\vec{y}$ must describe every final-state object (up to and including a remnant nucleus).
At the near detector, one can reconstruct events and measure its final state kinematics, $\vec{x}$.
These measured kinematics are related to the \emph{true} final state kinematics, $\vec{y}$ by the detectors response, $R(\vec{y},\vec{x})$ (which also encodes inefficiencies).
The measured event spectrum at the near detector is then given by:

\begin{equation}
N_{ND}(\vec{x}) = \int \phi(E_{\nu}) \sigma(E_{\nu}, \vec{y}) R(\vec{y},\vec{x}) d E_{\nu}.
\end{equation}

As can clearly be seen, the neutrino energy is a hidden variable which is necessarily integrated over, due to its inaccessibility by experiments.
The cross section, which critically includes the nuclear model, is the component that links a neutrinos energy to the measurable final state.
At the far detector, assuming no differences in detector response or neutrino energy spectrum, the equivalent rate is perturbed by oscillation effects:

\begin{equation}
N_{FD}(\vec{x}) = \int \phi(E_{\nu}) P(E_{\nu}) \sigma(E_{\nu}, \vec{y}) R(\vec{y},\vec{x}) d E_{\nu},
\end{equation}

where $P(E_{\nu})$ is the oscillation probability for a neutrino of a given energy.
It should be straightforward to see the benefits of a near detector - measurements of $N_{ND}$ allow the reduction of uncertainties in both the neutrino flux and the nuclear model, which then appear in the far detector prediction with smaller uncertainties.
However, due to the nature of this integral over neutrino energies, a near detector does not reduce the oscillation analysis to a simple rescaling.
Additional differences in detector response, and geometrical differences to the neutrino flux further complicate the matter, and when considering appearance measurements (i.e. measurements of a new flavor in the beam) the cross section also changes between flavours due to lepton mass effects.
A large and growing body of work over the past several years highlights how mis-modeling of the nucleus
(the nuclear model) could lead to unacceptably large systematic uncertainties or, worse, biased measurements in current and future oscillation experiments~\cite{Ankowski:2016bji,Mosel:2016cwa,Coloma:2013tba,Coloma:2013rqa}.  This is discussed further in section \ref{expt_motive}.

It is important to realize that the neutrino-nucleus interaction is usually the least understood component of a detector's response to neutrinos.
Understanding the subtleties of the nuclear model and its effects on what neutrino experimentalists measure in their detectors can only be accurately performed with the input of theorists specializing in this topic.

The following is a list of challenges facing the community.
For some of them a strategy of how to address them seems already clear, while others require a wider discussion among experts in the field.
After this general list, we describe the challenges in more detail, broken down into particular topics and interaction channels.

\begin{itemize}
\item Significant improvements of nuclear models by theorists are essential and include:
\begin{itemize}
\item The development of a unified model of nuclear structure giving the initial kinematics and dynamics of nucleons bound in the nucleus.

\item Modeling neutrino--bound-nucleon cross sections not only at the lepton semi-inclusive cross section level, but also in the full phase space for the exclusive channels that are kinematically allowed.

\item Improving our understanding of the role played by nucleon-nucleon correlations in interactions and implementing this in MC generators avoiding double counting.

\item Improving models of final state interactions,
which may call for further experimental input from other communities such as pion-nucleus scattering.

\item Expressing these improvements of the nuclear model in terms that can be successfully incorporated in the simulation of neutrino events by event generators.

\end{itemize}

\item Continued investment in event generators is critical, as the link between theory and experiment.  The challenges here include:
\begin{itemize}
  \item New theoretical models must be rapidly incorporated into the required generators for experiments to use them effectively, and this step is resource-limited.
  \item Uncertainty quantification is often not done for specific models and must be done once models are in generators.
  \item Even for models that do provide quantified uncertainties, in some cases these uncertainties can be reduced by correlating with other model elements and fitting to global data.  Such global fits come with their own challenges, for example how to handle data that disagree.
  \item Theoretically well-grounded models exist for ``low-energy'' interactions where a nucleon description applies, and ``high-energy'' interactions where a parton description applies.  In between, generators must handle how to smoothly transition between these, providing sufficient degrees of freedom to provide realistic uncertainties.
  \item Currently there exist several event generators with different frameworks but usually the same underlying physics, leading to duplication of effort.  Development of standardized interfaces or a common framework would significantly reduce the work required.
  \item Some high-dimensional models have efficiency challenges when using standard monte carlo techniques.
\end{itemize}

\item More measurements of neutrino-nucleus scattering are required, across a range of energies, nuclear targets, and final states.  While a wealth of data now exist, there are some clearly missing elements:
\begin{itemize}
  \item There are very few measurements that include final state neutrons.  Neutrons tend to represent considerable missing energy in neutrino interactions, and for non-isoscalar nuclei it is not necessarily safe to assume isospin symmetry and extrapolate from measurements of protons
  \item Measurements of low-energy protons can be highly sensitive to the nuclear effects that need to be understood.  These lower energy protons test the cababilities of many existing detectors.
  \item Pion kinematics in multi-pion final states should be measured, as these events tend to be in the so-called `transition region' where neither nucleon nor quark models can be employed to describe interactions.
  \item There are now some measurements which combine lepton and hadron measurements and their correlations (such as kinematic imbalance measurements).  These shoud be pursued further.
\end{itemize}

\item Additional measurements of electron-nucleus and hadron-nucleus scattering can provide unique inputs that neutrino scattering alone cannot provide.  Again more data is needed in this realm, with additional difficulties such as:
\begin{itemize}
  \item Theoretically connecting electron scattering and neutrino scattering must be done through a model which accounts for differences in the fundamental scatter processes, as well as details such as radiative corrections.
  \item No electron scattering experiment provides 4$\pi$ coverage, although there do exist high-coverage experiments, some phase space will never be covered.
  \item Intranuclear scattering occurs in a nuclear potential, whereas hadron beams are characterized in a vacuum, making these measurements slightly different to the process occuring in a nucleus.
  \item Neutral pion beams cannot be created, and neutron beams are challenging to use, and as such intranuclear scattering of these neutral particles is difficult or impossible to constrain with these methods.
\end{itemize}

\item Measurements of cross sections need to follow best practices, with careful attention to model-dependence and uncertainty correlations.  Key challenges include:
\begin{itemize}
  \item It is usually impossible to eliminate absolutely all model dependence from a measurement, so while this is an admirable goal, practical decisions must be made that weigh this against other concerns.
  \item Assessing correlations between a measurement, and an existing older measurement from the same experiment, or measurements from other experiments that use the same beam, is extremely challenging and sometimes impossible.  But these correlations are critical when assessing the level of agreement between generators and models.
  \item Published measurements should come with the most complete data releases possible, ideally using standardized formats.
  \item Evaluation of uncertainties with correlations usually assume guassianity.  This is not always a safe assumption, and experiments should at least confirm whether this is valid.  Handling correlated non-gaussian uncertainties is complex and resource-intensive.
\end{itemize}

\item Finally, reducing neutrino beam flux uncertainties will reduce neutrino cross section measurement uncertainties, particularly the normalization uncertainty.  This will require:
\begin{itemize}
  \item Dedicated hadron production measurements, ideally on replica neutrino beam targets
  \item Additional measurements of hadron scattering on relevant materials in neutrino beamlines
  \item In-situ measurements of neutrino-electron elastic scattering, inverse muon decay, and coherent neutrino scattering, which can all provide additional flux constraints
  \item Loop and radiative corrections to the theoretical calculations of such processes at least to match the experimental precision of a few percent
\end{itemize}

\end{itemize}

In addition to these general challenges facing the community, there are more specific concerns for particular topics and interaction channels. These are summarized below in the form of  observations, problem description or recommendations. For a deeper insight, the reader is encouraged to consult the referenced sections of this paper.

\subsection{Challenges: The impact of neutrino-nucleus interaction physics in determining neutrino oscillation parameters}
Neutrino oscillations fundamentally depend on the baseline traversed by the neutrino and on its energy. In long-baseline experiments, the baseline is fixed, but the incident neutrino energy is generally not known on an event-by-event basis. 
In the case of atmospheric neutrinos, both the neutrino energy and direction (thus the baseline) aren't known, leading to further complications.
The neutrino is not directly observed; its presence and characteristics is instead inferred from its interaction products.  
To complicate matters further, different interaction mechanisms contribute to a given selection of events used in an oscillation experiment, with each mechanism having its own modelling challenges.

The choice of neutrino energy estimator is closely connected to the event selection strategy and to the interaction channels that contribute to the experiment's energy range. 
Long-baseline neutrino oscillation experiments typically select events according to their visible topology. These selections can be relatively inclusive, such as the CC$\nu_\mu$ and CC$\nu_e$ samples used by NOvA and DUNE, or more exclusive, such as the single-ring muon and electron samples used by T2K and Hyper-Kamiokande. 
Common neutrino energy estimators include the calorimetric method, which reconstructs the neutrino energy by summing energy deposits in the detector together with particle-identification information. There is also the kinematic method, which infers the neutrino energy under an assumed interaction hypothesis, for example a charged-current quasielastic (CCQE) interaction. 

The neutrino energy estimators are generally biased---that is, the estimators do not always perfectly reconstruct the actual neutrino's energy. Understanding the impact that interaction modelling has on these estimators is of central importance to the precision of the experiments.
For example, several processes often contribute to a given selection of neutrino events, and each process will have a different bias. Understanding the bias therefore includes understanding the composition of the processes contributing to the overall selection.
The neutrino energy bias also depends on nuclear effects, such as how much energy the initial state nucleon carried, how the hadronic final state may have been altered away from the free-nucleon expectation due to nuclear effects, what particles were in the final state and how much energy was carried away by it.

Biases in the neutrino energy estimator only change the overall shape of the spectrum.
The contributions from different interaction mechanisms at a given neutrino energy fundamentally impacts the normalisation of the spectrum. 
For instance, consider an experiment which has a baseline and neutrino energy tuned to maximise muon neutrino disappearance with perfect neutrino energy reconstruction. The expectation would be to observe no muon neutrino events at all at the energy of maximal disappearance, since the muon neutrinos will have all oscillated into electron and tau neutrinos. 
However, those electron and tau neutrinos can interact via neutral current channels, where one of the outgoing particles, for example a $\pi^\pm$ may be mistaken for a $\mu^\pm$, and the event can therefore enter the muon selection as a background. This means that the depth of the disappearance spectrum is directly impacted by the amount of neutral-current interactions, and how often these risk being mischaracterised as charged-current interactions.
Similarly, understanding differences between electron and muon neutrino interactions directly affects the conclusions about the oscillations between the two flavours: did the experiment see more than expected electron neutrinos at its far detector because of a larger oscillation from muon neutrinos, or because of a larger electron-neutrino interaction cross section?

There are numerous critical challenges to overcome to maximise the sensitivity to neutrino oscillations and the physics reach of the future programs---which includes measurements of the extent of CP conservation in neutrinos, the neutrino mass ordering, and sub-percent precision on the dominant mass-squared splitting:
\begin{itemize}
    \item Flavor appearance measurements directly depends on understanding the cross section difference between the appearing flavor and some reference flavor. 
    For DUNE and \HK, this means knowing the differences between muon and electron (anti-)neutrino cross sections.  
    With extremely limited data on electron (anti-)neutrino cross sections, and a near detector that measures a different electron neutrino flux to the far detector, this will always have some reliance on theory to understand the effects, such as lepton mass and radiative corrections.

    \item Neutrino oscillations fundamentally change the flux of a given neutrino flavor in energy, and thus the interactions that are measured at the near detector does not match those at the far detector. 
    Therefore, the evolution of the neutrino interaction cross sections in energy is critical to understanding the details of neutrino oscillations.
    
    \item How nuclear effects depend on the nuclear target impacts the experiments in different ways. For T2K and HK, differences between carbon and water are important, whereas for DUNE it is instead important to understand the details of the nuclear effects on the larger, and less studied, argon nucleus. 

    \item Constraints on \dcp and the mass ordering are provided in part by measuring differences between neutrino and antineutrino oscillations. 
    At GeV-scale energies, the energy carried away by unmeasured particles, such as neutrons or low momentum charged particles, and by particles which are undetected due to interactions with the remnant nucleus, can differ greatly between neutrinos and antineutrinos and dramatically change the bias of the neutrino energy estimator. It is important to understand the prospects for semi-inclusive theoretical models that can accurately predict this missing energy. 
    Experimentally, programs to detect and measure neutron yields on the relevant targets are essential, as are measurements of missing energy, nuclear absorption, and low-energy particles.

    \item As DUNE and atmospheric experiments will observe a considerable rate of events with multi-pion final states, the ability to predict the strength of the interactions, and how to the interactions generate the lepton and hadronic systems, is a central challenge. The evolution of the final state through the transition from resonances to deep-inelastic scattering must be improved and the uncertainties better quantified.

    \item Near detectors are powerful in oscillation analyses, but have fundamental and practical limitations in the near-to-far extrapolation of neutrino events. 
    New experimental methods, such as NuPRISM~\cite{nuPRISM:2014mzw} which enables variable neutrino energy fluxes to enter the near detector, may provide additional information to this constraint, but it will not completely remove them.  It is therefore important to understand the corrections that are required, and their uncertainties.

    \item Oscillation analyses are often ``stress-tested'' with fake data studies, where aspects of a model are changed and the impact on measured oscillation parameters is determined. However, determining reasonable choices for such fake data samples is challenging, and input from the theory community is crucial.
\end{itemize}

\subsection{Challenges: The Nuclear Model}
With most neutrino experiments unable to use hydrogen targets, the interactions that are used invariably occur with medium to large nuclei, with nuclear masses anywhere from A=12 (carbon) to A=208(lead).
As the level of precision in neutrino experiments has increased, so has the need for more sophisticated treatments of the interactions with these complex nuclei.

Recent years have seen tremendous advancements in the amount and level of detail in the cross section data that are provided by experimental collaborations.
This presents strong challenges for theoretical modelling that must keep pace with experimental advances in order to secure smooth progress in the field.
Some key challenges in this field are

\begin{itemize}
\item It is very challenging to identify a single model that can describe every aspect of the nucleus in detail, but different kinematic regions and/or reaction channels are sensitive to different aspects of the nuclear model.  This can lead to a desire to use a different nuclear model in different kinematic regimes, however this leads to inconsistent treatments, and the transitions between these regions then need to be handled carefully.

\item It is challenging to model heavy nuclei (e.g. for ab-initio models), nuclei away from closed shells (e.g. for mean field models) and to establish the limits of non-relativistic approaches.

\item Nucleon-nucleon correlations are now known to play an important role for an accurate description of scattering cross sections, still their implementation in models is numerically challenging and subject to model dependencies. 
More data to constrain these models is highly desirable.

\item It is essential that progress in theoretical nuclear modelling finds its way to generators.  MC generators should include all essential nuclear effects needed for a state-of-the-art description of scattering cross sections.

\item In some cases, the inclusive response can be calculated more easily than calculating all exclusive components, however this can't then be used to simulate full events.  It is not clear how to use benefit from the detailed inclusive calculations to improve event generators.

\item Microscopic and sophisticated models that provide complete information about the final-state hadron(s) are needed, particularly for interactions which produce more than one final state hadron.
More exclusive data is needed to constraint the few existing models and to trigger new studies.

\item The kinematic region beyond the delta peak remains underexplored and is in desperate need of more theoretical efforts and more detailed data to constrain models, especially for the axial component.

\item A quantitative understanding of how uncertainties in input parameters and approximations in nuclear models affect the final cross sections is needed.
To date, a comprehensive estimation of the theoretical uncertainty in cross-section calculations has not been accomplished. A well-defined approach to better assess theory-related uncertainties is necessary.
\end{itemize}

\subsection{Challenges: Determining the Flux}
Crucial to the measurement of neutrino-nucleus scattering cross-sections, and for other measurements such as neutrino oscillations, is the flux of neutrinos.
As neutrino interactions are energy-dependent, correctly predicting measured event rates relies on knowing both the $\nu$-A cross sections, but also the energy-dependent neutrino flux (both its normalisation and its shape).
Due to the processes by which neutrino beams are made, this is highly non-trivial.

Currently, the systematic uncertainties on the a-priori flux predictions in neutrino oscillation experiments range from 5-10\% at energies near the oscillation maximum, and can be as large as 20\% in the tails.  These uncertainties are dominated by hadron scattering cross section uncertainties for processes that have large uncertainties or have not yet been measured.  Uncertainties associated with beam focusing can be as large as 5\% near the flux peak and can depend strongly on the neutrino energy.  Neutrino experiments with beam energies above 1-2 GeV are able to further constrain the flux normalization via the measurement of neutrino-electron scattering, which has $\sim1\%$ theoretical uncertainties since it is a pure leptonic interaction.  With significantly reduced theoretical uncertainties on certain neutrino-nucleus scattering process, additional ``in-situ'' constraints on the flux could be made from measurements of interactions with low hadronic recoil energies, coherent neutrino-nucleus scattering, and scattering off hydrogen.

Reducing the often-dominant $\sim10\%$ systematic uncertainties of the flux in $\nu$-A measurements is achievable via further, challenging measurements of:
\begin{itemize}
    \item hadron scattering cross sections for a variety of nuclear targets and beam momenta to cover the broad range of primary, secondary and tertiary hadron interactions occurring in accelerator beamline and the atmosphere,
    \item hadron production using real neutrino production targets and/or focusing horns,
    \item in-situ neutrino-electron, neutrino-proton and neutrino-nucleus scattering rates, especially if neutrino-nucleus scattering uncertainties are further reduced.

\end{itemize}
\subsection{Challenges: Generators}

Monte Carlo (MC) generators serve as a bridge between theoretical models and experimental measurements.
For future neutrino oscillation experiments it is of critical importance that they contain the best knowledge of neutrino-nucleus cross sections and nuclear effects.

Many MC generator challenges are closely tied to the fundamental physics they are trying to model - in those cases, they are discussed in other sections of this paper.
However, there are practical challenges that are general to MC generators, somewhat independent of the physics in question:

\begin{itemize}
\item The work of implementing existing models into generators has historically not been well-funded in part due to its inherent overlap between theory and experiment. Lack of funding leads to delays between models becoming available, and them being integrated into generators.

\item Further delays occur due to the need to integrate models into multiple generators which requires duplication of effort.  Standardization of interfaces would be a significant benefit in this regard, though this requires coordination.

\item All generators are tuned (or fit) to data, at some level.  However due to the large uncertainties and historically lack of information about correlations, these can depend not only on the data used, but also on choices made about how to handle cases where data disagree.  As new models and data are made available, re-tuning becomes a challenge.  Insipiration may be drawn from QCD global fits of parton distribution functions to produce a ``global'' tune.

\item speed and efficiencies of generators for high-dimensional calculations has recently become a problem. To address this shortcoming, many groups have begun to include machine learning (ML) techniques into their generators.
While ML techniques can provide a significant improvement in these areas, we have to ensure that the inclusion of them does not alter the underlying physics. For example, using
Generative-Adversarial Networks (GANs) to generate new events from a previously generated sample does not conserve the total cross section of the original sample. On the other hand,
using tools like Normalizing Flows (NFs) to more efficiently sample the phase space can be accomplished while guaranteeing no modification to the overall cross section.
\end{itemize}

\subsection{Challenges: Final State Interactions}

Under the short time approximation neutrino-nucleus interactions can be factored into a hard-scatter event where a neutrino interacts with one or multiple nucleons, and then final state interactions where the products of a hard-scatter propagate through, and re-interact with, the residual nucleus.
While much progress has been made, this is a fundamentally difficult process to describe, and there remain significant challenges.

\begin{itemize}
\item All existing models are based on semi-classical approaches, which cannot truly describe reality.  However, it is not clear where their limits are.
\item Almost all neutrino data has strong influence of FSI, and it is very challenging to identify measurements that are either independent of FSI, or independent of all other effects, meaning FSI is usually measured in conjunction with other nuclear effects.  However there are observables that isolate reasonably FSI well, such as
 the kinematic imbalance variables, 
 hadron-nucleus total reaction cross sections, and hadron transparency.
 \item Hadron-nucleus scattering data may be used to constrain elements of FSI, but are only an approximation to the processes ocurring after a neutrino scatter, where the nucleus tends to be in an excited state and hadrons may be off-shell.
 \item Particular attention needs to be applied to the emission of low energy nucleons and light ions.  This has been included in various models of hadron-nucleus interactions but need to be added to neutrino generators.  New data from liquid argon detectors may be able to directly constrain these elements of FSI models.
\end{itemize}

\subsection{Challenges: Electron-nucleus Scattering}

In the factorization commonly used to describe neutrino-nucleus interactions, uncertainties can be divided into: (i) the description of the elementary cross sections for neutrino scattering on nucleons, and (ii) the modeling of nuclear effects.

While near-detector measurements provide invaluable information on neutrino interactions, there are intrinsic limitations on what can be learned from them primarily due to the need to average observables over a broad flux.
In view of the similarities of neutrino and electron interactions with atomic nuclei, precise electron-scattering measurements offer a unique opportunity to greatly reduce the nuclear uncertainties of the neutrino cross sections, leaving only their axial contributions to be determined from the near-detector data. 

In order to leverage the similarities between neutrino-nucleus and electron-nucleus scattering, we require are a consistent treatment of neutrino and electron interactions in a simulation, and an abundance of relevant electron-scattering data.  There remain challenges in both of these components, for example:

\begin{itemize}
\item Neutrino generators need to be used to simulate electron scattering in a consistent manner, using the same underlying models and parameters.

\item Radiative corrections which are a large effect to electron scattering need to be calculated precisely and applied.

\item Due to the mass differences, electron scattering and neutrino scattering will always measure slightly different regions of energy and momentum transfer and this can only be accounted for using a model with consistent treatment.

\item Electron-scattering data for the targets used in the long-baseline program and in the kinematic regimes relevant to it are required, to reduce the need for extrapolations in nuclear size and beam energy.

\item Increasingly exclusive measurements are key to further reduction of uncertainty.  Neutrino experiments should evaluate which modeling elements have the most important uncertainties such that electron scattering experiments can make the most targeted measurements.
\end{itemize}

\subsection{Challenges: Pionless Interactions}
\label{over:CCQE}

The charged current quasielastic (CCQE) reactions
$$
\nu_\mu n\to \mu^- p,\qquad \bar\nu_\mu p\to \mu^+ n
$$
are one of the simplest signatures to observe, and the lowest threshold process, making them very important.
In the T2K experiment, and future \HK experiment, this interaction type is used as the primary signal process at as the far detectors which are only sensitive to the lepton.  In these events, neutrino energy can be inferred \textit{kinematically} under the assumption that the neutrino interacted with a stationary, free nucleon.
Unfortunately, this simple picture doesn't truly apply when the target nucleons are bound in a nucleus, so it becomes critical to understand the impact of the nucleus.

Experimentally, one can define a ``CCQE-like'' interaction based on the final state.  As the true CCQE interaction produces no pions, a ``pionless'' final state is the simplest proxy.
However, this definition will necessarily include other reaction mechanisms which cannot occur on a free nucleon, such as two-particle two-hole interactions (where the interaction occurs simultaneously with a pair of nucleons), and events which produce pions that are then absorbed in the nucleus.
These events may be distinguished (at least partially) by the details of the hadronic final state, but final state interactions produce substantial smearing out of the hadron kinematics (and multiplicities) making this challenging.
The non-CCQE contributions to pionless final states come with substantial uncertainties as they can only be measured in neutrino-nucleus interactions.

Even for experiments where the CCQE reaction is subdominant, the clean experimental topology remains extremely useful for studying nuclear effects which play a role in all topologies, such as the nuclear initial state, proton and pion rescattering in the nucleus, and correlations between nucleons in the nucleus.

Theoretically there are major challenges for this reaction channel:
\begin{itemize}
\item Predictions of hadronic final states in 2p2h interactions, which include the correlation between lepton, leading, and sub-leading hadrons
\item Careful assessment of potential interference and/or double counting of effects when considering multi-nucleon interactions, correlations in the initial state, and FSI
\item Development of a model/theory that remains applicable at low-energy transfers but retains the full hadronic final state predictions associated with models that rely on the impulse approximation.
\end{itemize}

However there are sme ways in which these challenges can be addressed at least in part by experimental measurements:
\begin{itemize}
\item Measurements of proton production down to lower energies (below the Fermi momentum) would probe FSI in a way currently inaccessible.
\item Measurements of multi-proton final states are likely to be unique probes of 2p2h events, and should be possible in current- and next-generation near detectors
\item Measurements of other hadronic components in the final state - neutrons and nuclear fragments - also provide unique information about FSI that measurements of only protons cannot achieve.  This is only realistically possible in pionless final states (in the near term, at least) due to detector limitations
\end{itemize}

\subsection{Challenges: The Resonance Region}
\label{over:RES}
Describing meson production in neutrino interactions is exceptionally challenging, particularly in the GeV energy range relevant for current and planned accelerator-based neutrino experiments. At these energies, neutrino beams interact with nucleons within a nuclear environment through various processes. The production of mesons, mainly pions, occurs primarily through the decay of nucleon resonances (nucleons in an excited state) but also by other mechanisms. Accurately describing meson production in nuclei requires knowledge of Hadronic, Particle and Nuclear Physics. The most important challenges include:

\begin{itemize}
\item  Establishing the domain of validity of the present meson production models and, if necessary, extending it to cover future neutrino experiments. 

\item Encompassing both resonant and non-resonant interactions for meson production throughout the kinematic region from the lowest $Q^2 < 1.0~(\text{GeV}/c)^2$ to the hadron-nonperturbative quark transition region ($Q^2 > 1.0~(\text{GeV}/c)^2$) as well as the $W$ regions significantly beyond the pion production threshold. 

\item Utilizing all available data in a unified analysis encompassing pion, photon, electron, and neutrino-induced reactions is essential for developing realistic models and determining the nucleon transition form factors for each process.

\item Understanding the flexibility of phenomenological models in the resonance region and evaluating their systematic uncertainties that is currently lacking in most of the theoretical models. 

\item Modeling pion production in the nuclear environment. This involves integrating lepton-nucleon models into a nuclear framework and accounting for various nuclear effects. Increased experimental data on pion electro-production on nuclei would be helpful.

\end{itemize}

\subsection{Challenges: Shallow and Deep-Inelastic Scattering Region} 
The description of inclusive lepton scattering in the transition region between resonance excitation and deep-inelastic scattering (DIS) is a subject of continuing study. This region, referred to as shallow inelastic scattering (SIS), will contribute a significant fraction of events for DUNE, and contributes through feed-down in both signal and background estimates for lower-energy experiments.

Deep inelastic scattering refers to a region of phase space with high energy and momentum transfer, which can be considered neutrino-quark scattering events.
This can be described primarily through perturbative QCD and rely on parton distribution functions (PDFs).

Shallow inelastic scattering~(SIS), on the other hand, is two different regions. The first region is the predominantly lower-$Q^2$ non-resonant meson production region. The second region begins at the somewhat higher $Q^2$ threshold of interactions within the nucleon and proceeds via increasing $Q^2$ through the non-perturbative QCD SIS interactions such as multi-quark interactions.   
Within this "multi-quark" scattering SIS region the nonperturbative QCD corrections such as Target Mass Correction (TMC) and dynamic higher twist (HT) effects become important.  

Since these SIS non-perturbative Higher Twist (HT) effects play a significant role for the typical kinematics of modern (anti)neutrino experiments, the possibility of using quark-hadron (QH) duality to describe this region has been considered to provide a connection between the average value of interaction strengths in the quark-gluon description of the DIS formalism at high $Q^2$ and in the pion-nucleon description in the region of resonance excitation at lower $Q^2$. However, the application of QH duality in neutrino scattering has proven challenging and is still being investigated.

In studies of charged lepton nucleus DIS there is evidence from experimental measurements as well as theoretical studies that the quark parton distribution function for the nucleons bound in nuclei (nPDF) differs from the quark PDFs in the free nucleon. These partonic nuclear effects demonstrate themselves even down into the SIS region.

A comprehensive understanding of these kinematic regions requires concerted efforts in the following directions:
 
\begin{itemize}
\item Develop and optimize a consistent framework describing the transition from resonance production through SIS to DIS.  Determine the impact of higher-twist contributions and clearly establish the kinematic boundaries within which the DIS formalism for structure functions and cross sections remains valid.

\item Carry out new, precise measurements of differential and total neutrino and antineutrino cross sections on a broad range of nuclear targets, with extensive Bj-$x$ and $Q^2$, coverage while systematically investigating the interplay among Fermi motion, nuclear binding, meson-exchange currents, nuclear shadowing, off-shell effects, and other nuclear mechanisms .

\item More studies are required to better understand the low-$x$ differences between nuclear effects observed in electromagnetic and weak DIS structure functions and cross sections.

\end{itemize}

\subsection{Challenges: High Energy Neutrinos}
Historically, a variety of experiments operated to measure neutrino interactions at accelerator facilities with beam energies around 100~GeV. In recent years, a new program emerged to measure neutrino interactions at TeV energies and beyond using neutrinos produced at the LHC or from natural sources. These offer exciting new potential for discovery, but several challenges have been identified: 

\begin{itemize}
  \item The first generation of LHC neutrino experiments are small and statistically limited. Larger experiments operating during the high-luminosity LHC era, such as those proposed in the context of the Forward Physics Facility, are needed to explore the physics potential. 
  \item A more precise understanding of forward particle production rates and the associated neutrino fluxes at the LHC is needed.
  \item Above the reach of accelerators, the naturally-ocurring flux of neutrinos is low, requiring larger detectors.  Ideally, upgraded detectors will be reasonably fine-grained, to better separate energy from hadronic cascades from energy from muon energy loss.
  \item Neutrino interaction modeling in generators must be updated with higher order corrections, mass effects, and modern PDFs
  \item Updated modeling of parton showers and hadronization must be used in generators, and this will likely require neutrino-specific tuning for the energy range of interest.
\end{itemize}

\subsection{Challenges: Strange Particle Production}

The production strange particles in neutrino and antineutrino interactions offers some unique and complementary information, compared to other neutrino interactions.  However, they also come with unique challenges:

\begin{itemize}
    \item Flavor-changing processes with strangeness production are Cabbibo suppressed and are then further suppressed due to energy thresholds for strange quark production. High-statistics data are needed to investigate the details of these processes.
    \item Due to the strange quark mass, strange hadrons are more massive and have higher detection thresholds than non-strange counterparts.  At low momentum transfers, these particles are challenging to observe and measure without new detector technologies (for example gaseous time projection chambers).
    \item Generators need to be updated to include all possible channels for strange particles interactions and their rescattering with appropriate nuclear potential.
    \item Due to the challenges above, many experiments have never attempted to measure strange particle production, further limiting the available data.  Data preservation efforts allowing for reanalysis would be valuable.
\end{itemize}

\subsection{Challenges: Coherent Processes}
Coherent - or diffractive in the case of a free nucleon - scattering occurs when the initial nuclear state remains unchanged after an interaction.
This can be elastic (known as CE$\nu$NS) - in which case the signature is extremely small nuclear recoil - or inelastic.  In inelastic coherent interactions, significantly more energy is available in the final state.  For example, charged-current coherent pion production:

$$
\nu_l + A \rightarrow l^- + \pi^+ + A
$$

Importantly, neutrino interactions can coherently produce photons, or other heavier mesons, in place of a pion, and these interactions can also occur via the neutral current.

Coherent production of neutral pions or photons can be significant backgrounds to $\nu_e$ signals in an oscillation experiment, or to the similar topology of $\nu$-$e$ elastic scattering which can be used as a flux constraint.
However, coherent interactions as a signal process are very useful as there is negligible energy transferred to the nucleus meaning there should be very good neutrino energy resolution for these events.
Additionally, there are ways in which production of BSM particles may be coherently enhanced in neutrino scatters, making them an interesting probe for new physics.

Specific challenges related to coherent interactions are:
\begin{itemize}
  \item PCAC models rely on good pion-nucleus scattering data on the relevant nuclei, which doesn't always exist.
  \item Microscopic models can be computationally expensive and limited to particular energy ranges.  Work to improve the efficiency and applicability is important.
  \item Although the reduced rates at low and medium $E_{\nu}$ have restricted measurements of many coherent processes, coherent $\pi^+$ production, including $A$-dependence, within this energy range has been measured and exhibits tension with current models. 
  \item When considering coherently-enhanced BSM physics searches, integration of these models in generators can be slow.
\end{itemize}

\newpage
\section{The Impact of Neutrino--Nucleus Interaction Physics on Oscillation Physics Analyses}
\label{expt_motive}
\subsection{Introduction}
Long-baseline neutrino oscillation experiments generally encompass baselines, $L$, of 100-1000 km and neutrino energies, $E_\nu$, between 0.5-5 GeV.
They estimate the neutrino oscillation parameters by studying the disappearance of muon neutrinos and the appearance of electron neutrinos in a beam of predominantly muon neutrinos, and how these phenomena depend on the neutrino energy. Typically, the appearance and disappearance are studied for both neutrinos and anti-neutrinos.
The range of $L/{E_{\nu}}$ and the oscillation channels which are measured make long-baseline experiments sensitive to the two mixing angles $\theta_{23}$ and $\theta_{13}$, the mass-squared splitting $\Delta m^2_{32}$ ($\Delta m^2_{13}$),  and the charge-parity (CP) violating phase $\delta_{CP}$, in the Pontecorvo-Maki-Nakagawa-Sakata (PMNS) formalism of neutrino oscillations~\cite{Pontecorvo:1968fh, Maki:1962mu}. 
The experiments are also sensitive to the neutrino mass ordering, that is, if $\Delta m^2_{31}> 0$ (normal ordering, ``NO'') or $\Delta m^2_{31}<0$ (inverted ordering, ``IO''). 
The two currently running long-baseline experiments, T2K~\cite{T2K:2011qtm} and NOvA~\cite{NOvA:2004blv,NOvA:2007rmc}, are seeing weak hints of CP violation and normal mass ordering on their own~\cite{T2K:2025yoy,T2K:2023smv,NOvA:2025tmb,NOvA:2023iam,NOvA:2021nfi} and in recent joint analyses~\cite{T2K:2024wfn,T2K:2025wet}.
However, significant improvements are required to decrease both statistical and systematic uncertainties to definitively establish if neutrinos violate CP symmetry, and which neutrino mass ordering is realized.
The two future long-baseline experiments, DUNE~\cite{DUNE:2020lwj} in the US and Hyper-Kamiokande (HK)~\cite{Hyper-Kamiokande:2018ofw} in Japan, will address this by collecting orders of magnitude more events with high performance detectors. Both experiments are currently under construction, and are expected to commence data-taking within the next five years. 
Their goals are to: 
\begin{itemize}
    \item measure the ``atmospheric'' neutrino oscillations (primarily governed by $\theta_{23}$ and $\Delta m^2_{32}$) with improved precision compared to previous long-baseline and atmospheric experiments, below the 1\% scale;
    \item determine the neutrino mass ordering;
    \item definitively measure if neutrino oscillations violate CP symmetry, regardless of how large it is;
    \item test the three-flavor neutrino oscillation paradigm, through probes such as unitarity and CPT symmetry.
\end{itemize}

This upcoming high-statistics, precision era, makes strict demands on the measurements. Accurate and well-understood neutrino energy reconstruction is essential for disentangling the effects of different oscillation parameters.
Although many oscillation parameters can be constrained using muon- and electron-neutrino data alone, measurements of the corresponding antineutrino channels are crucial for resolving degeneracies, for example between CP violation and the neutrino mass ordering, because these effects manifest differently in neutrinos and antineutrinos. 
Moreover, certain tests of charge-parity-time (CPT) symmetry and of the PMNS framework are only possible through direct comparisons of neutrino and antineutrino oscillations. 
Finally, measuring oscillations across multiple neutrino energies is also important, as the energy dependence of the oscillation probabilities provides additional leverage for breaking parameter degeneracies. 

\begin{table}[hbtp]
    \centering
    \begin{tabular}{c|c}
    \hline \hline
         $\Delta m^2_{32}$& $2.511\times10^{-3}~\textrm{eV}^2/c^4$ \\
         $\sin^2 \theta_{23}$ & 0.572\\
         $\sin^2 \theta_{13}$ & 0.2203 \\
         $\Delta m^2_{21}$ & $7.41\times10^{-5}~\textrm{eV}^2/c^4$\\
         $\sin^2 \theta_{12}$ & 0.303 \\
         $\delta_\textrm{CP}$ & $197^\circ$\\
         \hline \hline
    \end{tabular}
    \caption{Values for the oscillation parameters used in this section, coming from the NuFit 5.2 analysis~\cite{Esteban:2020cvm} in the normal ordering, excluding SK atmospheric results. 
    For the cases in the inverted ordering, the simple transformation $\Delta m^2_{31}\rightarrow-\Delta m^2_{32}$ is applied (i.e. $\Delta m^2_{31}: 2.5851\times10^{-3}\rightarrow-2.511\times10^{-3}~\textrm{eV}^2/c^4$), with the $\Delta m^2_{32}$ in the inverted ordering changing accordingly so that the three-flavor relation is respected ($\Delta m^2_{32} = \Delta m^2_{31} - \Delta m^2_{21}$), with no changes to the other parameters.}
    \label{tab:osc_nufit52}
\end{table}

\subsubsection{The impact of the CP-violating phase and the neutrino mass ordering}
The upcoming precision experiments aim to definitively measure the CP violating phase and the neutrino mass ordering. Both effects give rise to similar effects in the electron (anti-)neutrino appearance spectrum, which can be highly degenerate for certain baselines and energies.
This section introduces the problem, showing the impact of oscillations on the spectrum for different baseline, using full three-flavour oscillation calculations and the NuFit 5.2~\cite{Esteban:2020cvm} parameter values, detailed in \autoref{tab:osc_nufit52}.
\begin{figure}[hbtp]
    \centering
    \includegraphics[page=1,width=0.49\textwidth,clip, trim={0mm 0mm 0mm 0mm}]{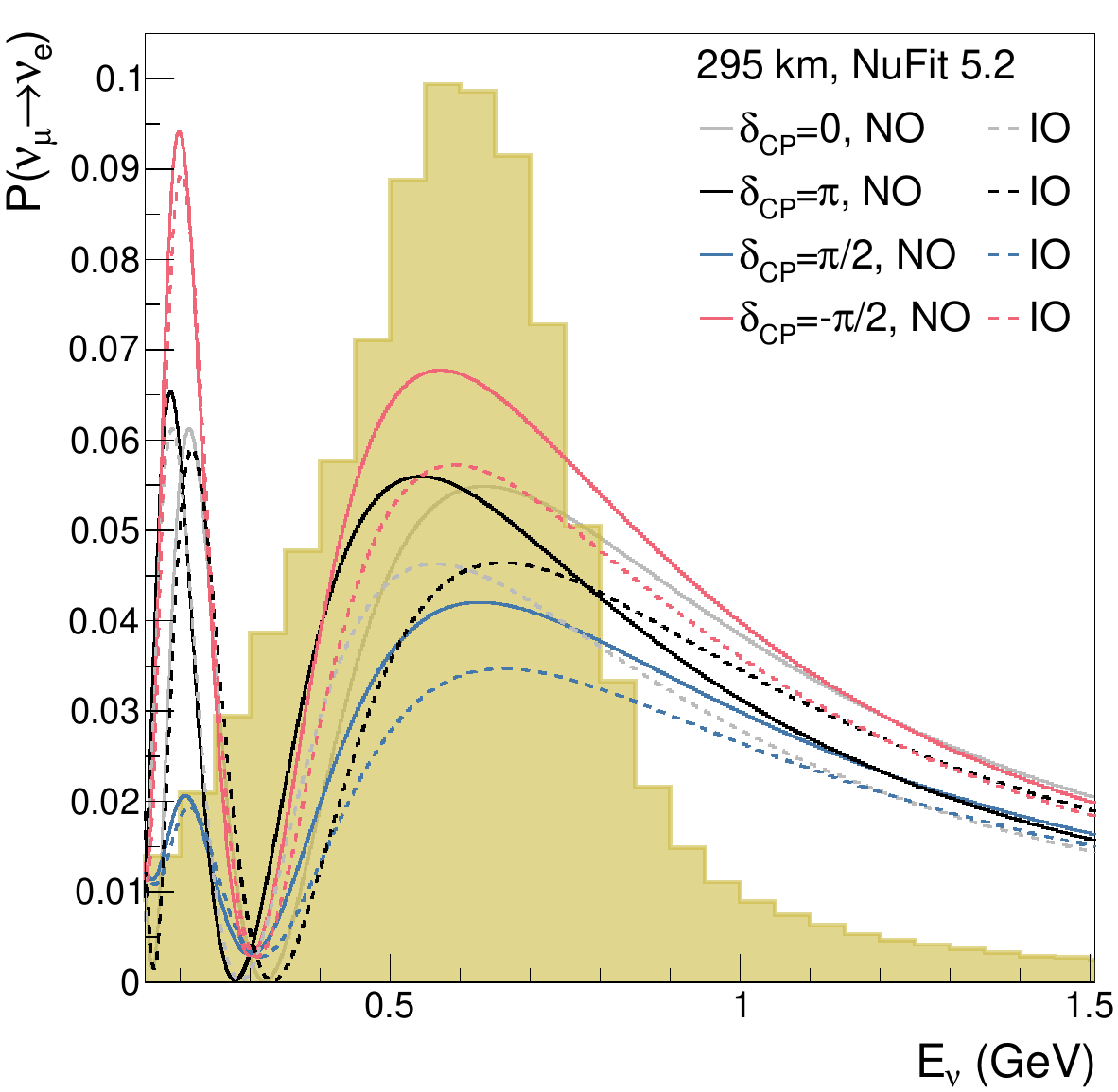}
    \includegraphics[page=1,width=0.49\textwidth,clip, trim={0mm 0mm 0mm 0mm}]{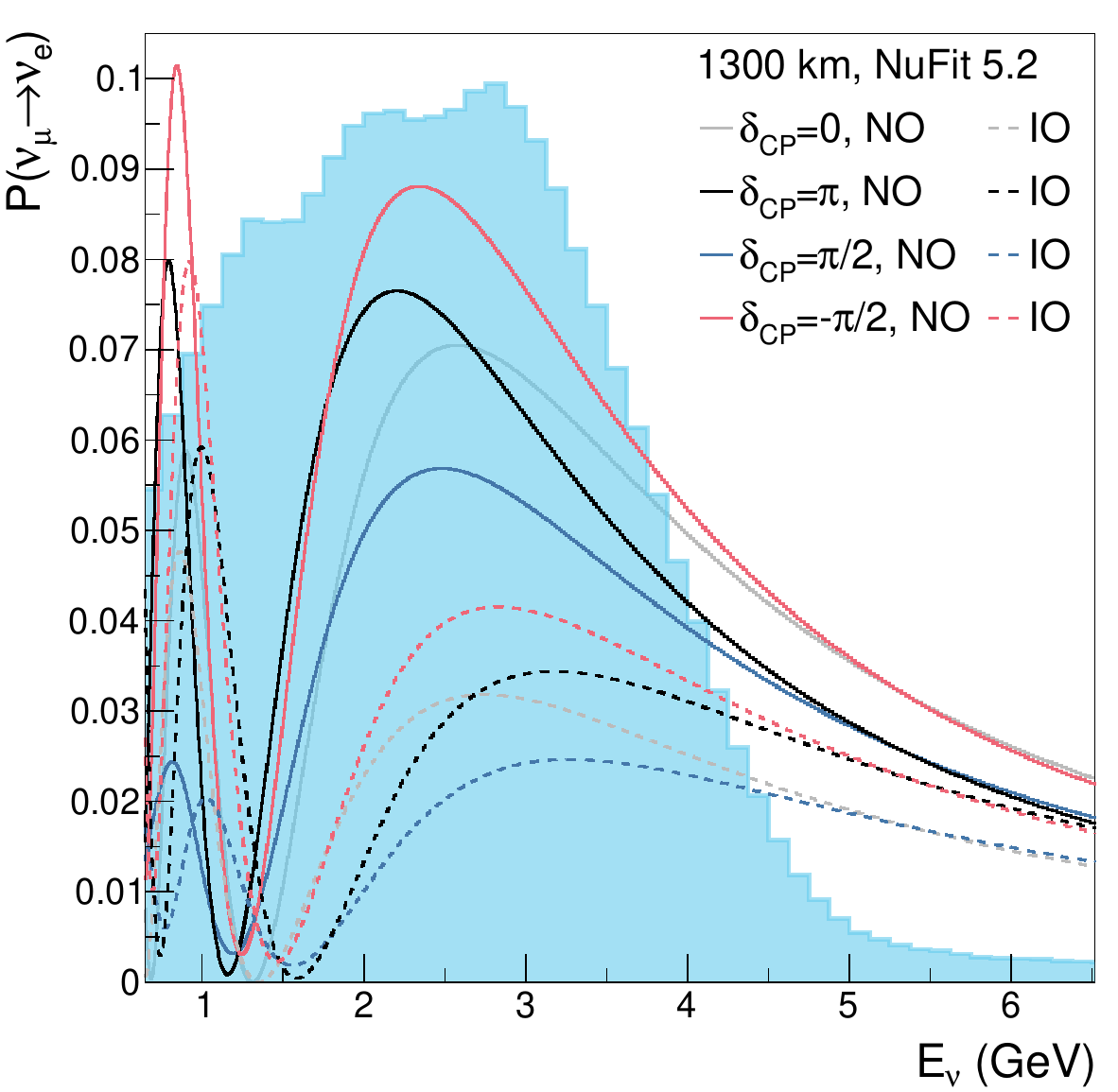}

    \includegraphics[page=1,width=0.49\textwidth,clip, trim={0mm 0mm 0mm 0mm}]{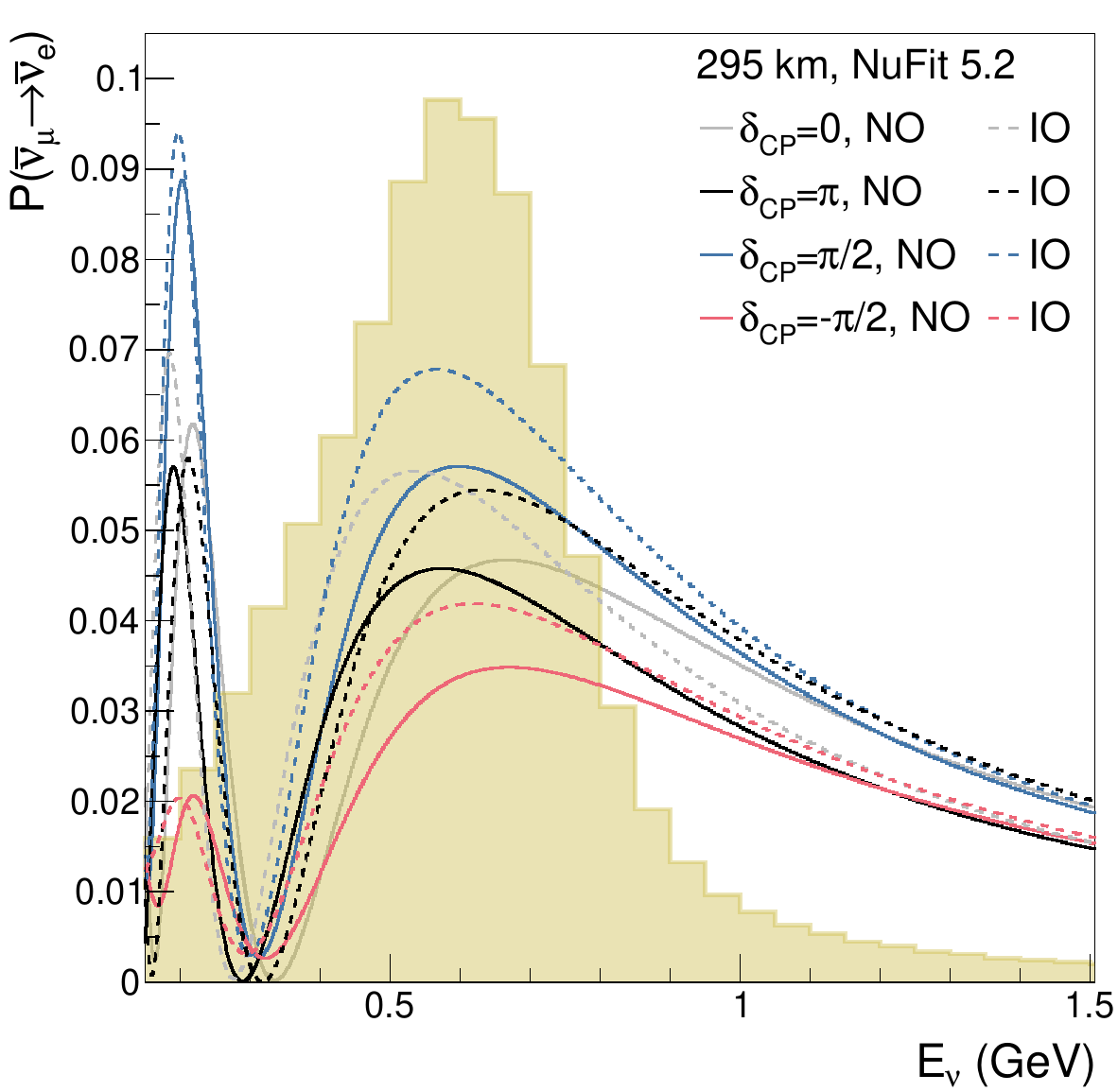}
    \includegraphics[page=1,width=0.49\textwidth,clip, trim={0mm 0mm 0mm 0mm}]{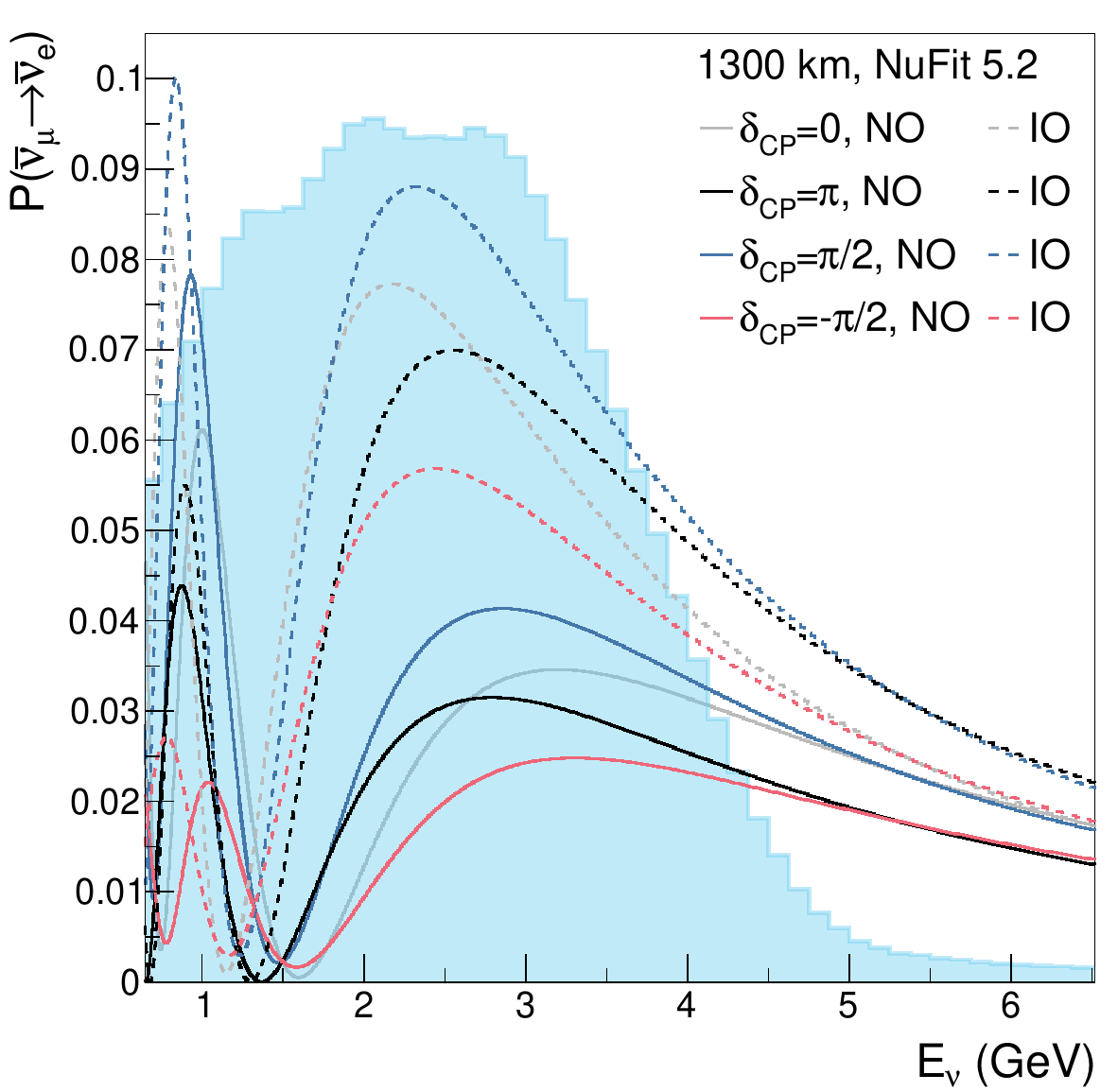}
    \caption{The effect of the CP violating phase and mass ordering scenarios on the electron neutrino appearance coming from muon neutrinos. The oscillation probability is shown for HK (left) and DUNE-like (right) baselines, for neutrinos (top) and anti-neutrinos (bottom), with the arbitrarily normalized neutrino fluxes for each experiment and appropriate species in the background. The CP-conserving points are $\delta_\textrm{CP}=0,\pi$ ($\sin\deltacp=0$), and the maximally CP-violating points are $\delta_\textrm{CP}=\pm\pi/2$ ($\sin \deltacp = \pm 1$).}
    \label{fig:osc_prob_numu_nue}
\end{figure}
The probabilities for muon (anti-)neutrinos to oscillate into electron (anti-)neutrinos (``appearance'')are shown in \autoref{fig:osc_prob_numu_nue} for different values of the CP violating phase and the two mass orderings for HK and DUNE, overlaid with their fluxes.
CP is conserved (CPC) when $\delta_\textrm{CP}=0,\pi$ and the two scenarios have similar impact on HK and DUNE: both CPC cases have comparable probabilities, but the position of the maximum shifts in neutrino energy between the two. The two CP conserving cases present the most notable degeneracy present for both experiments. 
Relative to the CPC cases, $\delta_\textrm{CP}>0$ causes a suppression of the appearing electron neutrinos, and an enhancement of electron anti-neutrinos, whereas $\delta_\textrm{CP}<0$ does the opposite. 
The figures demonstrate DUNE's ability to consistently separate the effects of the mass ordering from the CP-violating phase, whereas HK sees degeneracies between specific cases of the CP-violating phase and mass ordering.
There are two notable degeneracies for HK in the peak region: 1) the CPC scenarios ($\delta_\textrm{CP}=0,\pi$) in normal ordering are similar to the scenario of $\deltacp=-\pi/2$ in the inverted ordering; and 2) the scenario where $\deltacp=\pi/2$ in normal ordering is similar to the CPC scenarios in inverted ordering. The cases of inverted ordering with $\delta_\textrm{CP}=0,\pi/2,\pi$ are ruled out by current data with significances above $3\sigma$~\cite{T2K:2023smv,NOvA:2025tmb,T2K:2024wfn,T2K:2025wet}, making it likely that only the first degeneracy matters.
This degeneracy is lifted with other measurements of the mass ordering, from for example JUNO~\cite{JUNO:2015zny}, or atmospheric neutrinos, including HK and IceCube~\cite{IceCubeCollaborationP:2025rpl}.
Although the peak region is most sensitive to the oscillation parameters, the degeneracy between $\deltacp=0,\pi$ in the region of the flux peak exhibits considerably different behaviour at higher neutrino energy in the same flux for both experiments. 
Hence, experiments will need to study the oscillations throughout the neutrino energy spectrum to maximise their sensitivity to the oscillations parameters.
The oscillation probabilities also demonstrate the larger relative impact of the CP-violating phase on the rate of appearing electron (anti-)neutrinos has on HK compared to DUNE, and the much larger impact that the mass ordering has on DUNE compared to HK.

Other degeneracies, not shown here, exist between for instance \ssqthtwothree, \ssqthonethree and \deltacp~\cite{DUNE:2020jqi,T2K:2023smv,Hyper-Kamiokande:2025fci,NOvA:2023iam} due to their impact on the CP sensitive electron neutrino samples. These parameters are additionally constrained through other samples (e.g. long-baseline muon (anti-)neutrino disappearance) and external data (e.g. reactor electron anti-neutrino disappearance), which significantly reduces these degeneracies. 
Recent joint analyses combining T2K with NOvA~\cite{,T2K:2025wet} and T2K with SK atmospheric data~\cite{T2K:2024wfn} have demonstrated significant benefits of combining different experiments to lift degeneracies between oscillation parameters. A joint DUNE and HK analysis, including results from atmospheric neutrinos from IceCube and KM3NeT/ORCA and reactor neutrinos from Daya Bay and JUNO, will ultimately provide the most stringent constraints on neutrino oscillations in the GeV region. 
However, the success of such combinations profoundly depends on how the experiments design, evaluate and correlate their systematic uncertainties. This includes developing an interaction model which covers both experiments' neutrino energy regions and interactions on potentially different targets, which poses a formidable challenge. This is especially true for the high-statistics next-generation experiments where systematic uncertainties need to be small precise and well understood.

\subsection{Dependence on neutrino interaction modelling}
The impact of interaction modelling on neutrino oscillation measurements can be profound, and strongly depends on detector performance and design, the neutrino energy range, and analysis choices.
The vast field of neutrino interactions has been subject to multiple reviews in the last decade and a half~\cite{Formaggio:2012cpf,Alvarez-Ruso:2014bla,Mahn:2018mai,Jachowicz:2021ieb,SajjadAthar:2022pjt, Ruso:2022qes, Pandey:2025vpa}, with some focussing particularly on the impact on oscillation experiments~\cite{Mosel:2016cwa,katori:2016yel,NuSTEC:2017hzk,Dolan:2026nlr}.

The central requirement which enables long-baseline experiments to measure oscillations is reliably identifying neutrinos with a definite flavor, $\alpha$, and energy, $E_\nu$.
Measuring the neutrino energy is non-trivial, since the neutrinos in long baseline experiments are not mono-energetic and are not directly detected---their presence and energy is only inferred from the outgoing interaction products.
The primary means of characterizing the neutrino's flavour is by identifying the outgoing lepton from the neutrino's charged-current interaction in the detector.
These alone introduce a number of uncertainties related to neutrino interaction modelling.
For instance, an NC$1\pi^0$ interaction may produce a single visible electromagnetic shower, which can be mistaken for a single electron.
Including such an event in the electron-like selection biases the number of reconstructed electron neutrinos appearing in the far detector, and the estimation of the neutrino energy. This changes both the shape and normalization of the appearance energy spectrum, which directly impacts the inference of underlying neutrino oscillation parameters. 
Certain neutrino oscillation parameters are more sensitive to the overall normalisation of the spectra, such as $\delta_\textrm{CP}$, but misreconstructing the neutrino energy can have significant impact on parameters that oscillate in energy, for example $\Delta m^2_{32}$, and separating $\delta_\textrm{CP}=0,\pi$ scenarios, as detailed in \autoref{fig:osc_prob_numu_nue}.

\begin{figure}[hbtp]
    \centering
    \includegraphics[page=1,width=0.49\textwidth,clip, trim={0mm 0mm 0mm 0mm}]{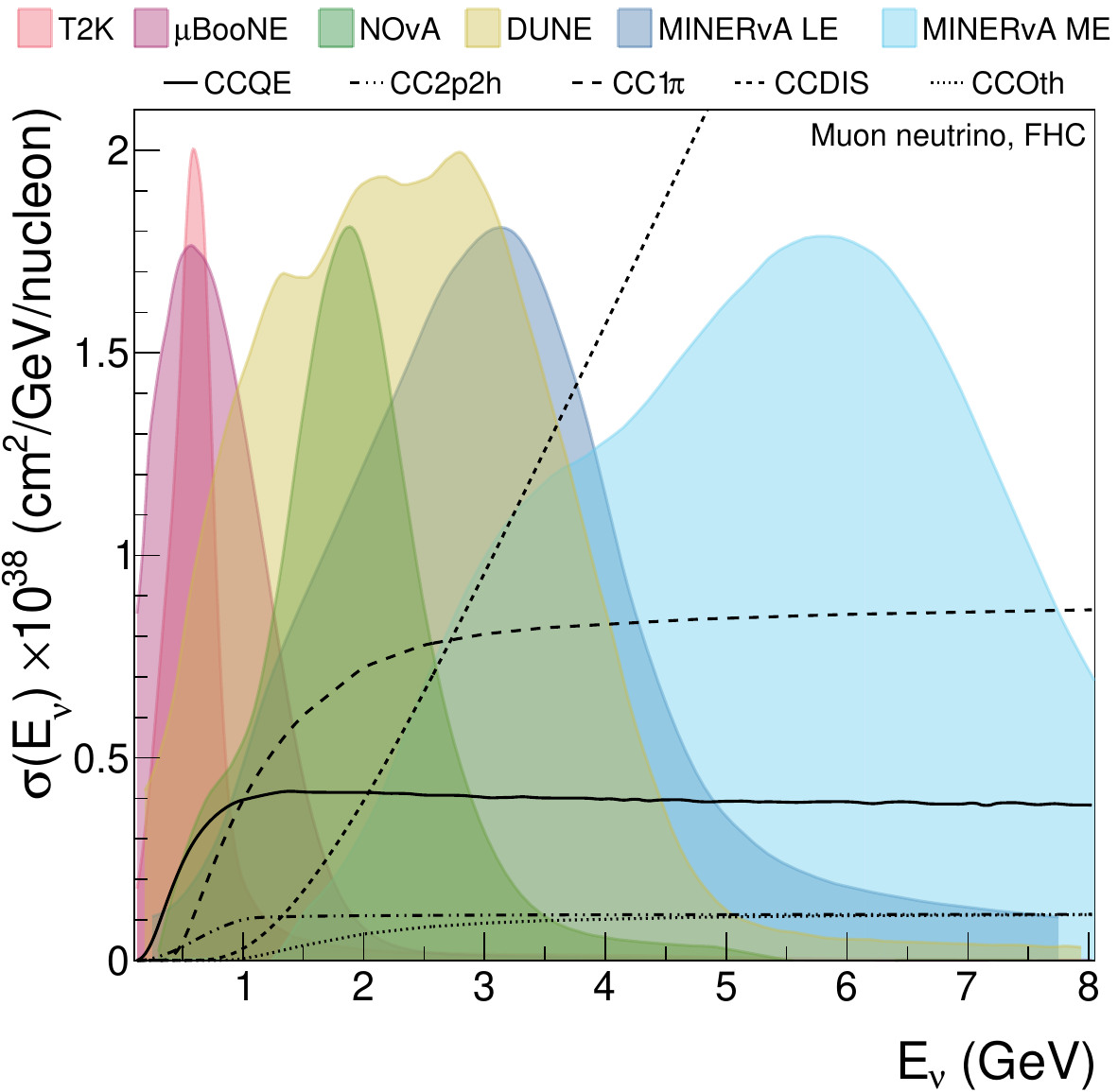}
    \includegraphics[page=1,width=0.49\textwidth,clip, trim={0mm 0mm 0mm 0mm}]{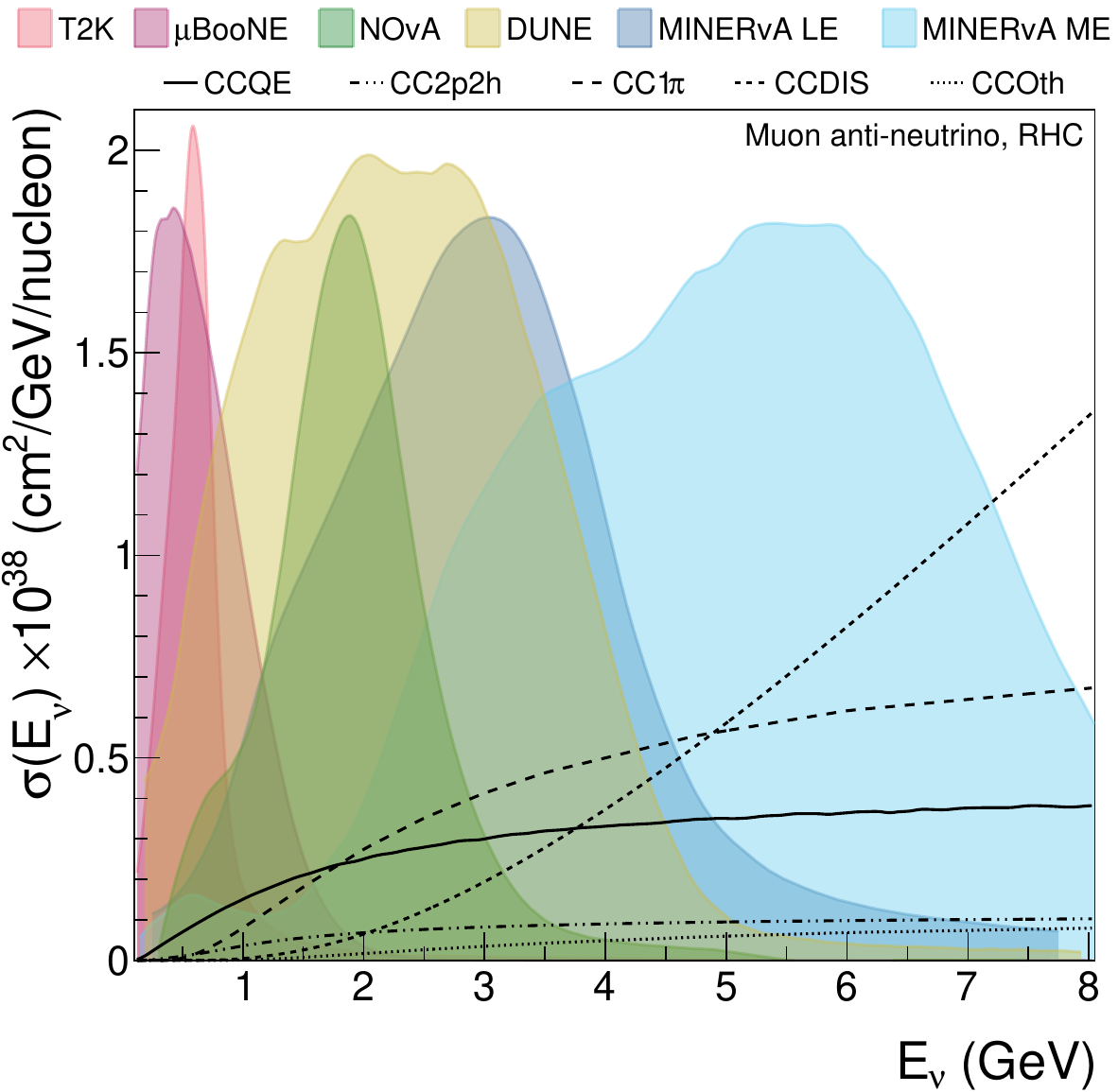}
    
    \caption{The energy of the neutrino fluxes in current and future GeV-scale oscillation and cross section experiments, overlaid with the interaction cross sections for different neutrino interaction channels from NEUT v5.4.0~\cite{Hayato:2021heg}. The fluxes and cross sections are shown for muon neutrinos (left) and muon anti-neutrinos (right), and the fluxes are each arbitrarily normalized. The MicroBooNE flux~\cite{MiniBooNE:2008hfu} is equivalent to the MiniBooNE flux, which is very similar to the SBND and ICARUS fluxes. The HK flux is equivalent to the T2K flux.}
    \label{fig:neu_flux}
\end{figure}

The energy range of 0.5-5 GeV renders the long-baseline experiments dependent on neutrino-nucleus interaction modelling to varying extents. 
\autoref{fig:neu_flux} shows the neutrino fluxes in the current and future long-baseline experiments with the cross sections for different types of interactions overlaid.
For instance, T2K and HK have a relatively narrow neutrino energy spectrum peaking at $E_\nu\sim0.6~\textrm{GeV}$, and is therefore more dependent on charge-current quasi-elastic (CCQE) modelling and initial-state nuclear effects.
NOvA also uses a narrow-band beam, but the peak energy is 2 GeV. This promotes the importance of a different set of interaction mechanisms than for T2K and HK, as there are roughly equal amounts of CCQE, resonant pion production, and soft inelastic scattering (SIS).
DUNE, on the other hand, will sample the neutrino beam on-axis and therefore have a broad energy spectrum, with significant flux between 0.5--5~\textrm{GeV}, making it sensitive to a wide range of interactions, including CCQE, resonant, SIS, and deep-inelastic scattering (DIS), and the interferences and transitions between them. 

The target material in the experiment's detectors also matters for accurately modelling neutrino-nucleus interactions, primarily from the myriads of nuclear effects.
For instance the NOvA, T2K, and HK experiments have opted for lighter nuclear targets, such as $\textrm{CH}_N$ and $\textrm{H}_2 \textrm{O}$, whereas DUNE and the short-baseline program at Fermilab~\cite{MicroBooNE:2015bmn}---encompassing SBND~\cite{SBND:2025lha}, MicroBooNE~\cite{MicroBooNE:2016pwy} and ICARUS~\cite{ICARUS:2004wqc,ICARUS:2023gpo}---use $^{40}\textrm{Ar}$ as the target in both the near and far detectors\footnote{One near-detector component of DUNE, SAND, will likely consist of a lighter carbon-based target~\cite{DUNE:2021tad}.}. 
Generally, neutrino interactions on heavier targets experience more collective nuclear effects and rescattering of the outgoing products due to the increased number of nucleons in the nucleus; this significantly complicates nuclear modelling.

\subsubsection{The role of the near detector}
\label{sec:near_det}
The neutrino beam is often characterized by using a near detector to mitigate the inherent challenges in long-baseline neutrino experiments. 
The objective is to measure the neutrinos \textit{before} their dominant oscillations, estimating the flavour content and kinematic distributions of the outgoing particles from the neutrino interaction, and applying that knowledge to estimate the appearance and disappearance spectra at the far detector.
In the case of perfectly matching near and far detectors, with identical acceptances and efficiencies for all particles, the far detector event spectra can be expressed as
\begin{equation}
    N_\textrm{FD}^\alpha = P(\nu_\alpha\rightarrow\nu_\alpha) \times
    N_\textrm{ND}^\alpha \times
    R_{\textrm{ND}\rightarrow\textrm{FD}}^\alpha
\end{equation}
\label{eq:osc_near_far}
where $N^\alpha_\textrm{FD}$ and $N^\alpha_\textrm{ND}$ are the number of neutrinos with flavor $\alpha$ at the far and near detectors, respectively;
$P(\nu_\alpha\rightarrow\nu_\alpha)$ is the neutrino oscillation probability of a neutrino with flavor $\alpha$ being a neutrino of flavor $\alpha$ after having travelled the baseline of the FD---which is calculated through the PMNS formalism and is dependent on the baseline, neutrino energy, and the oscillation parameters;
and $R^\alpha_{\textrm{ND}\rightarrow\textrm{FD}}$ is a constant near-to-far ratio, which accounts for the reduction of neutrinos at the far detector compared to the near detector.
This simplified approach---where $R_{\textrm{ND}\rightarrow\textrm{FD}}$ is only dependent on the neutrino beam divergence---is in practice never achieved: the acceptance at the near and far detectors never perfectly match, and there is often a complex dependence on the outgoing particles and their kinematics, which in turn depend on the incoming neutrino's energy and flavor.
Characterizing the degree of mismatch inevitably depends on the neutrino interaction model, which becomes central to reliably extrapolating measurements at the near detector to the far detector in the precision era.

\subsubsection{Matching selections}
\begin{figure}[hbtp]
    \centering
    \begin{subfigure}[b]{0.32\textwidth}
        \includegraphics[page=1,width=\textwidth,clip, trim={0mm 0mm 0mm 0mm}]{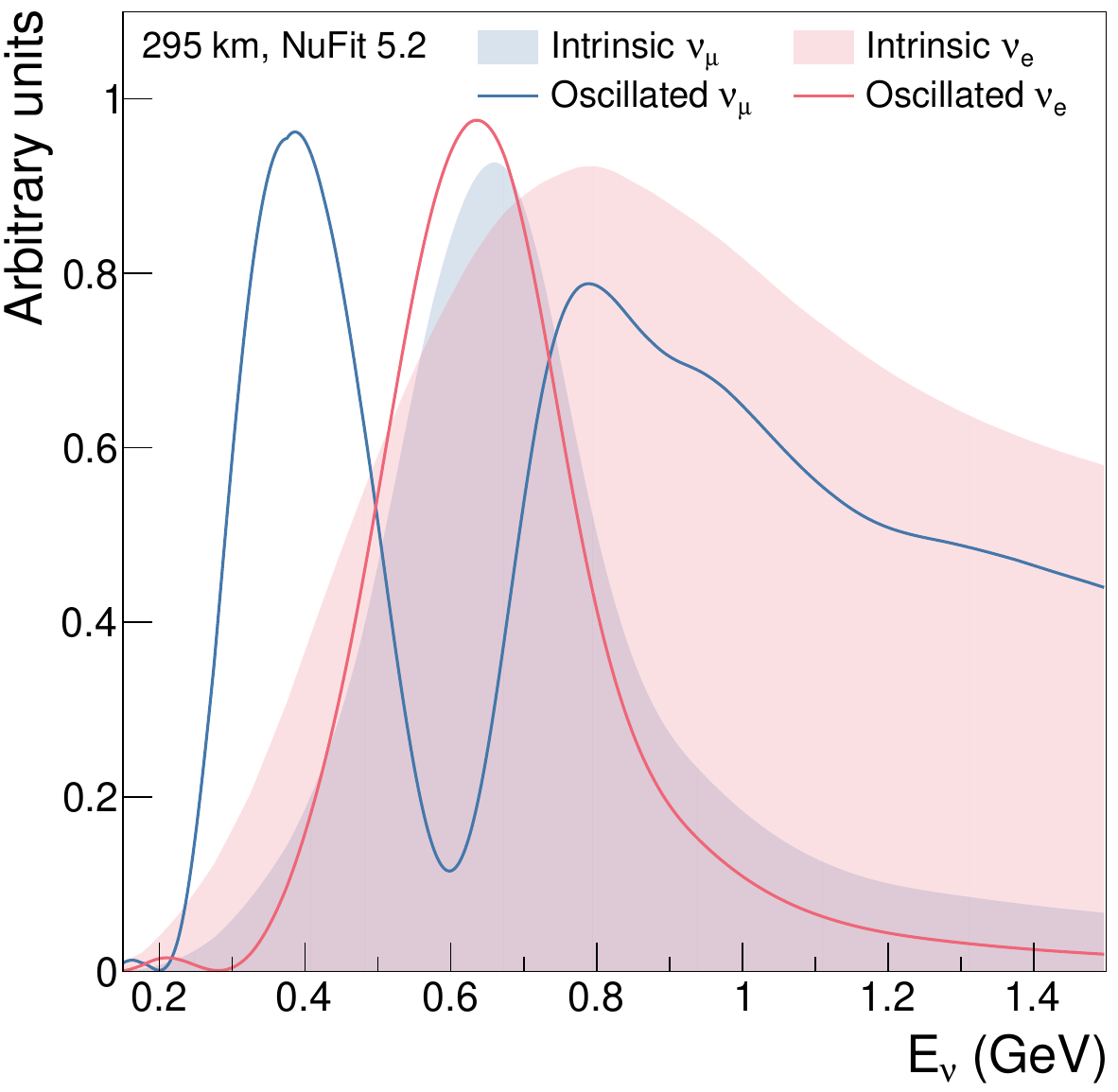}
    \end{subfigure}
    \begin{subfigure}[b]{0.32\textwidth}
        \includegraphics[page=1,width=\textwidth,clip, trim={0mm 0mm 0mm 0mm}]{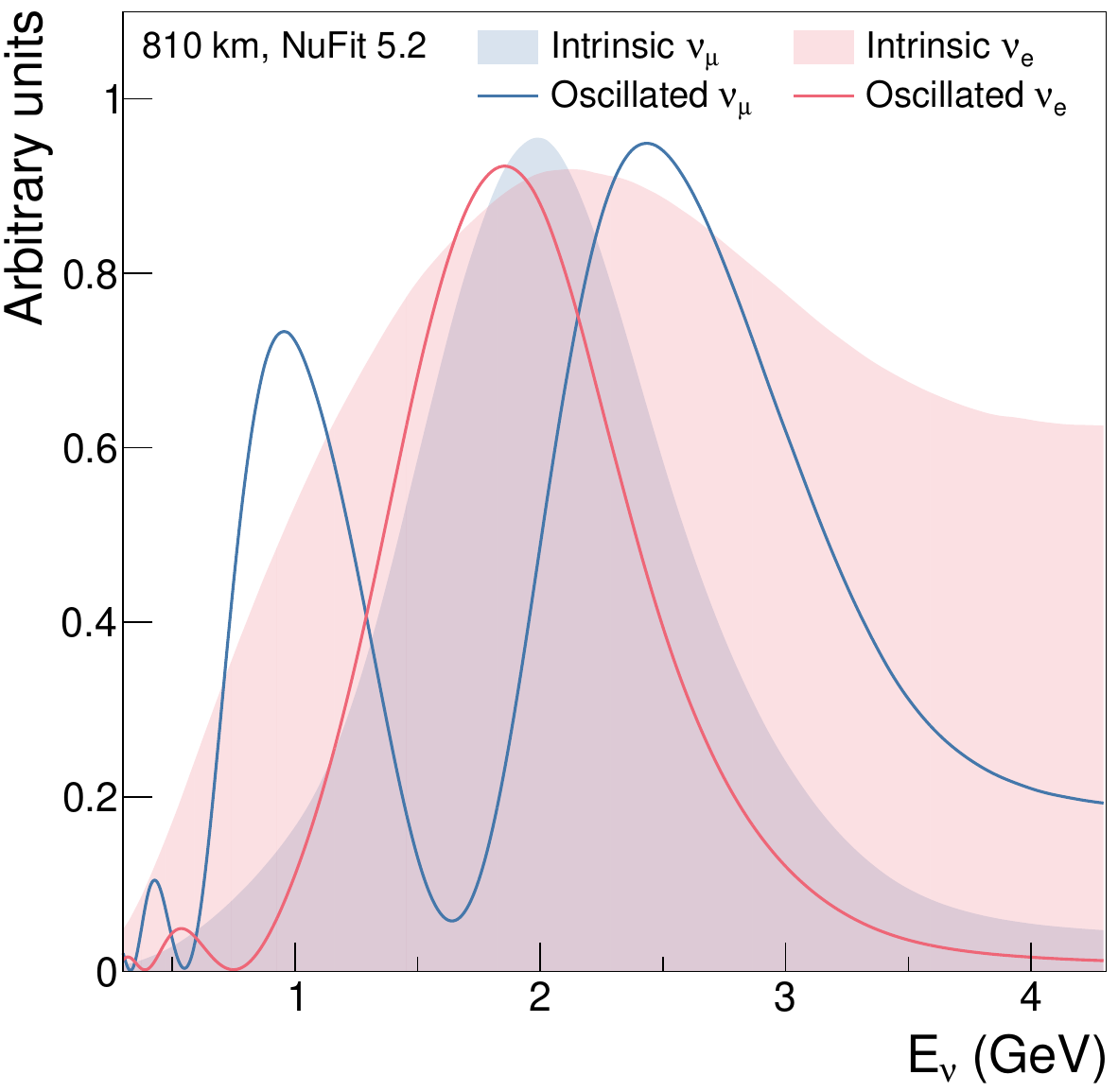}
    \end{subfigure}
        \begin{subfigure}[b]{0.32\textwidth}
        \includegraphics[page=1,width=\textwidth,clip, trim={0mm 0mm 0mm 0mm}]{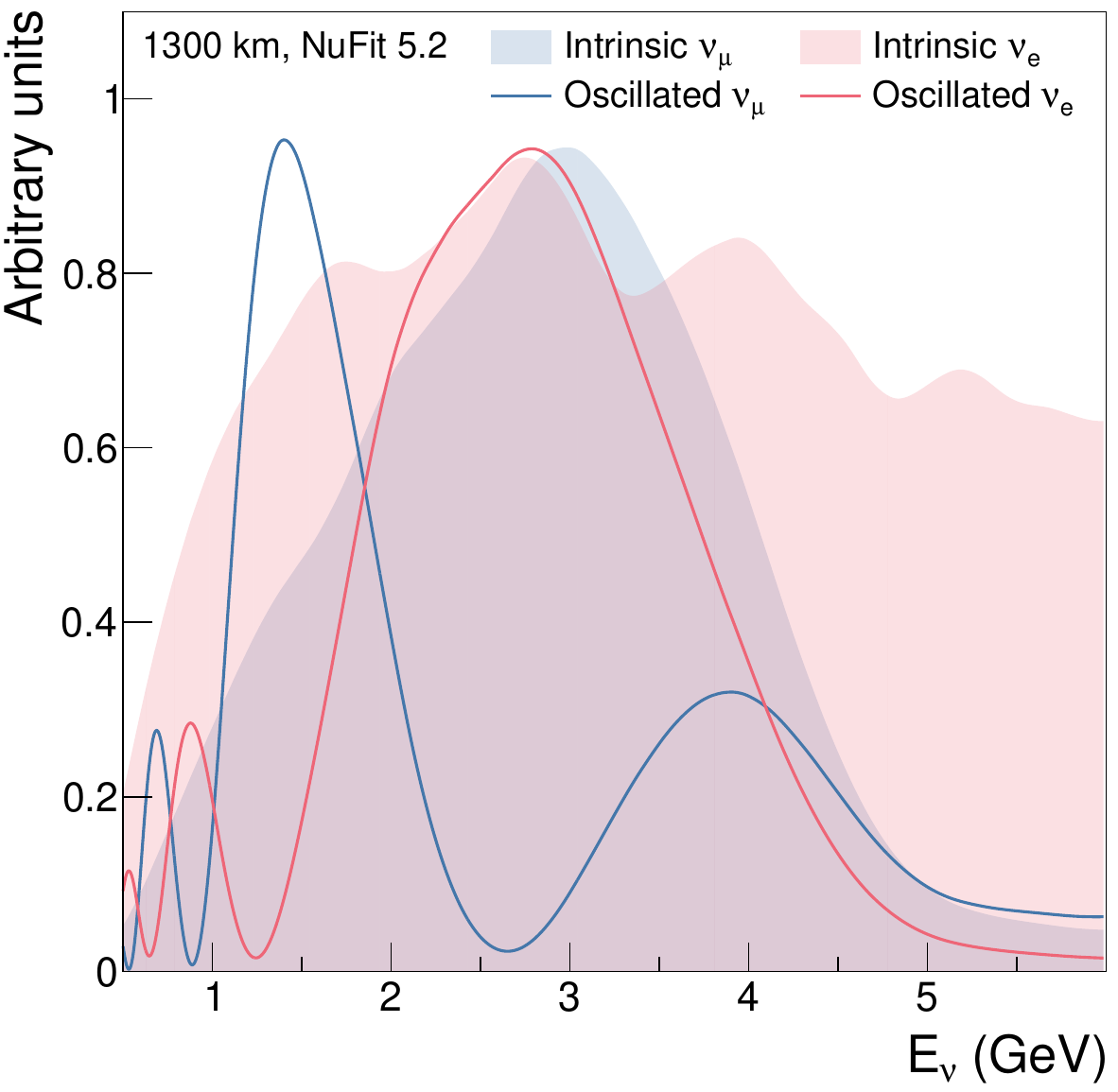}
    \end{subfigure}
    \caption{The expected event distributions, calculated as the neutrino flux times the cross section ($N=\Phi \sigma$), for intrinsic (filled) and oscillated (lines) muon (blue) and electron (red) neutrinos at the far detectors of T2K (left) NOvA (center) and DUNE (right). The event distributions are arbitrarily normalized, the oscillation probability is the full three-flavor probability with oscillation parameters from NuFit 5.2~\cite{Esteban:2020cvm}, and the cross sections are from NEUT~\cite{Hayato:2021heg}. No detector effects are included, and no selections are applied, apart from using the charged-current total cross section from NEUT.}
    \label{fig:t2k_nova_spectrum}
\end{figure}

The near detectors help in better understanding the neutrino flux and neutrino interactions by studying them in a high-rate environment. Experiments generally develop dedicated near-detector selections that approximately match the samples at the far detector which are used to infer the neutrino oscillation parameters.
However, imperfections arise due to the oscillations affecting the neutrinos in the far detector, altering the flavor composition for a given neutrino energy, which does not affect the neutrinos observed in the near detector. This impacts both the muon and electron neutrino selections.
Similarly, the electron-neutrino flux at the near detector---the ``intrinsic'' electron neutrino flux---is small and comes from a mix of muon, pion and kaon decays. However, at the far detector the electron neutrinos are largely from oscillated muon neutrinos, which mostly come from pion decays.
Thus, the electron neutrinos observed in the near detector have a different energy spectrum to the far detector, and cannot be directly used to inform predictions of the appearing electron neutrinos at the far detector without relying on an interaction model to extrapolate between the different regions of neutrino energy.
This difference is demonstrated in \autoref{fig:t2k_nova_spectrum} for T2K and NOvA's neutrino fluxes and baselines, which shows that the electron neutrinos appearing in the far detector have a notably different neutrino energy distribution to the unoscillated neutrinos measured in the near detector. The difference in peak position is between 0.2-0.3 GeV for both T2K and NOvA.
Neither T2K nor NOvA directly use electron-neutrino samples at their near detectors in the oscillation analysis; they are only used as cross-checks of the electron-neutrino interaction model, and the intrinsic component of electron (anti-)neutrinos in the beam.

Alternatively, \autoref{fig:t2k_nova_spectrum} also shows that the muon (anti-)neutrinos at the near detector have a much more similar energy spectra to the electron neutrinos appearing at the far detector. 
Here the model dependence instead enters by extrapolating measurements of muon (anti-)neutrinos at the near detector to electron (anti-)neutrinos at the far detector, which is non-trivial near the charged lepton production thresholds due to differences in the charged lepton masses\footnote{The threshold in turn depends on the interaction channel that the neutrino interacts through (CCQE, single pion production, etc).}. These differences generally become small, but not negligible, at higher energies.
Differences between electron and muon neutrinos can also come from the neutrino-nucleon interaction, for instance via radiative corrections~\cite{Day:2012gb,Tomalak:2021hec,Tomalak:2022xup}.
The uncertainty on electron-to-muon neutrino scaling is currently around 2--3\% from the neutrino-nucleon interaction, and is the dominant contributor to cross-section uncertainties for electron (anti-)neutrino selections. 
These are relatively small compared to the statistical uncertainties of the electron (anti-)neutrino samples on T2K and NOvA, which are around 10\%. However, for DUNE and HK, which each expect over 10 times more electron (anti-)neutrino events, the differences between electron and muon neutrino interactions significantly affects the ability to measure CP violation in neutrinos.
Studies of the impact of the nuclear model~\cite{Martini:2016eec,Ankowski:2017yvm,Nikolakopoulos:2019qcr,Gonzalez-Jimenez:2019qhq,Dieminger:2023oin,Martini:2023kem} suggest that a flat uncertainty is inappropriate, and the energy dependence should be accounted for. This generally means the uncertainty is largest near the thresholds and when the neutrino energy is comparable to the energy scale of nuclear effects, and decreases with increasing neutrino energy. 
Furthermore, the uncertainty may be underestimated below neutrino energies of 1 GeV, but is highly correlated for electron neutrinos and anti-neutrinos.
\autoref{fig:t2k_nova_spectrum} concluded that the event spectrum for the intrinsic electron neutrinos have long and wide tails, extending far beyond the energies of the appearing electron neutrinos. 
Since the differences between electron and muon neutrinos are large at lower energies and smaller at higher energies, measurements of the cross section of the intrinsic electron neutrinos do not necessarily probe the region that has the largest uncertainty, nor the region that has the biggest impact on understanding the appearing electron neutrinos' cross section.
An improved understanding of the $\nue/\numu$ cross section ratio, the $\nuebar/\numubar$ cross section ratio and their correlations, has been demonstrated to have profound implications on the sensitivity to CP violation in HK~\cite{Hyper-Kamiokande:2025fci} and atmospheric neutrinos~\cite{T2K:2024wfn}.

The issue of mismatching selections due to oscillations is not only important for electron neutrinos, but is also present for muon neutrinos. 
This is because for experiments with an $L/E$ ratio tuned to maximizing muon neutrino disappearance, the muon neutrino flux at the near detector peaks where the far detector flux has a minimum, also shown in \autoref{fig:t2k_nova_spectrum}.
Furthermore, the region just below and above the first oscillation dip often has a significantly lower flux than the peak region in the near detector, but is a crucial region for precisely measuring $\Delta m^2_{32}$ and $\sin^{2}\theta_{23}$ at the far detector. Isolating these two regions in the near detector for benchmarking the neutrino cross section is challenging, requiring reliable neutrino energy reconstruction, amongst others.

\subsubsection{NOvA}
\begin{figure}[hbtp]
    \centering
    \includegraphics[width=0.5\textwidth]{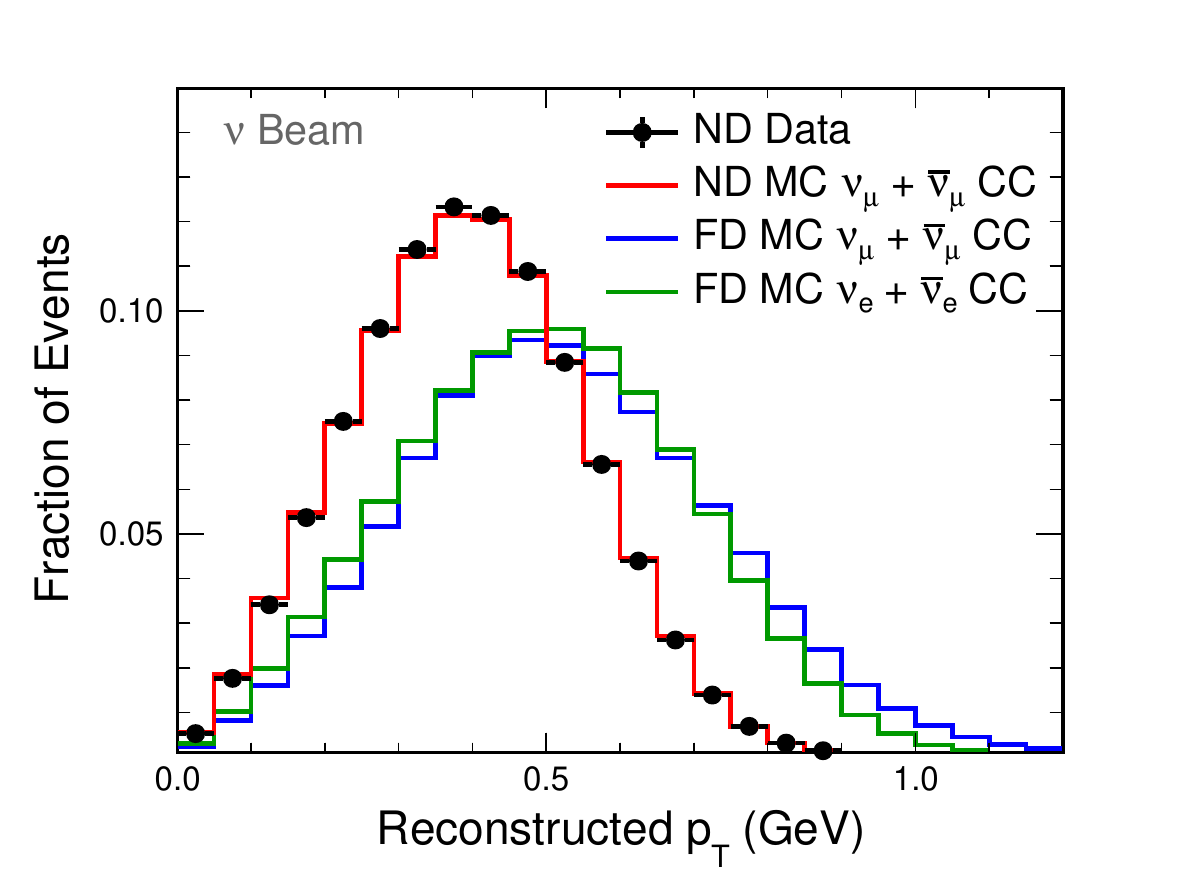}
    \caption{The fraction of events in reconstructed transverse lepton momentum for the NOvA near and far detectors, showing a mismatch in phase space due to detector geometry. Reproduced from Ref.~\cite{NOvA:2021nfi}.}
    \label{fig:nova_pt}
\end{figure}

NOvA utilizes a CC-inclusive selection at the far detector for both electron and muon neutrinos, and estimate the neutrino energy primarily through calorimetry of the outgoing particles, with muon energies estimated by range. 
NOvA's near detector is functionally identical to the far detector, and the near-detector selections are similar to the far-detector selections.
Although NOvA's detectors are designed to be functionally identical, the near detector is significantly smaller, which leads to a mismatch in their acceptances in transverse muon momentum, shown in \autoref{fig:nova_pt}. 
Hence, events in the NOvA far detector which have poor overlap with the events in near detector---for instance those with high transverse momentum---have a larger model dependence, or remain unconstrained by the near-detector analysis which leads to larger uncertainties.

Oscillation analyses on NOvA~\cite{NOvA:2025tmb,NOvA:2021nfi,Shanahan:2021jlp} have considerably enhanced the 2p2h cross section at the near detector to account for a noticeable under-prediction in the number of CC-inclusive events~\cite{NOvA:2020rbg}, most evident for events with low hadronic activity. 
Such a discrepancy is unlikely to be entirely due to 2p2h mismodelling, as the region is sensitive to a number of nuclear effects which are challenging to pinpoint in an inclusive selection; for example the absorption and rescattering of hadrons in the nucleus, the amount of low energy pions and protons, which can be impacted by nuclear effects such as Pauli blocking and in-medium resonance effects, and the modelling of larger nuclear resonances (e.g. CRPA).
NOvA primarily use the ratio between the observed data and simulation at the near detector after this enhancement to scale the simulation in bins of reconstructed neutrino energy and transverse lepton momentum, which partly mitigates wrongly assigning cross section strength to specific interaction modes. This is then used to inform the appropriate scaling of events at the far detector.

\begin{figure}[hbtp]
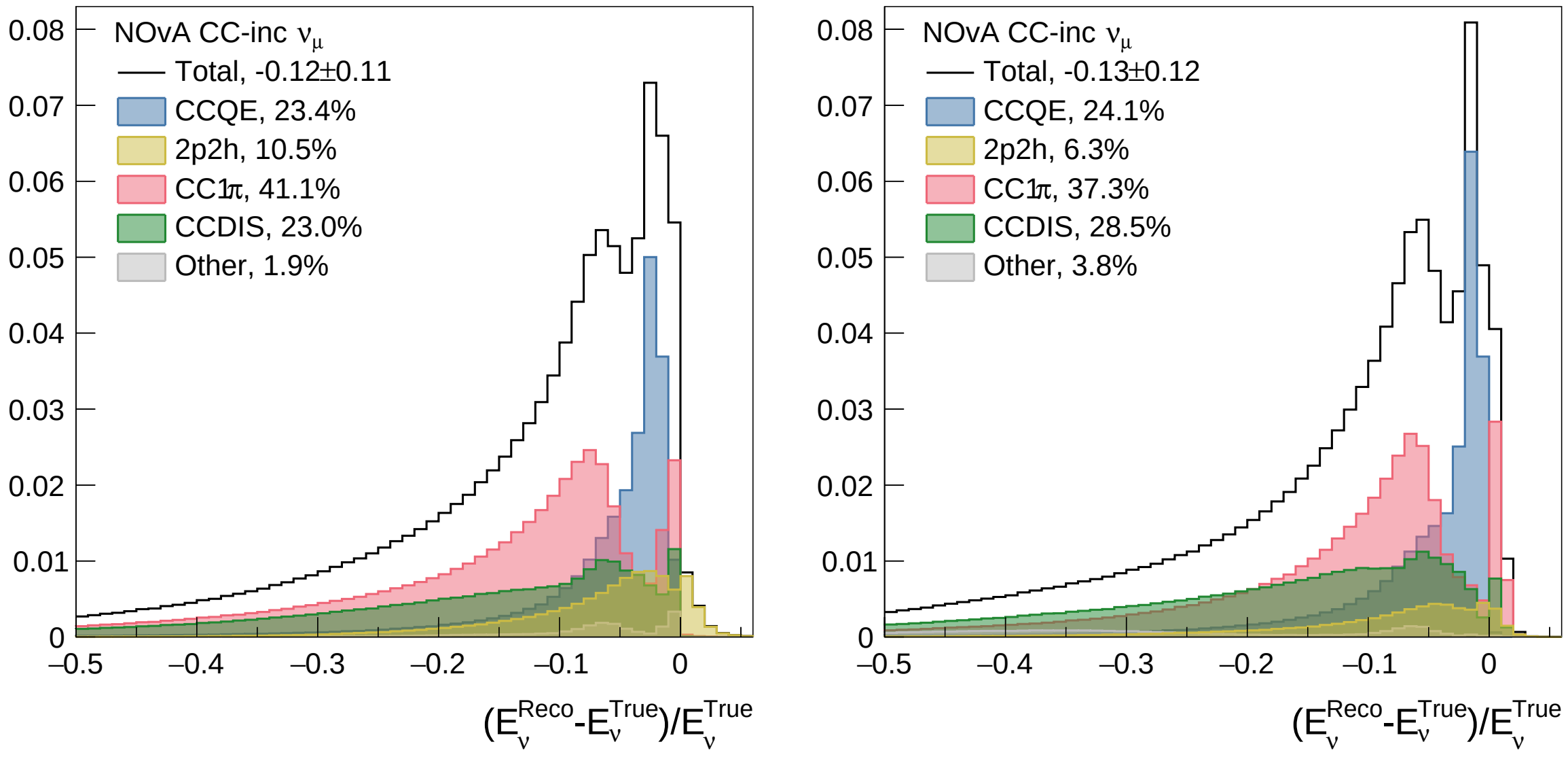

    \centering
    \begin{subfigure}[b]{0.47\textwidth}
        \includegraphics[page=2,width=\textwidth,clip, trim={0mm 0mm 0mm 0mm}]{Figures/osc_effects/NOvA_bias_genie.pdf}
    \end{subfigure}
    \begin{subfigure}[b]{0.47\textwidth}
        \includegraphics[page=2,width=\textwidth,clip, trim={0mm 0mm 0mm 0mm}]{Figures/osc_effects/NOvA_bias_neut.pdf}
    \end{subfigure}
    \caption{Biases of a NOvA-like calorimetric neutrino energy estimator using \autoref{eq:calo} with the \texttt{GENIE\ G18\_02a} (left), and \texttt{NEUT v5.4.0} (right) neutrino interaction generators. Interactions are generated on a CH target with the NOvA muon neutrino flux and a CC-inclusive selection. Particle thresholds of 200, 200, and 500 MeV/$c$ are applied for muons, pions and protons respectively to roughly emulate detection thresholds. The y-axes are area normalised.} 
    \label{fig:nova_bias}
\end{figure}

In a calorimetric analysis such as those on NOvA and DUNE, particle identification and missing energy can introduce significant biases.
Energy carried away by neutral particles, energy lost to nuclear breakup, re-interactions of outgoing particles as they propagate through the nucleus, and energy lost to overcoming binding energies directly bias the energy estimator.
Mistaking the type of outgoing particle can also bias the estimator, as the rest-mass energy needs to be added when particles are created, but not when they are the product of a re-scatter: for example, to be unbiased a pion needs its rest mass added to the calorimetric energy estimator, whereas a proton does not since it is ejected from the nucleus, not created.
The calorimetric energy estimator is generally expressed as
\begin{equation}
    E_\nu^{\textrm{calo}} = E_l + \left(\sum_{i=\pi^0,\gamma} E^i\right) + \left(\sum_{i=p,\pi^\pm} T^i\right)
    \label{eq:calo}
\end{equation}
\noindent where $E_l$ is the outgoing lepton energy, $E^i$ is the energy of the $i$th neutral pion and photon, and $T^i$ is the kinetic energy of the $i$th proton and charged pion.
Particle detectors that are able to separate protons from pions with suitable low particle tracking thresholds can improve the neutrino energy estimator by including the $\pi^\pm$ in the total energy term instead of the kinetic energy term.
An example of the bias from a calorimetric estimator using two interaction generators, GENIE and NEUT, are shown in \autoref{fig:nova_bias}. Whereas the overall shape of the estimator bias is similar for the two generators, there are important differences in the details, especially when the bias become large: for example the relative importance of 2p2h and CC$1\pi$ or the shape of the DIS distribution, which differ in part due to how the two generators predict neutrons.
Such biases are corrected for in analyses, and the size of the correction depends on the underlying interaction model.

NOvA's near detector is not magnetized, so is not able to reliably determine the charge of the primary lepton on an event-by-event basis. Although the purities of the muon neutrino and anti-muon neutrino beams are high and similar (93\% and 92\% purity, respectively), the larger muon-neutrino cross section means the wrong-sign background is larger in anti-neutrino mode compared to neutrino mode. 
Mistaking an electron anti-neutrino for an electron neutrino directly impacts the CP violation searches and mass ordering determination, which are sensitive to the differences between neutrino and anti-neutrino oscillations.
Although the flux simulation is used to predict the amount of muon (anti-)neutrinos passing through the detector, the cross-section model is relied on to simulate their respective rate of interactions.

\subsubsection{T2K}
The T2K experiment have opted for an analysis that topologically selects the dominant neutrino interaction topology in its energy spectrum, the charged-current interaction with no pions, $\textrm{CC}0\pi$. Here, the neutrino energy is estimated using only the kinematics of the outgoing charged lepton, assuming a four-legged ``CCQE-like'' interaction, where the neutrino interacts with a stationary initial-state nucleon, and the outgoing system is assumed to be a single nucleon and a charged lepton,
\begin{equation}
    E_\nu^\textrm{CCQE} = \frac{2 m_N E_l - m_l^2 + m_{N'}^2 - m_N^2}{2 \left( m_N - E_l + p_l \cos{\theta_{\nu,l}}\right)}\label{eq:ccqe_calo}
\end{equation}
where $m_N$ is the initial-state nucleon mass, corrected for an average binding energy, $m_{N'}$ is the final state nucleon mass, $\theta_{\nu,l}$ is the angle between the incoming neutrino and the outgoing charged lepton, and $m_l, E_l, p_l$ are the outgoing charged lepton's mass, energy and momentum, respectively.

\begin{figure}[hbtp]
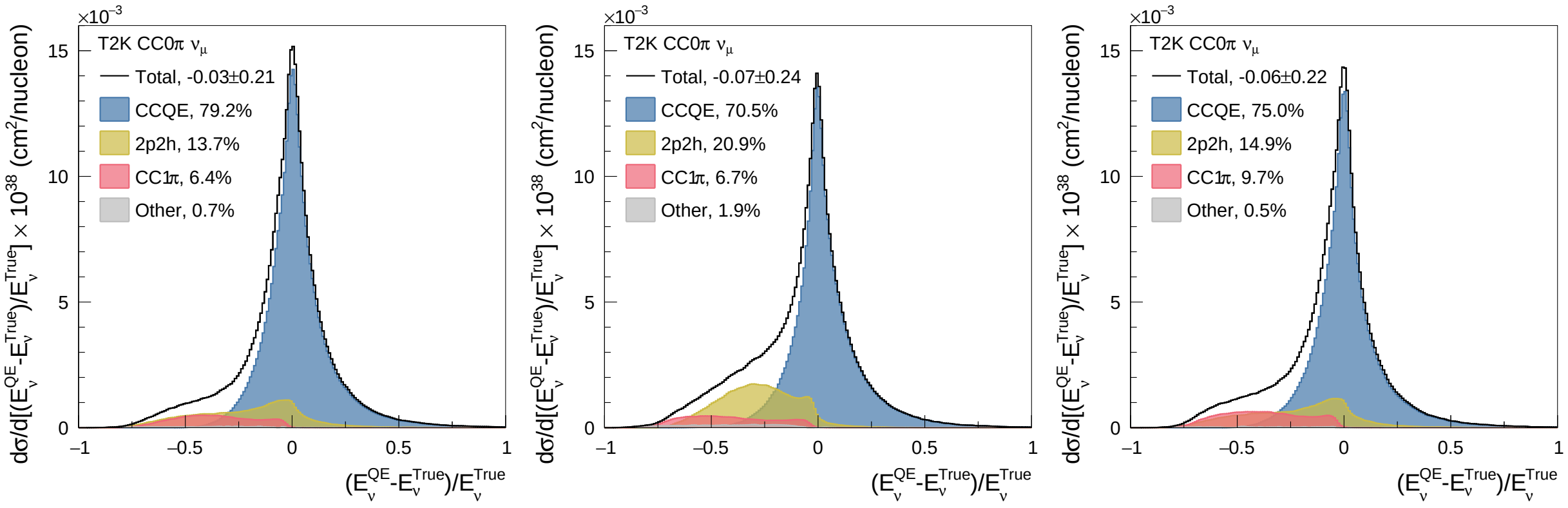

    \centering
    \begin{subfigure}[b]{0.32\textwidth}
        \includegraphics[page=2,width=\textwidth,clip, trim={0mm 0mm 0mm 0mm}]{Figures/osc_effects/T2K_CC0pi_bias_nuwro.pdf}
    \end{subfigure}
    \begin{subfigure}[b]{0.32\textwidth}
        \includegraphics[page=2,width=\textwidth,clip, trim={0mm 0mm 0mm 0mm}]{Figures/osc_effects/T2K_CC0pi_bias_G18_02a.pdf}
    \end{subfigure}
    \begin{subfigure}[b]{0.32\textwidth}
        \includegraphics[page=2,width=\textwidth,clip, trim={0mm 0mm 0mm 0mm}]{Figures/osc_effects/T2K_CC0pi_bias_neut.pdf}
    \end{subfigure}
    \caption{Biases of a T2K-like kinematic CCQE neutrino energy estimator in \autoref{eq:ccqe_calo}, using the NuWro 19.01 (left), \texttt{GENIE\ G18\_02a} (middle), and \texttt{NEUT v5.4.0} (right) neutrino interaction generators. Interactions are generated on a CH target with the T2K muon neutrino flux and a CC$0\pi$ selection. The prediction for \texttt{GENIE\ G18\_10a} is not shown because it closely matches that of NEUT.}
    \label{fig:t2k_bias}
\end{figure}

For T2K's CC$0\pi$ selection, the CCQE interaction dominates with a fractional contribution of 70--80\%, depending on the generator used. 
Two-particle-two-hole interactions (2p2h) contributes between 15--20\%, and there is also a 5--10\% contribution from CC$1\pi$ interactions where the pion is missed, either through detector effects or final-state interactions in the nucleus. These neutrino interaction channels all bias the energy estimator differently.
The bias of the energy estimator relative to the true neutrino energy for a T2K flux and CC$0\pi$ selection is shown in \autoref{fig:t2k_bias} for three different popular choices of generator, broken down by the interaction modes. Both the normalization and shape of the channels significantly differ between each other.
The balance of CCQE, 2p2h CC1$\pi$ and others is different for the generator models, as is the total cross section at a given bias.
Most notably, the ``Empirical'' 2p2h model in \texttt{GENIE G18\_02a} is considerably larger and has a different shape to the Nieves et al. (``Valencia'') calculation, used in both NuWro 19.01, \texttt{GENIE G18\_10a}, and NEUT.
NEUT also has a larger contribution of single pion interactions, coming from a larger probability to produce low-momentum pions, and for those pions to be absorbed during final-state interactions.

Separately, the initial-state motion of the struck nucleon smears the resolution of the neutrino energy estimator, and effects from removal energy and Pauli blocking sculpts the CCQE contribution away from being symmetric around 0.
These effects bias the estimator away from the true neutrino energy, and the amount of bias is necessarily predicted through the neutrino interaction model.
Comparing the NOvA and T2K energy estimators, T2K's estimator does not need to be as aggressively corrected as NOvA's (4\% vs 13\%), but the resolution of the estimator is half as good (24\% vs 12\%).

T2K, and to some extent HK, have additional complications to NOvA and DUNE from the different near and far detector designs.
The far detector target is water and the near detector (ND280) is primarily plastic scintillator (CH), with about a quarter of the target mass being passive $\textrm{H}_2\textrm{O}$. 
The use of the near detector data in the oscillation analysis relies on using interactions on a hydrocarbon to predict interactions on water, which depends on the neutrino interaction model and more specifically the nuclear model.
There are also notable differences in performance between the detectors: for example SK rarely detects protons due to their high Cherenkov threshold, and has difficulties separating muons and pions. On the other hand, ND280 observes protons down to $\sim450~\textrm{MeV}/c$, and has good muon-pion separation due to the magnetic field and multiple pion-tagging methods.
Additionally, due to the detectors' geometries, SK has a near-isotropic angular acceptance in reconstructing particles, whereas ND280 has excellent forward acceptance, and worse acceptance for high and backward angles. 
Any mismodelling of high-angle and backwards-going particles would likely not be seen in the near-detector analysis, but are present in the far-detector analysis.
T2K's near-detector upgrade~\cite{T2K:2019bbb} aims to address the acceptance differences, but the detectors will still have different particle tracking thresholds and a different detector material. 
The Intermediate Water Cherenkov Detector (IWCD)~\cite{Zhu:2023nsv,Scott:2016kdg} addresses many of these concerns for HK, matching the target, technology and geometry of the near detector to that of the far detector, but is unable to detect protons and neutrons.

T2K has since 2017 introduced lower-statistics samples from sub-dominant interaction topologies at the far detector, such as the single-ring electron-like single-Michel-electron (1R$e$1de) selection~\cite{T2K:2017rgv}, and the two-ring muon-like selection~\cite{T2K:2025yoy}; both dominated by single-pion production events. 
The samples are relatively small: the former contributes 14 events compared to the 94 in the dominant 1R$e$0de selection~\cite{T2K:2023smv}, and the latter contributes 134 events compared to the 318 in the dominant 1R$\mu$ selection~\cite{T2K:2025yoy}.
However, the higher neutrino energy required to produce 2-ring events means the sample's lowest energies are around $E_\nu\sim1~\textrm{GeV}$, so renders it less sensitive to $\nu_\mu\rightarrow\nu_\mu$ oscillations compared to the 1R$\mu$ selection. 
As a result, the modelling of single-pion production is becoming increasingly important in T2K analyses, although the CC$0\pi$ modelling continues to dominate the uncertainty budget~\cite{T2K:2024wfn,T2K:2025yoy,T2K:2023smv}.

\subsubsection{Atmospheric neutrinos}
\label{subsec:atmospheric}
Although not strictly ``long-baseline'', atmospheric neutrinos are included in this discussion since the region relevant to oscillation physics lays in the GeV region. 
Importantly, combinations of accelerator, reactor and atmospheric neutrinos may be crucial to unlocking precision measurements of CP violation and the neutrino mass ordering~\cite{T2K:2024wfn,Hyper-Kamiokande:2025fci,Super-Kamiokande:2023ahc,Esteban:2024eli,IceCubeCollaboration:2024ssx,Scott:2020gng,KM3NeT:2021ozk,KM3NeT:2024ecf}.
Hence, atmospheric neutrino experiments also share many of the challenges faced by the long-baseline experiments when interaction physics is concerned, and addressing those challenges will have significant impact on the overall precision of the atmospheric neutrino oscillation measurements.

\paragraph{Introduction}
Atmospheric neutrinos span a vast range of energies, from 0.1 GeV to the TeV-scale~\cite{Gaisser:2002jj,IceCube:2015mgt}, and a similarly wide range of baselines. 
Neutrinos that are ``down-going'' in the detector are created in the atmosphere directly above ($\cos\theta_{z}\sim1$) and travel 10s of kilometers, whereas upwards-going neutrinos travel further than the diameter of the earth, around 13,000 kilometers. 
This means that neutrinos that emanate from below the horizon ($\cos\theta_z<0$) and have energies between 0.1--50 GeV have similar $L/E$ to the long-baseline experiments, and therefore undergo those same oscillations.
Additionally, atmospheric neutrinos that cross through the earth experience significant ``matter effects''~\cite{Wolfenstein:1977ue,Mikheyev:1985zog}, which causes prominent distortions in the neutrino energy spectra, resulting in notable sensitivity to the neutrino mass ordering.

Whereas the neutrino direction and baseline is precisely known for accelerator long-baseline and reactor neutrino experiments, an atmospheric neutrino's direction can only be inferred from its outgoing interaction products. 
Additionally, their exact baseline is unknown since they can be produced anywhere in the atmosphere, which is roughly between 5 to 50 km~\cite{Honda:2015fha}.
This causes uncertainties in estimating both the interacting neutrino's energy (discussed in the previous section), and its baseline, complicating the ability to precisely resolve the $L/E$, and therefore also pin down the parameters governing the oscillations.
Furthermore, unlike accelerator and reactor neutrinos, where the initial state is either muon or electron neutrinos respectively---and those neutrinos can be measured in-situ by near detectors---the initial neutrino flavor composition for atmospheric neutrinos is a model-dependent mixture of \numu and \nue, with a small \nutau component, with large contributions from the anti-neutrino flavours.

\begin{figure}[hbpt]
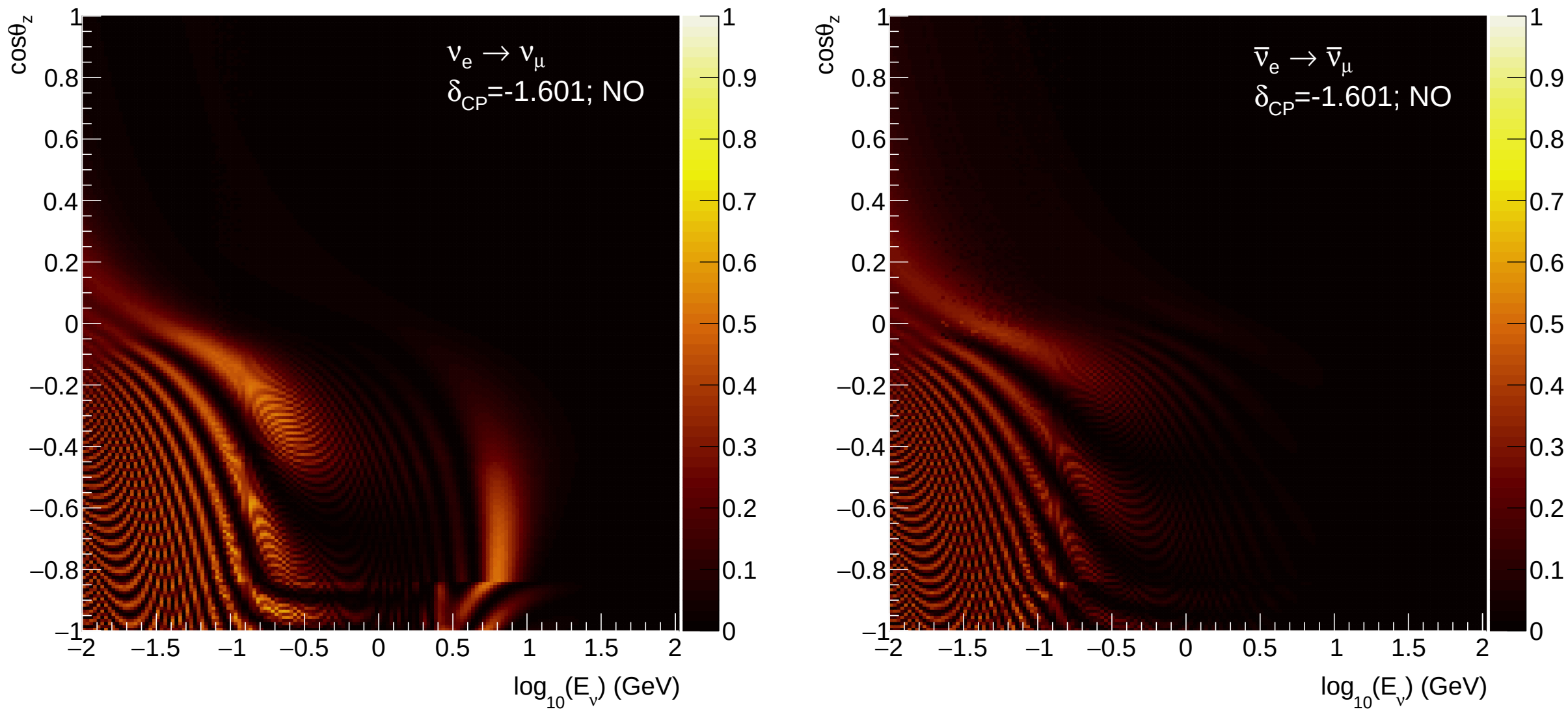

    \centering
    \begin{subfigure}[b]{0.49\textwidth}
        \includegraphics[width=\textwidth, trim={0mm 0mm 0mm 14mm}, clip, page=30]{Figures/osc_effects/Oscillograms_avg_dens_CPU.root_Oscillograms_avgdens_dcp-1601_IH.root.pdf}
    \end{subfigure}
    \begin{subfigure}[b]{0.49\textwidth}
        \includegraphics[width=\textwidth, trim={0mm 0mm 0mm 14mm}, clip, page=6]{Figures/osc_effects/Oscillograms_avg_dens_CPU.root_Oscillograms_avgdens_dcp-1601_IH.root.pdf}
    \end{subfigure}
    \caption{Atmospheric neutrino oscillations in neutrino energy and zenith angle (``oscillograms''), including earth matter effects, for $\nue\rightarrow\numu$ (left) and $\nuebar\rightarrow\numubar$ (right), in the normal mass ordering and $\deltacp=-1.601$.}
    \label{fig:oscillograms_abs}
\end{figure}

\paragraph{Atmospheric neutrinos, CP violation and the mass ordering}
The neutrino oscillation spectra as a function of zenith angle $\theta_{Z}$---defined as the angle between directly overhead and the incoming neutrino direction---and neutrino energy, commonly referred to as an ``oscillogram'', is shown in \autoref{fig:oscillograms_abs} for electron (anti-)neutrinos oscillating into muon (anti-)neutrinos.
Although atmospheric experiments do not have a near detector, the $\cos\theta_z>0.2$ region (the ``down-going'' region) is invariant to oscillations over a wide range of energies. 
The region is therefore used to constrain systematic uncertainties, for instance related to the atmospheric neutrino fluxes, detector and reconstruction, and neutrino interactions, effectively acting as a ``near detector''\footnote{This is why experiments occasionally report an ``up-down asymmetry'': taking the ratio cancels many of the systematic uncertainties, and compares the region which sees oscillations to the region which doesn't.}.

\begin{figure}[hbtp]
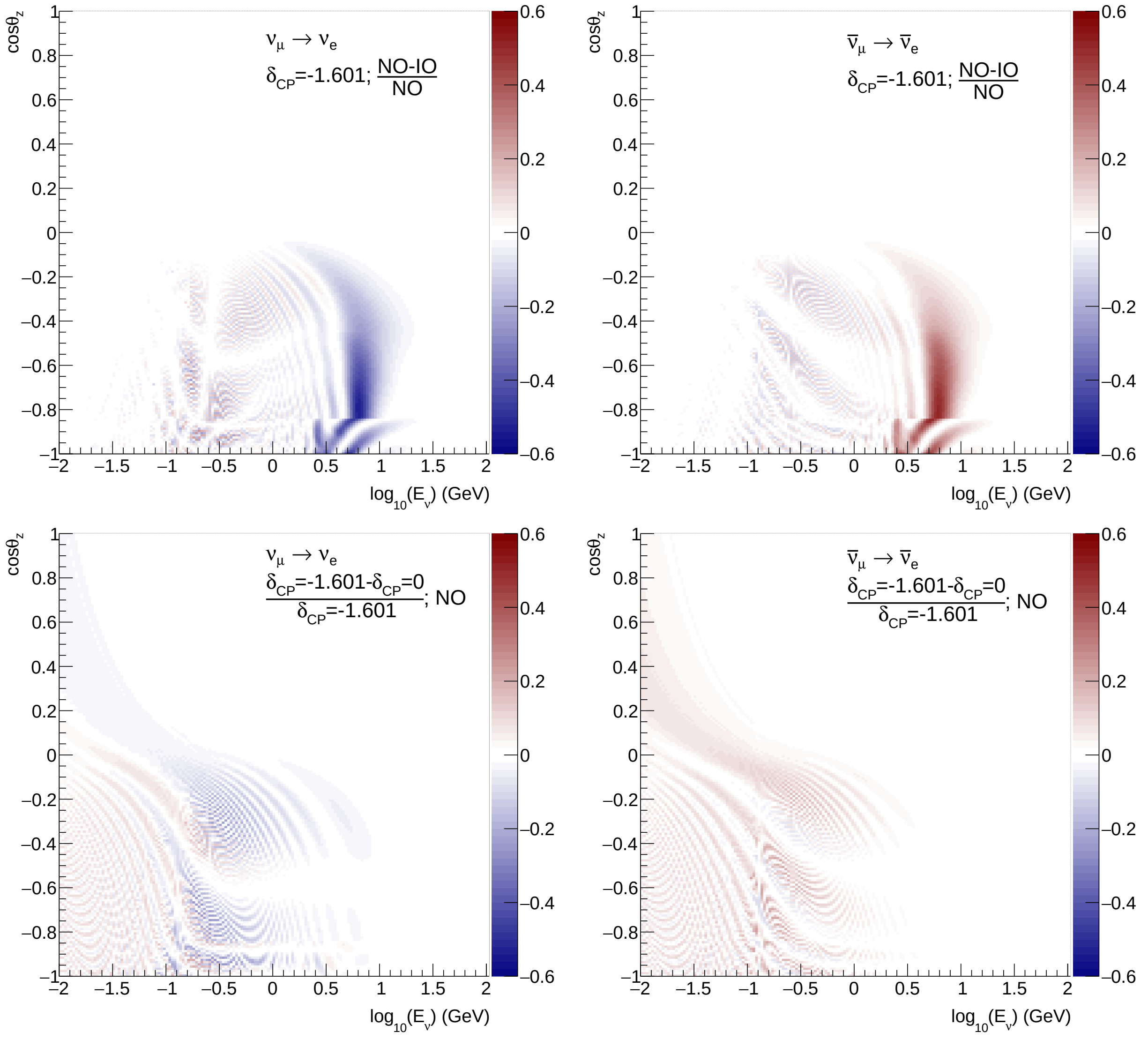

    \centering
    \begin{subfigure}[b]{0.49\textwidth}
        \includegraphics[width=\textwidth, trim={0mm 0mm 0mm 14mm}, clip, page=41]{Figures/osc_effects/Oscillograms_avg_dens_CPU.root_Oscillograms_avgdens_dcp-1601_IH.root.pdf}
    \end{subfigure}
    \begin{subfigure}[b]{0.49\textwidth}
        \includegraphics[width=\textwidth, trim={0mm 0mm 0mm 14mm}, clip, page=17]{Figures/osc_effects/Oscillograms_avg_dens_CPU.root_Oscillograms_avgdens_dcp-1601_IH.root.pdf}
    \end{subfigure}

    \begin{subfigure}[b]{0.49\textwidth}
        \includegraphics[width=\textwidth, trim={0mm 0mm 0mm 14mm}, clip, page=41]{Figures/osc_effects/Oscillograms_avg_dens_CPU.root_Oscillograms_avgdens_dcp0.root.pdf}
    \end{subfigure}
    \begin{subfigure}[b]{0.49\textwidth}
        \includegraphics[width=\textwidth, trim={0mm 0mm 0mm 14mm}, clip, page=17]{Figures/osc_effects/Oscillograms_avg_dens_CPU.root_Oscillograms_avgdens_dcp0.root.pdf}
    \end{subfigure}
    \caption{Relative differences in probabilities for $\nu_\mu\rightarrow\nu_e$ (left) and $\bar{\nu}_\mu\rightarrow\bar{\nu}_e$ (right) oscillations for atmospheric neutrinos in zenith angle and neutrino energy, for varying the mass ordering (top) and $\deltacp$ between -1.601 and 0 radians (bottom).}
    \label{fig:oscillograms_diff}
\end{figure}

\autoref{fig:oscillograms_abs} shows an important prominent feature between 3--10 GeV for upwards-going $\nue\rightarrow\numu$ oscillations, which is absent for $\nuebar\rightarrow\numubar$ oscillations in the normal ordering.
This feature is from the aforementioned ``matter effect'', in which the electron (anti-)neutrinos interact with electrons in the earth as they propagate through it. This produces a strong resonant effect for electron neutrinos in the normal ordering not present for electron anti-neutrinos, with the opposite applying in the inverted ordering.
To better illustrate the impact of the mass ordering and \deltacp in the $\numuany\rightarrow\nueany$ oscillation channels, \autoref{fig:oscillograms_diff} shows the relative differences in oscillation probabilities for different oscillation parameter scenarios.

The feature at 3--10 GeV is the primary mean by which atmospheric neutrino experiments constrain the mass ordering: if the normal ordering is realized, there should be a deficit of upwards-going \nue between 3--10 GeV which is not present for \nuebar; if the inverted ordering is realized, it is instead true for \nuebar.
Similarly, the muon and tau (anti-)neutrino rates around 3--10 GeV should drastically increase in the normal(inverted) ordering due to the ``disappearing'' electron (anti-)neutrinos.
Measuring the mass ordering with atmospheric neutrinos is therefore dependent on a) measuring the rate of neutrinos with energy between 3--10 GeV, b) establishing the different neutrino flavor fluxes for neutrinos and anti-neutrinos, and c) reliably inferring the sign and flavor of the interacting neutrinos.
The wide neutrino energy spectrum means interactions from atmospheric neutrinos involve a complex mixture of neutrino interaction channels, requiring a similarly complex set of uncertainties. 
The details and impact of this has not been studied in detail for the high statistics telescopes, and may limit the analyses' constraining power.
Studies of KM3NeT's model in their first analysis of data found a low p-value~\cite{KM3NeT:2024ecf}, $p=1.2\%$, suggesting that the model can not describe the data---but it is not clear if this is due to interaction modelling or other missing physics. 
Experiments with more developed uncertainty models have found better p-values: IceCube cite $p=26\%$~\cite{IceCubeCollaboration:2023wtb}, SK observed $p=88\%$~\cite{Super-Kamiokande:2023ahc}, and the joint T2K+SK analysis found $p=58\%$~\cite{T2K:2024wfn}.

\paragraph{Challenges in neutrino energy and direction reconstruction}
Due to the complex interaction channel mixture and the poorly known neutrino direction, the neutrino energy can not be reliably estimated by using the kinematics of CC$0\pi$ events, as done on T2K.
Instead, atmospheric neutrino experiments generally reconstruct the ``visible energy'' in the detector, which is similar to the calorimetric energy, and has relatively poor resolution and maps less well to the neutrino's energy.
Importantly, it inherently depends on the interaction model from estimating the number of particles below threshold, the energy losses from nuclear effects, and the energy carried away by neutral particles\footnote{Some analyses only measure the outgoing lepton's energy, and use the model-dependent connection between the interacting neutrino and outgoing lepton to make a connection back to the neutrino energy; this generally performs worse than the visible energy, especially for neutrino energies below a few GeV.}.
The resolution of the zenith angle is limited by the resolution of the neutrino's direction, which again is reconstructed using the outgoing interaction products, and therefore suffers from similar issues as the neutrino energy reconstruction.

The direction estimators are generally expected to be better in liquid-argon detectors than in Water Cherenkov detectors due to the ability to reconstruct low-momentum particles. 
This simple picture is complicated by the presence of nuclear effects when energies are below a few GeV, and that the amount of energy that the lepton carries away strongly depends on the neutrino energy: at low energy the hadronic side can carry away significant fractions of energy, meaning the lepton direction is only weakly correlated with the neutrino. At higher energies, the lepton direction is strongly correlated with the neutrino direction.

DUNE~\cite{DUNE:2026yly}, SK~\cite{Suzuki:2019jby}, IceCube~\cite{IceCubeCollaborationP:2025rpl}, KM3NeT/ORCA~\cite{KM3NeT:2021ozk} and JUNO~\cite{Yang:2023rbg} have resolutions of about $20-35^\circ$ for interactions around 1 GeV, $5-20^\circ$ at 10 GeV, and $2-10^\circ$ at 100 GeV, which strongly depends on detector design.
Hence the fast oscillations spanning widths of $\mathcal{O}(0.1~\textrm{GeV})$ in energy and $\cos\theta_{Z}\sim0.1$ in zenith angle are generally not resolved by the experiments, and the average effect is instead observed. 
Importantly, the corrections necessary to reconstruct the neutrino direction and energy from the observed outgoing products depend on many aspects of interaction modelling, and will inevitably benefit from related developments.

\paragraph{Unique challenges in atmospheric neutrinos}
Contrary to long-baseline experiments, the incoming neutrino species are not precisely known; there are significant contributions from muon and electron (anti-)neutrinos.
The effects of the CP violating phase and the mass ordering is markedly different for atmospheric neutrinos and anti-neutrinos, and the sensitivity to these effects therefore depends on the ability to separate neutrino and anti-neutrino interactions in the detector.
No running or planned atmospheric neutrino observatory is magnetized, so inferring the sign of the outgoing charged lepton is done indirectly.
Separating neutrinos and anti-neutrinos is in principle partly possible through kinematics, since for a given neutrino energy, anti-neutrinos a more likely to produce a more energetic outgoing lepton and less energy in the hadronic system. 
This can however only separate events statistically, i.e. a given observed event has a probability of being a neutrino or an anti-neutrino, rather than being able to definitively state that an event was from a neutrino or anti-neutrino interaction.
The specifics of this method depends profoundly on the neutrino interaction model which, due to the energy range, is a complex mixture of CCQE, resonance production, soft inelastic scattering, and deep inelastic scattering, and differs for neutrinos and anti-neutrinos, shown previously in \autoref{fig:neu_flux}.
Furthermore, correctly identifying the primary outgoing lepton can be challenging at these energies for low-granularity detectors, or Cherenkov detectors.
Whereas the matter effect causes a significant change to the (anti-)neutrino oscillograms dependent on the mass ordering, the realized sensitivity once the detection capabilities are accounted for is considerably weaker~\cite{IceCube:2019dyb,Super-Kamiokande:2023ahc}.

The impact of \deltacp on the oscillograms is demonstrated in the bottom half of \autoref{fig:oscillograms_diff}. 
\deltacp has a sizable effect in the $E_\nu\sim0.1-3~\textrm{GeV}$ region, with roughly opposite effects for muon (anti-)neutrinos oscillating into electron (anti-)neutrinos. It also impacts the \numu and \numubar disappearance spectrum in the same region.
Again, the frequencies of the oscillations are significantly faster than detector resolutions in both neutrino energy and zenith angle, so the average effect is measured rather than the intricate fast oscillations.
In this energy region, the impact of \deltacp is a decrease(increase) in the \nue(\nuebar) appearance, and an increase(decrease) in the \numu(\numubar) disappearance.
Hence, similar to the mass ordering sensitivity, measurements of \deltacp from atmospheric neutrino observatories also depend on the flavor identification, energy resolution, and separating neutrinos from anti-neutrinos, albeit at a lower energy than the mass ordering sensitivity.

Finally, the atmospheric experiments often detect neutrinos with energies above tau lepton production threshold. 
Atmospheric tau neutrinos are not part of the atmospheric neutrino flux below $10^{5}~\textrm{GeV}$~\cite{Honda:2015fha}, but their appearance is measurable, meaning more oscillation channels can be probed compared to accelerator-based experiments.
As yet, tau-enhanced samples have not been directly used in atmospheric oscillation analyses; instead, tau neutrino interactions enter as backgrounds to the main muon and electron (anti-)neutrino samples. This is because the charged-current tau neutrino interactions produce hadronic showers, in which one of the many outgoing particles in the shower may be mistaken as the primary lepton in a electron or muon (anti-)neutrino interaction.
As a result, the tau neutrino interaction cross section is an important uncertainty for the mass ordering constraint, as are details of the single pion, SIS and DIS cross sections, and the ratio of NC to CC interactions~\cite{Super-Kamiokande:2023ahc,IceCubeCollaboration:2024ssx, IceCubeCollaborationP:2025rpl}.
Techniques to isolate tau-enhanced samples have been used by SK~\cite{Super-Kamiokande:2017edb}, IceCube~\cite{IceCube:2019dqi, IceCubeCollaborationP:2025rpl}, and JUNO~\cite{Meighen-Berger:2023xpr}, and studies suggest atmospheric tau neutrinos may be reconstructed in the DUNE far detector~\cite{PhysRevD.102.053010, DeGouvea:2019kea}. 
These developments suggest that the tau-neutrino cross-section uncertainty may be experimentally constrained below the 10\% level in the next decade.
A recent summary on tau neutrinos can be found in Ref.~\cite{MammenAbraham:2022xoc}.

\subsection{Impact of neutrino interactions on long-baseline oscillation measurements}
Past, present and future long-baseline neutrino experiments all rely on a neutrino interaction model to interpret their data to varying extents.
As discussed above, the uncertainties inherent to these models directly impact the uncertainties on the inferred oscillation parameters.
\autoref{tab:uncert} presents the uncertainties on the key oscillation parameters and channels from the two currently operating long-baseline experiments, T2K and NOvA\footnote{The numbers for T2K and NOvA are not necessarily directly comparable because they are calculated at different values of the oscillation and systematic parameters. They also use a different base model, and thus a different set of systematic uncertainties. The numbers do, however, guide the relative importance of interaction modelling compared to the overall systematic and statistical uncertainties.}.

{\def\arraystretch{1.3}
\begin{table}[hbtp]
    \centering
    \begin{tabular}{l | c c c | c c c }
    \hline \hline
         Quantity & \multicolumn{3}{c|}{NOvA}   & \multicolumn{3}{c}{T2K}   \\
         \cline{2-7}
                  & Syst. & Int. model & Stat. & Syst. & Int. model & Stat. \\
         \hline
         \ssqthtwothree ($\times10^{-2}$)                                    & $_{-2.5}^{+0.8}$ & $_{-0.5}^{+0.3}$ & $_{-3.3}^{+2.2}$ & $\pm2.1$ & --- & $\pm4.7$ \\
         \dmsqtwothree $(\times 10^{-5}~\textrm{eV}^2)$     & $_{-2.8}^{+2.4}$ & $_{-1.1}^{+0.6}$ & $_{-5.5}^{+4.3}$ & $\pm2.1$ & --- & $\pm5.3$ \\
         \deltacp (rad.)                                    & $_{-0.19}^{+0.06}$   & $_{-0.07}^{+0.05}$ & $_{-0.87}^{+0.21}$     & $\pm0.29$ & --- & $\pm1.12$ \\

        $\nue, \nuebar$ events (\%)                                & 3.6, 3.7 & 3.0, 2.3 & 10.7, 17.4 & 4.7, 5.9, 14.3 & 3.2, 3.1, 4.2 & 10.3, 25.0, 26.7 \\
        $\numu, \numubar$ events (\%) & 6.7, 7.0 & 2.5, 2.1 & 6.6, 9.8 &3.0, 4.0 & 3.1, 3.0& 5.6, 8.5 \\
        \hline \hline
    \end{tabular}
    \caption{The uncertainty on key oscillation parameters in the two currently running long-baseline neutrino oscillation experiments, T2K and NOvA. The uncertainties are split into systematic (``Syst.'') and statistical (``Stat.'') sources, and NOvA additionally provides the uncertainty from different classes of systematics~\cite{miranda_elkins_2020_4253085}. The interaction model uncertainty for T2K can be larger than the overall systematic uncertainty, since the latter includes (anti-)correlations between flux and interaction parameters, whereas the former does not, which can decrease the overall impact on the number of events. The third number for T2K in the \nue,\nueany events category refers to the FHC 1Re1de sample, dominated by charged-current \nue single pion production interactions where the pion falls below Cherenkov threshold. }
    \label{tab:uncert}
\end{table}
}

In these analyses, NOvA and T2K have similar total uncertainty from systematics on \ssqthtwothree ($^{+0.8}_{-2.5}\times10^{-2}$ and $\pm2.1 \times10^{-2}$ respectively) and \dmsqtwothree ($^{+2.4}_{-2.8} \times10^{-5}~\textrm{eV}^2$ and $\pm2.1\times10^{-5}~\textrm{eV}^2$ respectively). 
Similarly, directly comparing these to the statistical uncertainty, it's slightly above half for NOvA ($^{+2.2}_{-3.3}\times10^{-2}$ for \ssqthtwothree and $^{+4.3}_{-5.5}\times10^{-5}~\textrm{eV}^2$ for \dmsqtwothree) and slightly below half for T2K ($\pm4.7\times10^{-2}$ for \ssqthtwothree, $\pm5.3\times10^{-5}~\textrm{eV}^2$ for \dmsqtwothree). 
Considering the uncertainty contributions to \deltacp in NOvA, the uncertainty from the interaction model (0.05 rad) is the dominant contributor to the total (0.06 rad) for the upper edge of the 68\% confidence interval, although statistics contribute a much larger uncertainty overall (0.21 rad). Hence, NOvA's \deltacp constraint is dominated by statistical uncertainty in the \nueany samples, but as more data is collected, neutrino interaction uncertainties, notably on \nueany/\numuany differences, are increasingly important. Qualitatively similar, this has also been observed by T2K when analysing the impact of alternative model choices~\cite{T2K:2023smv, T2K:2025yoy}.

The neutrino interaction model is a significant contributor for both experiments when predicting the number of events at the far detectors.
Uncertainties in neutrino interaction modelling contribute an uncertainty of 2.3--4.2\% depending on the selection and experiment for the electron (anti-)neutrino measurements. 
This is primarily due to the 2--3\% uncertainties on the differences between \nueany and \numuany, and the lack of high-statistics \nueany samples at the ND constraining such systematics, which applies for both T2K and NOvA.
Although the interaction model is consistently a larger contributor for T2K (3.1--4.2\%) compared to NOvA (2.3--3.0\%), it is a smaller contributor to the overall larger systematic uncertainty for T2K, whose statistical uncertainty is also larger.
For the muon (anti-)neutrino samples, the overall systematic uncertainty is smaller for T2K than NOvA, and is dominated by the neutrino interaction model for T2K.
For NOvA, the largest systematic uncertainties for muon (anti-)neutrinos are detector calibrations and neutron-related uncertainties, which also sees contributions from neutrino interaction modelling.
These are being studied with dedicated test-beam experiments, after which neutrino interaction systematics become important.
In NOvA's \numu sample, the total systematic uncertainty (6.7\%) is on a similar scale to the statistical uncertainty (6.6\%), whereas for T2K the statistical uncertainty is approximately double the systematic uncertainty. 

Overall, this suggests that the final analyses of T2K and NOvA will be limited by systematic, rather than statistical, uncertainties, of which neutrino interactions play an significant part---especially important for their final joint analysis.
This is particularly true for the ``disappearance'' parameters \ssqthtwothree and \dmsqtwothree.
Additionally, neutrino interaction modelling uncertainties make a formidable contribution to the overall systematic uncertainty for the measurements of the CP violating phase, although they are generally more limited by statistical uncertainty overall.
This presents a significant challenge to both DUNE and Hyper-Kamiokande, which will collect orders of magnitude more data compared to NOvA and T2K, meaning neutrino interaction uncertainties have to be better understood to fully realise the physics potential of these experiments.

\subsection{Mitigation of the impact}
As discussed, the uncertainties from neutrino interactions is becoming a dominant source of uncertainty in oscillation experiments.
Current experiments try to mitigate for these by making dedicated measurements that can reduce or better understand such uncertainties, and quantify the impacts of mismodelling using techniques such as fake data studies.

The concept of ``fake-data studies'' (sometimes referred to as ``simulated-data studies'') introduces setting an alternative model---potentially not covered by the uncertainties in the oscillation analysis---as the fake ``data''. This studies the biases on the extracted oscillation parameters were an alternative model (not included in the experiment's simulation) to perfectly describe the data.
The fake data can be made by applying weights from the alternative model to the base simulation, or by generating new simulated events. This fake data is either directly fitted to, as done by T2K, or used to build the near-to-far extrapolation templates, as done by NOvA.
The fake-data procedure is also repeated at the far detector, where predictions of the event spectra are made using the alternative model.
The result of the near-detector analysis---which fit the model to the fake data---is used as the base simulation, with new values for the systematics parameters and their covariances, which may wrongly predict the extrapolation to the far detector. 
Neutrino oscillations are applied to the fake-data prediction at the far detector.
The bias is firstly evaluated by overlaying the extrapolated and the fake-data prediction at the far detector: this studies the bias in the extrapolation on the event counts.
Secondly, the potentially biased extrapolated model, which used the results from the near-detector fake-data analysis, is used to fit the oscillation parameters to the fake-data spectrum generated directly by the alternative model. 
A set of biased oscillation parameters are extracted, and the oscillation parameters' central values, uncertainties, and $\Delta \chi^2$ surfaces, are used to evaluate the extent of the bias.
These metrics are then used to make statements on the robustness of a particular result, for instance how much the confidence interval on $\Delta m^{2}_{32}$ should be inflated, or if statements on excluding CP conservation (e.g. $\sin\delta_\textrm{CP}=0$) at a certain confidence level is robust to specific modelling choices. 
Generally speaking, this is a rather conservative approach since it assumes that each tested model is equally plausible. 
On the other hand, the tests are not complete: they fundamentally depend on the choice of which models that are tested, and the evaluation of what constitutes a ``reasonable'' model. This often requires confronting the alternative model with existing measurements. If the appropriate measurements don't exist, new measurements have to be made, which may take years of development. 
\begin{figure}[hbtp]
    \centering
    \includegraphics[page=1,width=0.49\textwidth,clip, trim={6mm 0mm 5mm 5mm}]{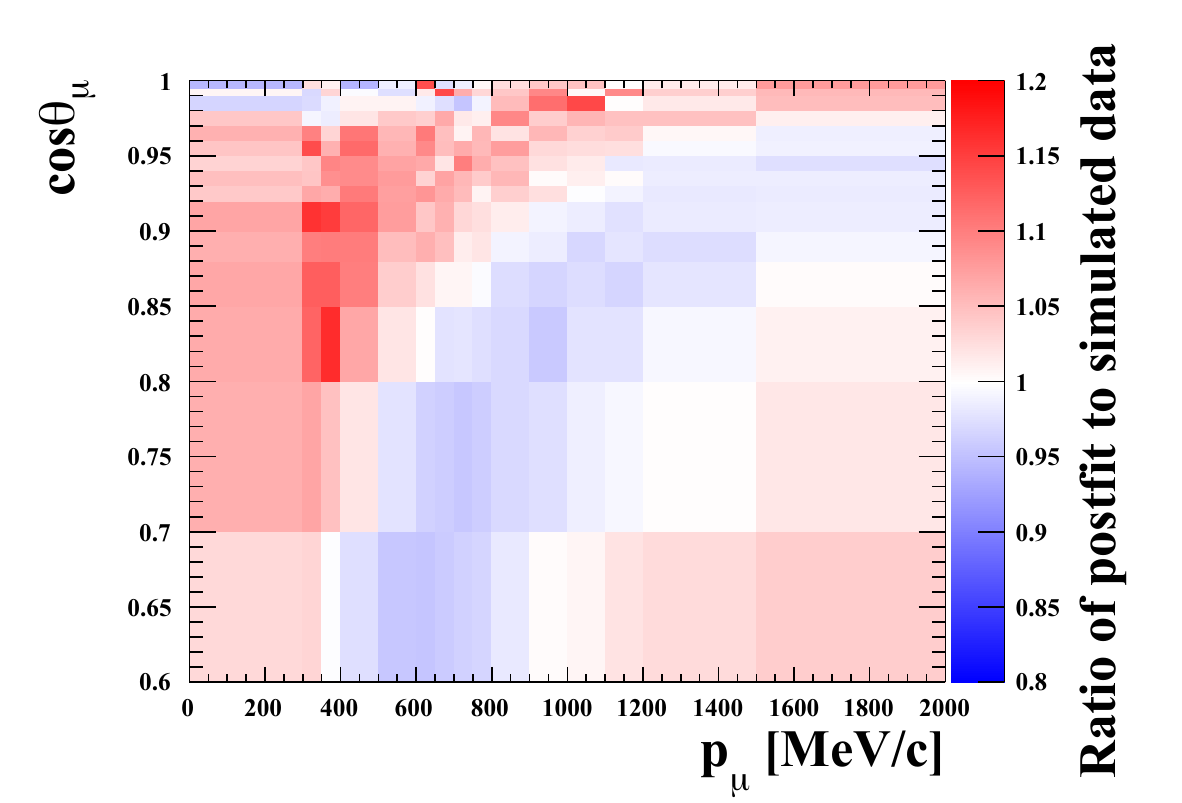}
    \includegraphics[page=1,width=0.49\textwidth,clip, trim={10mm 3mm 15mm 10mm}]{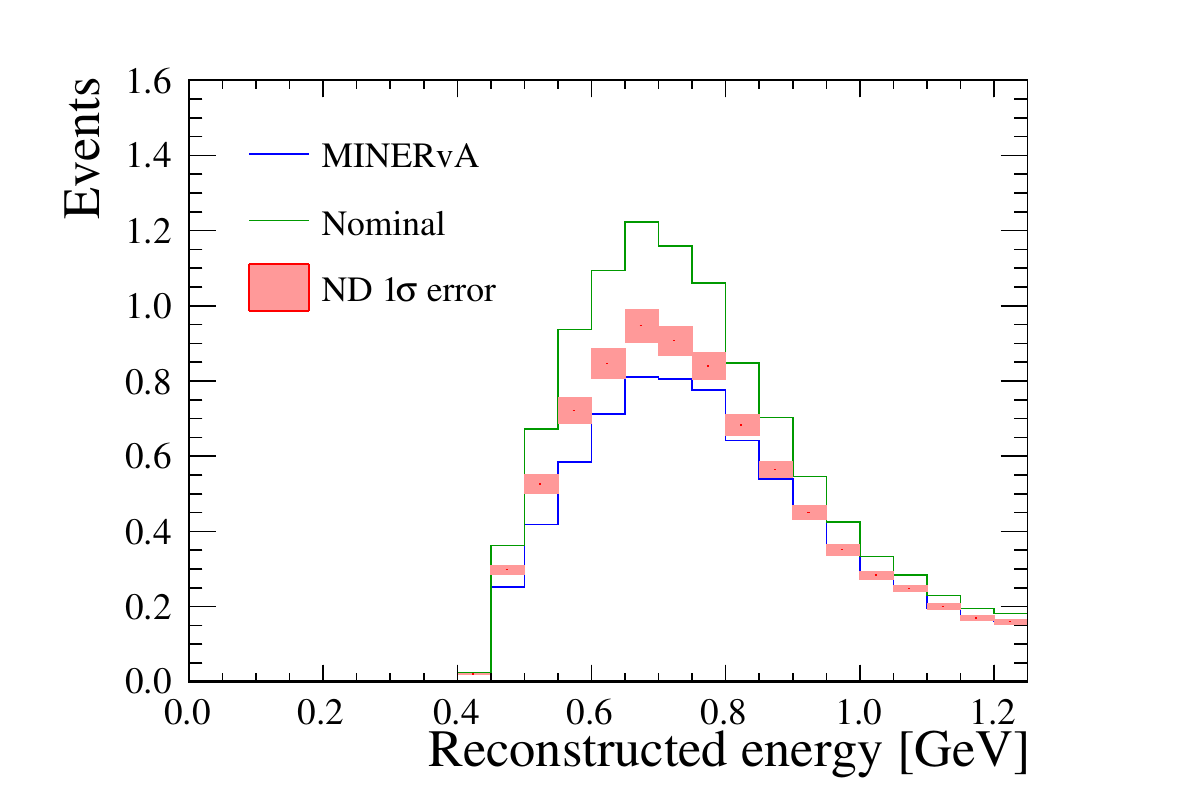}
    \caption{The impact of the low-$Q^2$ single-pion production fake-data study on T2K, using fake-data generated from MINERvA~\cite{MINERvA:2019kfr}, shown at the near (left) and far (right) detectors. Figure is reproduced from Ref.~\cite{T2K:2023smv}}
    \label{fig:fakedata}
\end{figure}

An example demonstrating this from T2K's analyses is shown in \autoref{fig:fakedata}.
This fake data set includes a low-$Q^2$ suppression of single-pion production events found from analyses of MINERvA data~\cite{MINERvA:2019kfr}.
The prediction after fitting to the near detector fake-data is divided by the fake-data itself to show the fractional bias.
This example significantly biases the single-pion dominated samples at T2K, but---due to low statistics of those samples compared to the large CCQE-dominated samples---it currently has little impact on the extracted oscillation parameters.
In future experiments with higher statistics, effects like this will play a large role if the uncertainties for single pion production aren't reduced.
The example also demonstrates the conservative and subjective nature of fake-data studies: the low-$Q^2$ suppression was needed to make the GENIE v2.12.10 prediction describe bubble chamber data and MINERvA data at the same time, but subsequent studies using the generator used by T2K (NEUT) found that such suppression was not needed~\cite{Avanzini:2021qlx}.

Whereas fake-data studies do little to directly avoid model dependence, the technique does guide the extent of the model dependence in a result, and can aid the community in which directions model development is needed. 
This is particularly important in the current experimental climate, where neutrino interaction theory is not fully understood, models are being developed at a fast pace, and their impact on oscillation analyses needs to be urgently assessed as it may impact conclusions on, for example, CP violation in neutrinos and the neutrino mass ordering.

The future DUNE and HK experiments will partly address some aspects of the model-dependence by utilizing the PRISM~\cite{nuPRISM:2014mzw,DUNE:2021tad} technique for the muon (anti-)neutrinos. 
PRISM provides measurements of neutrinos with correlated spectra, using near-detector data from different off-axis angles.
In principle, this allows a linear combination of near detector measurements to provide an effective flux that matches the oscillated far detector flux, which addresses some of the concerns raised in this section.
This approach significantly lessens dependence on the interaction model in the near-to-far extrapolation, but its impact has not been fully demonstrated. Many of the issues discussed in this section such as mismatched in detector acceptances, still apply to a PRISM analysis.

\subsection{Summary of current challenges}
As detailed in the previous sections, the challenges facing neutrino interaction modeling for long-baseline oscillation measurements are experiment-dependent.
The neutrino flux, interaction target, detector technology, reconstruction methods, and selections all play important roles in determining which aspects of the modelling impact the measurements, and to what extent.
But generally speaking, issues surround the model-dependence of reconstructing the neutrino energy, the lepton flavour, and the lepton charge, in the GeV neutrino energy range.
Model-dependency also arises from an imperfect extrapolation from the near detector to the far detector: for instance the particle acceptance is different due to geometry or detector design, one flavour of neutrino may be used to predict another flavour via a model (including anti-neutrinos), and oscillations mean that the neutrino energy at the near and far detectors are different, necessitating a model-dependent extrapolation.

The T2K and HK experiments face similar challenges due to analogous detector designs and analysis methods. 
First and foremost is understanding the relative contributions of CCQE, 2p2h, and CC$1\pi$ via Delta excitations to the CC0$\pi$ topology, which are the dominant contributions in the energy region below 1 GeV. 
Understanding the small but important NC$1\pi^\pm$ and NC$1\pi^0$ backgrounds are also critically important for precision measurements of muon neutrino disappearance and electron neutrino appearance, respectively.
Separately, the introduction of smaller statistics samples---where the signal is instead dominated by pion production---has exposed that low-momentum pions in both electron and muon samples are poorly described. 
For T2K, the impact of such samples are limited by their statistical power, but for HK the samples are roughly equivalent in size to T2K's final data set and will see much larger impact from the uncertainties related to pion production.
Since both the mass ordering and CP violating phase are estimated by measuring appearing electron (anti-)neutrinos, the modelling of their cross sections and the difference relative the muon (anti-)neutrinos (which are measured at the near detector) are centrally important. This is especially true in the low-energy region where the differences can be large due to threshold effects.
Due to the near and far detector designs, T2K and HK need to understand differences between interactions on $^{12}\textrm{C}$ (the predominant near-detector target) and $^{16}\textrm{O}$ (the far-detector target), and how to reliably extrapolate between them.
Also related to the detector design is the different acceptance at the near detector---which has a lower efficiency to reconstruct high-angle and backward events---and the far detector---which has generally good reconstruction regardless of angle, but can not reconstruct protons due to high thresholds.

T2K and HK's near-detector upgrade programmes intend to address many of these issues. For instance, the recently upgraded near detector ND280~\cite{T2K:2019bbb} is pursuing measurements detailing nuclear effects with a much improved high-angle and backward acceptance. In addition, the IWCD~\cite{Zhu:2023nsv,Scott:2016kdg} will measure a large quantity of interactions of electron (anti-)neutrinos and interactions on water, with similar acceptance to the HK far detector and at various off-axis angles to widen the neutrino energy spectra and precisely study the evolution in neutrino energy~\cite{nuPRISM:2014mzw}. 
But importantly, the IWCD will not be able to measure detailed nuclear effects in the same way as ND280 due to the high proton tracking threshold of Water Cherenkov detectors. To a lesser extent, this also applies to the ability to track low-momentum muons and pions.

The NOvA experiment is at a higher energy with a more inclusive selection compared to T2K, uses a calorimetric energy reconstruction, and functionally identical near and far detectors. 
This renders it more sensitive to the modeling of energy depositions and correct particle identification to reconstruct the neutrino energy, at an energy which sees sizable contributions from CCQE, 2p2h, resonant single pion production, and soft inelastic scattering. Whereas CCQE, 2p2h, and resonant interactions in the Delta resonance region has seen notable development in the last decade, the modeling of higher resonances and the transition to soft and deep inelastic scattering needs further study.
Energy depositions from neutrons, and more generally ``missing energy'', is a leading systematic in NOvA.
The different interaction mechanisms produce notably different particle multiplicities and kinematics, impacting energy depositions and the neutrino energy estimator.
Missing energy may be related to the excess that NOvA sees at low available energy in the CC-inclusive selection, which is a region sensitive to a wide range of nuclear effects.
The ND and FD also have different reconstruction efficiencies to events with high transverse momentum (i.e. large neutrino-muon angle). This stems from the different absolute sizes of the two detectors, rather than a different detector material or design, which is additionally present for T2K. Similarly, this introduces dependence on the interaction model as the analysis uses events in the ND, at a predominantly forward angle, to predict the higher-angle events in the FD.

DUNE shares many challenges with NOvA, where the reference analysis uses a calorimetric energy reconstruction and a CC-inclusive selection.
To complicate this, DUNE has a much wider neutrino energy spectrum than NOvA, making it more sensitive to the resonant-SIS-DIS transition region.
A capable near-detector design and the PRISM technique helps relieve some of the issues from having a wide neutrino energy spectrum, possibly allowing dedicated near-detector measurements of specific topologies as a function of energy, which can help decouple effects. However, the particle acceptance of the first phase of the near detector design has similar issues to NOvA and T2K: its near detector is significantly smaller than the far detector, and thus does not measure particles at higher angles, causing a mismatch in acceptance.
Neutrino interactions in DUNE will be on dense liquid $^{40}\text{Ar}$, which has significantly stronger nuclear effects compared to the lighter targets previously used by many long-baseline neutrino oscillation experiments. 
Similarly, modeling particle multiplicity, notably neutrons but also the amount of hadron absorption and low-momentum hadrons that may be missed, in such a large nucleus remains a significant challenge, and directly impacts the neutrino energy estimator.
The SBN programme at Fermilab and ProtoDUNE at CERN are studying interactions on $^{40}\text{Ar}$ before DUNE comes online to address many of these challenges, although measurements by SBN are in a different neutrino energy range, which means it will not be able to measure the same resonant-SIS-DIS transition region as in DUNE.

Finally, measurements of atmospheric neutrinos may provide critical information to determine the neutrino mass ordering and CP violation. They too have notable interaction modeling challenges ahead---many of them shared with the long-baseline experiments.
The impact of \deltacp is concentrated at low neutrino energies below a GeV. The challenge is focused on how accurately such events can be reconstructed, and measurements are particularly sensitive to the modeling of electron (anti-)neutrino interactions. 
The effect of the neutrino mass ordering is instead visible at energies between 3--10 GeV, rendering it sensitive to the resonant-SIS-DIS transition region---similar to DUNE---and the separation of different lepton flavors.
A unique challenge for atmospheric experiments is that measurements of \deltacp and the mass ordering heavily benefits from in-situ methods to differentiate between neutrinos and anti-neutrinos, which profoundly depends on how well interaction physics in the region is understood.
Another unique challenge is the large uncertainty on the tau neutrino interaction cross section when concerning the mass ordering determination in atmospheric experiments, as tau neutrinos constitute a challenging background at high energies.

\newpage
\section{Nuclear Models}
\label{nuclear}
Neutrino interaction cross sections present one of the main sources of uncertainty in the oscillation analyses for long-baseline experiments, that use nuclei to detect neutrinos in their near and far detectors.  Progress towards a better understanding of oscillation phenomena hence  necessitates detailed studies of the nuclear effects impacting the interaction of the lepton with one or more nucleons embedded in the nuclear medium.
The broad spectrum of neutrino energies impinging on the target nuclei and the various interaction channels involved, makes this a challenging and exciting program.
The heavier target nuclei used in new detectors  and the improved detection capabilities that enable the experimental study of more exclusive process, require  further efforts in the  modeling of the nucleus for theory to keep up with experimental progress.

Over the last few decades, several nuclear models have been adopted as the cornerstone of reliable descriptions of neutrino-nucleus interactions~\cite{Ruso:2022qes}. Whereas some were custom-made, aiming to provide predictions for cross sections needed in long-baseline oscillation experiments, others were inherited from photon and electron scattering studies. The choice of a model ultimately depends on the interplay between the desired level of sophistication, numerical efficiency, and accessibility for the research community. The latter aspect is particularly crucial for experimental analyses, as models must not only be implemented in Monte Carlo event generators but also equipped with physically-justified degrees of freedom, allowing us to avoid speculations and estimate systematic uncertainties meaningfully~\cite{Ruso:2022qes}.

Describing the scattering of leptons on nuclear targets in the intermediate energy region presents a challenge due to the differing nature of the particles involved. While neutrinos or electrons can be considered elementary spin-$\sfrac{1}{2}$ Dirac particles, nuclei are complex many-body systems that require phenomenological modelling. 
The nuclear models presented in this section are derived in different formalisms, employ different assumptions and introduce various approximations. 
Already the description of the groundstate nuclear wavefunction is a challenging task which has been approached in various ways: from the most simplified model of the Fermi gas (Sec.~\ref{section:LDA}), through mean-field models which employ various effective nuclear potentials (Sec.~\ref{sec:mf}), to more sophisticated ab initio models which retain information on the nuclear entanglement (Sec.~\ref{sec:abinitio}). 
The description of electroweak nuclear responses, introduces yet an additional layer of complication, since the final nuclear state is an unbound system with possibly non-nucleonic degrees of freedom.
The level of nuclear correlations that needs to be considered, depends, to some extent, on the energy regime under study~\cite{Walecka2023}. 
At low energy transfers (at the order of 100 MeV and below), interaction cross sections become increasingly sensitive to nuclear structure details. This is where nuclear correlations (sometimes referred to as the long-range correlations) play an essential role in capturing the complicated nuclear many-body dynamics. While in this regime the ab initio methods bring valuable results, some collective phenomena are also addressed in a more phenomenological way by the RPA description of the Valencia and Lyon groups (see Sec.~\ref{section:LDA}).
For intermediate energy-momenta, the quasi-elastic mechanism starts playing a significant role. In this region, nuclear responses are dominated by genuine one-body processes, allowing for a good mean-field description. 
This inspired the spectral function formalism which--assuming the impulse approximation--relates the cross-section to the spectral functions (Sec.~\ref{section:SFA}). These can be obtained by different theoretical methods or can be extracted from experimental data.
Interestingly, in this particular reaction channel the scaling phenomenon was observed which led to establishing a phenomenological lepton-nucleus scattering model, described in Sec.~\ref{sec:scaling}. To understand the details of this regime, however, the interplay of a single- and multiple-nucleon knockout processes has to be addressed. While this phenomenon has been described within various models, it is worth noting that the short-time-approximation has been proposed from the ab initio community specifically to retain the correlations of two-nucleon final states (see Sec.~\ref{section:STA}).
At higher energies, the nuclear models are challenged further to account for the multi-nucleon knockout processes, as well as production of mesons, mostly pions. These reaction channels involving various hadrons in the final state, go beyond the scope of traditional nuclear modelling. It becomes increasingly difficult to consistently account for final state interactions of outgoing hadrons. Due to the growing phase-space of final particles the calculations become also increasingly expensive.
Some models derive their efficiency from an intrinsically inclusive description of the process, often integrating out the nucleon degrees of freedom and thus strongly reducing the computational cost. Therefore, with a growing need for describing exclusive final states in neutrino-nucleus scattering processes, a suitable model should strike an optimal balance between the level of detail and computational performance.

The following sections provide an overview of the status and future challenges for the main nuclear modelling techniques currently under active development and use in this field. 
We examine the specifics of each model, their strengths and limitations focusing on the relevance to neutrino physics. 
More details on specific applications to particular channels of neutrino-nucleus scattering processes will be provided in the dedicated sections of this paper.

\subsection{Ab initio techniques}
\label{sec:abinitio}
In the recent years both the advancements in high-performance computing and theoretical developments have allowed to establish a new family of theoretical methods which solve the nuclear many-body problem in a systematically improvable manner, allowing for precise nuclear structure calculations with a robust uncertainty quantification.
These ab initio (``from first principles'') techniques aim at describing the structure and dynamics of atomic nuclei in terms of the individual interactions among protons and neutrons, which are treated as ``fundamental'' degrees of freedom. The nucleus is modeled as a collection of $A$ non-relativistic nucleons whose dynamics is dictated by the Hamiltonian which constraints two- and three-nucleon interactions. 

Traditionally, phenomenological nucleon-nucleon interactions have been constructed by including the long-range one-pion exchange interaction, while different  schemes are implemented to account for intermediate and short-range effects. Highly-realistic interactions~\cite{Wiringa:1994wb,machleidt2001} of this kind, such as the Argonne $v_{18}$ (AV18) potential~\cite{Wiringa:1994wb}, involve a number of parameters that are determined by fitting experimental data. In the recent years, due to the seminal works of Weinberg~\cite{Weinberg:1978kz,Weinberg:1990rz,Weinberg:1991um}, we have witnessed the tremendous development and success of Hamiltonians based on chiral Effective Field Theory~\cite{epelbaum2009} ($\chi$EFTs) which allow for order-by-order calculation respecting the broken symmetry pattern of QCD.

Within a broad landscape of ab initio approaches, several theoretical methods were used to calculate electroweak nuclear responses in the regime of light and medium-mass nuclei. The main strength of these calculations is the treatment of both initial and final nuclear wavefunctions on the same footing. This is particularly advantageous at low and intermediate momentum transfers where the nuclear effects become important and the more approximated methods fail to capture the complexity of nuclear responses.
The general strategy in these calculations consists in accessing first the integral transform (typically Euclidean or Lorentzian) of the response, and then inverting it. The inversion procedures introduce an uncertainty which can be accounted for.

\begin{figure}[ht]
\centering
\includegraphics[width=\textwidth]{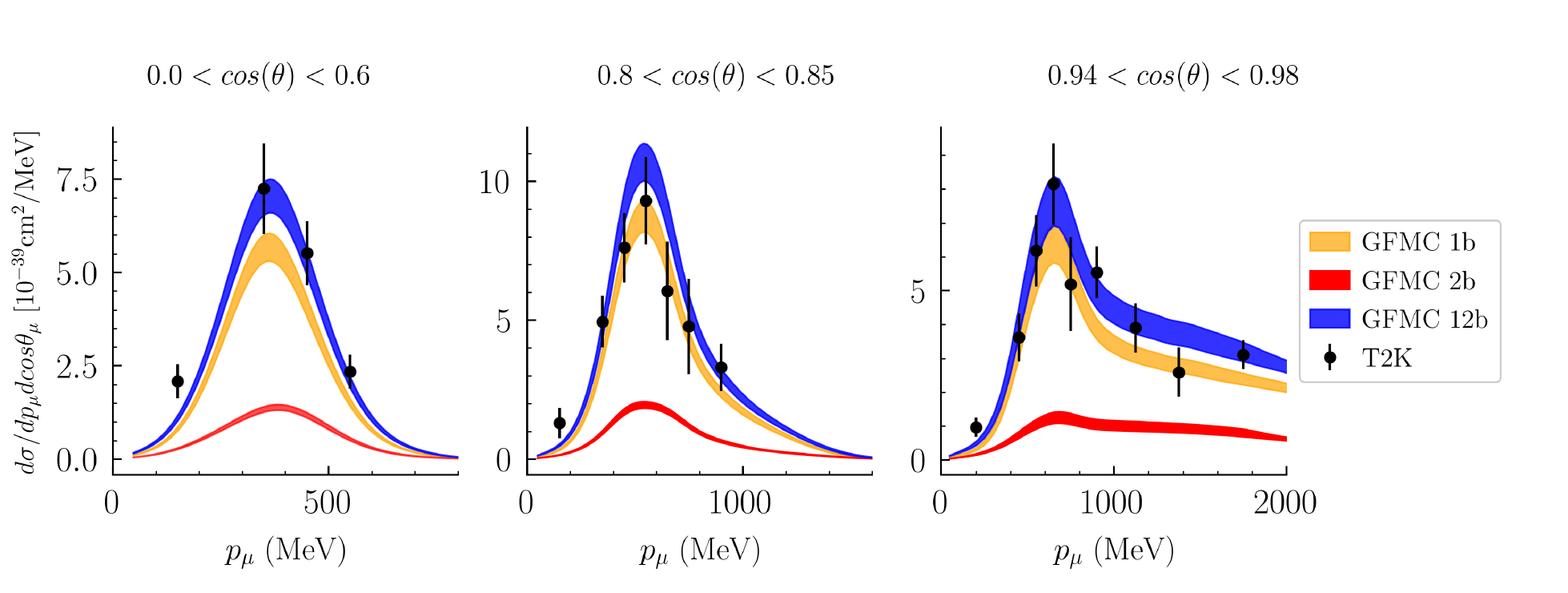} 
\caption{ Comparison with T2K data, adapted from Ref.~\cite{Simons:2022ltq}.
Breakdown into one- and two-body current contributions of the $\nu_{\mu}$ flux-averaged differential cross sections for T2K in three bins of $\cos\theta_{\mu}$.  The GFMC predictions refer to one-body (1b) contributions in orange, two-body (2b) contributions in red, and the total (12b) in blue. The two-body results include interference effects only.  The D2 Meyer et al. $z$ expansion results for $F_{A}$ are used in both cases~\cite{Meyer:2016oeg}.}
\label{fig:T2Kab}
\end{figure}

In the range of light nuclear systems, the many-body problem (whose complexity grows exponentially with the number of nucleons) can be solved in a virtually exact way. Among ab initio techniques applied in this regime, the Green's function Monte Carlo (GFMC) has been employed to perform calculations of inclusive electron- and  neutrino-scattering~\cite{Lovato:2016gkq}
on $^4$He and $^{12}$C, which turned out to be in excellent agreement with experiments in the quasielastic region. Various studies have been performed employing the GFMC framework. Recently, the sensitivity to different axial form factor parametrizations was investigated in Ref.~\cite{Simons:2022ltq}. Selected angular bins for T2K kinematics are presented in Fig.\ref{fig:T2Kab}. The uncertainty bands reflect the propagated errors in the axial nucleon form factor, obtained using the $z$-expansion, with coefficients fitted to deuterium bubble chamber data from Ref.\cite{Meyer:2016oeg}. In addition, Ref.\cite{Nikolakopoulos:2023zse} recently evaluated relativistic corrections to GFMC predictions for flux-averaged neutrino cross sections. The impact of these corrections on the T2K results displayed in Fig.~\ref{fig:T2K} is minor and generally remains within the uncertainty bands arising from the axial form factor.

\begin{figure}[ht]
\centering
\includegraphics[width=6cm]{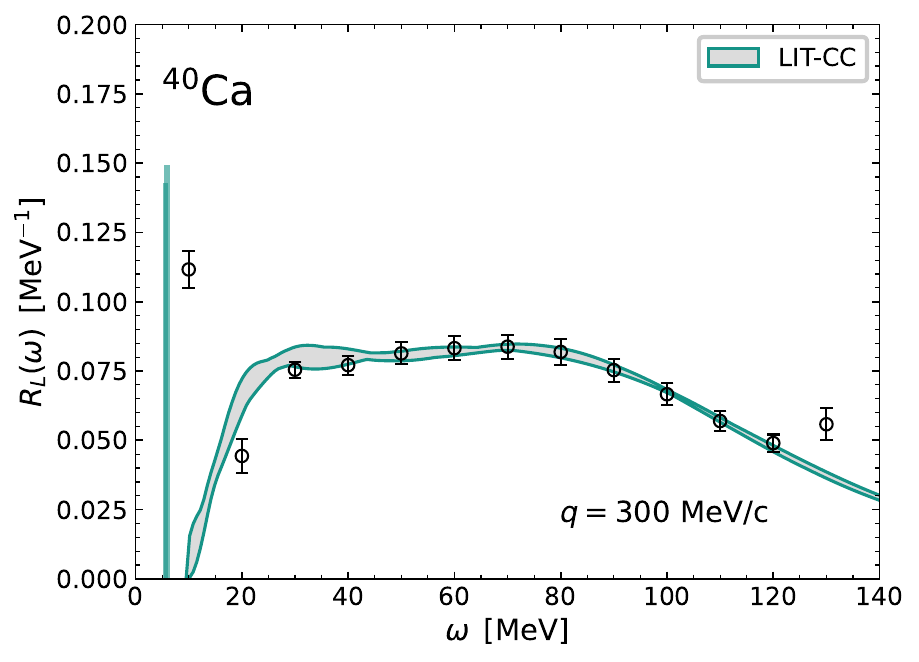}
\includegraphics[width=6cm]{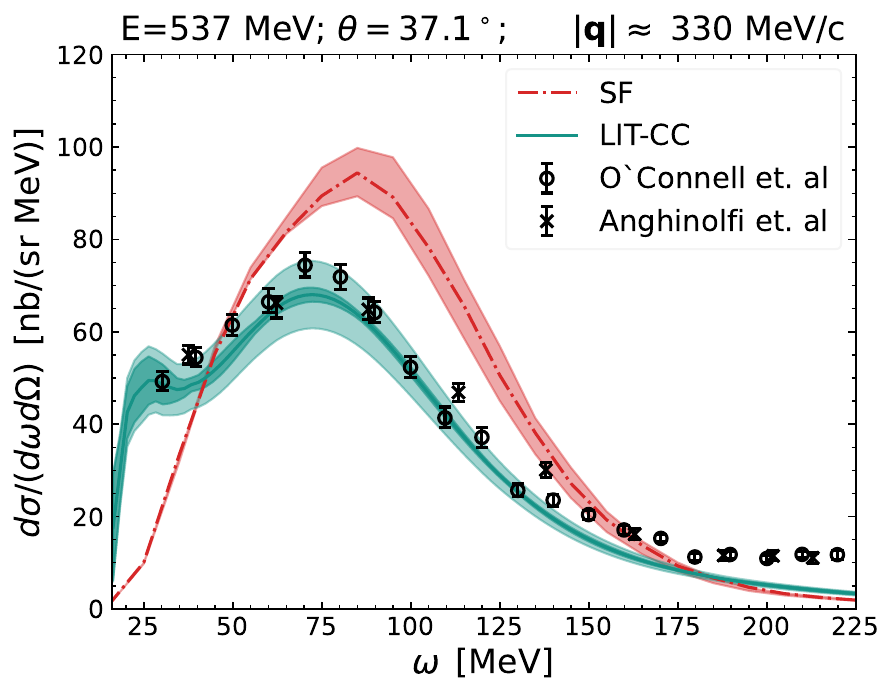}
\caption{Coupled cluster using Lorentz integral transform (LIT-CC) results for the longitudinal response on $^{40}$Ca (left panel) and the inclusive electron scattering on $^{16}$O (right panel). For the latter, a comparison with the spectral function (SF) is presented. }
\label{fig:CC}
\end{figure}

For heavier nuclei, the Coupled Cluster framework~\cite{Hagen:2013nca} has been used to deliver the first results for electron and neutrino scattering on $^{16}$O~\cite{Acharya:2024xah}, as well as electron scattering on $^{40}$Ca~\cite{Sobczyk:2021dwm,Sobczyk:2023sxh}, see Fig.~\ref{fig:CC}. These encouraging results, showing a remarkable agreement with the data, have proved the capability of ab initio methods to provide calculations for medium-mass nuclei of interest for long-baseline experiments. They also showed that nuclear Hamiltonians derived in $\chi$EFT are suitable to deliver electroweak responses for the momentum transfers of the order of $400$ MeV. 
The Coupled Cluster calculations allow to scrutinize various sources of theoretical uncertainty, most importantly stemming from the many-body expansion and the nuclear forces. The $\chi$EFT framework allows to estimate both the truncation error of the chiral expansion and the uncertainty propagated from the low energy constants which are fitted to experimental data.

Apart from the fully consistent calculations of nuclear responses, various ab initio methods were employed to deliver more approximated results within factorization schemes, such as spectral functions or short time approximation. These are described in the next two sections, ~\ref{section:SFA} and~\ref{section:STA}.

\textbf {Advantages and Limitations:} Ab initio methods offer a major advantage by enabling rigorous assessment of theoretical uncertainties. This is achieved through the use of different nuclear interactions, careful propagation of their associated uncertainties, and evaluation of the errors intrinsic to the many-body techniques. Moreover, these approaches allow for a fully consistent treatment of final-state interactions, aligned with the description of the initial state.

Their limitations include their non-relativistic formulation, restriction to inclusive observables, and the inability to incorporate explicit pion degrees of freedom in a straightforward manner.
They can nonetheless be employed to constrain more approximate methods in the quasi-elastic region and at low to moderate momentum transfers.

\subsection{Spectral function approaches}\label{section:SFA}
The spectral function (SF) formalism relies on the factorization of the nuclear final state~\cite{Rocco:2018mwt}, enabling to reach higher energies and momentum transfers, as it includes relativistic effects in both the kinematics and in the interaction vertex. In contrast with ab initio techniques, the SF approach can access exclusive channels and covers the relativistic regime. However, while based on a similar treatment of the initial target state, factorizing the final state involves additional approximations, which are only valid at large momentum transfer. 
The hole SF carries the probability of leaving the residual system with a specific recoil momentum ($\mathbf{k}$) and excitation energy ($E$) ans it is commonly formulated as
\begin{equation}
    S_h (E,\mathbf{k}) = \sum_n \left| \langle R_{A-1}(-\mathbf{k}) | a_\mathbf{k} | I_A \rangle \right|^2 \delta(E - E_n + E_I)
    = \frac{1}{\pi} \mathrm{Im} G_h (E,\mathbf{k}) ,
\label{eq:sf_def}
\end{equation}
where $n$ denotes all possible nucleons one can remove while making a transition between the initial ($I$) and final ($R$) hadronic states, and $G_h$ is the Green's function describing hole dynamics in the nuclear medium. This framework allows us to explore various nuclear models by comparing their spectral functions and the effects they include beyond this picture.

In various calculations, the spectral function of Eq.~\eqref{eq:sf_def} is rewritten as a sum of a mean field and a correlation term. The mean field piece contains the shell structure with nucleons occupying orbitals obeying the Pauli principle and contributes to the low $k$ and $E$ region. On the other hand, the correlation term comes from pairs and triplets of interacting nucleons with low center-of-mass momentum but relative momenta above the Fermi momentum. 
Many prescriptions of the spectral function for finite nuclei are available from a combination of fits to $(e,e'p)$ cross sections and theoretical calculations. The spectral function of Benhar et al. \cite{Benhar:1994hw} obtains the mean field piece combining shell model calculations with a quenching of the spectroscopic factors and an energy width of the different peaks obtained from exclusive electron scattering data. While the correlation piece is computed within the Correlated Basis Function (CBF) theory for symmetric nuclear matter~\cite{Benhar:1994hw,Benhar:1989aw}. The local density approximation (LDA) is used to extrapolate the correlation piece to finite nuclei by convoluting the correlation component of the nuclear spectral function $P_{\mathrm{NM}}^{\mathrm{corr}}$ with density profile of the nucleus $\rho_{A}(\mathbf{R})$~\cite{VanNeck:1994ge}.

In recent years, various ab initio techniques have also been employed to obtain the spectral function of different nuclei. Below, we provide a succinct overview of recent efforts.
Quantum Monte Carlo Techniques have been adopted in Refs.~\cite{CLAS:2022odn,Simons:2022ltq,Andreoli:2021cxo,Lovato:2023khk} to obtain the spectral function of nuclei with A=3, 4 and 12. Within this approach the mean field contribution has been obtained by Fourier transforming the variational Monte Carlo (VMC) radial overlaps for the relevant transitions~\cite{Mecca:2019lxd,overlaps_web}. The correlation contribution is obtained from the VMC two-nucleon momentum distributions of Ref.~\cite{nkk_web}, but in order to isolate the contribution of short-range correlated nucleons, cuts in the relative momentum of the pairs, requiring that the overall normalization and shape of the one-nucleon momentum distributions are correctly recovered have to be performed. These QMC spectral functions have been used as input for the impulse approximation scheme and compared with electron scattering data in the quasielastic region for a variety of kinematics~\cite{Andreoli:2021cxo,Betancourt:2023uxz,Lovato:2023raf}.

The Self Consistent Green's Function (SCGF)~\cite{Barbieri:2016uib} and the Coupled Cluster~\cite{Hagen:2013nca} approaches have also been recently used to obtain nuclear spectral functions in the range of medium-mass nuclei. The complexity of these approaches exhibits a polynomial scaling with the number of nucleons, allowing to describe medium size nuclei such as A=40. In these approaches, the calculation of the spectral function does not require any separation into the mean-field and correlation parts since the overlap between the initial and final hadronic states in Eq.~\eqref{eq:sf_def} can be directly calculated.
In order to obtain the spectral function of finite nuclei, the SCGF method involves an iterative calculation of the Green's function's imaginary component, which is directly related to the one-body spectral function.  In Ref.~\cite{Rocco:2018vbf} the inclusive electromagnetic cross sections of $^4$He and $^{16}$O, obtained using the associated SCGF spectral functions, have been compared with the experimental data.  In order to subtract the center of mass contribution from the SCGF trail wave functions and expectation values, a new Metropolis Monte Carlo calculation has been adopted. 
Recently, a new method to construct spectral functions for the many-body system based on Chebyshev polynomials expansion of the Gaussian integral kernel~\cite{Sobczyk:2022ezo} was proposed. This method used in conjunction with Coupled Cluster theory was benchmarked on $^4$He and $^{16}$O leading to a good agreement with electron-scattering data~\cite{Sobczyk:2023mey}. 
Both approaches, in principle, are limited to describing closed-shell nuclei and necessitate some generalizations to address open-shell ones. To this aim, the SCGF approach utilizes Gorkov theory~\cite{Soma:2011aj,Soma:2012zd,Barbieri:2021ezv} and has successfully employed the spectral functions of $^{12}$C and $^{40}$Ar to generate electron- and neutrino-scattering predictions that correctly match experimental data~\cite{Rocco:2018mwt,Barbieri:2019ual}. For the Coupled Cluster method, treating the ground state of open-shell nuclei might be achieved by employing charge-exchange or particle-attached/particle-removed techniques. 

In models based on the spectral function (SF) formalism, the outgoing nucleon in the quasi-elastic region is typically described as a plane wave. As a result, the residual A–1 nucleons interact among themselves, but interactions between the struck nucleon and the remnant are not taken into account. To compare with exclusive observables, one possible approach to include these missing effects is to propagate the outgoing nucleon using semi-classical intranuclear cascade models, such as those implemented in event generators (see Sec.\ref{sect:fsi}).
However, it is important to note that these cascades do not affect inclusive observables due to their semi-classical nature. To account for rescattering effects in inclusive processes, the procedure developed and discussed in Refs.\cite{Benhar:2006wy,Benhar:2013dq} is commonly adopted.

Within this procedure, the propagation of the knocked-out particle in the mean field generated by the spectator system is modeled by introducing the real part of the optical potential $U$ derived from the Dirac phenomenological fit of Ref.\cite{Cooper:1993nx}. The latter is used to modify the free energy of the outgoing particle during its propagation through the mean field generated by the spectator system. Multiple scatterings experienced by the struck particle as it propagates through the nuclear medium are accounted for by folding the plane-wave response with a folding function derived from a generalization of Glauber theory~\cite{Benhar:1991af}.

Lastly, various LDA models, described in Sec.~\ref{section:LDA}, can be recast to the spectral function formalism for the nuclear matter. The nucleons' propagators, both for the hole and particle states, can be modified to account for in-medium effects. While for the Fermi gas model the propagators correspond to free states, GiBUU introduces a mean-field potential which alters the nucleon's effective mass, while the Valencia group calculates nucleon's self-energy. All these approaches allow to describe the final states interactions within the same formalism. However, due to the LDA approximation, they lack information on the nuclear shell structure.

\textbf {Advantages and Limitations:}
This approach enables a consistent description of different reaction mechanisms. Various nuclear models can be employed to compute the spectral function, allowing for a systematic study of model dependence. The main limitation, however, is that the final states are currently treated as plane waves, as there is currently no framework, beyond the LDA approaches, that incorporates final-state interactions in a fully consistent way. Estimating the associated uncertainty in a rigorous manner remains a nontrivial task.

\subsection{Short Time Approximation}\label{section:STA}

The Short Time Approximation (STA)~\cite{Pastore:2019urn} algorithm finds its roots in ab initio techniques but exploits a factorization scheme to retain two-body physics, namely two-body correlations and associated two-body currents. Despite limiting the description of the scattering process to interactions of the probe with pairs of correlated nucleons, the STA is found to be in good agreement with both GFMC predictions and experimental data for electron scattering from the alpha particle, the trinucleon systems, and $^{12}$C~\cite{Pastore:2019urn,Barrow:2020mfy,Andreoli:2021cxo,Andreoli:2024ovl,Andreoli:2025pvm}. This approach encompasses interference effects between one- and two-body current contributions and provides nuclear response densities, which are expressed in terms of the relative and center of mass energies of the struck nucleon pair.  The STA method can accommodate fully-relativistic kinematic and currents~\cite{Andreoli:2025pvm}, enabling  detailed information on the kinematic variables associated with the hadronic final states. While the current results are limited to light and medium size nuclei, Auxiliary Field Diffusion Monte Carlo can be used to extend the predictive power of the STA to larger nuclei.

\textbf {Advantages and Limitations:}
In the STA approximation, final-state interactions between the two emitted nucleons and the residual (A–2) system are included. What is neglected, however, are the interactions between the emitted nucleons and the remnant nucleus.

\subsection{Mean-field approaches}
\label{sec:mf}
Independent-particle shell-model (IPSM) approaches describe the ground state nucleus as a set of independent-particle nucleon wave functions that are solutions of the mean-field equations. In this way, a good part of the nuclear dynamics is incorporated, basically, the shell structure, which includes realistic predictions of binding energies, momentum distributions and nuclear densities. The IPSM provides not only predictions about the initial state but a quantum mechanical framework to describe some reaction channels. In particular, the channels for which the impulse approximation work well, like quasielastic and single-pion production, can be efficiently modelled by describing the knocked-out nucleon as a scattering solution of the same wave equation used for describing the ground state, i.e., using the same mean-field potentials. This consistency in the description of initial and final state results in accounting for Pauli exclusion and nucleon distortion (i.e.~elastic final-state interactions) in a natural and quantum mechanical way.  

The limitations of the IPSM are also evident. For example, due to the absence of explicit long- and short-range correlations, in the IPSM nucleons exist in shells with full occupancy. However, reduced occupancies are observed in nature~\cite{Kramer89, Volkov90, Garino92, Udias93, Holtrop98, Dutta03, Fissum04, Egiyan06, Yasuda10, Giusti11, Atkinson18, Duer18, JeffersonLabHallA:2022cit, JeffersonLabHallA:2022ljj}, therefore, the use of spectroscopic factors lower than one is needed to reproduce experimental data accurately.
Addressing some of these beyond-mean-field effects can be partially done in an effective and relatively simple way by incorporating a phenomenological spectral function that, essentially, accounts for the spectroscopic factors and the widening of the deeper energy levels~\cite{Gonzalez-Jimenez22, Franco-Patino22, Franco-Patino24}. 
But also, IPSM can be extended to incorporate some of these additional nuclear effects in a more fashion, microscopic and dynamical way, e.g. long-range correlations in a continuum random phase (CRPA) model \cite{pandey:2014tza}, or the influence of short-range correlations and meson-exchange currents on 1p-1h processes \cite{vancuyck:2016fab, vancuyck:2017wfn, Franco-Munoz23}. Additionally, two-nucleon knockout and pion production processes can also be studied within a mean-field framework~\cite{vancuyck:2016fab,vancuyck:2017wfn,Garcia-Marcos2023}.

\begin{figure}
    \centering
   \includegraphics[width=0.7\linewidth]{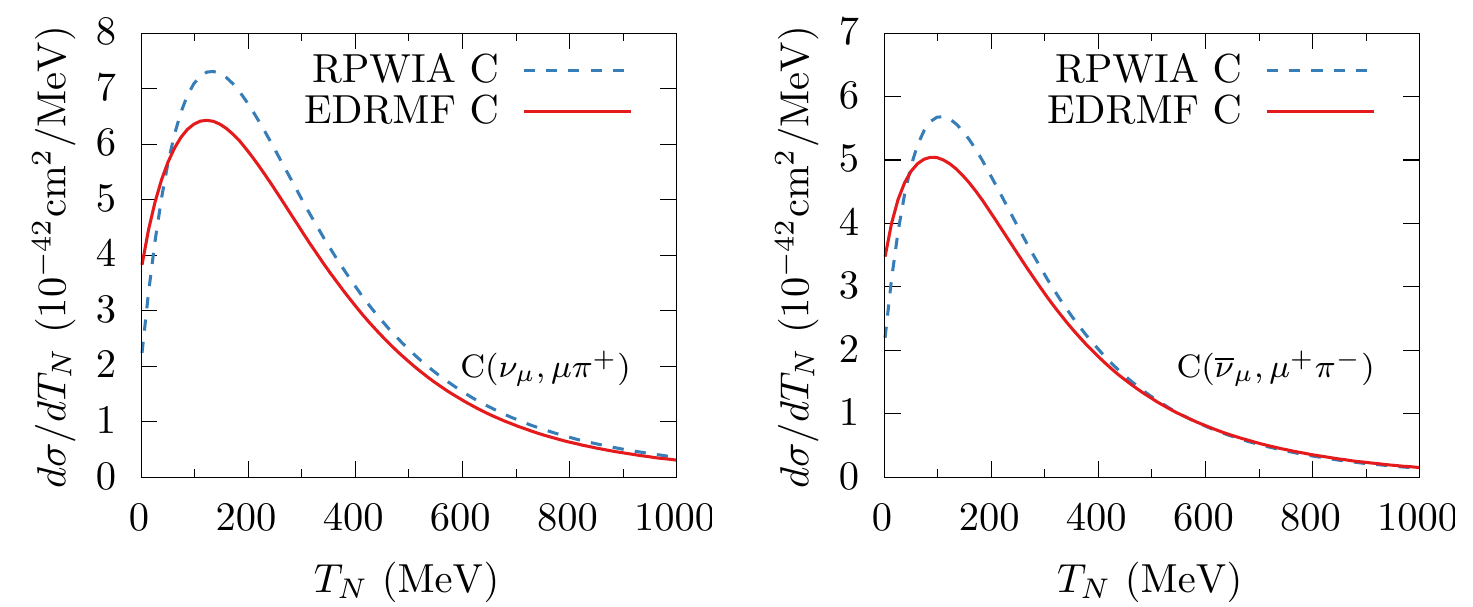}
    \caption{Comparison of RDWIA (relativistic distorted wave impulse approximation) and RPWIA (relativistic plane wave impulse approximation)  cross sections for charged pion production on $^{12}$C, folded with the MINERvA flux, shown as a function of the kinetic energy of the outgoing nucleon ($T_N$)~\cite{Nikolakopoulos:2021nmb}. It is clear that especially for small $T_N$, the influence of including nucleon distortion is sizeable, shifting the strength to lower energies and reducing the peak.}
    \label{fig:RMF-pions}
\end{figure}

 When real potentials are used, the flux (or probability) is conserved and these approaches provide reliable predictions of the inclusive cross section~\cite{Maieron03, Meucci09, Kim07, Butkevich07, Pandey16, Gonzalez-Jimenez20, Bodek24}.  
 If the goal is to predict exclusive cross sections, i.e., all particle 4-vectors need to be determined, including those of the outgoing nucleons, then one needs to make use of phenomenological complex optical potentials~\cite{Udias93, Udias01, Giusti11, Butkevich12, Gil21}. Those are usually extracted from fits to elastic nucleon-nucleus scattering data~\cite{Cooper93,Cooper:2009zza}. A slightly different approach is the the Relativistic Green Function~\cite{Capuzzi91, Meucci09, Ivanov16b}, which uses full complex optical potentials to make predictions of the inclusive cross section; in this case, the flux that was migrated to inelastic channels, due to the imaginary part of the potential, is recovered by a formal summation over those channels.

By nucleon distortion, or `elastic FSI', we refer to the fact that the struck nucleon is under the influence of a mean field while it is propagating out of the nucleus. The impact of this on the cross-section distributions is quite relevant, producing an important redistribution of the strength that is needed to match the inclusive and exclusive experimental data. It is important not to confuse this elastic FSI with the inelastic FSI, the latter could, eventually, be taken into account by a classical cascade approach. 
We stress here that the inclusive cross section is not affected by the cascade process, therefore, the model that describes the elementary vertex should give a reasonable inclusive response, and for that, all the nuclear effects mentioned above, including nucleon distortion, are necessary. 
Whereas inclusive cross sections obtained in an mean-field model can be fed as input to a generator using tables, as e.g. done in \cite{Dolan:2021rdd, Dolan:2019bxf},
it is actually preferable to use a model that also provides detailed information on the final hadrons, so they can be propagated in the generator’s cascade more consistently.

The mean-field approaches discussed here are good candidates for these two tasks. The price is that as soon as one moves beyond the plane-wave approach, the factorization of the cross section, explained in section \ref{section:SFA}, is broken. The use of distorted waves allows for a more accurate description of the nuclear dynamics, but it also means that these models are computationally more demanding, which poses new challenges to incorporate them into MC generators. Some work has been done to study the fit of these approaches in MC event generators \cite{Nikolakopoulos:2022qkq, Nikolakopoulos24b} and the implementation of such `full-kinematic' models in generators is in progress~\cite{McKean25, ElBaz25}.

\subsection{Local density approximation based models}
\label{section:LDA}

Within the local density approximation (LDA), the nucleus is viewed locally as the nuclear matter of a given density of neutrons and protons $\rho_{n,p}(r)$. The density profile of each nucleus can be either retrieved from experimental data or calculated theoretically. It is worth noting that within this description, the information about the nuclear shell structure is smeared out. Therefore, calculating semi-exclusive observables such as cross-sections measured in \textit{(e,e'p)} experiments are beyond the scope of these models. Nevertheless, they found an important place in modelling neutrino-nucleus interactions in long-baseline experiments.

The chief example of an LDA model is the Fermi gas (FG) which introduces yet another simplification by treating nucleons as particles bound by a constant binding energy. The global FG treats the nucleus as a sphere of nuclear matter of constant density, while for the local FG the density has a radial dependence, $\rho_{n.p}(r)$. In the Fermi gas model the nuclear density is directly related to the Fermi momentum. Nucleons in the groundstate are on-shell particles described by the momentum distribution which for the global FG is a simple step function, while for the local FG it has an additional $r$ dependence. Similarly, the knocked-out nucleons have momenta above the Fermi level, in accordance with the Pauli exclusion principle. This model is still widely used in the Monte Carlo event generators, also to model dynamical mechanisms other than quasi-elastic reaction.
In spite of its simplicity, the Fermi gas describes the bulk properties of scattering cross sections, as has been analyzed in the past for the quasi-elastic dynamics~\cite{Sobczyk:2017mts}. However, the approach is known for various deficiencies. The constant binding energy is not enough to predict precisely the position of the quasi-elastic peak. Moreover, the model breaks at lower energy-momentum transfer region where a more refined treatment of nuclear dynamics is essential even for the description of inclusive cross sections. In this regime, the LFG is known to largely overestimate the strength. This behaviour was scrutinized for the quasi-elastic in case of electron scattering where a lot of experimental data is available. These theoretical deficiencies remain when the model is applied to other physical mechanisms (multinucleon knockout or resonance production).

Various theoretical models based on the LDA have been proposed and successfully used to describe neutrino-nucleus cross-sections. They are based on the Fermi gas description incorporating additional information on the nuclear dynamics. The LDA approach is appealing for the Monte Carlo generators, since it naturally allows to model the neutrino-nucleus scattering event as a two-step process. Namely, the interaction vertex at the position $r$ directly serves as a starting point for the intra-nuclear cascade (see Sect.~\ref{sect:fsi}) which itself follows the LDA when propagating nucleons through the nucleus.
Within the physical model used by the GiBUU event generator, the nucleons are submerged into a relativistic mean-field potential which is both density and momentum dependent, leading to the in-medium modification of nucleons' dispersion relations~\cite{leitner:2006ww}. Importantly, both the initial and final nucleons are treated consistently and--since the potential is valid in the relativistic regime--it can be used for the wide range of energies relevant for long-baseline experiments. 

Other Monte Carlo event generators widely employ the Valencia model~\cite{Nieves:2004wx}, which is based on the local FG. It was initially proposed and benchmarked for nuclear interactions with several probes: pions, real photons, electrons, and finally neutrinos~\cite{Nieves:1991ye,Carrasco:1989vq,Gil:1997bm}. 
The approach accounts for some nuclear medium polarization effects sensitive to the collective degrees of freedom of the nucleus through the inclusion of random phase approximation (RPA).
These corrections bear some resemblance with the polarization experienced by a probe charge inside of an electron gas. Within the model, an infinite series of particle-hole and $\Delta$-hole excitations, which interact via an effective spin–isospin non-relativistic potential, is summed up. These collective phenomena influence mostly the low-energy part of the nuclear response, leading to its quenching and redistribution towards higher energies.
Additionally, the Valencia model also accounts for in-medium treatment of nucleons by introducing their complex self-energy. This can be also understood in terms of the spectral functions for nuclear medium which introduce the collisional broadening (through the imaginary part of the self-energy), as well as changes the dispersion relations of nucleons in nuclear medium~\cite{Nieves:2017lij}. 

The nuclear description employed by the Valencia group bears much resemblance to the Lyon model~\cite{Martini09} which includes RPA in a similar manner, leading to consistent predictions for the quasi-elastic excitations. Initially, the results of both group for multi-nucleon knockout were quite different~\cite{Morfin:2012kn,katori:2016yel}. However, in a recent Valencia reexamination of the neutrino multinucleon cross sections this discrepancy has been resolved~\cite{Sobczyk:2024ecl}, leading to a good agreement with the inclusive MiniBooNE and T2K data.

\textbf {Advantages and Limitations:}
These approaches are computationally inexpensive and -- apart from the FG model -- they have been successful in various comparisons to the experimental data. They are also well suited for existing Monte Carlo event generators.

Their main limitation is the simplified treatment of nuclear dynamics which affects the exclusive observables. It is also not straightforward to estimate theoretical uncertainties.

\subsection{Superscaling phenomenology}
\label{sec:scaling}

Scaling is a phenomenon observed in several areas of physics and constitutes a fundamental part of the scientific methodology. In studies of lepton-nucleus reactions~\cite{West:1974ua,Donnelly:1998xg,Donnelly:1999sw}, it has been shown that when FSI effects are negligible, the QE cross section can be written in terms of a factor containing a single-nucleon cross section times a specific function, usually defined as the scaling function, that embodies the nuclear dynamics. Indeed, within some specific approaches, this function is related to the momentum distribution of the nucleons in the nucleus (or, more generally, with the spectral function)~\cite{Caballero:2010fn,Antonov:2011bi}. At intermediate and large momentum transfers, above 400-500 MeV/c, this function scales; i.e., it becomes a function of only one variable, namely, the scaling variable $\psi$, defined in terms of the energy and momentum transfers ($q$, $\omega$), and largely independent of the momentum transfer. Further studies show that this phenomenon can also be observed for different nuclei. The superscaling phenomenon occurs when a simultaneous independence on $q$ and on the nuclear target is observed in the analysis of scaling functions for lepton-nucleus reactions~\cite{Donnelly:1998xg,Donnelly:1999sw}.

Note that the RFG approach is a natural starting point for these studies as, in spite of its simplicity, it fulfills exactly this superscaling behavior. Within the RFG model, the QE cross section in terms of the lepton kinematics can be analytically calculated as shown in~\cite{amaro2020} and the nuclear responses related to the different components of the nuclear tensor can be roughly described in terms of the single-nucleon hadronic respones multiplied by the RFG scaling function~\cite{amaro:2004bs,megias_thesis}.
Although within this approach a universal scaling function emerges from the calculation; scaling functions can also be extracted from experimental data or different nuclear models. The general procedure used to define the scaling functions consists of constructing the inclusive cross section, or nuclear response functions, using either a theoretical model or experimental data, and dividing them by the appropriate single-nucleon hadronic responses $G_K$, arising from the nuclear hadron tensor and computed within the RFG model, i.e.,
\begin{eqnarray}
\displaystyle f(\psi)=\displaystyle k_F\frac{\frac{d^2\sigma}{d\Omega d\omega}}{\sigma_{Mott}(v_LG_L+v_TG_T)}\,,
\end{eqnarray}
where $k_F$ is the Fermi momentum, $\sigma_{Mott}$ is the Mott cross section and the $v_K$ terms are the lepton kinematic factors that are constructed via the leptonic currents. The $L$ and $T$ terms in the previous expression are related, respectively, to the longitudinal and transverse components of the leptonic and nuclear tensors with regard to the direction of the momentum transfers to the nucleus (see~\cite{amaro:2004bs,megias_thesis} for details).

\begin{figure}
\includegraphics[scale=0.3,clip,angle=0]{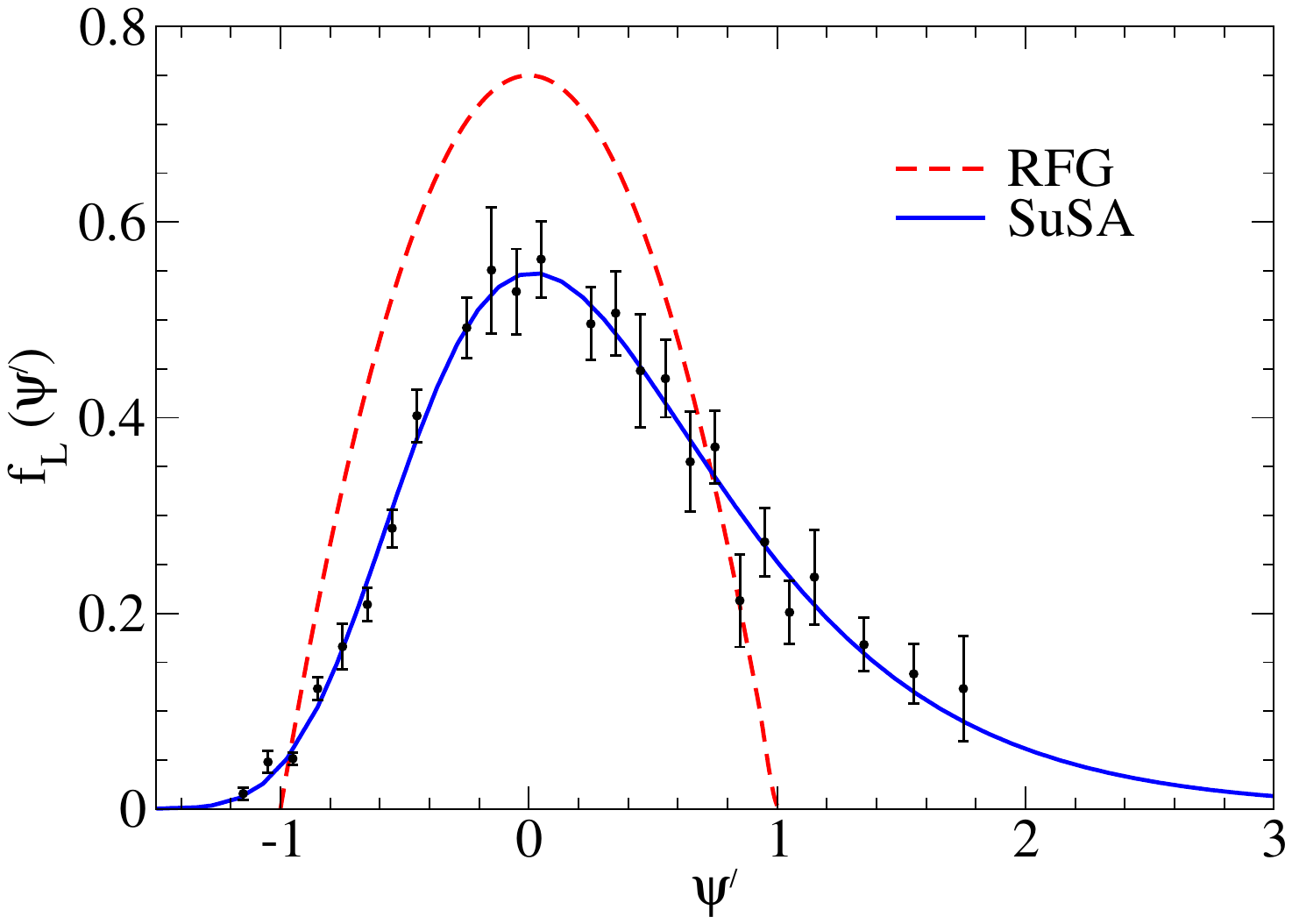}          \includegraphics[width=0.45\textwidth,angle=0]{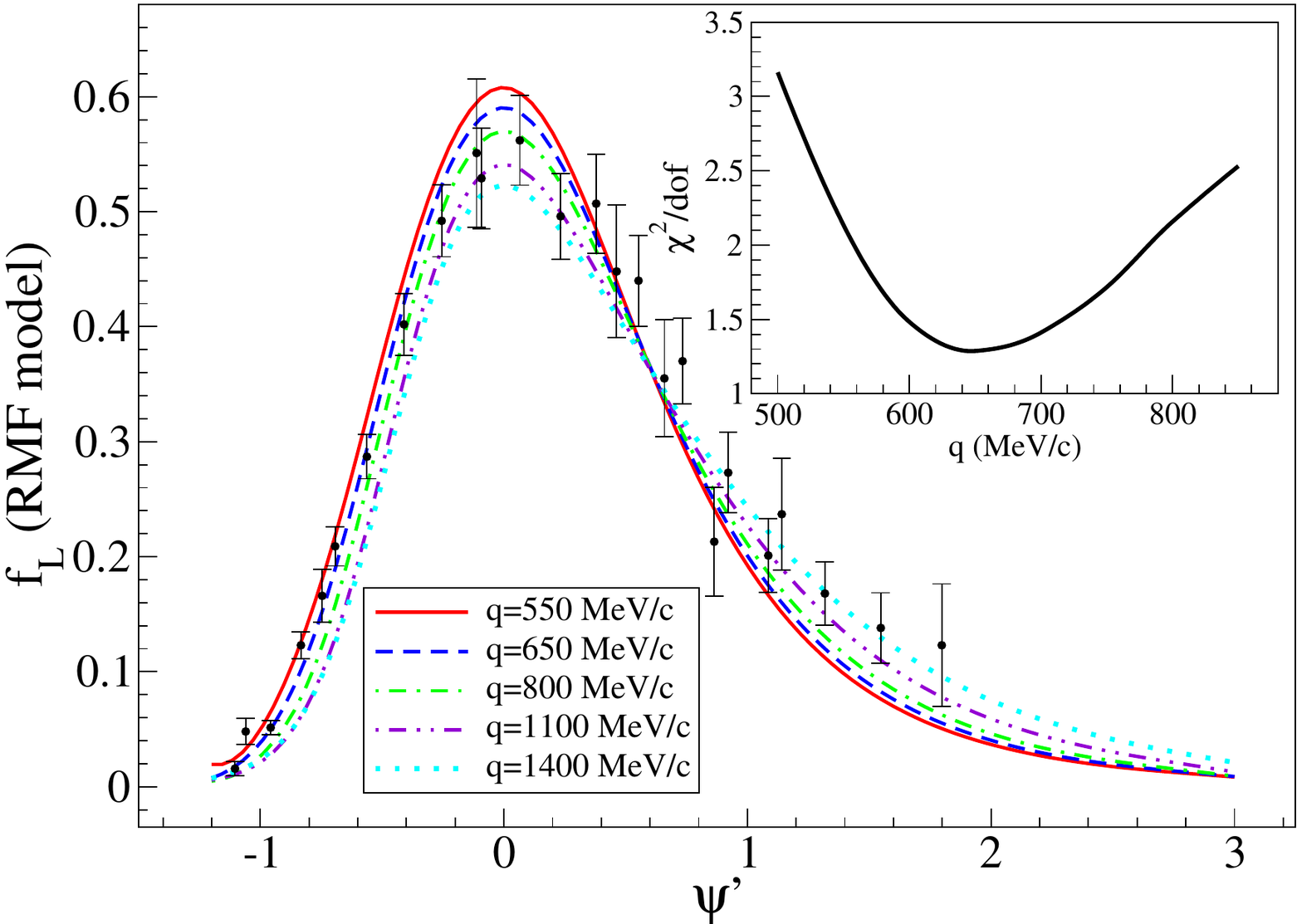}
\caption{ (Left panel) Averaged experimental $f_{SuSA}(\psi^\prime)$ and $f_{RFG}(\psi^\prime)$ scaling functions versus $\psi^\prime$ in the QE region together with a selection of $(e,e')$ longitudinal scaling data ($f^{ee'}_{L, exp}$) on $^{12}$C obtained from~\cite{Jourdan:1996aa}. (Right panel) Longitudinal scaling function for $(e,e')$ data~\cite{Jourdan:1996aa} computed within RMF for different $q$ values. In the inner smaller plot a reduced-$\chi^2$ analysis is presented versus $q$. The minimum $\chi^2$ is around $q=650$ MeV/c.
}\label{fpsiplaf} 
\end{figure}

This analysis has been applied to extensive electron scattering data on $^{12}$C, among other nuclei, where the experimental longitudinal scaling function has been compared with the RFG one and also with the SuSA (SuperScaling Approach) one, where in the latter the scaling function is obtained as a fit of the data~\cite{amaro:2004bs,Amaro:2006pr}. The comparison, only for the longitudinal (L) channels, can be observed in Fig.~\ref{fpsiplaf} (left panel) where the simple RFG model cannot describe the nuclear dynamics exhibited by the data. However, this scaling behavior is well reproduced by sophisticated nuclear models such as the relativistic mean field (RMF) one~\cite{susav2qe} as shown in Fig.~\ref{fpsiplaf} (right panel). Consequently, the RMF-scaling functions were adapted into the RFG framework described in previous equations in the so-called SuSAv2 (SuperScaling Approach version 2) model~\cite{susav2qe,megias:2016lke}, which has shown enormous success in describing not only electron reactions with several nuclei~\cite{megias:2016lke,Barbaro:2019vsr,Gonzalez-Rosa:2022ltp} but also CC neutrino cross-section measurements in a wide range of kinematics~\cite{megias:2016fjk,Megias:2018ujz}, also speeding up complex RMF calculations and easening its extension to other nuclei in a straightforward formalism. Thus, these scaling analyses must be taken as a solid benchmark for assessing the validity of nuclear models prior to their application in the study of electron and neutrino scattering data.

Unlike RMF models, the SuSAv2 approach only predicts lepton kinematics although work is in progress to also describe final-state hadrons by relying on factorization approaches combined with information from the RMF theory.
These superscaling approaches originally developed for the QE regime have been extended in further works~\cite{megias:2016fjk,Megias:2014qva,Barbaro:2017hkz,Barbaro:2019vsr,Gonzalez-Rosa:2022ltp,Gonzalez-Rosa:2023aim,PhysRevD.111.073002} to other channels, namely the 2p2h-MEC one and the inelastic regime, where the scaling rules can differ but the great accordance with data is still preserved. Some of these models have been implemented in MC event generators for the QE and 2p2h regimes~\cite{Dolan:2019bxf,PhysRevD.103.113003,clas:2021neh} and the inelastic channels will soon be available.

\subsection{Challenges and open questions}

While significant progress has been made in understanding the impact of nuclear modelling on neutrino interactions, and focused development on the most important elements, a perfect solution is not expected to be possible and so there remain some significant challenges.

\begin{itemize}
\item In general there is limited \textbf{uncertainty quantification} for all models.  Given the approximations generally needed, this uncertainty quantification is challenging but necessary.
\item As is discussed in Chapter \ref{generators}, there are challenges integrating new nuclear models into Monte Carlo generators, especially those that break the standard factorisation of initial state, interaction, and final state interactions.
\item There remains no general method to predict specific exclusive channels without approximations. 
\item Covering the entire phase space: the range of energies explored by accelerator neutrino is so vast that effective field theories are not expected to be valid throughout the energy interval explored. Combining an accurate description of nuclear structure and dynamics with electroweak interactions over such a broad energy range requires approximations. 
\item Machine learning techniques are currently being explored to extract properties of nuclei and to develop data-driven models capable of predicting cross sections.
\end{itemize}

\newpage
\section{Determining the Neutrino Flux}
\label{flux}
\subsection{Introduction}
Conventional neutrino beams, used in modern long-baseline oscillation experiments, provide naturally broad neutrino energy spectra from the decays of short-lived mesons.
Proton beams impinged on a solid target produce a spectrum of mainly pions and kaons, which are focused with magnetic horns and directed to a decay region where they decay to produce neutrinos.
Atmospheric neutrinos are produced through very similar processes, with the exception of magnetic focusing.

\begin{figure}
    \centering
    \includegraphics[width=0.37\linewidth]{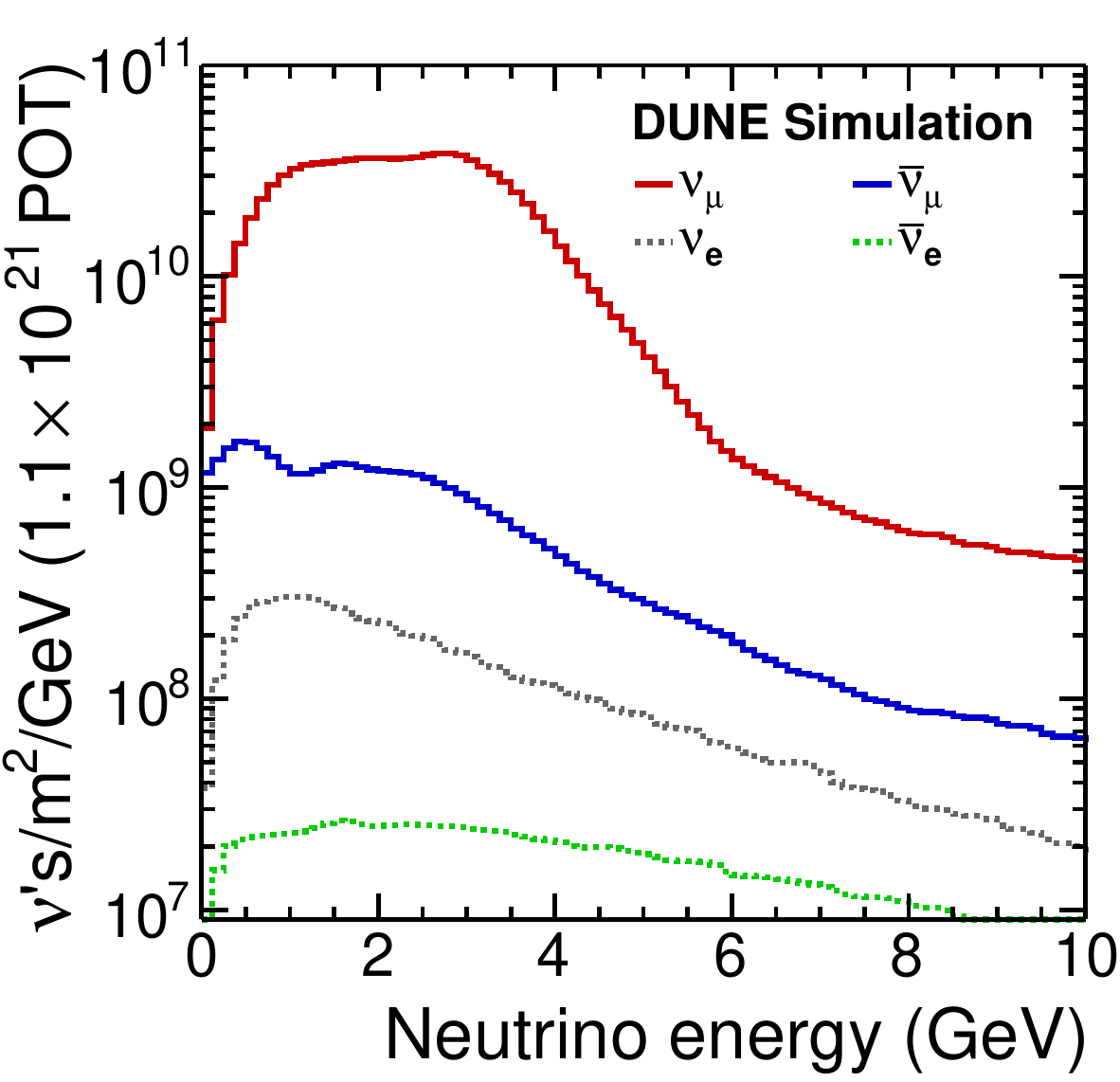}
    \includegraphics[width=0.50\linewidth]{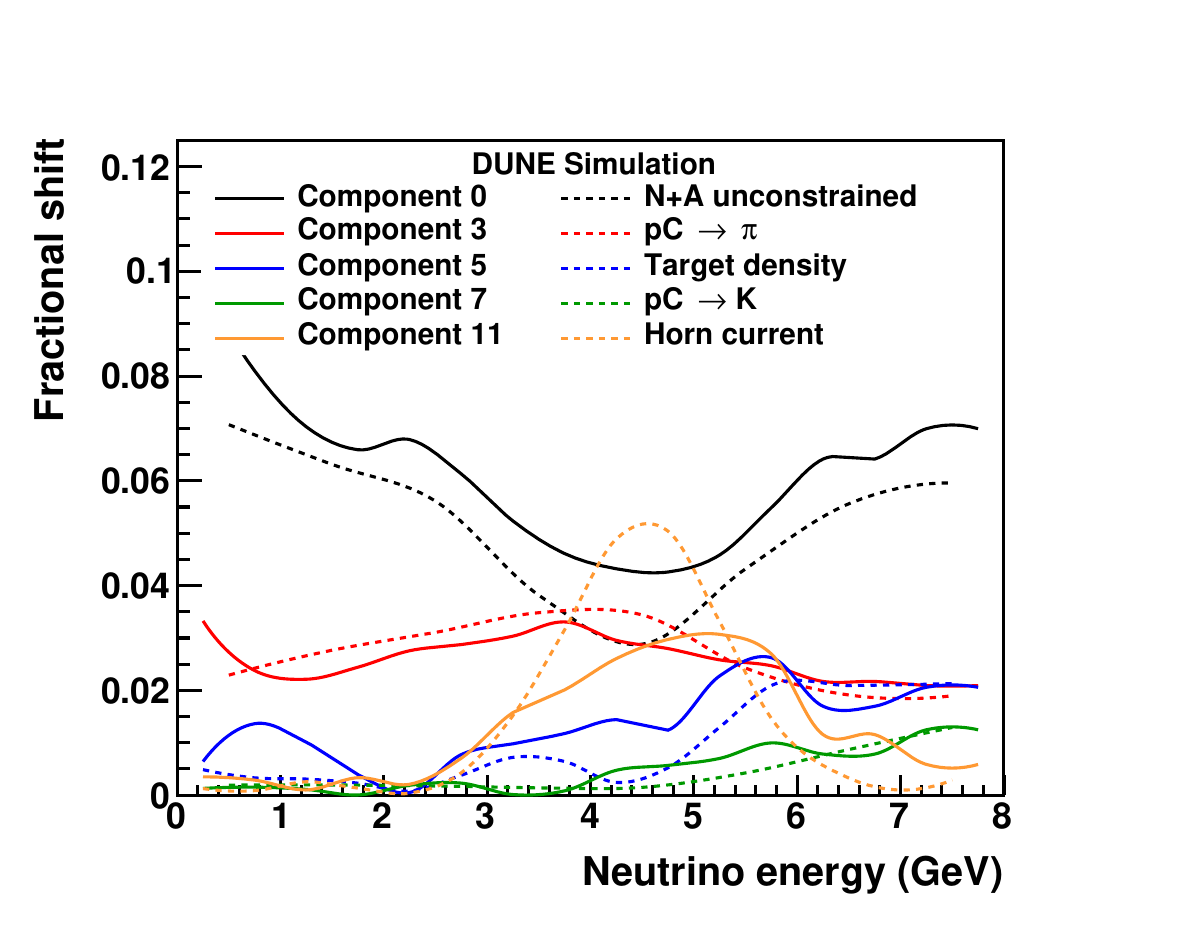}
    \caption{LBNF neutrino-mode neutrino flux (left) and uncertainties (right) as a function of neutrino .
    Figures taken from~\cite{DUNE:2020jqi}}
    \label{fig:DUNEFDFluxUncertainty}
\end{figure}

Flux uncertainties are often a dominant systematic uncertainty in neutrino-nucleus cross section measurements, limiting their precision.
This is because a-priori flux predictions rely on simulations of hadron production and scattering across a broad range of energies and nuclear targets, for which there are large uncertainties (some estimated as high as 40\%\cite{MINERvA:2016iqn}).
These uncertainties are particularly large for secondary and tertiary interactions involving particles with energies below 15~GeV.
The predictions for accelerator-based fluxes must also include very sophisticated descriptions of the materials and geometry of the production target and focusing horns, and the proton beam profile at the target.

Currently, hadron production uncertainties are the dominant contribution to the flux uncertainty, at the level of 10\%.
Hadron production uncertainties primarily impact the flux normalization, with a smaller impact to the shape of the flux.

Shape uncertainties in the flux (eg, as a function of neutrino energy) can be as large as 5-6\%, and arise due to a combination of hadron production and beam focusing uncertainties.  This can be seen in Fig.~\ref{fig:DUNEFDFluxUncertainty} which shows the muon-neutrino flux (left), and hadron production and a subset of beam-focusing  uncertainties (right) for the DUNE near detector in the LBNF neutrino mode.

Since flux uncertainties are dominated by hadron production uncertainties, a global effort has been underway to measure hadron scattering and hadron production cross sections with relevant beam energies, beam particles and nuclear targets for the past two decades.  
Initial efforts by the NA49~\cite{NA49:2006oyk}, HARP~\cite{HARP:2009drm} and MIPP~\cite{MIPP:2014shj} collaborations reduced flux uncertainties by about a factor of 2, from 20-30\% to 10-15\%.
More recent and future measurements by the NA61/SHINE~\cite{NA61SHINE:2023bqo} and EMPHATIC~\cite{EMPHATIC:2019xmc} collaborations aim to further reduce hadron production uncertainties by another factor of 2 or more.

For many experiments, a component of the flux cannot be constrained by existing data, with predictions relying on simulations such as FLUKA and GEANT4.
Existing measurement campaigns aim to not only reduce uncertainties but also expand the phase space to provide more direct data constraints and reduce the reliance on simulation which provides more robust predictions and uncertainties.

\subsection{Hadron Spectra from Real Neutrino Production Targets and Focusing Horns}

The most straightforward approach to reducing neutrino beam flux uncertainties is to measure the spectra of hadrons downstream of actual production targets and focusing horns, as the decays of these particles is very well understood.
Ideally this measurement would be in-situ, but this is very impractical for a number of reasons.
Firstly, detectors cannot operate in the high-radiation environment of a modern neutrino beam, meaning the intensity would need to be reduced for a measurement campaign with detectors removed afterwards.
General access to these beamline components during operation is extremely limited, and there is rarely space available for such detectors.
For these reasons, ex-situ measurements are preferred, allowing for greater flexibility and control over the measurements, at a cost of needing additional or spare production target, horns and power supplies.

\begin{figure}
    \centering
    \includegraphics[width=0.65\linewidth]{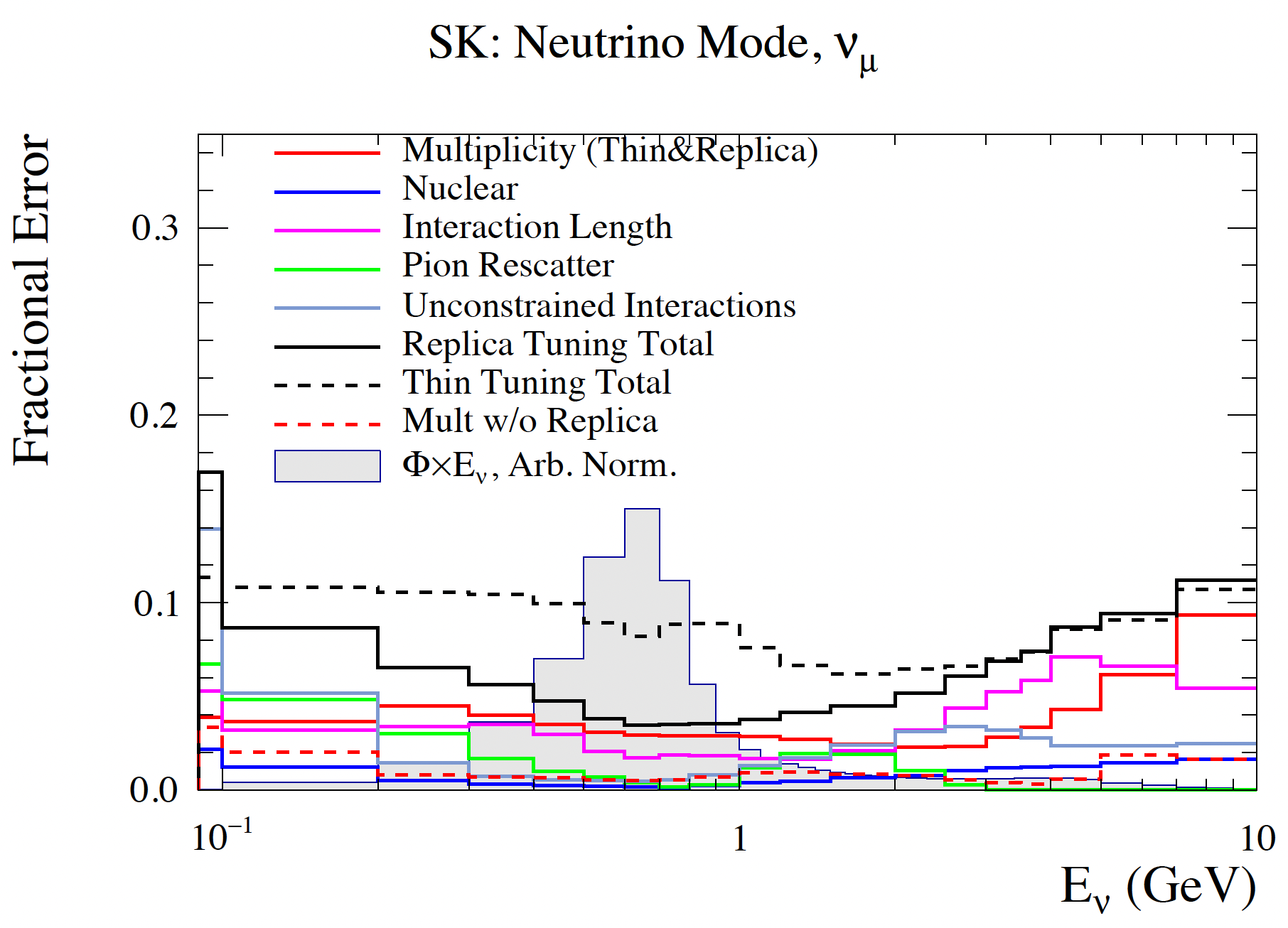}
    \caption{The hadron interaction model uncertainties evaluated on the T2K flux prediction. The uncertainties have been calculated for the flux constrained with a combination of NA61 2009 thin-target and replica-target data~\cite{Vladisavljevic:2019nfm}.}
    \label{fig:T2KFluxBeforeAfterReplica}
\end{figure}

Measurements by MIPP\cite{MIPP:2014shj} and NA61/SHINE\cite{NA61SHINE:2012vyd} of hadron spectra downstream of replica production targets have demonstrated the feasibility of constraining the flux to within 5\% at the flux peak.  Figure \ref{fig:T2KFluxBeforeAfterReplica} shows the impact such measurements have on reducing overall T2K flux uncertainties, where hadron production measurements were made using a replica T2K target in the NA61/SHINE experiment and used to constrain the flux prediction.  NA61/SHINE has also collected data using a replica NuMI target, and plans to collect data with a replica LBNF target in future.

The measurements using replica neutrino production targets will significantly reduce the flux uncertainties for current and future neutrino experiments.  The next step in further reducing a-priori flux prediction uncertainties is to account the impact of the focusing horns and other beamline elements.  The EMPHATIC collaboration has a goal of making the first measurement of hadron spectra downstream of a neutrino production target and focusing horn by placing the table-top-sized spectrometer on a motion table downstream of the focusing horn, and making measurements across the face of the horn.  A proof-of-principle measurement downstream of an unpowered horn is planned for 2026.  Measurements with a powered horn will require additional funding for engineering and upgrades to the horn power supply and EMPHATIC detectors.

\subsection{Hadron Scattering and Production Uncertainties on Thin Targets}

\begin{figure}
    \centering
    \includegraphics[width=0.45\linewidth]{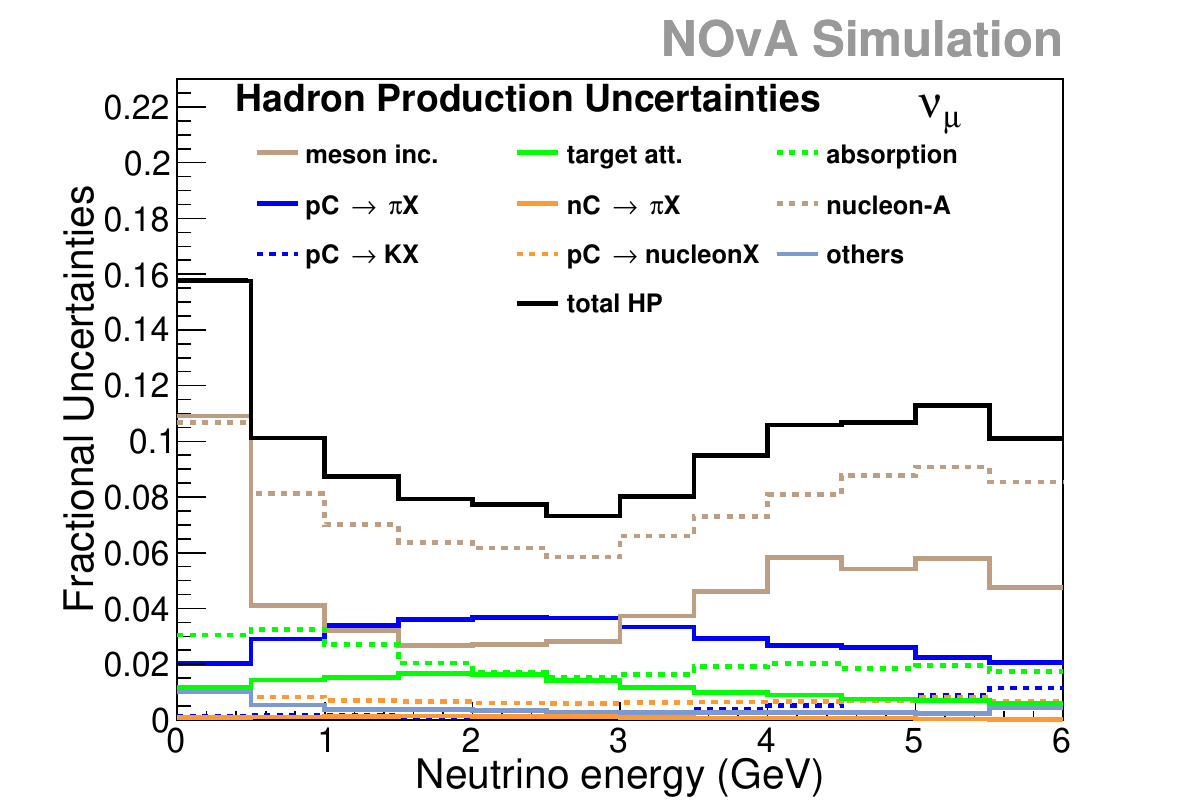}
    \includegraphics[width=0.45\linewidth]{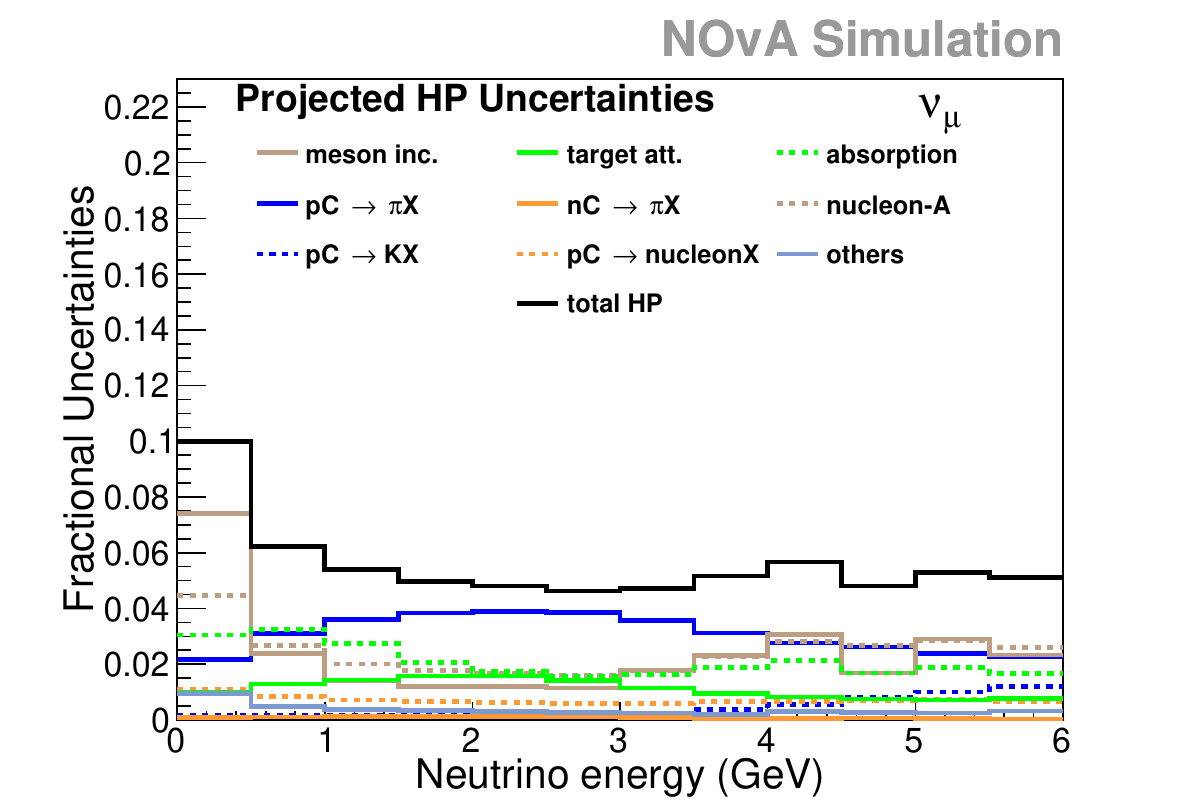}
    \caption{Current (left) and potential (right) flux uncertainties with modest reductions in uncertainties in processes are unconstrained by external data\cite{EMPHATICWP}.}
    \label{fig:NOvAFluxBeforeAfter}
\end{figure}

Hadron production cross section measurements typically take many years to both collect and analyse the data.  Direct measurements of hadron spectra from real production target and horns are typically not possible years in advance of an experiment, and often not for several years of beam operation, as spare targets and horns are not available at that point in time.  Therefore, hadron scattering and production cross section measurements on different materials, beam particles and beam momenta are needed to robustly predict the flux from a first-principles simulation.  These cross section measurements are typically referred to as ``thin-target'' measurements, since the nuclear targets are typically much less than an interaction length thick (usually 2-5\% of an interaction length), making secondary interactions within the target negligible.  These measurements are also applicable to atmospheric flux predictions.

The MINERvA, NOvA and DUNE collaborations currently use thin-target measurements as constraints on their flux predictions.  The left-hand plot in Fig~\ref{fig:NOvAFluxBeforeAfter} shows the current hadron-production fractional uncertainty on the neutrino flux as a function of neutrino energy in the NOvA off-axis near detector, broken down by the types of hadron interactions in the target and horn.  The prediction is constrained primarily by NA49 data of 158 GeV/c protons striking a thin carbon target\cite{NA49pCpi158}.  Work is underway to include additional measurements by the NA61/SHINE and EMPHATIC experiments as additional constraints.

(AF added) To constrain secondary interactions in the production target and surrounding material (horns, decay volume, etc), thin target measurements are needed at a variety of energies, for a variety of materials, and for all secondary hadrons (primarily pions, kaons, protons, and neutrons).  Such an expansive program is challenging due to the number of different datasets needed, which is why replica measurements have such value.

\subsection{Atmospheric neutrino flux uncertainties}

Atmospheric neutrino fluxes are subject to additional uncertainties beyond those present at accelerators.  In addition to the hadronic production uncertainties, there is also uncertainty due to the limited knowledge of the incident proton/heavier nuclei beam, uncertainties in how pions and kaons interact, and even from the atmospheric density, which affects the likelihood of pions and kaons decaying before they interact.   These are issues even for relatively low-energy neutrinos, since high-energy cosmic-rays also produce low-energy neutrinos. 

The cosmic-ray energy spectrum and nuclear composition (protons vs. heavier nuclei) have been measured by many experiments.  The composition has a significant effect on neutrino production.  At energies up to about about 10-50 TeV, there are direct measurements of energy spectrum and the composition, using satellites and balloons \cite{Tomassetti:2023xph}, although the composition measurements are incomplete at the higher energy end of this range.  At higher energies, we are forced to use indirect measurements, such as measuring the depth of shower maximum, or the muon content measured by ground-based arrays \cite{Supanitsky:2022zcw}.    

The hadronic uncertainties are broadly similar to those that pertain to accelerator experiments, but there are additional contributions. Cosmic-ray air showers occur in a long cascade.  After a high-energy incident primary cosmic-ray interacts in the upper atmosphere, the lower energy particles will propagate downward, undergoing further interactions\cite{Gaisser:2002jj}.  At sea level, the atmosphere is about 13 hadronic interaction lengths thick.  Because of the long cascade, the cross-sections for $\pi^\pm$, $K^\pm$, $K_S$ and $K_L$ are all important, along with their interaction characteristics.  These cross sections have not been measured at TeV energies or above.   At high energies, prompt neutrinos may be a significant fraction of the flux.  Unfortunately, the cross-section for prompt neutrino production is not well known \cite{Bai:2022xad}.  

The final new ingredient in the atmospheric neutrino flux is the atmosphere, which affects the competition between $\pi$ and $K$ decay vs. interaction (the same phenomena happens for charmed hadrons at high energies).  The probability of $\pi,K$ interaction before decay (producing another generation of the shower) increases nearly linearly with the meson energy, and depends on the atmospheric density profile, including its seasonal dependence \cite{IceCube:2025wtk}.  This competition affects the zenith-angle dependence of the flux, with neutrinos from $\pi,K$ clustered near the horizon, while prompt neutrinos are more isotropic.  Neutrinos from $K_L$ are intermediate between these two classes \cite{Gaisser:2014pda}.  

Fortunately, data can be used to constraint these uncertainties; the atmospheric neutrino spectrum can be measured in the energy range  above about 100 GeV, where oscillation effects can be  
neglected, up to energies of about 1 PeV for $\nu_\mu$ \cite{IceCube:2014slq,ANTARES:2021cwc} and about 100 TeV for $\nu_e$ \cite{IceCube:2015mgt}. At higher energies, astrophysical neutrinos dominate.  
Unfortunately, the prompt neutrino flux has not yet been seen \cite{IceCube:2023mrq}.

Atmospheric muon measurements can help constrain neutrino flux calculation \cite{IceCube:2015wro}.  Modern calculations use fits to multiple data sets, both muon and neutrino, to minimize the uncertainty in atmospheric neutrino flux predictions \cite{Yanez:2023lsy}.  

Looking ahead, a wide range of data should contribute to gradually lowering the uncertainties on the atmospheric neutrino flux. 

\subsection{In-situ Constraints from Neutrino Scattering}

Processes such as neutrino-electron, coherent pion production, and ``low-$\nu$'' scattering can provide constraints on the flux normalization, and with enough statistics and energy resolution, the shape of the flux can also be constrained.  In nearly all cases, improvements to the theoretical predictions of the cross section will be needed to take advantage of these measurements.

By far the ``cleanest'' constraint on the flux is obtained from neutrino-electron scattering.  Because this is a purely leptonic, predominantly neutral-current process, the theoretical uncertainties on the cross section are at the percent level \cite{TomalakHill:2020nue}.  Therefore, a measurement of the rate of these interactions can be converted into a constraint on the total neutrino flux.  
One challenge of this measurement is that the cross section of this process is very small, about 1\% that of the charged-current neutrino-nucleus cross section.  It is a small signal with extremely large backgrounds.  However, the backgrounds are rejected via requirements on the forward angle of the final-state electron.  
MINERvA has demonstrated the feasibility of these measurements to constrain the flux normalization uncertainty to 3-4\%\cite{minerva:2019hhc}, and DUNE expects to eventually be able to further reduce the uncertainty to 1\% \cite{Marshall:2019vdy}.  
It remains to be seen if this technique can be applied to atmospheric neutrinos, but if it can be demonstrated, it would be a useful additional handle in constraining the atmospheric neutrino flux in a manner independent of nuclear interaction modeling.

Anti-neutrino scattering off hydrogen is a process that is be predicted with small uncertainties via lattice-QCD calculations \cite{latticeQCD:2019}.  Therefore a cross section measurement can be used as a constraint on the flux. Pure hydrogen targets are experimentally challenging, although a resurgent R\&D effort is underway at Fermilab \cite{Alvarez-Ruso:2022exy}.  MINERvA has demonstrated that \cite{minerva:2023avz} interactions off hydrogen can be kinematically separated from interactions off carbon nuclei in a high-statistics sample.  The technique has also been proposed to be done in DUNE with much higher statistics, but it is important to continue to reduce the theoretical predictions of the cross section.

Coherent pion production has a very simple topology: a 2-particle final state including the lepton and pion.  Although the cross section for this process is quite small, approximately a factor $10^{-3}$ smaller than \mm{$\sigma^{\mathrm{Res}}_{CC}$},
these interactions will be cleanly identifiable with the highly capable detectors that will be used in future experiments.  The neutrino energy in these interactions can also be determined with very high resolution.  Therefore this process could be used to constrain the flux as a function of neutrino energy.  However, for this process to useful to constrain the flux, the theoretical uncertainties on the process must be significantly reduced from their current level ($\sim20\%$).

Finally, low-$\nu$ scattering refers to interactions where there is low energy-transfer to the nucleus, such that the cross section is fairly constant as a function of neutrino energy.  This approach has been used by several experiments to constrain flux uncertainties, including MINERvA \cite{minerva:2021mpk}.  However, as pointed out in \cite{Wilkinson:2022dyx}, this approach has large uncertainties from modelling nuclear effects, especially for neutrino energies below 5 GeV.  Improvements and reductions of these modelling uncertainties would certaintly make this approach very viable for very high statistics samples.

\subsection{Summary and prospects}
Reduced flux uncertainties will lead to more precise measurements of almost all interaction processes in future, and has a direct impact on the precision of oscillation experiments.
In order to achieve substantial reduction, it is important to use all of the tools available.
This will require:

\begin{itemize}
\item Dedicated measurements of hadron production from replica targets, ideally with powered focusing horns in place.
\item A broad range of thin target hadron scattering measurements with higher precision and larger phase space coverage than existing data.
\item Highly capable detectors able to identify neutrino-electron elastic scattering, coherent events, and neutrons from $\bar{\nu}$-hydrogen scattering, for in-situ direct flux constraints.
\item Theoretical developments to reduce the a-priori theoretical uncertainties on $\nu$-e elastic scattering, coherent scattering, and quasi-elastic scattering on hydrogen.
\end{itemize}

With these developments, there is a clear path to few-percent flux uncertainties.
This would be sufficient for most oscillation measurements as a near detector constraint is able to effectively reduce the flux uncertainties further. 
Reduction below a few percent would be extremely challenging, likely requiring new ideas and/or technology such as tagged beams.

\newpage
\section{Neutrino Event Generators}
\label{generators}
Simulations of interaction physics processes play a critical role for neutrino
experiments just as they do for the high-energy and nuclear research
communities more broadly~\cite{Campbell2022,Diefenthaler2022}. Such simulations
are provided by software tools called Monte Carlo event generators. These
computer codes take information about the experimental setup as input and use
probabilistic methods to produce \textit{events}: full predictions of the final
state for individual simulated interactions. Experiments rely upon high-quality
event generator models to design and execute physics measurements. Estimation
of the efficiency and purity with which a physics signal can be experimentally
isolated from uninteresting background is an example of one of the key tasks in
any data analysis for which a reliable event generator is essential.

Several modern event generator codes are available for simulations of neutrino
scattering at accelerator energies.
GENIE~\cite{Andreopoulos:2009rq,GENIE:2021npt} is most commonly used in
experimental production workflows, especially at Fermilab.
NEUT~\cite{Hayato:2009zz,Hayato:2021heg} has been developed for use by J-PARC
experiments such as T2K. Both of these experimentally-focused generators
provide substantial technical infrastructure needed to interface with
simulations of neutrino beam production and to evaluate model uncertainties.
NuWro~\cite{Golan:2012rfa} and GiBUU~\cite{Buss:2011mx,Mosel:2019vhx}, on the other
hand, are developed primarily as an aid for theoretical investigation of the
interaction physics. These codes place greater emphasis on the quality and
self-consistency of the physics implementation but lack some features needed
to pursue an experimental analysis. Finally, the ACHILLES~\cite{achilles}
event generator is an emerging effort that combines aspects of the two previous
categories. While the code is written mostly by theorists and prioritizes a
consistent physics treatment, the long-term development plan for ACHILLES
involves full support for its use in an experimental context.

While there are strong connections to the ways in which event generators are
used in other areas, accelerator-based neutrino experiments are faced with
unique physics modeling needs that directly translate to requirements on the
interaction simulation. Many of these are driven by the need to use the
contents of the final state to infer the incident neutrino energy, which is
unknown on an event-by-event basis due to the wide energy range covered by
relevant beamlines. All standard strategies for neutrino calorimetry are
sensitive to nuclear modeling details (such as neutron emission and
intranuclear pion absorption) that are theoretically complicated to
describe~\cite{AlvarezRuso2018}. Recent work using electron beam data to
benchmark the energy reconstruction techniques used for neutrinos has revealed
substantial event generator mismodeling that must be addressed via
problematically large systematic uncertainties~\cite{e4nuNature}. Ultimately,
the high precision needed to achieve the design goals of next-generation
accelerator-based neutrino oscillation experiments will only be achieved if
considerable refinements to nuclear modeling (contemplated in the other
sections of this document) are incorporated into an event generator usable in
an experimental analysis.

Beyond the necessary improvements to theoretical understanding of
neutrino-nucleus scattering and related phenomena, there are additional
challenges for the neutrino event generator community that arise in three
areas. Organizational difficulties remain a significant hindrance to progress,
both in obtaining sufficient support for the work and in directing the existing
development efforts most effectively. Challenges impacting the usability,
maintainability, and performance of neutrino event generators as software
products will require investment to ensure the future success of the field.
Finally, there are also various issues related to the physics content of
event-generator-based interaction models and their use by experimental
collaborations. Specific needs in each of these categories are discussed in the
subsections below.

\subsection{Organizational challenges}
\label{sec:generator_org}

  \paragraph{Funding issues from multiple agencies within multiple countries.} 

Although the technical and scientific issues for neutrino event generators are
significant and will require concerted effort to address, organizational
challenges are presently the most intractable. A chronic problem faced by the
field is the difficulty in securing funding to support generator development.
One of the major causes of this situation is the inherently cross-cutting
nature of the work along multiple axes. For example, the most compelling
motivations for creating precise simulations of neutrino interactions come from
open questions in high-energy physics, but much of the needed modeling
expertise comes from the nuclear community. The traditional rigid boundaries
between these two subdisciplines are necessarily breached by neutrino event
generators, but funding mechanisms have been slow to adapt in response.
Research focused on improving generators thus falls far too often into an
illusory ``no man's land'' between categories that are less scientifically
distinct than current policy entails.

A similar state of affairs exists due to the competing demands that must be met
by a successful event generator. A mixture of theoretical, computational, and
experimental competence is required to address users' needs for high-quality
model predictions from a simulation that runs efficiently and is usable for
data interpretation if not the actual execution of an analysis. Different
facets of an event generator are valued by theory (rigor, self-consistency),
computation (performance, code design, usability), and experiment
(completeness, flexibility). As a result, attempts to gain support for the full
breadth of needed activities from any one area are unlikely to succeed. The
present lack of dedicated resources necessitates piecemeal funding strategies
that promote inefficiency, fragility, and fragmentation of expertise when
greater unity is required to pursue common goals.

Facilitating collaboration across national borders in addition to disciplinary
ones remains critical for event generator work as it has been for high-energy
physics more broadly for decades. Many of the most active theory groups
studying neutrino-nucleus scattering are based in Europe, while the largest
precision neutrino experiments are hosted by laboratories in Asia and North
America. Designing and implementing an effective global strategy for the field
will require interacting with government organizations from multiple countries
and obtaining support for joint projects, workshops, and international travel
between collaborating institutions.

\paragraph{Promoting career advancement for junior physicists involved in generator advancement. }

Enabling robust participation in generator development by a diverse and
energetic cohort of junior physicists is a related need that requires increased
attention. Research activities related to event generators, despite being
necessary and physics-laden, are poorly incentivized by the prevailing system
of scientific career advancement. In particular, event generator development is
generally not as valued by hiring committees, funding agencies, etc. as model
building (theory) and data analysis (experiment). There is no dedicated funding
stream for students and postdocs pursuing physics simulation work of this kind,
and senior scientific positions for which it can be the primary focus are
exceptionally rare. This situation makes it unattractive and inadvisable for
junior scientists to invest the time needed to become expert contributors to
event generators. It also represents a serious threat to the continuity of
generator knowledge at the very time when it is most sorely needed for neutrino
physics; the small and aging community of neutrino event generator authors is
likely to further contract in the coming years unless a substantial corrective
investment is made to strengthen the career pipeline.

\paragraph{Issues related to reduced generator support on physics results. }

The meagerness of support currently available for neutrino event generators has
practical consequences for the analysis strategies used by contemporary
experiments. Both strong incentives to produce physics results quickly as well
as a dearth of collaborators with appropriate expertise encourage quick-fix
solutions to generator modeling problems with dubious validity and reusability
outside of a specific context. An example that has begun to draw appropriate
scrutiny from the theory community~\cite{Coyle2022} is the use of ad hoc
generator model tuning by experiments like NOvA~\cite{Acero2020} and
MicroBooNE~\cite{MicroBooNE:2021ccs} to mitigate observed tensions with data.
While some degree of tuning is usually necessary given the limitations of the
available neutrino interaction models, the scope and robustness of the tuning
studies pursued so far have been severely resource-limited. As a consequence,
these tuning exercises can lead to results that are difficult to interpret
theoretically and to apply to other experiments. There is also a danger of
performing a generator tune that might cover up an exotic new physics
signal~\cite{Coyle2022}. Properly supporting greater participation in similar
future efforts, especially by theorists, will help to ensure the best possible
choice of tuning degrees of freedom, design of the parameter fitting campaign,
and physics impact of the resulting interaction model.

The unavailability of designated generator-focused resources also pushes the
field toward a haphazard development model driven more by the idiosyncratic
physics interests of groups funded for other reasons than a systematic overall
research program. Independent approaches and ideas are valuable for innovation,
and simply merging all existing efforts into a single generator code and
collaboration is unrealistic and undesirable. However, a clearer community
leadership structure and more well-defined priorities, with special attention
paid to input from the major experimental collaborations, would reduce
duplication of effort and help to ensure that improving neutrino interaction
simulations does not become a rate-limiting barrier for achieving new
discoveries. Recent discussions hosted at ECT*~\cite{ectstar2018,ectstar2019}
and Fermilab~\cite{fnalTools2020,Barrow:2020gzb,EGs2023} have been a productive
step toward greater coordination of effort among event generator groups and
other stakeholders. However, further clarification of the consensus vision and
sufficient, well-organized work towards its timely execution will only
materialize if generator development is supported at a level commensurate with
its decisive importance to the future of neutrino physics.

\subsection{Software challenges}
\label{sec:generator_software}

The embodiment of neutrino interaction models in simulation codes suitable for
general use imposes software engineering requirements far beyond what is
typically needed to produce a theory publication. Achieving a satisfactory
description of the scattering physics is the primary goal, but generator
authors must also provide efficient performance, portable and maintainable
code, clear documentation, and often a variety of supporting software tools in
order to meet users' needs. Delivering these items requires significant
technical expertise and effort which, despite being persistently undervalued
relative to other contributions, is nevertheless crucial. Expanded investment
in ``long-term development, maintenance, and user support'' of event generators
and similar scientific software was a major recommendation of the 2021 Snowmass
Comptutational Frontier report~\cite{elvira2022} and would do much to help
confront present concerns. Many of the required tasks to maintain neutrino
event generators as software products are a poor fit for ordinary university
research groups. Laboratory support for development and maintenance of
generator software infrastructure, in addition to investigation of related
scientific issues, will thus be essential going forward.

At present, computing performance of neutrino event generators is generally a
sub-leading concern in experimental workflows compared to event reconstruction
and simulation of the detector response. However, as the theoretical
sophistication of generator models and their uncertainty treatments continues
to grow, this situation is likely to change. A historical precedent exists for
experiments at the Large Hadron Collider (LHC), which have had to begin careful
consideration of event generator optimization in recent years to remove a
dominant bottleneck~\cite[Sec.~4.7.1]{Campbell2022}. Analyses from the NOvA
experiment have already benefited from running large-scale calculations of
systematic uncertainties, including those arising from its GENIE-based neutrino
interaction model~\cite{Acero2020}, using high-performance computing (HPC)
resources hosted by the National Energy Research Scientific Computing Center
(NERSC)~\cite{Sousa:2019nhd,NormanNERSC}. Although the future is likely to
place significant computational demands on neutrino event generator
performance, at present there is little anticipatory effort to address them.

A related task with potential benefits both to the computing efficiency and
modeling complexity achievable with neutrino event generators is the
opportunity to adapt to new computing techniques. Machine learning is widely
used in the analysis of neutrino data~\cite{Psihas2020} and in physics
simulation for LHC experiments~\cite{Butter2023}, but its application to
neutrino event generators is largely unexplored. Development to optimize
particle transport and detector simulations to execute on GPUs is well
underway~\cite{Tognini2022}, but none of the major neutrino event generators
has yet been refactored to maximize performance in such a workflow. Although
still in early stages of exploration, quantum computing algorithms also appear
promising for simulations of neutrino scattering~\cite{quantum,Roggero:2019myu}
and merit further study.

A more mundane but compelling need arises due to a lack of standard interfaces
between neutrino event generators and associated software tools. At present,
there is no common data format for storing simulated neutrino interaction
events, and the major generators share only the most basic conventions, such as
the particle ID codes~\cite[sec.~44, pp. 661--664]{pdg2020} defined by the
Particle Data Group (PDG), for representing the scattering physics. A concrete
proposal for such a format, based upon the HepMC3~\cite{Buckley2021} data
structures developed for LHC simulations, has recently been
released~\cite{Gardiner:2023ejq}, but its actual adoption by generator groups
and experiments will require further discussion and technical work.

There is also no universal interface between neutrino event generators and the
simulation software used to model beam production and the detector geometry in
experimental production environments. This missing infrastructure forces
experiments to largely rely on the physics models available in the single
generator most compatible with the remainder of the software stack. Widespread
recognition of the value of using multiple generators for experimental analysis
has motivated ad hoc approaches for injecting events from one generator into a
simulation chain intended for use with another (see
Ref.~\cite[Sec.~2.3]{GENIE:2021npt} and Ref.~\cite{Soplin:2023nxg}). However, design
and implementation of a standardized interface would overcome the fragility of
present strategies and lower the barrier for delivering new model calculations
to experimental collaborations.

Beyond the generators themselves and the experimental frameworks that use them,
software that enables detailed model comparisons and parameter tuning to
neutrino cross-section data will become increasingly important to achieving
precision in the coming years. For the LHC community, the Rivet
toolkit~\cite{bierlich:2019rhm} provides automated access to over 900 data sets
together with analysis tools to manipulate them and compare them to multiple
generators. The only comparable open-source tool for neutrino physics,
NUISANCE~\cite{Stowell:2016jfr}, seeks to play a similar role but lacks
dedicated support and is at a much earlier stage of development. As the
experimental neutrino cross-section literature continues to grow at an
accelerating pace, investment in further technical
enhancements~\cite[Sec.~3.2.2]{Campbell2022} to \mbox{NUISANCE} or a similar
tool, including standardization of data releases~\cite{DataStandard}, will
enable generator improvement efforts to leverage the latest data in a timely
manner.

Solving these and similar software problems will require coordination across
the entire neutrino generator landscape. Dedicated meetings explicitly devoted
to relevant technical discussions and hands-on work will be invaluable for
progress. In organizing these, the neutrino community will be following the
strong precedent of the ``PhysTeV'' series of meetings held in Les Houches,
France by collider physicists. The meeting series originally led to an
agreement on a standardized interface~\cite{Boos2001generic} for information
passing between event generators that was subsequently expanded to event file
formats~\cite{Alwall2007, Butterworth2010tools, Andersen:2014efa} and
conventions for supersymmetric physics~\cite{Allanach2004, Allanach2006}. These
``Les Houches Accords'' proved valuable for facilitating physics simulations
for the LHC, and recent meetings in the series~\cite{Bendavid2018, Amoroso2020,
Huss2023, leshouches2023} have retained proceedings related to Monte Carlo
generators as a major focus.

\subsection{Physics challenges}
\label{sec:generator_physics}

\paragraph{Scope of the problem}
Neutrino scattering theory is a highly multifaceted subject, with many
subtleties and open questions even when a single reaction mode or nuclear
effect is considered in isolation. Modeling efforts developed by nuclear and
high energy theorists therefore usually have a restricted scope (e.g. a single
process restricted to a specific energy range) to make pursuit of a detailed
understanding more tractable. This strategy is prudent and productive. However,
because the experimental program requires a practical and comprehensive
simulation of the interaction physics, computational efficiency and full
coverage of relevant processes and nuclides must unavoidably be prioritized
over rigor. An especially stringent demand is imposed by the broad range of
neutrino energies involved in accelerator-based experiments; beam fluxes for
MicroBooNE and DUNE are significant within 0.3--2~GeV and 1--20~GeV,
respectively. Generator authors are thus obliged to implement simultaneous
treatments of many relevant processes, including those where theoretical
complications, such as RPA effects in quasielastic interactions
(Sec.~\ref{CCQE}) and higher-twist corrections in deep inelastic scattering
(Sec.~\ref{sisdis}), are considerable. The resulting simulation, despite a
preference for theoretical consistency among stakeholders, is nevertheless
often a \textit{Frankenstein's} patchwork of model ingredients, with validity
caveats that may not always be stated or even fully understood.

\paragraph{Relationship to experiments}
Experiments benefit greatly from the models available in generators but can
always use more. Simulation capabilities for hypothetical processes beyond the
Standard Model are especially limited in the established neutrino generators,
and a major goal of the new ACHILLES effort is to facilitate their development
with maximum efficiency~\cite{Isaacson:2021xty}. To fully meet experimental
needs, an implementation of a theoretical calculation must be provided together
with relevant modeling uncertainties. When such uncertainties are unavailable,
either because the model authors have not yet quantified them or because the
generator code does not yet include them, makeshift techniques for estimating
uncertainties must be relied upon instead. The need for well-justified
uncertainties in generators thus further complicates the already formidable
physics modeling demands required for precision neutrino experiments.

A related issue that complicates prioritization of model improvements in
generators is the way in which experiments use data-driven constraints on the
simulation to achieve precise measurements. As described in
Sec.~\ref{sec:near_det}, oscillation analyses typically make use of near
detector data to mitigate mismodeling of the beam flux and neutrino
cross-section physics. In cases where a near detector is unavailable,
measurements of neutrino interactions in channels that are not of primary
interest, known as \textit{sidebands}, can be used to achieve a similar model
constraint. A recent example is the use of $\nu_\mu$ data in MicroBooNE to
constrain a GENIE-based prediction of the $\nu_e$ interaction
rate~\cite{microboone:2021tya}. In either of these cases, the relationship
between the \textit{prior} modeling uncertainties, which reflect a combination
of the present theory landscape and its generator implementation, and the
\textit{posterior} uncertainties, which are informed by new data and
non-trivial analysis procedures, is difficult to anticipate. Theory and
generator development effort to reduce the subset of uncertainties that can be
well-constrained by such techniques may thus have a small impact on ultimate
oscillation sensitivity, while previously sub-leading contributions may become
dominant post-constraint. The question of which interaction model uncertainties
are most important to address is therefore highly analysis-dependent, and only
continuing dialogue between experimental collaborations, generator developers,
and theorists will allow the community to fulfill needs for model improvements
in an effective way.

\paragraph{Relationship to theory}
Realistic treatments of the nuclear processes simulated by event generators are
often computationally demanding. High-quality models of neutrino scattering
thus present practical difficulties when embodied in Monte Carlo codes intended
for general use. To achieve the necessary simulation performance, generator
developers are often obliged to use approximations and software techniques that
imperfectly capture the physics content of the models being implemented. A
concrete example is the hadronic tensor framework used in GENIE as the basis
for incorporating several models of QE and 2p2h
interactions~\cite{schwehr:2016pvn, Dolan:2019bxf, GENIE:2021npt,
electronsforneutrinos:2020tbf, Barrow:2020mfy, Dolan:2021rdd,Russo:2025oph}. Within this
framework, inclusive cross sections are evaluated using precomputed tables of
nuclear responses, which are provided as external input and interpolated as a
function of the lepton kinematics. The outgoing hadronic system is attached to
the event by imparting the sampled leptonic four-momentum transfer to one or
two initial-state nucleons~\cite{sobczyk:2012ms, Katori:2013eoa}. This strategy
avoids the significant and often prohibitive computational burden of evaluating
the nuclear responses within the generator itself. However, while it is capable
of reproducing inclusive model predictions with high fidelity, this approach
has notable limitations in both physical correctness and usability. Because
hadronic degrees of freedom are integrated out of the tabulated nuclear
responses, correlations between the lepton and nucleon kinematics are ignored
in a way which leads to noticeable mismodeling of the
latter~\cite{Nikolakopoulos:2023pdw}. The input parameters used to generate the
tabulated responses are also not directly accessible in the generator itself.
This omission prevents straightforward model variations in a way that makes
related tuning and uncertainty quantification more difficult. It may also
become an obstacle to future model refinements if the external code used to
generate the response functions becomes unavailable.

Even for model calculations executed entirely inside a generator code with full
access to the underlying parameters, tradeoffs between correctness and
computational efficiency are also a significant concern for uncertainty
quantification. This is especially true for large-scale simulation campaigns
run by experiments, in which the neutrino generator is one of many software
tools used to process each event. Brute-force creation of high-statistics
samples of Monte Carlo events for each model variation of interest would easily
exceed the computing budget of the largest experimental collaborations. To
estimate relevant systematic uncertainties, neutrino experiments instead rely
upon a numerical technique called \textit{reweighting}~\cite{Dobson:2009zzb};
for a given set of simulation outputs, the impact of a model variation may be
estimated by assigning a new statistical weight to each original event.
Technical infrastructure provided by some neutrino generators allows automated
calculation of the weights corresponding to modifications of numerous physics
parameters. While reweighting of this kind is powerful and standard for
experiments, it is ill-suited for estimating some uncertainties. In particular,
the method is inappropriate for model variations that allow new regions of
phase space to be populated; events in regions not represented in the original
sample cannot be reweighted into existence. As neutrino experiments continue to
pursue higher precision, efficient evaluation of such non-reweightable
uncertainties will likely become important, and new tools and techniques must
be developed in response.

\subsection{Open challenges, future prospects}
This section on event generators has described various issues in detail.  The situation is complicated because as a bridge between theory and experiment it seems land firmly in neither effort.  Thus, funding and recruitment of young people is tricky.  The overall status can be described as too small but definitely improving.  A significant advance came about 15 years ago when four postdocs were hired in US and UK with the goal of contributing to event generator work at the 50\% level.  In addition, the level of activity from young researchers in experiments significantly increased at the same time.  This was highly successful at the time and at least two of those young people is now in a permanent position. Nevertheless, a shortfall in workers is still evident.  Until experiments regard the issues as important, most of the problems will continue.

The breadth of the effort in event generators is a central issue. Currently, there are about five programming groups devoted to this effort; each is briefly described in Sect.`\ref{fsi}.  With such a small group of workers, the need for such a large set of efforts with no central guiding leadership can be questioned.  Workshops have been held at ECT (Trieste), Fermilab, and INT (Seattle) with minimal accomplishments.  A notable advance was the HepMC3 effort to unify event formats~\cite{Gardiner:2023ejq}.  The rapid recent increase in technological capabilities of Machine Learning has provided a special challenge for optimal usage that is still evolving.  Collider experiments had a similar problem at the beginning of LHC running; CERN led the effort to standardize and prioritize the existing efforts.

\newpage
\section{Final State Interactions}
\label{fsi}
\label{sect:fsi}
\subsection{Introduction}
Final state interactions are an essential ingredient in modeling neutrino interactions and the difficulty in that task is reflected in the uncertainties in each measurement. Good control of FSI effects is necessary to correctly estimate the energy of each interacting neutrino and its uncertainties. The mean free path of all hadrons (pions, nucleons, hyperons, kaons...) inside nuclei is of order of femtometers and they typically experience FSI, especially for targets as heavy as argon.  These interactions mask the results of any principal interaction, e.g. QE or RES.  Any outgoing hadron can change charge (charge exchange, CEX), energy, or direction (inelastic scattering, INEL) as a result.  Recoil nucleons must then be separately tracked.  Even more complicated possibilities such as pion absorption (ABS) (the pion disappears and 2 or more baryons are emitted) and pion production (PIPRO) (at least one additional pion is produced) are also seen.  All generator codes used in experiments must include all these effects for a large variety of targets at a wide range of energies.  Since the effects are too complicated for a fundamental model, all codes provide numerous ways to adjust parameters or swap models.

Most FSI models (GENIE, NuWro, and NEUT) are based on semi-classical intranuclear (INC) cascade models.  They follow the basic pattern of seminal papers by N. Metropolis et al.~\cite{Metropolis:1958wvo}, which proposed an algorithm of the space-like propagation of particles. In this approach, the hadrons propagate through a continuous medium in straight sections of the length not exceeding the distance between consecutive interactions. A probability for an interaction to occur is determined with a Monte Carlo algorithm using the textbook non-interaction probability formula
$P (\Delta x) = exp(-\frac{\Delta x}{\lambda})$,
where $\lambda = \frac{1}{\rho\sigma}$ is a mean free path expressed in terms of the local nuclear density $\rho$ and  cross section $\sigma$ for the relevant total hadron-nucleon interaction.  While Metropolis et al used free hadron-nucleon cross sections, currently MC codes employ more realistic approaches including nuclear medium effects.  All codes choose the nucleon positions, momenta, and binding energies according to prescriptions based on studies of electron scattering.  For example, the model of Salcedo~\emph{et al.}~\cite{salcedo:1987md} which employs a density dependent $\pi N$ cross section is often used by all three codes. Their model uses the local Fermi-gas model for nucleon densities and incorporates medium effects on the $\Delta$ self energy.  Pandaripande and Pieper~\cite{pandharipande:1992zz} have a similar model for NN scattering that is used in GENIE and NuWro. INCL++~\cite{cugnon:2016ghr,mancusi:2014eia} has a more sophisticated version of medium effects; it is an option in GENIE and in private versions of NuWro.
\textsc{Achilles} defines a different impact parameter algorithm inspired by LHC event generator modelling of multiparton interactions~\cite{Sjostrand:1987su}.  GiBUU~\cite{Buss:2011mx} is another code which has improved physics models.  Although it is a self-consistent set of models which have a very good nuclear model, it is still based on a semi-classical description.  GiBUU depends on modelling various kinds of underlying data (same as the other codes), but has a more complete and separate choice of formulae.  It is often displayed with other generators for comparison with experimental data.  Although it is on the whole a better description of the wide body of relevant data, differences between it and the models of this study aren't large.

The traditional validation study of FSI models uses hadron-nucleus measurements, esp. the total cross sections for specific channels.  There is a good sample of channel total cross section data for pions but much less for nucleons and kaons.  On the other hand, there is a significant body of data for inclusive scattering of pions, neutrons, and protons that is not often used. A useful compilation of these data and numerous comparison to simulations is available~\cite{iaea-spall}. There is concern that these data do not correspond  to exactly the same physics situation as comes with neutrino experiments.  This has produced a new interest in hadron transparency measurements.  A new measurement from CLAS\cite{CLAS:2025fqh} at proton $|\vec{q}|$ from 0.8 to 1.4 GeV/c has provided clarity for both experiment and interpretations.  Traditional GENIE models are significantly larger than data and problems are traced to the nuclear model; the spectral function is a much better match to data than traditional LFG models.  

\subsection{Model Descriptions}
\label{sec:models}
\subsubsection{NuWro}
NuWro is a neutrino-nucleon/nucleus Monte Carlo event generator under development at Wroclaw University since 2005. It is optimized for use in accelerator-based neutrino oscillation experiments in the few-GeV energy region. NuWro covers neutrino energy region from $\sim 100$~MeV to $\sim 100$~GeV and it includes several interaction mechanisms giving predictions for all available phase space.
For scattering off free nucleons the interaction mechanisms include quasi-elastic process, hyperon production, single-pion production and more inelastic channels (DIS). In the case of nuclear target additional reaction channels are: meson exchange current (MEC),  coherent pion production, and scattering off atomic electrons.
In general, neutrino interactions are modeled in the impulse approximation scheme in which primary interaction on bound nucleon or nucleon pair is followed by final state interactions (FSI) of hadrons modeled with semiclassical INC model. NuWro allows for a variety of choices for theoretical interaction model parameters and form factors, and also for several alternative models to describe nuclear effects. The models are: local and global Fermi gas, hole spectral function, and effective momentum and density dependent potential. The NuWro version 19.02.2 used in this study uses local Fermi gas. DIS events are modeled with Bodek-Yang model for the inclusive cross section and hadronization is done with PYTHIA6. Resonant pion production is optimized for the Delta resonance region with form factors obtained as a joint fit to the old ANL and BNL experimental data.  Quasi-elastic process is described in the Llewellyn-Smith scheme with dipole axial form factor with axial mass $M_A=1.03$~GeV. MEC events are described by the Valencia model.

The most important ingredient of the cascade is a model for in-medium cross sections which adjust free hadron-nucleon cross sections.
NuWro uses theoretical models for pion~\cite{salcedo:1987md} and nucleon~\cite{pandharipande:1992zz} FSI. The cascade terminates when all the moving hadrons leave the nucleus or do not have enough kinetic energy to move through the energy barrier of 7 MeV. The remnant nucleus is typically left in an excited state.

NuWro allows for tests and comparisons with several FSI parameters to be changed at user’s convenience. E.g. the values of the mean free paths can be scaled by a constant factor. The maximal value of the stepping $\Delta x$ can be changed from the default value of 0.2 fm. There are various options for a parameterization of free differential nucleon-nucleon cross sections used in deriving in-medium $\sigma$’s. For DIS events NuWro offers several options to model formation time effects.  One can change default values of $\Lambda$ and $\Sigma$ nuclear potential in the hyperon FSI model.

\subsubsection{GENIE}

GENIE~\cite{Andreopoulos:2009rq} is widely used in experiments because it has a variety of interfaces to geometry and flux packages.  It has a variety of models from which to choose~\cite{GENIE:2021npt} and associated estimates of uncertainty.  It has models that work in the approximate neutrino energy range of 100 MeV-100 TeV for all nuclei.
The physics interaction processes used in this study are from v3.4 with the AR23 tune recently established for use by DUNE and SBN collaborations.  It contains a mixture of models.  The nuclear structure model is common to all normal modes; here, the Local Fermi Gas momentum distribution with an empirical NN correlation tail is used.  The $\nu$-nucleon models include the Nieves~\cite{Nieves:2004wx} QE model with the z-expansion axial form factor~\cite{Meyer:2016oeg} and SuSAv2~\cite{megias:2016lke,Dolan:2019bxf}
2p2h model.  For meson production, the Berger-Sehgal models are used for $N^*$ resonance production~\cite{berger:2007rq} and coherent pion production~\cite{Berger:2008xs}.  Modern treatments have a boundary between resonance and DIS processes of $W=1.95$ GeV while older models use 1.7 GeV.  At higher inelasticities, the Bodek-Yang DIS model~\cite{Bodek:2002vp} with hadronization coming from the AGKY~\cite{Yang:2009zx,GENIE:2021wox} model of $\nu$ multiplicity data. There is a smooth transition to the Pythia model at higher W.

The primary goal in this section is to study FSI.  GENIE also has 4 FSI models~\cite{GENIE:2021npt}.  These include the hN2018 (which is very similar to what is in NuWro), INCL++, and GEANT4 models.  Here, the empirically-based $hA2018$~\cite{Dytman:2009zza} is chosen.  It employs mostly standard stepping in space as described in the introduction to this section.  Hadron-nucleon cross sections all come from the SAID partial wave analysis results~\cite{said}.  The choice and execution of interaction is based on data as much as possible.  Thus, it can be labeled as data-driven.  The nuclear model is the same as is used for the neutrino interaction vertex - local Fermi gas with a correlation tail.  A nuclear medium model correction~\cite{pandharipande:1992zz} is used for nucleon FSI.  Pauli blocking has not been implemented.

For pions, a significant supply of data for a variety of targets is available for the inelastic, charge exchange, and absorption channels in addition to the total reaction cross section.  A spline dependent on pion energy and nucleus for these data is implemented for these channels.  Total cross section data for pion production is not measured for nuclei and the single nucleon values with appropriate factors are used.  Since proton-nucleus total cross section data is limited to the total reaction cross section, calculations must be used for the choice of channel.  For both pions and nucleons, kinematics are chosen through quasielastic models as this is known to dominate the response.  Although these choices are simple, agreement with a large variety of data is seen.

\subsubsection{NEUT}

NEUT~\cite{Hayato:2021heg} is a neutrino--nucleus scattering simulation program library, designed primarily to be used in atmospheric and accelerator neutrino oscillation and nucleon decay experiments.
To satisfy the demands, various neutrino interaction models are implemented in the range of 100 MeV to a few TeV.
In simulating neutrino interactions with nucleus, NEUT not only takes into account nuclear effects of the initial target nucleon and final state hadrons, but re-interactions of hadrons produced in the nucleus.
The FSI simulation works independently and can then be used to simulate e.g. nucleon decay events.

NEUT is able to simulate final state interactions of pions, eta, kaons and nucleons. Dedicated custom semi-classical cascade models are implemented for each hadron.
The density-dependent prescription by Salcedo~\emph{et al.}~\cite{salcedo:1987md} is used for pion interactions for momenta smaller than 500~MeV/c and free $\pi N$ cross sections at higher momenta.
The mean free paths are further adjusted to reproduce various pion--nucleus scattering data sets~\cite{PinzonGuerra:2018rju}.
The determination of kinematics uses $\pi N$ partial wave analysis results~\cite{Rowe:1978fb} with medium corrections of Seki~\emph{et al.}~\cite{Seki:1983sh}.

Nucleon scattering model in NEUT is based on the work
by Bertini~\emph{et al.}~\cite{Bertini:1972vz}, for MECC-7.
It uses the local Fermi-gas nucleus model and the scattered
nucleon is required to have larger momentum than the
local Fermi surface momentum ($p_F^{surf}$), which depends
on the density.  It uses the nuclear density function
extracted from the experiments.
In calculating the kinematics, effective nucleon mass
($M_N^{eff}$) is used instead of the on-shell mass
($M_N^{free}$). $M_N^{eff}$ is defined as follows:
\begin{equation}
  M_N^{eff} = \sqrt{(M_N^{free} - 8\mathrm{MeV/c^2})^2-(p_F^{surf})^2}.
\end{equation}

The pion production channels are based on a simple isobar model by Lindenbaum~\emph{et al.}~\cite{Lindenbaum:1957ec}
and all pions are assumed to be produced from the decay of resonance. The produced pions go through the pion rescattering simulation in NEUT.
In the past, most of the detectors using NEUT were insensitive to the low energy nucleons or do not have high resolution. Therefore, current version of NEUT does not simulate the low energy nucleon rescattering,\ below several tens of MeV.
\jgm{JGM: The Bodek-Yang model and the important NEUT model for DIS multiplicites of two or more are significant and should be included}
\subsubsection{\textsc{Achilles}}

The \textsc{Achilles}~\cite{Isaacson:2020wlx,Isaacson:2022cwh} code is an emerging effort in the landscape of neutrino event generators.
The approach taken tends to be motivated by the development of event generators for
the Large Hadron Collider (LHC).
For this reason, the approach to the simulation pipeline differs from those in
the traditional neutrino generators. The focus of the code is to develop a modern theory-driven event
generator with a focus on appropriately modeling physics, while also providing the flexibility for
extension by those interested in BSM~\cite{Isaacson:2021xty} or new nuclear models. There is also a focus on properly
handling theoretical systematic uncertainties and tuning.

Since \textsc{Achilles} is a relatively new effort, the generator only has quasielastic scattering and resonance production implemented.
This is handled with the Spectral Function formalism~\cite{Benhar:2006wy}.
Recently, resonance production followed by one pion emission and subsequent FSI~\cite{Isaacson:2025cnk} have been added.
There are ongoing efforts to extend \textsc{Achilles} to include MEC and DIS interaction modes in the near future.

The approach to intranuclear cascades in \textsc{Achilles} differs from those in other generators in a few
key ways.  Similar to treatment of multiparton interactions at the LHC~\cite{Sjostrand:1987su}, interaction models interpret the nucleon-nucleon cross-section as a physical size.
All struck nucleons are propagated in time and
the probability of interaction is calculated in two possible ways~\cite{Isaacson:2020wlx}.  The first option, the \textit{Cylinder} approach, is to give the nucleon to have a radius squared of $\sigma_{NN}/\pi$, as it propagates. An interaction occurs if a potential struck nucleon falls
within the cylinder traced out by the nucleon during the time step.
The second option, \textit{Gaussian}, is to use  $\sigma_{NN}/\pi$ to define the variance of a Gaussian.
In both cases, the probability is calculated using the impact parameter of the propagating and background nucleons.
Furthermore, the momentum of the struck nucleon is selected using a local Fermi Gas, with a density consistent
with the hard interaction nuclear modeling. The locations of the nucleons are fixed at the start of the
cascade and distributed either according to a mean field distribution or obtained using Green Function
Monte Carlo methods~\cite{Wiringa:2000gb, Lovato:2013cua}. The two different distributions allow a direct study of the effect NN  correlations.
In medium effects are also included~\cite{Isaacson:2022cwh}, by both propagating using an optical potential~\cite{Cooper:2009zza} and
modifying the nucleon-nucleon interactions using a density-dependent model~\cite{pandharipande:1992zz}. The handling of pions in the cascade can
also be handled through two different approaches~\cite{Isaacson:2025cnk}.
In the first approach called "Virtual Resonances" is using the DCC model~\cite{Nakamura:2015rta} for pion-nucleon interactions, and the Oset-Scaledo model for pion absorption~\cite{salcedo:1987md}. The second approach called "Propagating Resonances" directly
propagates Deltas through the cascade. Absorption is handled by a set of processes in which the pion-nucleon interaction excites a Delta, followed by a Delta-nucleon to nucleon-nucleon interaction. The interactions with the Delta are handled using models originally developed within GiBUU~\cite{Buss:2011mx}.
The inclusion of both models provides a theory-choice uncertainty associated with how resonances are handled within the cascade.

\subsection{FSI Challenges}
\label{sec:fsi-challenges}
Various challenges apply to FSI.  Primary among them are neutrals, esp. neutrons and $\pi^0$ and they become the main subject of this study. Experimental estimations of transparency are tricky as it is hard to avoid theoretical assumptions. Still, its importance is recognized for its potential role in validating FSI.

\subsubsection{Neutral particle production}
Neutrino detectors have great difficulty with neutrons because they leave no ionization track and travel significant distance before interacting.  Detection of $\pi^0$ is easier because the dominant decay is to two $\gamma$'s which then produce observable showers.  However, their extremely small lifetime means there can be no $\pi^0$ laboratory beam and it is impossible to measure $\pi^0$-nucleus cross sections directly.

Experiments to date must often rely on Monte Carlo simulations to estimate the contribution of neutral particles. MC modeling relies on isospin symmetry.  However, Coulomb effects are hard to include.  Thus, the MC codes must rely on uncertain models and hadron-nucleus data. To explore this issue, a simulation study of neutral particles was performed in TENSIONS19~\cite{Avanzini:2021qlx} that employed NEUT, NuWro, and GENIE generators using the MicroBooNE flux for an argon target.  The fraction of energy in the final state due to neutrals was largely flat in neutrino energy at roughly 30\% but relatively varied by $\pm$50\% between the generators.

Neutrons are copiously produced but can only be seen through their interactions in the detector, which produce charged particles or gammas at some distance from the neutrino vertex~\cite{MINERvA:2019wqe}. Event generators use neutron scattering data to simulate neutron FSI but have almost no confirmation of their existence in neutrino interactions.  Studies of the neutron multiplicity in MINERvA~\cite{MINERvA:2019wqe} show that the simulations are in reasonable agreement.  Much more experimental evidence is needed.  The TENSIONS19 publication also showed the fraction of energy due to neutrons in the final state as a function of $\nu$ energy for a MicroBooNE sample from various event generators~\cite{Avanzini:2021qlx}.  Here, the average fraction is roughly 25\% but significant deviations among the event generators were again seen.  Projects nearing completion will significantly improve the ability to measure neutrons in neutrino experiments~\cite{ANNIE:2023yny,T2K:2019bbb}

$\pi^0$ mesons are produced in resonance events in primary interactions, also due to FSI through charge exchange reactions. In practice, validation comes from inclusive $\pi^0$ measurements in various neutrino experiments.
Reasonable agreement with these data is often seen.  However, these data are typically subject to large statistical uncertainty and they have not yet had a strong impact on the generators.
There is also some available data for $\pi^\pm$- nucleus charge exchange cross section which can be used to benchmark MCs pion FSI modeling, although experimental uncertainties are quite large.

\subsubsection{Transparency}

Traditionally, FSI codes are validated against hadron-nucleus cross sections with emphasis on channel-dependent (exclusive) total cross sections, e.g. the total reaction cross section $\sigma_{reac}$ and the total absorption cross section $\sigma_{abs}$ for pions.   Transparency is the probability that a hadron which comes out of an interaction escapes the nucleus without FSI when averaged over all locations in the nucleus.  This matches the conditions for hadrons produced by probes such as electrons and neutrinos. Therefore, it more directly measures what is needed for FSI validation as $\sigma_{reac}$ is dominated by interactions at the periphery of the nucleus when the mean free path is very small, e.g. pions at energies near the peak of the $\Delta(1232)$ resonance.

Existing data is heavily weighted toward $\sigma_{reac}$.  Transparency data exists for a few targets for protons but but none for neutrons and there is no pion transparency data for pions less than 1 GeV kinetic energy.  More and better transparency data is needed to validate models.

Recent studies~\cite{Niewczas:2019fro,Isaacson:2020wlx,Dytman:2021ohr} have shown existing models describe the transparency data.  Both $\sigma_{reac}$ and transparency observables were studied in Ref.~\cite{Dytman:2021ohr}.  They showed that results for the two observables are closely related but transparency is much more sensitive to nuclear medium corrections, e.g. NN correlations in the ground state.  This is most visible for low energy nucleon emission and pions at energies near the $\Delta(1232)$ resonance.  To date, no tension has been seen because existing data doesn't have enough precision.

Transparency is very sensitive to the conditions of the {\it first} rescattering interaction.  Therefore, it depends very much on the stepping mechanism and the nuclear medium around each nucleon.
While NuWro, GENIE, and NEUT use a definition of closest approach in 3 dimensions, \textsc{Achilles}~\cite{Isaacson:2020wlx} uses a more sophisticated empirical approach motivated by the simulation of multiparton interactions at hadron colliders~\cite{Sjostrand:1987su}.  In addition, \textsc{Achilles} uses a superior nuclear structure code.  Further study is needed to disentangle all these effects.

\subsubsection{Low energy nucleon production}
\label{FSI:lowe}
Many detectors are insensitive to low energy nucleons (e.g. kinetic energy less
than about 40 MeV).  Therefore, MC codes are required to supply estimates of
this missing strength. The danger is that the uncertainty in these corrections
is large because this is where most generator models show significant problems.
The Liege intranuclear cascade model (INCL++)
model~\cite{cugnon:2016ghr,mancusi:2014eia} has a much better treatment of
nuclear structure than previous models in GENIE, NuWro, and NEUT. It delegates
de-excitation handling to a few supplementary codes, with
ABLA~\cite{Kelic:2009yg} being the most studied by both generator groups.
De-excitation products considered include gamma-rays, nucleons, and few-nucleon
clusters. It is now an alternative model in GENIE~\cite{GENIE:2021npt} and
NuWro~\cite{Ershova:2022jah} and the effects can be seen in
Ref.~\cite{Dytman:2021ohr}. GENIE also interfaces with the Bertini casacade
model from Geant4~\cite{Wright:2015xia} as an alternative treatment that
includes de-excitation processes.

Although there is general agreement between models of transparency for high energy protons (where the nuclear corrections are small), there is significant deviation for low energy protons.  INCL++ is seen to have significantly better agreement with proton transparency data.  It is clear that better treatments of nuclear structure and reaction mechanisms are available and they should be adopted as much as is reasonable.

\subsection{FSI Study Investigating Challenges}
To elucidate the concerns discussed above (Sect.~\ref{sec:fsi-challenges}), a dedicated study was done.  For the model choices discussed in Sect.~\ref{sec:models}, large samples were generated for each event generator.  The unoscillated $\nu_\mu$ DUNE flux was used.  While GENIE, NEUT, and NuWro all generated events from all CC and NC process, \textsc{Achilles} only includes the quasielastic process.  For this study, the NuHepMC~\cite{Gardiner:2023ejq} format which is based on the HepMC3~\cite{Buckley:2019xhk} standard used in collider and heavy-ion collisions was used.
Then, NUISANCE was modified to accept this format. This provides a streamlined method of investigating new approaches without having to modify the analysis at all.  Although the observables studied are not available in experiment, they directly show properties of the FSI code in each event generator.

\begin{figure}[ht]
    \centering
\includegraphics[page=5,width=0.5\textwidth]{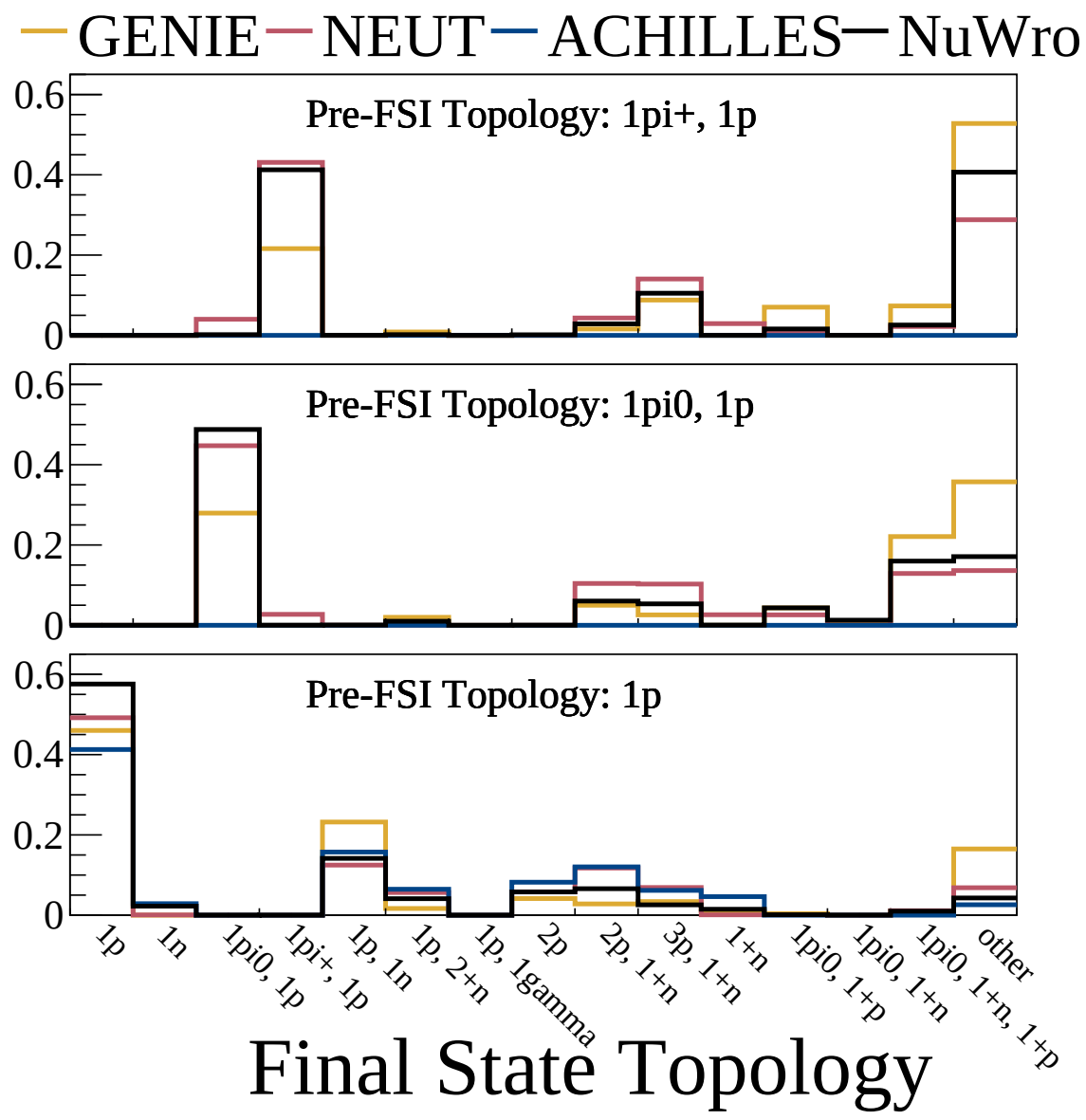}
    \caption{Results from the study of $\nu_\mu ^{40}Ar$ interactions using the DUNE unoscillated flux. The figure shows the topology smearing produced by FSI for three {\it preFSI} topologies of interest. Each pane shows the fraction of {\it postFSI} topologies produced from a given {\it preFSI} topology for the four models in the study. }
    \label{fig:fsi-channel}
\end{figure}

The first study was of neutral particle emission.
To focus on FSI effects, each simulation was divided into smaller samples, each of which were determined by the {\it preFSI} particles, i.e. those that come out of the primary $\nu$ interaction.  These events are plotted in Fig.~\ref{fig:fsi-channel} according to topology in the final state; significant differences are seen between the simulations.  Then, the neutral particle spectra in the final state are plotted according to particular {\it preFSI} combinations. To avoid further dependence on the initial state, these spectra were normalized to the GENIE sample.

\begin{figure}[h]
     \centering
     \begin{subfigure}{.65\textwidth}
       \centering
       \includegraphics[page=1,width=\textwidth]{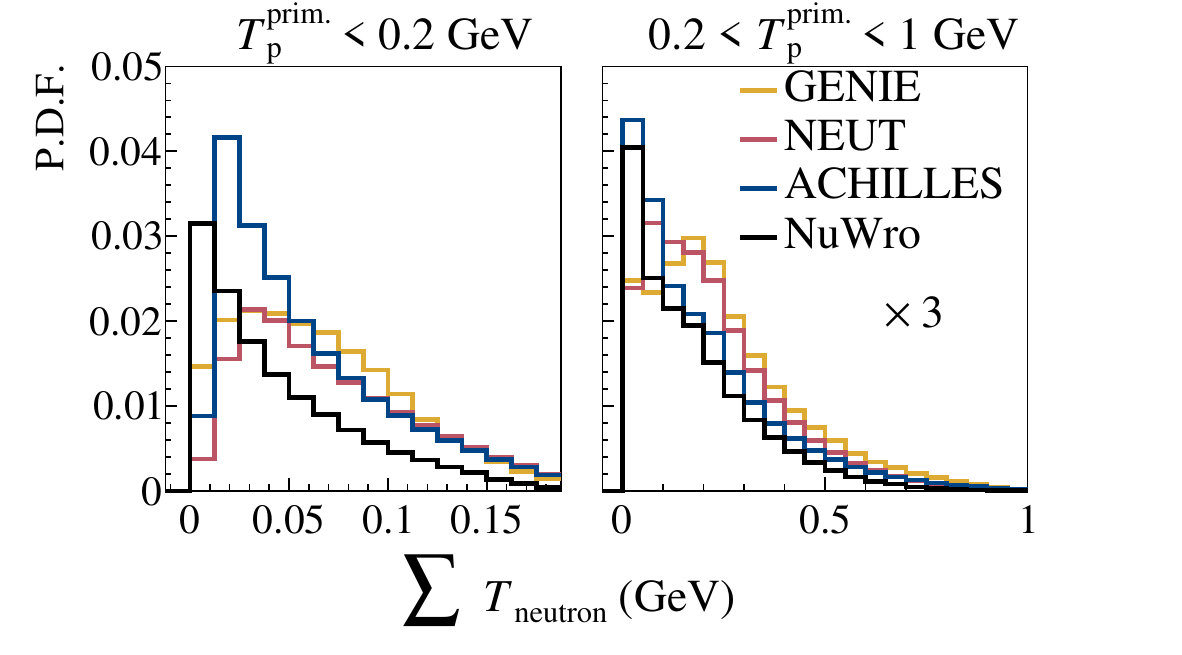}
       \caption{ }
       \label{fig:fsi-n-from-p}
     \end{subfigure}
     \begin{subfigure}{.34\textwidth}
       \centering
       \includegraphics[page=4,width=\textwidth]{Figures/FSI/fsi-allplots.pdf}
       \vspace{-0.75em}
       \caption{ }
       \label{fig:fsi-p-trans}
     \end{subfigure}
     \caption{These figures show the effect of FSI on a sample of muon neutrino--${}^{40}$Ar hard interactions, simulated with the unoscillated DUNE flux, that produce a 1$\mu$1proton {\it preFSI} topology. Four models are shown, where any differences between the hard scatter models are factorised out by a reweighting of the {\it preFSI} proton momentum spectra to match the GENIE prediction. (a) The total neutron KE spectrum {\it postFSI} in two bins of {\it preFSI} primary proton KE. (b)The fractional proton transparency. Those FSI simulations that neither change the particle topology nor scatter the proton by more than $5^\circ$ contribute to the numerator of the fractional transparency. The denominator contains all simulated events with a 1$\mu$1proton {\it preFSI} topology.}\label{figs:fsi-p}
\end{figure}

Fig.~\ref{fig:fsi-n-from-p} shows the integrated neutron kinetic energy spectra that come from initial states with a single proton (no neutrons and no pions).  This focuses on the QE interaction which is available in all codes in this study.  Since the physics changes significantly at proton kinetic energy (KE), $KE_p=200$ MeV~\cite{Dytman:2021ohr}, the data is divided into two plots for {\it preFSI} protons above and below that value.  The total outgoing neutron KE spectra are somewhat similar (within about 25\%) for $T_P^{preFSI}>200$ MeV, reflecting the dominance of single scattering processes.  However, the total neutron KE distributions for lower energy $\it preFSI$ protons show significant deviation among the models.  This is because nuclear medium effects such as nucleon-nucleon (NN) correlations, Pauli blocking, and spallation effects become important.  Experiments sensitive to these neutrons should take note of the significant model dependence seen here.

\begin{figure}[h]
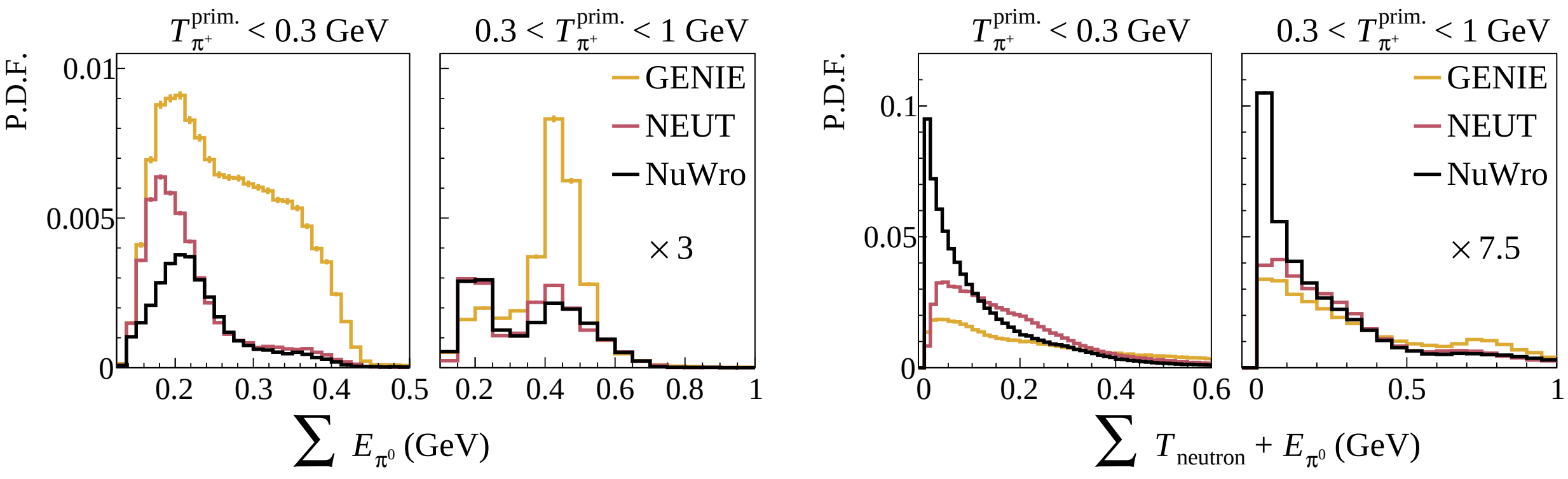

     \centering
     \begin{subfigure}{.49\textwidth}
        \hspace{-3.5em}
    \includegraphics[page=2,width=1.125\textwidth]{Figures/FSI/fsi-allplots.pdf}
       \caption{}
       \label{fig:fsi-pi0-from-pip}
     \end{subfigure}
     \begin{subfigure}{.49\textwidth}
       \hspace{-3em}
       \includegraphics[page=3,width=1.125\textwidth]{Figures/FSI/fsi-allplots.pdf}
       \caption{}
       \label{fig:fsi-pi0n-from-pip}
     \end{subfigure}
     \caption{These figures show the effect of FSI on a sample of muon neutrino--${}^{40}$Ar hard scatters, simulated with the unoscillated DUNE flux, that produce a 1$\mu$1proton1$\pi^{+}$ {\it preFSI} topology. Four models are shown, where any differences between the hard scatter models are factorised out by a reweighting of the {\it preFSI} 2D proton/$\pi^{+}$ momentum spectra to match the GENIE prediction. (a) The total neutral pion energy spectrum after hadron transport in two bins of {\it preFSI} primary $\pi^{+}$ KE. (b) The total neutron KE and neutral pion energy spectrum after hadron transport in two bins of {\it preFSI} primary $\pi^{+}$ KE.}\label{figs:fsi-pip}
 \end{figure}

The second study focuses on the RES process by selecting {\it preFSI} events with exactly 1 $\pi^+$ and 1 proton.  This corresponds largely to $\Delta(1232)$ excitation which is expected to be prominent in the DUNE data sample.  Fig.~\ref{fig:fsi-pi0-from-pip} shows the final state total $\pi^0$ KE spectra for $\it preFSI$ $\pi^+$ KE above and below 300 MeV.  This focuses on charge exchange interactions and shows deviations in its implementation among the event generators.  Fig.~\ref{fig:fsi-pi0n-from-pip} shows the sum of total neutron and $\pi^0$ KE for the same conditions.  Here, the spectra are dominated by low energy particles, likely due to neutrons similar to what was seen in Fig.~\ref{fig:fsi-n-from-p}.  Thus, it is apparent that the neutral energy comes largely as low energy neutrons independent of the $preFSI$ properties.

The last study was of proton transparency for $\nu_\mu$ interactions in argon.  The same samples were used, but now
{\it preFSI} protons are checked for being part of the final state.  If the {\it preFSI} proton is in the final state unchanged, it is considered to have avoided interaction and is included in the calculation of transparency.  Fig.~\ref{fig:fsi-p-trans} shows the results as a function of $KE_p$ for argon.  As in Ref.~\cite{Dytman:2021ohr}, this is a Monte Carlo calculation which is not determined for any specific experimental configuration other than the requirement that the proton scattering angle be larger than 5$^0$.  Unlike Ref.~\cite{Dytman:2021ohr}, there is no overlap between NEUT and GENIE at high $KE_p$.  It is interesting that although both NuWro and \textsc{ACHILLES} consider similar added nuclear medium effects, they have quite different predictions.  At low $KE_p$, there is a divergence of predictions similar to what is seen in Ref.~\cite{Dytman:2021ohr}.  Since these codes tend to agree better with $\sigma_{reac}$, this divergence is likely the result of choices in nuclear structure.

\subsection{Open challenges, future prospects}

Studies presented here are interesting because the distributions are designed to be solely dependent on FSI modeling. This is valuable because experimental results are seldom as direct.  These comparisons show significant variation and experimental measurements can only check the transparency result.  Even there, these simulations use no specific geometry and the main value is to show model dependence.  Our conclusion is that FSI uncertainty is even larger than commonly considered.  Tuning can make these disagreements go away but could cause problems elsewhere. Only improvements in theory can solve problems consistently.

FSI will always be difficult because of the problems associated with many body theory involving strong interactions.  Theoretical nuclear physicists have made excellent progress in developing nuclear wave functions in recent years and applying them to inclusive lepton scattering. There has been some work using relativistic optical potentials for proton elastic scattering.  Recent work in models for nucleon-nucleus scattering has started to be introduced into neutrino codes but more work in this direction is needed.

The general problem remains complicated.  As a result, there will always be uncertainty in treatments and recourse to fitting better and more hadron data is the best choice.  The primary data source is still total reaction cross section where data is of reasonable quality.  Total channel dependent cross sections are less common (particularly for nucleon probes) and uncertainties are often large.  Hadron-nucleus inclusive scattering data are plentiful;  coverage is often large and uncertainties are smaller than what comes with channel dependent total cross sections.  Treatments of medium corrections have improved dramatically and that is the best way to improve models.  Methods for this have been developed in the electron and hadron scattering communities and transfer to the neutrino scattering community is ongoing.

Data specifically related to neutral hadrons are less plentiful and not usually of high quality because neutral beams are hard to develop.  Nevertheless, there are many recent publications for neutron interactions, much of it at low energies.  These data should be used more.  The $\pi^0$ has a short lifetime and therefore beams are not possible.  Although various production measurements exist (see Sect.~\ref{sec:RES}), more data for $\pi^0$ production in any interaction is needed.  Electron interactions are closely related to neutrino interactions and data for electroproduction of hadrons is strongly desired.  This includes hadron differential cross sections and transparencies.  Most existing electron scattering data uses nucleon or light ion targets.  More attention to heavier targets, especially targets used in neutrino experiments, is sorely needed.  Interest is growing through the e4nu experiment which already has data and the planned LDMX experiment.  These opportunities are discussed in Sect.~\ref{eA}.

Another emerging topic in FSI modeling, largely driven by experimental searches
for rare processes, is particle emission via de-excitations of the remnant
nucleus following the cascade. This step has traditionally been neglected in the
FSI simulations used for GeV-scale neutrino interactions, but interest is
growing (see Sect.~\ref{FSI:lowe}). Liquid argon experiments have recently
demonstrated sensitivity to MeV-scale gamma-rays~\cite{ArgoNeuT:2018tvi} and
both $\beta$ and $\alpha$ decay products~\cite{MicroBooNE:2023sxs}.
De-excitation $\gamma$-rays~\cite{SKptoKaonWithGamma} and
neutrons~\cite{JUNOpnuK} have also been studied as a means of background
rejection in proton decay searches using water- and scintillator-based
detectors. Beyond the official interfaces to INCL++/ABLA present in GENIE and
NuWro, there have also been multiple published studies~\cite{Hu2022,
Cheng:2020aaw, Abe:2023iwq} coupling the TALYS de-excitation
model~\cite{koning2023talys} to one or more neutrino event generators. One of
these implements an open-source software tool called NucDeEx which is intended
for future integration with NEUT. Because TALYS is a deterministic code, all of
these implementations require inputs evaluated in advance of event generation
itself. A native Monte Carlo implementation of a de-excitation model is also
available in the MARLEY event generator~\cite{marleyPRC, marleyCPC} which
focuses on low-energy (tens-of-MeV) neutrino interactions. Overall, this aspect
of FSI modeling is still in an exploratory phase, but interest and progress is
currently accelerating~\cite{Andringa:2023aax, Castiglioni2020}.

Additional data can of course improve the situation.  Ongoing measurements with electron and neutrino experiments have a mixture of initial state and final state information that must be disentangled.  New hadron beam experiments (LARIAT and ProtoDune) have new data which is directly related to FSI issues and publications are eagerly anticipated.

\newpage
\section{Electron-Nucleus Scattering as Input to Neutrino-Nucleus Scattering}
\label{eA}
\subsection{Introduction}

While electron and neutrino interactions are different at the primary vertex, the underlying nuclear physics probed by electrons and neutrinos is intimately connected to each other. The ground-state properties of the initial nucleus and the effects of final-state interactions, both being governed by the strong interactions in the nucleus, are the same in the case of electron and neutrino interactions. Neglecting isospin breaking effects, the vector current in neutrino and electron interactions are components of an isotriplet of vector current operators, hence the vector form factors that enter in neutrino interactions are given by an isospin rotation of the electromagnetic ones, and are constrained by electron scattering measurements of protons and neutrons~\cite{osti_4093914}. The axial responses are unique to neutrinos or polarized electrons.

Electron scattering off the nucleus, governed by quantum electrodynamics, has an advantage over proton or pion scattering, dominated by strong forces. The electromagnetic interaction is well known within quantum electrodynamics and is weak compared to hadronic interaction. Hence, the interaction between the incident electron and the nucleus can be treated within the Born approximation, i.e. assuming the single-photon exchange mechanism. Therefore, various aspects of nuclear structure and dynamics influencing the neutrino-nucleus cross section can be studied in electron scattering. Any model or event generator that cannot reproduce electron scattering data ignores significant constraints and should not be trusted to provide robust results for neutrino scattering.

In typical electron scattering experiments, the incident beam energy and direction is known with good accuracy, hence the transferred energy, $\omega = E_i - E_f$, and momentum,  $\vec{q} = \vec{k}_i - \vec{k}_f$, can be precisely determined by measuring the outgoing lepton kinematics, $E_f$ and $\vec{k}_f$. 
This allows one to measure the final state as a function of energy and momentum transfer, which the nuclear response directly depends on and are not accessible to neutrino measurements.
To this end, electron-nucleus scattering experiments provide vital complementary information to test and assess the accuracy of different nuclear models and event generators intended to be used in neutrino experiments.

In the past several decades, a wealth of high-precision electron-scattering data has been collected at the inclusive level for a~variety of nuclei, ranging from $^{2}$H to $^{238}$U,
at several facilities including Bates, Saclay, Frascati, DESY, SLAC, NIKHEF, and JLab~\cite{revmodphys.80.189}. 
The ability to vary the beam energy and measure the outgoing electron (hence determining the energy and momentum transferred to the nucleus), combined with the advancement in high-intensity electron beams and high-performance spectrometers, resulted in investigating processes ranging from quasi-elastic (QE) scattering and resonance excitation to deep inelastic scattering (DIS) with great precision. A~number of those datasets were further utilized to separate the longitudinal and transverse response functions through the Rosenbluth separation. Several decades of experimental work have provided sufficient testbed to assess the validity of different theoretical frameworks, and propelled the theoretical progress staged around nuclear ground state properties, nucleon-nucleon interactions, form factors, etc. A web archive of accumulated data is maintained at the link given in Ref.~\cite{Benhar:2006er}.
A significant number of inputs to neutrino generators come from this inclusive data (nuclear binding energies, Fermi momenta, and vector form factors, for example).

Additionally, the process of proton knockout induced by electron scattering on a nuclear target, $A(e,e'p)$, gives a unique possibility for detailed
investigations of nuclear structure. Detecting the scattered electron in coincidence with the knocked-out proton, one can analyze the energy spectra of the shell-model
states, obtain nearly direct information on the momentum
distributions of the individual shells, as well as observe
deviations from the single-particle model originating from
short-range correlations between the nucleons in the nucleus~\cite{Jacob:1962lny}.

Studies of atomic nuclei through $(e,e'p)$ scattering have been performed over more than five
decades in several laboratories worldwide~\cite{Benhar:2016imd}.
In particular, experiments performed at Saclay and NIKHEF
confirmed overall validity of the shell-model description of atomic
nuclei, but also analyzed deviations from it, such as partial
depletion of the occupied states due to short-range correlations~\cite{Mougey:1980gu,deWittHuberts:1990zy}.

Over the past two decades, experiments performed in JLab reported
$(e,e'p)$ measurements for variety of complex
targets~\cite{Garrow:2001di,Rohe:2003bwf,JeffersonLabHallA:2004pgx,JeffersonLabHallA:2022cit,JeffersonLabHallA:2022ljj}---ranging from
$^{12}$C to $^{197}$Au---and for light
nuclei~\cite{JeffersonLabHallA:2004tej,JeffersonLabHallA:2004tad,CLAS:2007tee,HallA:2011gjn,JeffersonLabHallA:2019qqn}. The studies performed for argon and titanium~\cite{JeffersonLabHallA:2022cit,JeffersonLabHallA:2022ljj} are of particular importance to DUNE and other argon-based neutrino detectors.

\begin{figure}
  \centering
  \includegraphics[width=0.8\textwidth]{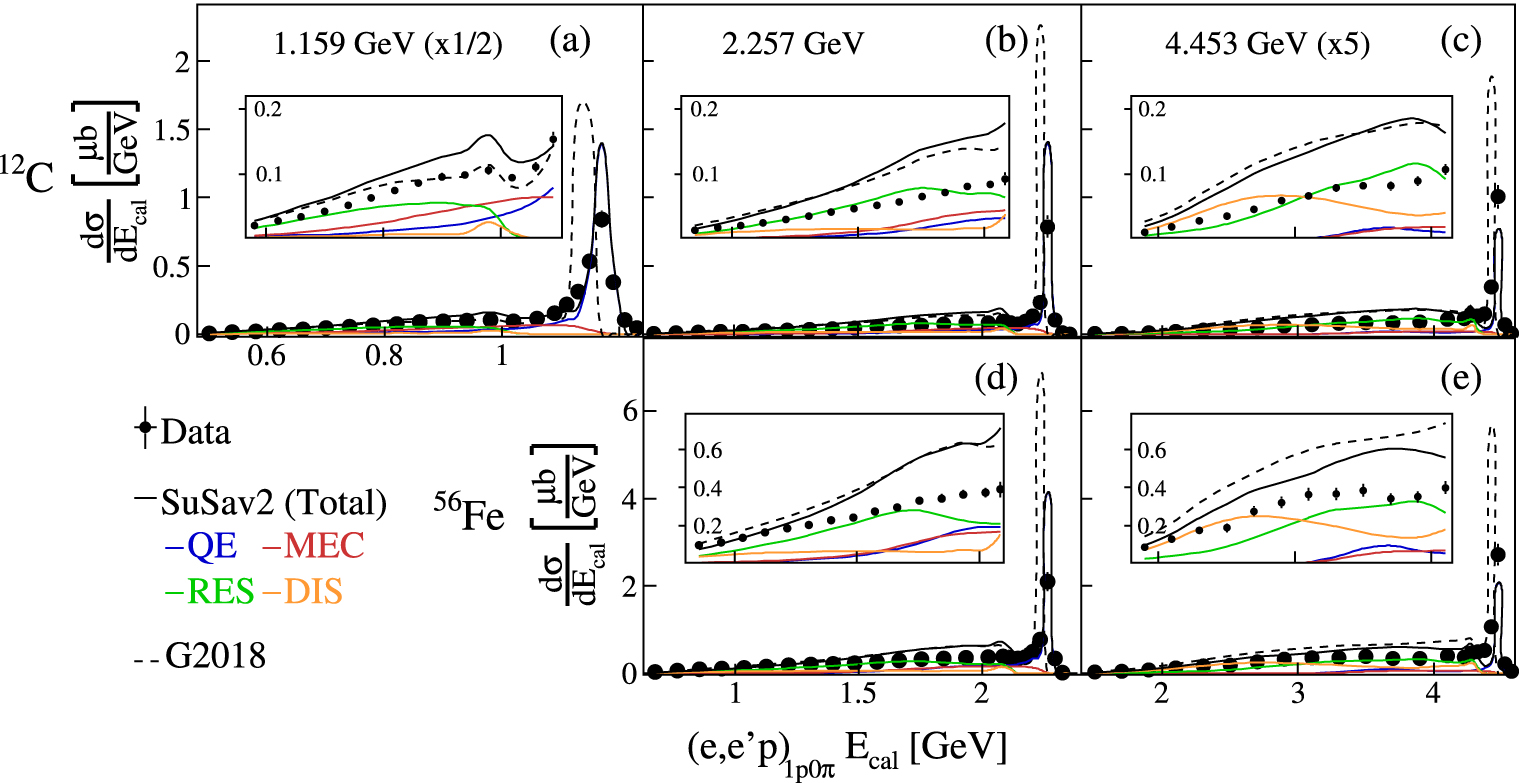}
  \caption{The 1e1p0$\pi$ cross section as a function of the reconstructed energy for CLAS data, compared to multiple predictions, for multiple beam energies on carbon (top) and iron (bottom) targets.  While the main peak location and width are quite well predicted, the tails are often wrong by a factor of two (taken from \cite{clas:2021neh}).}
  \label{fig:CLAS_data}
\end{figure}

Beyond inclusive $(e,e')$ and semi-inclusive $(e,e'p)$ measurements, there are now opportunities to measure complex hadronic final states in electron scattering experiments.
This allows one to compare fully exclusive final state predictions from event generators with data in the case that the beam energy is precisely known.
Fig. \ref{fig:CLAS_data} shows an example of this, using final states with a single proton and no pions.
The data is reconstructed under a `neutrino hypothesis' whereby the incoming beam energy is estimated from the sum of the electron and proton energies.
Due to the monoenergetic beam, this leads to a narrow peak at approximately the beam energy, but with a long tail.
This is a critical direct test of neutrino energy estimators, and this initial data suggests that considerable work is needed to correctly predict even this simplest topology.
There remain numerous opportunities for further measurements in this area which will be detailed in the next section.

\subsection{Current and Future Experimental Efforts}

A large body of high precision electron-nucleus cross sections exists, covering to some degree many nuclei and spanning a~broad kinematic region, corresponding to different reaction mechanisms. While available electron-scattering data provide important information, most measurements focus on carefully-selected kinematics to study effects of interest, and provide limited (or zero) information about the final state hadronic system.
Current and proposed measurements are tied closely to what is required for the neutrino program, in terms of expanding kinematic reach, addition of relevant nuclei, and providing information on final states of the hadronic system.
The first of these measurement programs has already demonstrated considerable challenges for models.
While past measurements have primarily been firmly in the realm of nuclear physics, these new endeavours require cross-community engagement between nuclear and high energy physics.

In Table~\ref{tab_escattering}, we present a summary of the current and planned electron-scattering experiments, primarily motivated by the needs of the accelerator-based neutrino experiments. They include complementary efforts that cover a broad range of kinematics, and carry a varied level of particle identification and other detection capabilities. For more information on details of individual experiments, we refer readers to a recent Snowmass white paper~\cite{Ankowski:2022thw}.

\begin{table}
\centering
\begin{tabular}{|c|ccc|}
\hline
\textbf{Collaborations}	& \textbf{Kinematics}	& \textbf{Targets} &  \textbf{Scattering}\\
\hline
{\bf E12-14-012 (JLab)} & $E_e$ = 2.222 GeV & Ar, Ti & ($e,e'$) \\
{(Data collected: 2017)} & $15.5^\circ\leq{\theta_e}\leq21.5^\circ$ & Al, C & $(e, e'p)$ \\
\hline
{\bf e4nu/CLAS (JLab)} & $E_e$ = 1, 2, 4, 6 GeV& H, D, He, & ($e,e'$) \\
{(Data collected: 1999, 2022)} & ${\theta_e} > 5^\circ$  & C, Ar, $^{40}$Ca, & $e, p, n, \pi,\gamma$ \\
 &  & $^{48}$Ca, Fe, Sn & in the final state \\
\hline
{\bf LDMX (SLAC)} & $E_e$ = 4.0, 8.0 GeV & & ($e,e'$) \\
  {(Planned)} & ${\theta_e} < 40^\circ$ & W, Ti, Al & $e, p, n, \pi, \gamma$ \\
 &  &  &  in the final state \\
\hline
{\bf A1 (MAMI)} & 50 MeV $\leq E_e \leq 1.5$ GeV & H, D, He & ($e,e'$) \\
 (Data collected: 2020) & $7^\circ\leq{\theta_e} \leq 160^\circ$ & C, O, Al & 2 additional \\
 (More data planned) &  & Ca, Ar, Xe & charged particles  \\
\hline
{\bf eALBA} & $E_e$ = 500 MeV & C, CH & ($e,e'$) \\
 {(Planned)} & ~~~~to few GeV & Be, Ca & \\
\hline
\end{tabular}
\caption{Current and planned electron scattering experiments whose goals are tied to neutrino program. Table taken from Ref.~\cite{Ankowski:2022thw}.\label{tab_escattering}}
\end{table}

\begin{figure}
\includegraphics[width=1.0\textwidth]{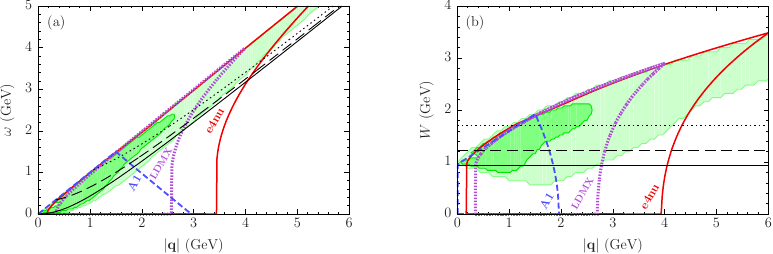}
\caption{Kinematic coverage of the ongoing and planned electron scattering (on targets including argon and titanium) experiments  presented in the (a) $(|\mathbf{q}|, \omega)$ and (b) $(|\mathbf{q}|, W)$ planes. The thin solid, dashed, and dotted lines correspond to the kinematics of quasielastic scattering, $\Delta$ excitation, and the onset of deep-inelastic scattering at $W = 1.7$ GeV on free nucleons. The light and dark shaded areas cover 68\% and 95\% of charged-current $\nu_{\mu}$Ar events expected in the DUNE near detector. Figure taken from Ref.~\cite{Ankowski:2022thw}.\label{fig_escattering}}
\end{figure}

The kinematic reaches of those experiments are presented in Fig.~\ref{fig_escattering}, where they are overlaid on the regions expected to contain 68\% (light-shaded) and 95\% (dark-shaded) of charged current interactions of muon neutrinos with argon in the DUNE near detector~\cite{DUNE:2021cuw}, as estimated using GENIE 3.0.6. The e4nu experiment at JLab~\cite{clas:2021neh} employs a broad range of energies and has the potential to study a significant part of the phase space relevant to DUNE. The 4-GeV (expected to increase to 8~GeV shortly after data-taking begins) beam energy of the LDMX experiment at SLAC~\cite{Ankowski:2019mfd} closely corresponds to the average neutrino energy in DUNE and the detector's large hadronic calorimeter will provide opportunities to measure neutron production across a range of kinematics.
Both e4nu and LDMX include  expect to be able to measure exclusive cross sections, using wide-angle trackers and calorimeters to measure the full hadronic final state.
The A1 experiment at MAMI covers a broad range of scattering angles---from $7^\circ$ to $160^\circ$---and beam energies---from $\sim$50 MeV to 1.5 GeV--- using three spectrometers.  This enables extensive studies of the quasielastic and $\Delta$ peaks, including semi-inclusive final state measurements.

Those programs, at various stages of operation, complement each other in the kinematic reach and the physics goals: the A1 and eALBA efforts are designed to probe the low-energy region and to focus on the QE scattering, while the JLab and SLAC programs can reach higher energies and can study inelastic processes.

A1 and eALBA will measure inclusive cross sections, the JLab experiment E12-14-012 measured both inclusive~\cite{Murphy:2019wed} and exclusive electron scattering cross section and measured spectral function of argon and titanium nuclei~\cite{JeffersonLabHallA:2022cit, JeffersonLabHallA:2022ljj}. It can measure protons with $-50.0^\circ\leq{\theta_p}\leq-39.0^\circ$. 

The goals of e4nu and LDMX are to measure exclusive final states. Thanks to the wide acceptance and low thresholds for detecting various particles, including neutrons, e4nu and LDMX will perform unprecedented studies of fundamental importance for neutrino experiments employing the calorimetric method of energy reconstruction. Specifically, LDMX is expected to measure forward-going particles with scattering angle up to $40^\circ$, with additional `tagging' (i.e. veto) capabilities up to around $80^\circ$. 
The e4nu collaboration utilises past data from the CEBAF Large Acceptance Spectrometer (CLAS) (1999)~\cite{CLAS:2003umf}, and data from its successor the CLAS12~\cite{Burkert:2020akg}. 
CLAS (CLAS12) does not measure particles with scattering angle below $15^\circ(5^\circ)$, but has a large solid angle coverage of $2\pi(3\pi)$ and can reach backward going particles up to $135^\circ$. CLAS was also able to detect protons with momentum above 300~MeV/c and charged pions above 150~MeV/c, while CLAS12 can detect protons and neutrons with momentum above 400 MeV/c and charged pions above 200~MeV/c.

These measurement programs offer an incredibly rigorous test of neutrino event generators, providing measurements of complete hadronic final states as a function of energy and momentum transfer to the nucleus, with small uncertainties.

\subsection{Expected challenges}

Due to the fact that essentially all neutrino cross section measurements are averaged over a broad flux, existing neutrino data is not usually able to uniquely determine the nuclear response.
Fortunately, the similarities of neutrino and electron interactions with atomic nuclei create an opportunity to greatly reduce the nuclear uncertainties, leaving only the axial contributions to the cross sections to be determined from the neutrino data. However, for this opportunity to be fully realized, an abundance of relevant electron-scattering data must be met with generators that include a consistent treatment of neutrino and electron interactions.
In doing so, the following challenges can be expected:

\begin{itemize}
    \item Long data-collection cycles

The long lead time for measurements---spanning proposal writing, data collection, analysis, and publication---necessitates that the community be adept at anticipating future needs.

\item Prioritizing measurements

Without a survey of the entire kinematic region of relevance, it is difficult to determine which measurements should be prioritized. However, interactions at the kinematics where various scattering mechanisms are significant can be expected to be the most challenging to model. In particular, the transition region from resonance excitation to deep-inelastic scattering deserves special attention.
Supporting general-purpose detectors that can make many measurements (such as CLAS and LDMX) provides for future flexibility in prioritization at the level of data analysis.

\item Defining sufficient data coverage

Currently, it remains an open issue to determine what data coverage is necessary, especially for semi-inclusive and exclusive measurements where no detector will be able to provide perfect kinematic coverage.
\item Performing increasingly exclusive measurements

It is of fundamental importance to go beyond the inclusive cross sections. 
Semi-inclusive and exclusive measurements will open new avenues of studies, both on the experimental and theoretical side.
Measurements will need to declare their kinematic coverage rather than correcting for unmeasured phase space, as neutrino experiments are increasingly doing.

\item Data interpretation

While it is reasonably straightforward to measure an electron scattering cross section, using any given data measurement to directly reduce uncertainties of neutrino experiments is non-trivial.
The only way this can be done in a general way is to fit model parameters to electron data and then use an internally consistent model at a neutrino experiment, but this relies on having self-consistent models which can't rely on parametrizations that only apply to either neutrinos or electrons.

\item Support for theory and Monte Carlo generator developments

The accuracy of the description of nuclear effects can only progress if the measurements go hand in hand with the developments in theory and their implementations in Monte Carlo generators. It is in the vital interest of the community to strongly support those efforts.  In particular, ensuring models included in neutrino event generators are able to also predict electron-nucleus cross sections in a consistent manner.

\end{itemize}

The cross section description is a fundamental ingredient of oscillation analyses, in particular due to the cross section linking the neutrino energy with the observable final state.
Therefore, it needs to be made abundantly clear to the entire neutrino community that, for the oscillation analysis to be reliable, the underlying cross-section description must be extensively demonstrated to be accurate in comparisons to both neutrino measurements and electron-scattering measurements.

 \newpage
\section{Quasi-Elastic, Quasi-Elastic-Like Scattering}
\label{CCQE}
\subsection{History}
Some of the first high-statistics data measuring pionless interactions on nuclei were from MiniBooNE~\cite{MiniBooNE:2010bsu} and NOMAD~\cite{NOMAD:2009qmu}.
At the time, standard predictions relied on modelling the nuclear target as a relativistic Fermi gas nuclear ground state using the Smith-Moniz formalism~\cite{Smith:1972xh}, dipole form factors, and simple cascade models for final state interactions.
These measurements made it clear that there were inconsistencies between the data and the best theory at the time.

The MiniBooNE CCQE-like measurement~\cite{MiniBooNE:2010bsu} was the first to measure flux-integrated topology-defined cross sections, mitigating some of the corrections to the data that rely on modelling of the process being measured.
Since then, T2K~\cite{T2K:2011qtm}, MINERvA~\cite{MINERvA:2013zvz} and many other experiments have produced similar measurements of the cross section of neutrino interactions with final state particle topologies that contain no pions. 

Over the past decade, significant strides have been made in understanding the limitations of neutrino interaction model predictions, using a combination of measurements and theory.
Using inclusive measurements (where a charged lepton is seen, and no pions are observed, but inclusive across other hadrons) and semi-inclusive measurements (requiring the presence of one or more protons, for example) it has been possible to better constrain the regions of kinematics phase space in which measurements disagree with predictions.
Much of the recent progress in cross section measurements and comparisons with models is discussed in~\cite{Avanzini:2021qlx}.
At the same time, theoretical predictions have incorporated much more advanced nuclear ground state models, multinucleon effects, and improved form factor predictions.  Initial steps have been taken to include correlations and other collective effects.

Significant strides have also been made in developing means to extract neutrino interaction cross sections that are as unbiased from the input model used and ensure as good uncertainty coverage properties as possible. The state of such work was summeraised at the recent \texttt{NuXTract} workshop~\cite{NuXTract}. Overall, this is manifest in: 
\begin{itemize}
\item the choice of topology-based signal definitions with phase space restrictions matching experimental particle tracking thresholds; 
\item the use of observables that are directly experimentally accessible (e.g. the four momentum transfer reconstructed under a CCQE hypothesis, $Q^2_{QE}$, rather than true $Q^2$, for example in~\cite{MINERvA:2015jih}); 
\item recent T2K and MicroBooNE efforts to provide results that are free from regularisation bias and associated uncertainty under coverage (see e.g.~\cite{T2K:2018rnz} as an example from T2K or~\cite{Tang_2017} used by MicroBooNE);
\item analyses performed in high dimensional spaces to mitigate biases from efficiency corrections (see e.g.~\cite{T2K:2018rnz}).
\end{itemize}

While these considerations are not exclusive to this final state, continued development and advocacy of analysis techniques and tools to ensure that our growing library of pionless cross section measurements remain as free from model dependence as possible will likely be of fundamental importance as the current challenges in model-measurement discrepancies are confronted with more precise predictions and higher statistics measurements.

\subsection{Inclusive pionless interactions}\label{pionless}
Inclusive pionless interactions are defined by the lack of a pion in the final state\footnote{Note that ``pionless'' is sometimes taken to mean mesonless. Most pionless, ``CC0$\pi$'' or ``CCQE-like'' measurements employ signal definitions that allow only a single lepton, nucleons and potentially low energy photons in the final state.}, but are inclusive over all other hadronic final states.
As such, the lepton kinematics are the only final state component that is common across all interactions.
Many microscopic calculations utilise symmetries which allow the complexity of an inclusive calculation to be significantly reduced compared to one that requires restrictions in the outgoing hadron kinematics (i.e. a semi-inclusive measurement)~\cite{Moreno:2014kia}.

\subsubsection{Theory}

Pionless interactions are highly sensitive to nuclear effects, especially at low energy and momentum transfers.
While many of these are discussed in section \ref{nuclear}, the inclusive cross section is sensitive to some particularly important aspects of the nuclear nuclear response:

\begin{itemize}
\item The inclusive cross section at {\bf low energy and momentum transfer} is sensitive to collective excitations of the nucleus, i.e. long-range correlations (LRCs), and the consistent description of initial and final-states~\cite{Nikolakopoulos:2019qcr}. For example, excitation of giant resonances (GR) is possible~\cite{RevModPhys.47.713,Snover:1986dg}.
The LRC are often taken into account in the Random-Phase approximation (RPA)~\cite{Martini09, Nieves:2017lij}. 
The GRs in particular are taken into account in the CRPA approach of Refs.~\cite{jachowicz:2002rr, Jachowicz:2019eul, pandey:2014tza}.
As discussed in section~\ref{nuclear}, RPA treatments differ in the set of 'bare' states and the residual interaction used~\cite{Martini:2016eec}.
Uncertainties underlying these RPA-like corrections (e.g. the parametrization of the residual interaction) are presently not quantified for neutrino interactions.

\item It is known that {\bf short-range correlations} (SRCs) lead to a depletion of mean-field strength, and a distinctive high-momentum tail in the nucleon momentum distribution~\cite{RevModPhys.65.817}. 
This leads to a reduction of strength in the quasielastic peak and a more pronounced tail in the inclusive response~\cite{revmodphys.80.189}.
These effects may be taken into account effectively through the single or two-nucleon spectral function as discussed in section~\ref{section:SFA}.
However, since these SRC pairs with large relative momentum lead predominantly to two nucleon knockout, it is important to understand the relation of SRCs to two-body currents (e.g. meson exchange currents) and their interference~\cite{PhysRevC.94.024611, PhysRevC.95.054611}.

\item { \bf The interference between one and two body currents} leads to non-negligible contributions in the transverse electromagnetic response in the region of the quasielastic peak. The importance of two-body currents in the transverse response has of course been established in ab-initio calculation~\cite{Lovato:2016gkq,PhysRevC.65.024002, Carlson:2014vla}, but different contributions to the response cannot be easily separated in this case.
Model calculations agree that the excitation of the delta plays an important role in the interference contribution~\cite{Amaro:2003yd, PhysRevC.55.338, VanderSluys:1995rp,Casale:2025zac, Casale:2025avv, Casale:2025wsg, Franco-Munoz23, Franco-Munoz:2022jcl, Lovato:2023khk}.
However, there is disagreement on the relative phases of delta and other contributions, and the possible role of correlations. 
A consistent treatment of resonance and pion-exchange terms, e.g. through $\Delta$-EFT~\cite{Piarulli:2016vel, Piarulli:2019cqu}, might resolve these discrepancies. 
Explicit calculations that include the axial current are more scarce~\cite{PhysRevC.95.054611, Casale:2025avv, Lovato:2023khk}.
Green's function Monte Carlo (GFMC) results for the weak interaction show that also the longitudinal axial-axial responses are affected by two-body currents~\cite{Lovato:2020kba, Lovato:2017cux}.

\item {\bf Interference between two-body currents in the delta region and pion-absorption:} A major contribution to the $2p2h$ response corresponds to MEC diagrams that involve the excitation of the delta resonance.
This contribution overlaps kinematically with single pion production off the delta, and may be related to pion-absorption on two nucleons. 
Hence, MEC and pion production mechanisms need to be treated consistently in order to avoid double counting of the inclusive rate. The possible interference and complementarity of MEC and pion absorption has to be clarified.
\end{itemize}

\subsubsection{Experiment}
\label{sec:qeinc}
For some detector technologies, inclusive pionless measurements are much easier from an experimental perspective than their semi-inclusive counterparts. A single muon or a muon with accompanying protons is a relatively clean signal without a great number of neutrino-induced backgrounds.
Moreover, the signal selection efficiency is expected to depend strongly only on two physics-sensitive kinematic variables (the outgoing lepton momentum and angle
) making efficiency corrections with minimal model dependence much easier than the semi-inclusive case. 

Measuring the inclusive cross section over a broad band beam with an uncertain normalisation suffers from the challenge that many microphysical effects can have similar impacts to the cross section predictions.  However, additional sensitivity to these different effects can be achieved by reporting correlations (or ratios, differences, sums or asymmetries of) different inclusive measurements. For example, a joint measurement of the neutrino and anti-neutrino inclusive pionless cross section has been shown to be sensitive to 2p2h modelling~\cite{Martini:2010ex,T2K:2020nqo}. Similarly, making the same measurement on different nuclear targets can probe the nuclear dependence of different nuclear effects, whilst making correlated measurements in detectors exposed to different neutrino beam energies can allow a model-independent probe of the energy dependence of cross sections (for example ``PRISM'' analyses~\cite{nuPRISM:2014mzw,T2K:2023qjb}).

Since the previous NuSTEC white paper in 2017 paper, a variety of new measurements have helped to develop a substantially more complete experimental picture of inclusive muon kinematics in pionless interactions over a wide range of neutrino energies, nuclear targets and neutrino species:
\begin{itemize}
\item T2K's contributions to this are five new measurements focusing on analyses which are correlated between: neutrinos and anti-neutrinos~\cite{T2K:2020sbd}, carbon and oxygen nuclear targets using two different detectors~\cite{T2K:2020jav,T2K:2020txr,T2K:2019ddy} and a joint on- and off-axis detector analysis~\cite{T2K:2023qjb}, in addition to a specialised measurement on a water target~\cite{T2K:2017qxv}.
\item There are another five such measurements from MINERvA corresponding to: neutrino and anti-neutrino measurements on a hydrocarbon target in their low energy ($<E_\nu>\sim$3.5~GeV)~\cite{MINERvA:2018vjb,MINERvA:2018hqn} and medium energy ($<E_\nu>\sim$6.0~GeV)~\cite{MINERvA:2019gsf,MINERvA:2022bno} beams in addition to a measurement on multiple nuclear target in the medium energy beam~\cite{MINERvA:2023kuz}.

\item Additionally, there is a recent measurement from MicroBooNE \cite{MicroBooNE:2025ooi}, which also shows tension with a variety of generator predictions in the forward region.
\end{itemize}

\subsubsection{Current model-measurement comparisons}

\begin{figure*}[h]
\begin{center}
\includegraphics[trim={0 0 2.3cm 1cm},clip,width=0.35\textwidth]{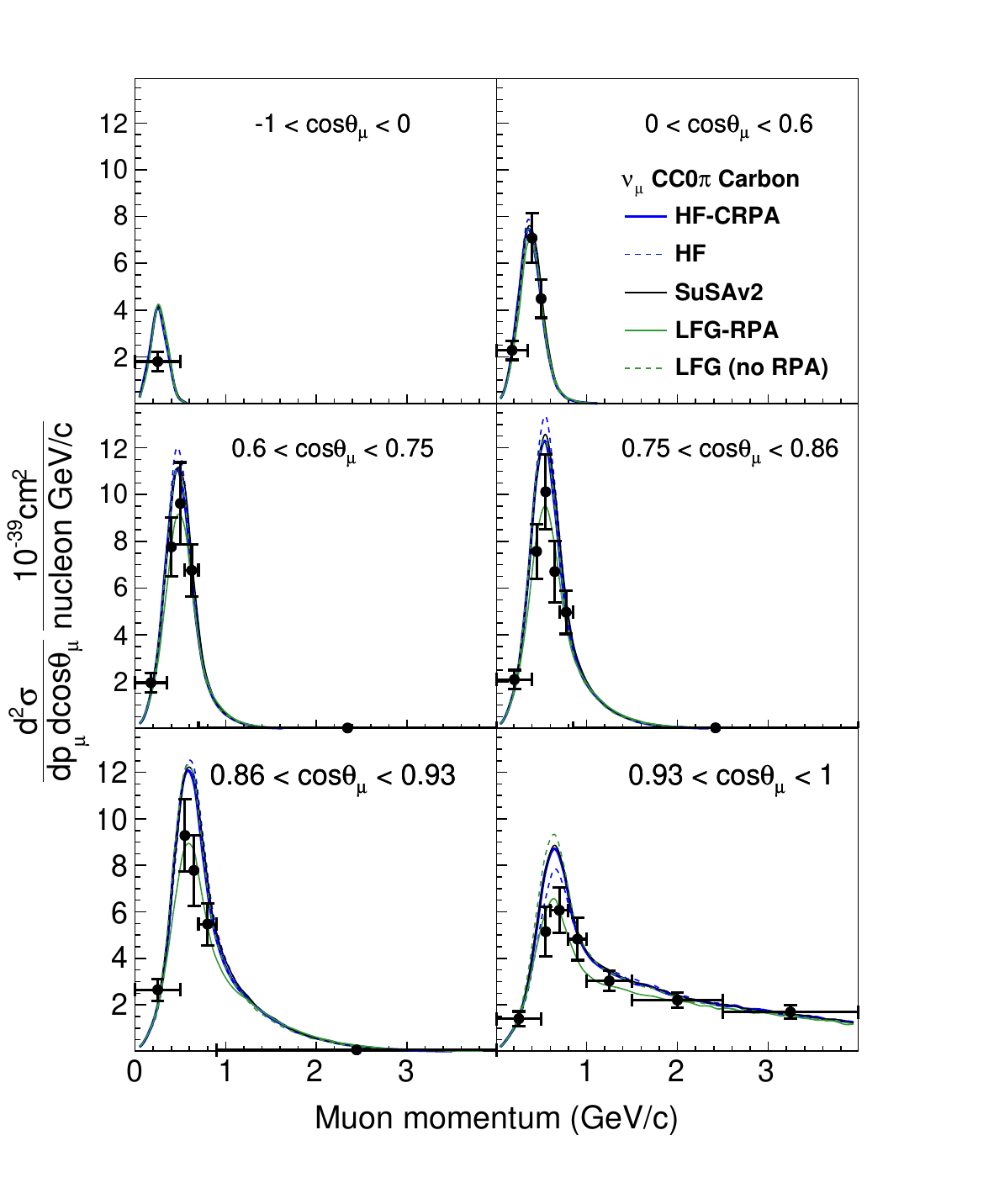}
\includegraphics[trim={3cm 0 4cm 0},clip,width=0.6\textwidth]{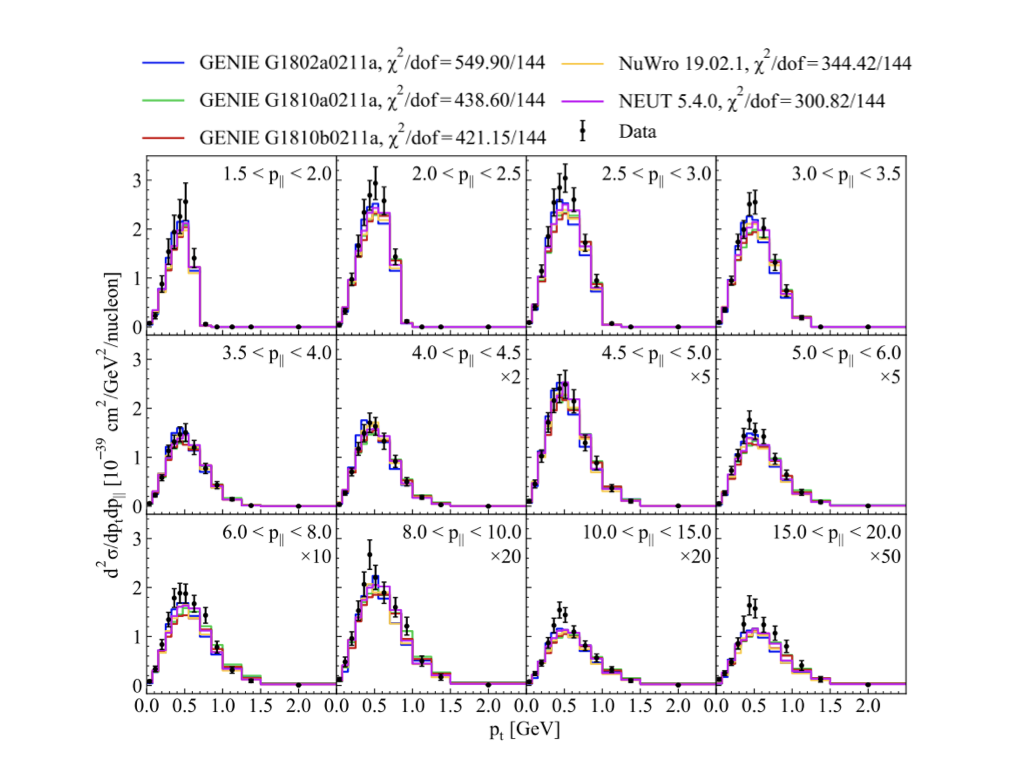}
    \caption{\textbf{Left:} A comparison of various model predictions with a recent T2K cross section measurement of CC0$\pi$ interactions on a carbon target~\cite{T2K:2020jav}, taken from Ref.~\cite{Dolan:2021rdd}. The only model not quantitatively rejected by the T2K measurement at more than three $\sigma$ 
is the LFG-RPA model from the Valencia group. \textbf{Right:} A comparison of various model predictions with a recent MINERvA cross section measurement of CC0$\pi$ interactions on a hydrocarbon target~\cite{MINERvA:2018hqn}, taken from Ref.~\cite{Avanzini:2021qlx}. The $\chi^2$ shown in the legend show a rejection of all considered models.}
    \label{fig:IncCC0piComp}
\end{center}
\end{figure*}

\begin{itemize}
    \item The new abundance of measurements should enable powerful constraints on models of pionless interactions. There are, however, important challenges in multi-measurement model comparisons due to correlations between them. In particular, the full sensitivity of a joint analysis across multiple  flavours, energy spectra or nuclear targets can only be realised if correlations are provided (for example, this permits the cancellation of many systematic uncertainties in derivations of ratios). This is especially important for the largely systematic-limited measurements from MINERvA. Without these correlations the power of all measurements cannot be effectively used.
    \begin{itemize}
        \item Each of T2K's new measurements provide the necessary correlations between the flavours, energy spectra or nuclear targets under consideration within each measurements, but correlations between measurements are not available.
        \item MINERvA does not generally provide correlations between measurements, but ratio measurements and accompanying covariance matrices are provided. Overall, this allows conclusions to be drawn from comparisons of models to each measurement separately, but any quantitative assessment of model agreements across measurements should be treated with caution (for more details, this issue is discussed in Ref.~\cite{MINERvA:2019kfr}).
    \end{itemize} 

    \item No model has been shown to describe available global data. Comparisons of the models to the new measurements from MINERvA and T2K can be found both in the experimental results paper, the recent ``TENSIONS'' white paper~\cite{Avanzini:2021qlx} and in a series of papers from theory groups~\cite{Lovato:2023khk, Sobczyk:2023mey, Dolan:2021rdd, Chakrani:2023htw, Tomalak:2022xup, Bonus:2020yrd, Barbaro:2021psv, Chanfray:2021tsp, Jachowicz:2021ieb, Martinez-Consentino:2023dbt, Kim:2022rfw, Amaro:2021sec}.
    In general, assuming a Gaussian interpretation of the $\chi^2$ quoted in model comparisons, no model has been shown give quantitatively good agreement (excluded with $p<0.01$) with any of the new measurements from MINERvA or T2K with the exception of the T2K joint measurement on carbon and oxygen.
    \begin{itemize}
        \item Note that this Gaussian uncertainty assertion present in cross section data releases from most experimental data releases is not necessarily correct and may be responsible for some of the very strong model rejections~\cite{DAgostini:1993arp,Chakrani:2023htw,Koch:2021yda}.
        \item Challenges with correct comprehensive assessment of flux shape uncertainties have also been raised~\cite{Koch:2020oyf}.
    \end{itemize}

    \item T2K measurements show broad qualitative agreement with most models, except for at forward angles where model predictions can differ dramatically~\cite{Wilkinson:2022dyx}. Compared to most models, T2K measurements at forward angles seem to require additional suppression, although it remains unclear whether this should be applied to the CCQE or non-QE contributions\footnote{a (planned) T2K measurement similar to Ref.~\cite{MINERvA:2022mnw} may be able to clarify this.}.
    One exception to the consistent model rejection by the measurements is that the Valencia model~\cite{nieves:2011pp} description of the T2K joint measurement on carbon and oxygen achieves good agreement even in the forward angle region thanks to a heavy suppression of the CCQE cross section from a strong RPA correction~\cite{T2K:2020jav,Dolan:2021rdd,Avanzini:2021qlx}\footnote{Similar conclusions have been obtained in Ref.~\cite{Russo:2025oph} by employing the RPA-based Lyon model.}. Conversely, MINERvA measurements seem to suggest a need for an \textit{enhancement} of model predictions in the most forward regions.
    These broad conclusions are concisely illustrated in Fig.~\ref{fig:IncCC0piComp}:
    \begin{itemize}
        \item A comparison to the carbon component of the T2K cross section measurement from Ref.~\cite{T2K:2020jav} shows that the only model able to describe the most forward region is ``LFG-RPA'' which is the aforementioned Valencia model, whilst the other models overestimate.
        \item  The MINERvA measurement from Ref.~\cite{MINERvA:2018hqn} is also shown compared to similar models (the G18\_10* configurations are very similar to LFG-RPA), which conversely shows an underestimation of the measurement from all models.
    \end{itemize}

    \item The poor model-measurement agreement does not appear to be a consequence of only using the central value model predictions and not considering the range of applicable modelling uncertainties. 
    \begin{itemize}
        \item A recent fit to the T2K joint measurement on carbon and oxygen with a Spectral Function baseline model is able to achieve a reasonable agreement once a set of uncertainties is considered~\cite{Chakrani:2023htw}, but the fitted model remains incompatible with other cross section measurements. 
        \item Recent fits ~\cite{MicroBooNE:2021ccs, GENIE:2022qrc} remain incompatible with the inclusive measurements they fit even after consideration of uncertainties.
        \item Overall, the complexity and computational cost of quantitatively evaluating agreement while considering model uncertainties means only a handful of such analyses are available.
        \item Ideally tools such as NUISANCE~\cite{Stowell:2016jfr} and GENIE's fitting functionality will be further developed to allow future fits to cross section measurements more tractable. 
    \end{itemize}  
\end{itemize}

\subsubsection{Prospects}

As described, inclusive pionless measurements provide powerful benchmarks that are sensitive to some of the most important nuclear physics effects that concern current and future neutrino oscillation experiments.
The provision of future correlated measurements are likely to provide sets of inclusive measurements which offer much more power to determine the cause of observed poor model-measurement comparisons. 

\begin{itemize}
\item 
T2K is expected to extend its correlated measurements with larger statistics, and improved acceptance thanks to its recently installed upgrade to its near detector~\cite{T2K:2019bbb}. 
Additionally, T2K will provide data from the WAGASCI water-based detector~\cite{Ovsiannikova:2016add}, as well as its on-axis detectors and ND280, providing two different targets, across three different energy spectra, for neutrino and anti-neutrino interactions.

\item MINERvA's data preservation efforts~\cite{Fine:2020snd} offer the opportunity for future reanalyses, which can provide a similar nucleus dependency but at higher energies than T2K.

\item MicroBooNE, SBND and ICARUS all offer the potential to provide exceptionally informative future inclusive pionless measurements. In particular MicroBooNE has recently presented plans to provide detailed correlations between future measurements similarly to T2K~\cite{MicroBooNE:2024tmp,t2k:2020lrr}. The similar energies of the BNB and T2K beamline provide the potential to study A-dependence, whilst ICARUS's simultaneous operation in the BNB and NuMI beamlines offer prospects for high-statistics measurements at two very different neutrino energy spectra to probe energy dependence. SBND will soon collect enormous data with excellent resolution to enable more granular measurements. 

\item Another intriguing prospect provided by SBND is the potential to use vertex position within the detector as a means to explore different energy spectra: SBND operates close enough to the neutrino beamline that the spectra is expected to differ significantly within it~\cite{del2021sbnd}. Such ``SBND-PRISM'' measurements offer an additional means of probing the energy dependence of cross sections with inclusive measurements inside a single detector.

\end{itemize}

Additionally, improved modelling of nuclear effects pertinent at the low energy transfers that PWIA breaks down are showing potential to reduce the uncertainty on the ratio of the muon to electron neutrino and antineutrino cross sections~\cite{Martini:2016eec,Ankowski:2017yvm,Nikolakopoulos:2019qcr,Gonzalez-Jimenez:2019qhq,Martini:2023kem,Dieminger:2023oin}, which is responsible for dominant systematic uncertainties in current and projected measurements of the CP-violating phase in neutrino oscillations.  This ratio is very challenging to determine experimentally.

\subsection{Hadronic final states}
\subsubsection{Experiment}
\label{sec:qesemiinc}
While inclusive measurements of lepton kinematics are informative, more detailed information can be gleaned by probing the hadron kinematics too.
This has become possible with the advent of fully active tracking detectors.
Recent data from T2K, MINERvA, and MicroBooNE probe the kinematics of protons produced in pionless interactions, as well as the correlation between proton and muon kinematics.
These correlations are generally sensitive to specific nuclear effects such as nuclear (Fermi) motion and FSI, and can be a unique probe of multinucleon interactions which are relatively unconstrained.
While electron scattering is able to provide constraints on many of these effects, neutrino data is still required because the axial coupling cannot be measured in electron scattering.

Depending on the specific detector used, the information available in semi-exclusive measurements can be different.
Ideally all experiments would measure the charged lepton and all hadrons, but usually information about neutrons is functionally absent (with some exceptions) and protons are only measured down to a detector-specific threshold.
The proton tracking threshold can have strong angular dependencies due to detector geometries, requiring complicated signal definitions to avoid model dependence.

As the statistics and resolutions of fully active tracking calorimeter detectors have increased, one challenge has been to determine which measurements are most sensitive to specific effects.
Early measurements often focused on simply measuring the kinematics of a muon and a proton \cite{MicroBooNE:2020akw,MicroBooNE:2023cmw,MINERvA:2013kdn,MINERvA:2014ypj}.
More recently, a number of strategies have been adopted to be sensitive to specific features.

\begin{itemize}
\item The most prominent of these is measurements of kinematic imbalance (often in the plane transverse to the neutrino direction, thereby removing the necessity to account for the unknown incoming neutrino energy)~\cite{lu:2015hea,Furmanski:2016wqo,T2K:2018rnz,munteanu:2019llq,minerva:2018hba, MicroBooNE:2023tzj, MicroBooNE:2023cmw, MicroBooNE:2023krv}.
In these analyses the effects of nuclear motion can be separated from FSI and MEC contributions to the cross section~\cite{lu:2015hea,Furmanski:2016wqo,Baudis:2023tma, Dolan:2018zye,Filali:2024vpy,Dolan:2018sbb}.
The most recent MicroBooNE measurements~\cite{MicroBooNE:2023krv, MicroBooNE:2023tzj, MicroBooNE:2023cmw, MicroBooNE:2023krv} separates these further by measuring multiple kinematic imbalance variables simultaneously (a technique also shown to be of particular interest to T2K's near detector upgrade~\cite{T2K:2019bbb,Dolan:2021hbw}), and including new imbalance measurements in the longitudinal direction.

\item An interesting analysis from MINERvA measures (in three dimensions) the muon kinematics (longitudinal and transverse) as well as the total kinetic energy of all protons~\cite{MINERvA:2022mnw}, which acts as a close approximation of the energy transfer. Such analyses are sensitive to the exact process that causes discrepancies in the forward regions of inclusive pionless analyses.

\item Some additional information about the contributions of MEC and FSI can be obtained by measuring the breakdown of proton multiplicities.  This is difficult to do in existing scintillator detectors but can be done using liquid argon.
A recent MicroBooNE measurement of the transverse momentum imbalance for events with two final state protons shows sensitivity to the normalisation of MEC events, as well as the nucleon-nucleon correlations in these events.
Still, it is difficult currently to infer how the MEC cross section depends on energy transfer, and how momentum is shared between nucleons.  Additional statistics in this channel would be invaluable in resolving this question.

\end{itemize}

While recent measurements contain a wealth of new information, a challenge remains in interpretation - with data on different targets at different energies, no generator can currently reproduce all the data.

\subsubsection{Theory}
In a semi-inclusive interaction, one or more nucleons are detected in coincidence with the scattered lepton, but the residual hadronic system is not characterized.
This is the situation in most neutrino experiments, since the total energy transferred to the nucleus is not known.
The semi-inclusive cross section can in principle be obtained by considering all the available exclusive channels, but such microscopic calculations are presently unavailable. 
The tool of choice in neutrino event generators is the intranuclear cascade model (INC). 
In this case a factorized form for the interaction matrix element is assumed. Schematically:
\begin{equation}
\label{eq:FSI_INC}
\lvert \langle \Psi_0 \rvert J^\mu \rvert \alpha \rangle \lvert^2 = \sum_{\beta}\lvert \langle \Psi_0 \rvert J^\mu \lvert \beta \rangle \rvert^2 \lvert \langle \beta \rvert \alpha \rangle \lvert^2.
\end{equation}
The production of a distinct final-state $\alpha$, is written as the product of a matrix element for an initial 'hard' process that produces a state $\beta$ with the probability for producing $\alpha$ from $\beta$. This probability is given by the INC.

The specific implementations and intricacies of INCs are discussed in section~\ref{sec:generator_physics}. 
Without considering the inner workings of the INC there are a number of questions that pose themselves:

\begin{itemize}
\item Eq.~\ref{eq:FSI_INC} can be read as a classical approximation that considers a specific set of intermediate states~\cite{Nikolakopoulos24b}.
Formal bounds on the validity of this approximation or possible correction terms have not yet been established.
\item Interference between different interaction mechanisms in this approach are dictated by the first factor. However the final-states $\alpha$ produced from different $\beta$ can overlap. 
For example $\beta$ might contain a single pion that gets absorbed, knocking out two nucleons. On the other hand, two nucleons might be produced in $\beta$ which do not reinteract. These processes are physically indistinguishable and should be treated consistently.
\item What the 'correct' decomposition in intermediate states is is not fully explored. The INC ($\lvert \langle \beta \rvert \alpha \rangle \lvert^2$) and theoretical calculations used as input ($\lvert \langle \Psi_0 \rvert J^\mu \lvert \beta \rangle \rvert^2$) are often formulated in terms of different states $\lvert\beta\rangle$. 
The INC should ideally be tailored to the input that can be provided and vice-versa.
\end{itemize}

\subsubsection{Current model-measurement comparisons}
In general, no model in neutrino event generators is able to describe all semi-inclusive pionless interaction measurements at the same time.
Despite this, comparisons of predictions to exclusive measurements tailored to lifting degeneracies between nuclear effects provides a very useful means to assess their limitations and to motivate systematic uncertainties.
The comparison of similar measurements from different experiments offers powerful additional handles on isolating the cause of generator mis-modelling.

\begin{itemize}
    \item Model comparisons to kinematic imbalance measurements from T2K, MicroBooNE and MINERvA~\cite{Dolan:2018zye,Bathe-Peters:2022kkj,Bourguille:2020bvw,Dolan:2018sbb,Filali:2024vpy} allows an exploration of how nuclear effects scale with energy on the same target (T2K to MINERvA) or at similar energies on different nuclear targets (T2K to MicroBooNE). Such meta-analyses expose potential issues with A-scaling, energy dependence and FSI modelling. 
    \item Recent work has confronted measurements with more sophisticated FSI models~\cite{Ershova:2022jah,Ershova:2023dbv,Niewczas:2019fro} to explore both measurements' sensitivities to alterations in FSI modelling (which is considerable) and event generator limitations. 
    \item Until very recently, most microscopic theory models have not been able to make  predictions of semi-inclusive measurements, but recent calculations from theory groups have used microscopic calculations of FSI under a distorted-wave-impulse-approximation to calculate the CCQE contribution to semi-inclusive experimental measurements~\cite{Dolan:2019bxf,Franco-Patino:2022tvv,Franco-Patino:2023msk,Franco-Patino:2023orj,Nikolakopoulos:2022qkq,Nikolakopoulos:2023pdw}. Also the 2p2h contribution to semi-inclusive processes is starting to be microscopically studied~\cite{Martinez-Consentino:2023hcx,Martinez-Consentino:2024mdr,Belocchi:2025eix,Rocco:2026kae}.
\end{itemize}

\subsubsection{Prospects}

Significant progress has been made over the past decade in understanding the impact of nuclear effects in pionless interactions.
Nevertheless some challenges remain:

\begin{itemize}
\item New measurements should be made that take advantage of increased statistics and lower proton thresholds to probe higher nucleon multiplicities.

\item It is important to ensure efficiency corrections are made in a way that doesn't rely on the details of the underlying modeling, and in the large phase space that exists in exclusive kinematics this can be challenging.  New detectors that have a flatter response as a function of angle will improve this.

\item Some experiments can take advantage of detectors that span off-axis beam angles to determine energy dependence directly.

\item MINERvA data on several nuclear targets has already probed the dependence on nuclear size, and more joint measurements that span nuclear targets would be invaluable.
In general, measurements with correlated uncertainties are necessary in order for these measurements to be most useful.

\item New scintillator detectors which have lower thresholds (such as the ND280 upgrade~\cite{T2K:2019bbb} or even lower threshold proposals~\cite{Franks:2023ttv}) can also make large contributions here by measuring the nucleon multiplicity. FSI and MEC are expected to scale differently with nuclear size, so comparing data on carbon and argon would be highly informative to model builders.

\item While measurements of kinematic imbalance have shown sensitivity to missing \textit{momentum}, different particles carrying away this momentum lead to very different missing energies.  Any sensitivity to neutrons and nuclear fragments can reveal details of missing neutrino energy in this channel which would be extremely challening to measure in other channels.
These measurements stretch the capabilities of existing detectors.

\item Detailed predictions of hadronic final states remain challenging to model, though some microscopic models have made progress here.  In general providing generators with sufficient degrees of freedom to be able to describe hadronic measurements across multiple energies and nuclei will be required to make use of the newer (and upcoming) data.
\end{itemize}

\subsection{Determining the single-nucleon response}
While the pionless cross section is highly sensitive to nuclear effects, there remain uncertainties related to the single-nucleon response in the absence of a nucleus.  However there has been some recent progress in this area.

\subsubsection{Theory}
\begin{itemize}
\item {\bf Axial form factor:} The $Q^2$ dependence of the axial form factor is a source of uncertainty which hampers analysis and interpretation of neutrino data. 

The most direct way to extract the axial form factor is by the measurement of neutrino-neutron or anti-neutrino proton scattering, e.g. in bubble chambers.
Historically, dipole form factors were used to fit bubble chamber data.
This introduces ad-hoc constraints on the parametrization, which leads to an underestimation of the uncertainty in comparison to a constraint-free $z$-expansion parametrization~\cite{Meyer:2016oeg}.
Analysis of deuterium data is in principle further complicated by final-state interactions, and the influence of two-body currents~\cite{shen:2012xz, PhysRevC.101.015505}. Attempts to extract the axial form factor should ideally consider the deuteron effects fully.

Theoretical avenues for the determination of the nucleon axial form factor are
\begin{itemize}
\item {\bf Pion electroproduction} amplitudes at threshold are related to the axial coupling to the nucleon, through low-energy theorems (LET) derived in ChPT~\cite{bernard:2001rs}. The original LET was derived using current-algebra techniques, and only valid in the chiral limit~\cite{Nambu:1962ilv,Nambu:1962lbq}. Several approaches have been used to extend the LET to deal with pion mass corrections~\cite{Nambu:1970jm, BENFATTO1972205, Furlan:1970cw, DOMBEY197365}. The result for the axial form factor depends on the particular method used~\cite{bernard:2001rs}. Recent works have included analyses performed with these different schemes, and treat the scheme-dependence as a source of uncertainty~\cite{PhysRevC.101.025501}. 
This uncertainty can in principle be overcome by a careful treatment in ChPT~\cite{bernard:2001rs}.

\item Recently, {\bf Lattice QCD (LQCD)} calculations of nucleon vector form factors~\cite{Park:2021ypf} which are in agreement with experimental analyses have been performed~\cite{PhysRevC.70.068202}. 
Several groups have computed the axial form factors in recent years, see Refs.~\cite{Meyer:2022mix,Ruso:2022qes} for reviews.
The results for the axial form factor from different analyses are found to be consistent with one another.
The resulting form factor has a more shallow slope than what is obtained from deuteron data. 
Further progress will require an assessment of the contamination of excited states in the lattice data.
\end{itemize}

\item  The {\bf Delta coupling} is assessed in more detail in the context of pion production in this paper. However, the axial couplings to the delta are an important source of uncertainty in the strength of MEC contributions. This uncertainty should be quantified.
Efficiently propagating this uncertainty to resulting observables remains a theoretical-technical challenge.
\end{itemize}

\subsubsection{Experiment}

Recent advancements in analysis methods and detector technology have presented the tantalising possibility of using pionless neutrino interactions to isolate anti-neutrino CCQE interactions on free proton targets. Such interactions are free of nuclear effects, and therefore offer the interesting opportunity to probe the details of neutrino-nucleon interaction physics (primarily the axial nucleon form factor) as well as a potential standard candle for constraining incoming neutrino flux. Such measurements can broadly be split into three categories: direct, subtraction and kinematic isolation measurements. 
\begin{itemize}
\item \textbf{Direct measurements:} 
The only direct neutrino-hydrogen data in existence comes from liquid hydrogen bubble chambers operated mostly in the 1970's and 80's.
New data from a modern liquid hydrogen bubble chamber in a contemporary neutrino beam has the potential to provide significantly more precise measurements crucial inputs for modelling of neutrino-nucleon interactions. Whilst technical challenges remain, an effort has begun to evaluate the prospects of operating a modern bubble chamber~\cite{Alvarez-Ruso:2022exy}.

\item \textbf{Subtraction measurements:} another means to access antineutrino-hydrogen interactions is via subtraction methods. The MINERvA experiment and the planned SAND DUNE near detector~\cite{Duyang:2018xcc,DUNE:2021tad} employ pure carbon as well as hydrocarbon target elements. In principle, the former offers a means to subtract the carbon component of the latter. 
However, the subtraction of the vast majority of the interactions ($\sim$6/7 for a CH target or $\sim$3/4 for a CH$_2$ target) on a hydrocarbon target necessarily implies larger statistical uncertainties than a pure hydrogen measurement, thus requiring larger target masses in an intense neutrino beam.
Beyond this, additional challenges are presented by the fact that only the hydrocarbon target can be an active detector element and therefore the kinematic acceptance of interactions on hydrocarbon and carbon components of the detector are likely to differ which introduces model dependence into the subtraction. Still, the SAND detector is proposed with a specific goal of realising such a subtraction analysis and expects to collect many millions of neutrino interactions, potentially making a subtraction analysis both viable and statistically compelling. 

\item \textbf{Kinematic isolation measurements:}\label{sec:tki} 
For composite targets that include hydrogen, it is in principle possible to kinematically separate interactions on hydrogen from those on other nuclei (first proposed for single pion production ~\cite{lu:2015hea}).
This relies on missing momentum from nuclear recoil, which doesn't occur for hydrogen interactions.
A detector that can measure neutron kinematics (using techniques such as those developed by the MINERvA collaboration~\cite{MINERvA:2019wqe}) can in principle kinematically isolate and measure antineutrino-hydrogen CCQE interactions.
The MINERvA experiment has put these approaches into a real-data analysis, producing the first inference of the axial form factor from experimental data since bubble chamber analyses~\cite{minerva:2023avz}.
Whilst most existing form factor models remain within MINERvA's current uncertainties, the analysis serves as a powerful first demonstration of the types of constraints future measurements can offer.
The ``Super-FGD'', recently installed as part of T2K's ND280 near detector upgrade, is particularly well suited for such measurements as its position and timing resolutions allow measurement of both neutron direction but also energy through time-of-flight.
Prospects for such measurements~\cite{munteanu:2019llq, Dolan:2021hbw, Baudis:2023tma} show that the additional capability can produce a dramatically improved purity.
The MINERvA analysis estimated $\sim$30\% purity and $\sim$10\% efficiency, while the most recent Super-FGD study~\cite{Baudis:2023tma} estimates a $\sim$90\% purity for the same efficiency.
\end{itemize}

Lastly, experiments that operate intense electron beams can probe the axial hadronic current, either through parity-violating electron scattering (PVES), or the electron-induced charged-current process.
The contribution of the axial form factor to the parity violating asymmetry is relatively low in PVES~\cite{Gonzalez-Jimenez:2011qkf}. The CC process is experimentally cumbersome as the hadron final-state is a neutron, but might be possible on the deuteron.
The sensitivity to the axial current in PVES is larger in pion-production reactions, but theoretical and statistical uncertainties on measurements are still quite large~\cite{G0:2011aa,G0:2012hkh,Mukhopadhyay:1998mn,JONES1980137}
The reaction rate and experimental signature of electron-induced CC pion production might make measurement of axial form factors feasible in high-intensity electron beams~\cite{Alvarez-Ruso:1997njo, Fatima:2024hlu}. 

\subsection{Challenges and open questions in pionless interactions}

Since the last NuSTEC white paper, measurements of pionless interactions, and developments in modelling them, have made significant progress. It is clear that suppression of the cross section at low energy transfers from physics beyond PWIA has a significant effect, and more sophisticated nuclear ground states and multi-nucleon interaction modelling are likely required to explain observations. Semi-inclusive analyses that probe correlations between the outgoing lepton and hadronic systems have proven particularly powerful.  However, overall no model is quantitatively compatible with global cross section measurements.

The future of pionless interactions should strive to supply measurements which can better isolate the source of the disagreements between models and measurements. This will involve making highly differential measurements that are correlated across energy spectra, target and flavour made using large statistics samples with high-resolution, low-threshold detectors. Pionless interactions are the best place to begin a campaign to make such comprehensive measurements, due to the relatively low dimensionality of the cross section and the relative ease with which the complete final state can measured.
Theoretical results are increasingly being compared in a systematic way to available data, and validated against electron-scattering data in an attempt to isolate how the various assumptions and differences in models lead to the observed discrepancies. 
Several open questions and challenges derived from the assessment of model-measurement discrepancies stand out:

\begin{itemize}
 \item 
 All currently implemented models for CCQE interactions in neutrino event generators have limited predictive power across a significant portion of the kinematic phase space where the cross section is expected to be significant. Some models are inclusive (i.e. come pre-integrated over the hadronic side of the interaction), such as SuSAv2, whilst others neglect most physics beyond the plane wave impulse approximation, such as spectral function implementations, or use nuclear ground states ruled out by electron scattering measurements (such as Fermi gas models). The development of a model with genuine predictive power over the entire phase space for CCQE interactions  relevant for neutrino oscillation experiments (including the scaling of the cross section between nuclear targets and neutrino flavours), and its implementation inside an event generator, is therefore a key challenge.
 
 \item
 A complete implementation of multinucleon interactions in generators is still pending. The adoption of the hadron-tensor approach simplifies both the numerical calculation and enables the easy adoption of multiple models with the same generator implementation scheme. However, within the current scheme inclusive 2p2h theory calculations are used, such that the outgoing nucleon kinematics need to be estimated in an \textit{ad-hoc} way during event generation, losing potential correlations between the outgoing lepton and the hadrons. The development and implementation of a 2p2h model with predictive power for outgoing nucleon kinematics is thus a key challenge. 

\item
Measurements of multi-proton events may provide data that is able to constrain correlations between lepton and hadrons in 2p2h processes.  Making these measurements with high statistics should now be possible, and should be pursued.

\item
 There is ongoing discussion about the possible double counting of cross section contributions by considering both microscopic two-body multinucleon models and the presence of initial nucleon correlations that appear in spectral functions. Whilst these two contributions arrive from different formalism, they may be accounting for elements of the same physics. Moreover, our understanding of theoretical predictions and discrepancies among them would benefit from a more careful treatment of interferences between these and other nuclear effects / interaction channels.
\item 
 Similarly to the multinucleon case, the predictive power of generators to predict the quite substantial non-QE and non-multinucleon contribution to the pionless cross section (i.e. interactions in which a pion is absorbed inside the nuclear medium) is limited. More details are given in \ref{sect:fsi}, but it is clear that FSI models differ greatly in their prediction of both the rate and shape of this component.
 \item
 More generally, one of the main problems faced by experimentalists is the rigidity of the theoretical models. Whilst an increasingly wide range of models are becoming available, most do not offer much to quantify their plausible variation. Experimentalists are thus forced to add heuristic uncertainties loosely covering known approximations made in the models and their disagreement with experimental measurements. Whilst such loose approaches are manageable as oscillation experiments remain largely statistics limited, they will become a limiting factor in both the sensitivity and robustness of oscillation measurements from next-generation experiments. A key challenge is therefore for the experiment and theory communities to come together to produce a comprehensive parameterisation to describe model's plausible variation. 
 \item
 The modelling of the axial vector form factor as a simple dipole needs to be revisited both theoretically and experimentally. 
\item 
  A detailed understanding and a rigorous reporting of potential model dependence in experimental cross section measurements is essential as we enter a very high-statistics era rich in a wide variety of neutrino interaction measurements. It is essential that analysers of cross section measurements are able to easily assess the scope for bias or undercoverage in uncertainties to mitigate the risk of either rejecting reasonable models or overconstraining parameters within them.
\end{itemize}

Efforts to confront these open questions and challenges are underway and are being confronted along three principal and complementary axes:

\begin{itemize}
    \item \textbf{New measurements:} current measurements show clear disagreement with models, but they are only beginning to indicate the source of the disagreement. The enormous statistics that have been or will be collected by current and future experiments such as MINERvA and SBND, alongside a host of newly introduced creative analyses techniques and observables of interest, permit increasingly high dimensional and informative measurements. Examples of such measurements have been found in analyses of transverse or similar variables, but other measurements isolating particularly interesting regions of kinematic phase space as a function of intuitive observables is likely to prove valuable. As these are increasingly provided in a way that mitigates bias and under-coverage in cross section extraction, and such that they can be correlated across different nuclear targets, neutrino flavour and energy spectra, they will permit increasing detailed analyses of where exactly our models are failing, to inspire new theory developments.
    \item \textbf{Better detectors:} in order to realise the new measurements which will permit a detailed analysis of the cause of model-measurement discrepancies, finer binned analyses with lower particle tracking thresholds are likely to be crucial. Recently developed and deployed detectors, such as SBND and T2K's ``Super-FGD'' represent considerable steps forward in detector technology and permit such analyses. They also open the door to measurements that would not easily be possible with current detectors such as those possible using ``SBND-PRISM'' or that require measurements of outgoing neutron kinematic energy.
    \item \textbf{Continued close collaboration with the nuclear theory community:} even if new measurements can isolate the cause of model-measurement discrepancies, remedying them necessarily requires a tight collaboration between neutrino-physics experimentalists and nuclear theorists.
\end{itemize}

\newpage
\section{Resonance Model}
\label{sec:RES}
\subsection{Introduction}
When the energy transferred to the nucleon increases, inelastic scattering resulting in meson production in the final state becomes possible. These processes often involve the excitation of baryon resonances, which then promptly decay to produce final state nucleons and mesons. At lower energy it leads to the production of a $\Delta(1232)$ in the final state, and with further increase in energy transfer higher mass resonant states are being produced. Presently, 18 resonances have been identified in the CM energy range $W<2$~GeV with different quantum numbers~\cite{pdg2020}. 

The threshold for meson production is defined by the creation of the lightest meson, the pion, at rest, i.e., when the invariant hadron mass $W=M_N + m_{\pi}$, where $M_N$ ($m_{\pi}$) is the nucleon (pion) mass. Similarly, the threshold for other meson productions occurs at $W=M_N + m_M$ where $m_M$ is the mass of the produced meson. Mesons with strange quark content will be discussed later in the section on strange particle production~\ref{strange}.

To understand resonance production in the weak sector induced by the (anti-)neutrinos, a model is required to describe the resonance dynamics. In this region of W, mesons are also created via the non-resonant interactions, which produce final states identical to resonance production resulting in non-negligible interference terms. As a result, a complete model in the resonance region must include all resonance and non-resonant interactions. Hence these overlapping interactions make the resonance region so complex!

Resonances can be excited by various probes, including electromagnetic, hadronic, and weak interactions. Isospin symmetry \footnote{Isospin symmetry also relates neutrino and anti-neutrino channels} connects the weak and electromagnetic (EM) vector currents, relating the weak vector  form factors in charged current (CC) and neutral current (NC) interactions to those in EM interactions. Since high-quality (anti-)neutrino data are limited, electromagnetic and hadronic scattering data therefore provide essential constraints for phenomenological descriptions of weak interactions in the resonance region. Furthermore, the Partially Conserved Axial Current (PCAC) relation links the axial-vector weak current to the pion scattering cross-section, providing a valuable constraint for axial form factors relevant to weak interactions. Accurate definitions of all form factors across various processes in the region are crucial for precise theoretical descriptions of meson production.

Theoretical models for single-pion production serve as essential inputs for models that extend beyond the impulse approximation. For instance, in hadronic two-body current interactions, the pion created by resonance decay or in background diagrams is a virtual particle that couples with a second nucleon. This interaction can result in the nucleon being knocked out, leading to two-particle-two-hole (2p-2h) processes induced by meson exchange currents (MEC)~\cite{VanOrden81,Ryckebusch94,DePace:2003spn,Martini09,Nieves11,Amaro20}. Alternatively, the pion may remain bound within the nucleus, contributing to two-body current effects in quasi-elastic (QE) scattering~\cite{Dekker:1994yc,VanderSluys95,Franco-Munoz23,Lovato:2023khk}. Therefore, accurate form factors in single-pion production processes are crucial for the development of advanced nuclear models.

Resonant pion production processes are a dominant interaction mode in the NOvA experiment and play a significant role in the oscillation analyses of DUNE, as illustrated in \autoref{fig:neu_flux}. These processes also serve as the main background for low-energy ($E_{\nu} <1\,\text{GeV}$) neutrino oscillation experiments such as T2K and Hyper-K, as the initially produced pion can be absorbed in final state interactions. Such events can be misidentified as quasielastic (QE) events, leading to biased reconstruction of the neutrino energy. In water Cherenkov detectors neutral current $\pi^0$
  production is a significant background in electron neutrino and antineutrino appearance searches, since the photons from $\pi^0$ decay can be misidentified as electron or positron signals.

\subsection{The domain of resonance production}\label{domain}

\begin{figure*}[t]
\begin{center}	 \includegraphics[width=0.49\textwidth] {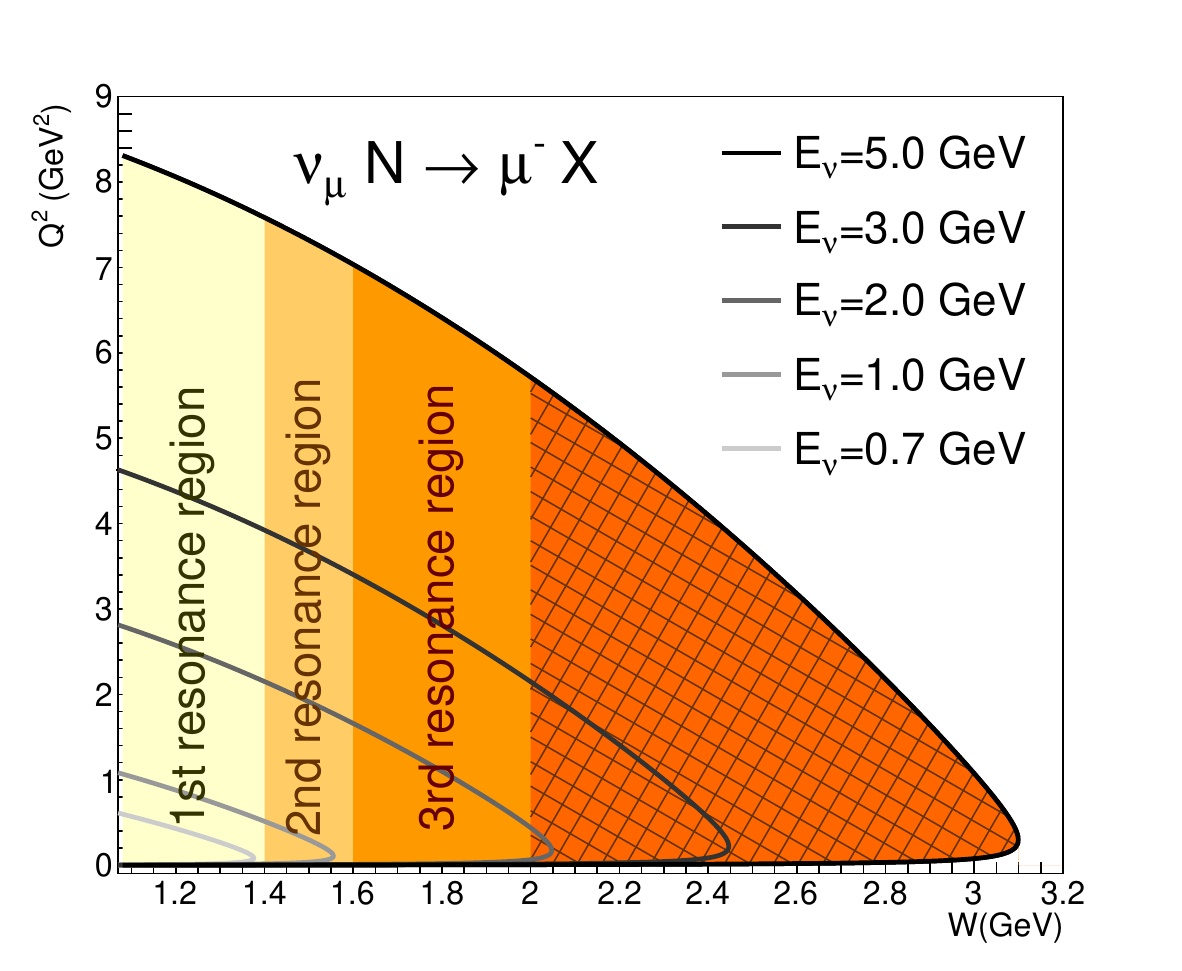}
\includegraphics[width=0.49\textwidth] {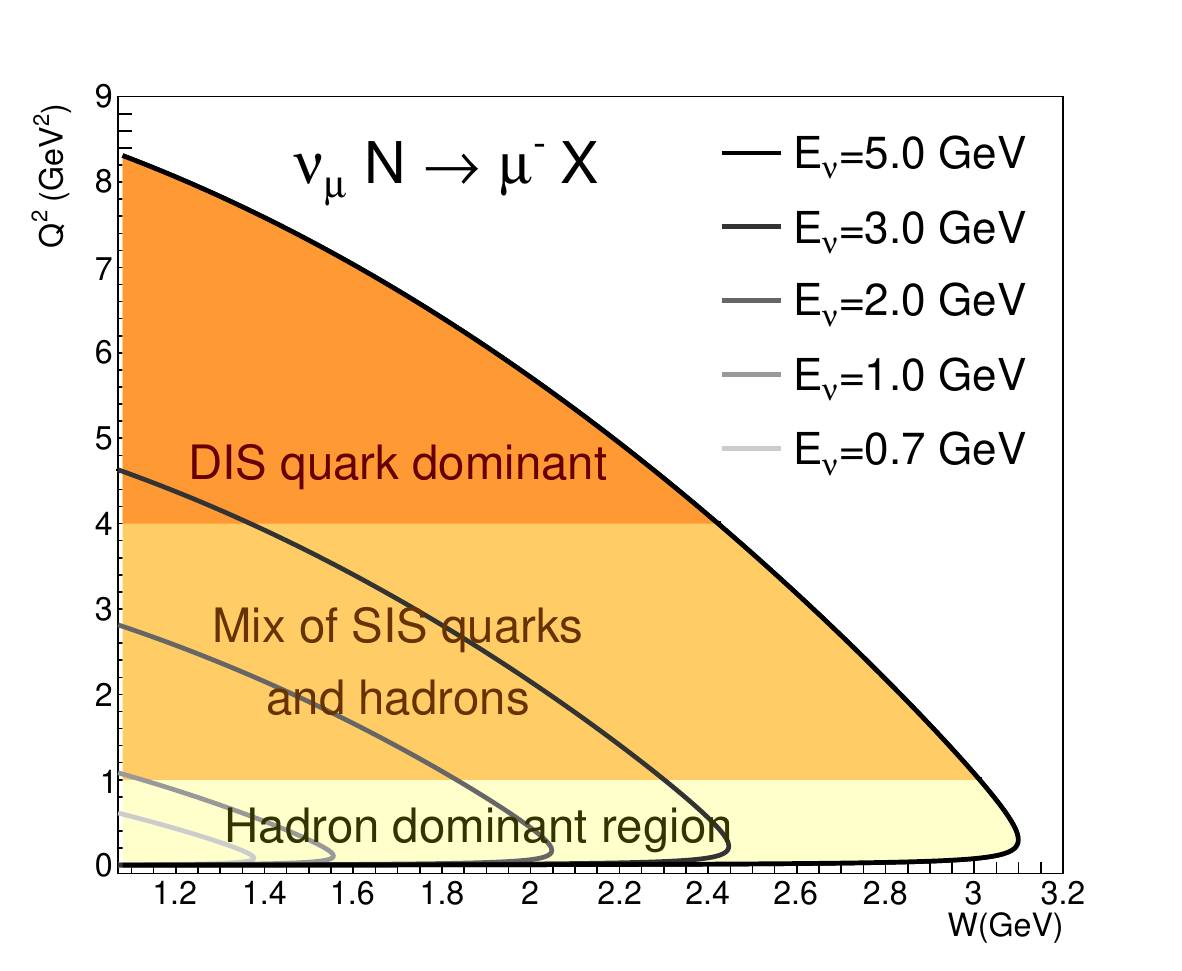}
    \caption{Allowed kinematic region for charged-current inelastic neutrino–nucleon scattering in the $(Q^{2},W)$ plane for several representative neutrino energies relevant to current and future accelerator-based experiments.
The left panel illustrates the evolution of the invariant hadronic mass $W$, highlighting the resonance region. While the resonance region extends up to $W = 2.0~\text{GeV}$, single-pion production can still occur at higher $W$ through non-resonant interactions. Consequently, for $E_{\nu}\lesssim 2$~GeV, the accessible kinematic region remains largely confined to the resonance domain.
The right panel shows the corresponding evolution in $Q^{2}$ according to ref. \cite{Muzakka:2022wey, MINERvA:2025hzq}: at low $Q^{2}$ neutrinos predominantly probe the nucleon as a whole, while at higher $Q^{2}$ they resolve individual quarks, with a broad transition region between hadronic and partonic degrees of freedom.}
    \label{PhaseSpace2}
\end{center}
\end{figure*}

Inelastic neutrino–nucleon scattering, including single pion production (SPP) channels, can be described using two independent kinematic variables, typically chosen as invariants: the four-momentum transfer ($Q^2$) and the invariant hadronic mass ($W$).
 The accessible kinematic region in terms of $Q^2$ and $W$
 for various neutrino energies relevant to current and future accelerator-based experiments is shown in Fig.~\ref{PhaseSpace2}. The neutrino energy spectra for T2K, Hyper-Kamiokande, and MicroBooNE peak below $E_{\nu}=1$ GeV, while NOvA operates primarily in the range of $1<E_{\nu}<3$ GeV, and DUNE extends up to $E_{\nu}<5$ GeV.
Consequently, for wide-band beams in accelerator-based neutrino experiments, interaction models must remain valid across a broad range of $Q^2$ and $W$.

At low invariant mass $W$ ($<1.4$~GeV), the target nucleon is excited to the $\Delta(1232)$ resonance, which promptly decays into a nucleon and a pion. The $\Delta$ resonance provides the dominant contribution in the resonance region and is the sole resonance present in the first resonance region ($1.08~\text{GeV}<W<1.4$~GeV). At higher $W$, the second resonance region ($1.4~\text{GeV}<W<1.6$~GeV) contains three isospin-$1/2$ resonances: the $P_{11}(1440)$ Roper resonance and the two negative-parity states $D_{13}(1520)$ and $S_{11}(1535)$. Beyond this, the third resonance region ($1.6~\text{GeV}<W<2.0$~GeV) includes several additional resonances, although their overall contribution to the cross section is smaller. In both the second and third resonance regions, mesons heavier than the pion, such as $\eta$, $K$, $\omega$, and $\rho$, can also be produced. 

The $Q^2$ dependence further characterises the scattering regime. At low $Q^2$ ($\lesssim 1$~GeV$^2$), the interaction is dominated by hadronic resonance production and non-resonant background processes, without quark-level interactions inside the nucleon. In the intermediate region ($1 \lesssim Q^2 \lesssim 4$~GeV$^2$), hadronic resonance production coexists with non-perturbative shallow inelastic scattering (SIS), often described as the quark–hadron transition region. At higher momentum transfers ($Q^2 \gtrsim 4$~GeV$^2$), perturbative QCD becomes applicable, and deep inelastic scattering (DIS) processes, dominated by quark-level interactions, prevail. 

\subsection{Status and prospects of theory}\label{res_nucleon_theory}
\subsubsection{neutrino-nucleon interactions}\label{neutrino-nucleon-SPP}
The underlying fundamental theory of meson production from a nucleon is quantum chromodynamics (QCD), and QCD
calculations provide important and reliable predictions in the perturbative (high-$Q^2$) domain. Furthermore, at sufficiently low momentum transfer, where the characteristic momentum is much smaller than the chiral symmetry breaking scale ($\Lambda \sim 1$ GeV), chiral perturbation theory (ChPT) provides a systematic effective field theory description of QCD in terms of hadronic degrees of freedom. For weak pion production, such studies have been reported in Refs.~\cite{Hernandez:2007qq, yao2018, yao2019}.

At the intermediate energies relevant for accelerator-based neutrino experiments, presented in Fig. \ref{PhaseSpace2}, neither perturbative QCD nor ChPT applies. Interaction models must therefore rely on phenomenological descriptions that respect the underlying symmetries and conservation laws of QCD. In this regime, both resonant and non-resonant amplitudes are expressed through $Q^{2}$-dependent form factors. Dipole form factors are commonly used, as they provide a reasonable description in the hadron-dominant region at low to moderate $Q^{2}$; however, they fail to reproduce data at higher $Q^{2}$ and lack a theoretical foundation for inelastic dynamics.

An alternative is provided by the meson dominance (MD) form factor model, originally proposed by Sakurai~\cite{Sakurai:1960ju} and Gell-Mann~\cite{Gell-Mann:1961jim}. In this approach, the interaction between leptons and nucleons is mediated through the exchange of virtual mesons carrying the appropriate quantum numbers. The MD model consistent with QCD constraints~\cite{vereshkov:2007xy} can successfully describe nucleon form factors across both perturbative and non-perturbative regimes~\cite{Stoler:1993yk}. This makes it a well-motivated choice for the kinematic region probed in accelerator-based neutrino experiments.

Complete phenomenological models must coherently include multiple resonances and non-resonant interactions. They must be carefully parametrized and constrained to respect fundamental theories and conservation relations using relevant data. This task is particularly challenging because the interactions overlap across kinematic regions that span both perturbative and non-perturbative regimes, where the hadronic degrees of freedom involve a mixture of nucleons, resonances, and partons. Moreover, systematic uncertainties associated with the adjustable parameters and their correlations must be rigorously evaluated. Nevertheless, most phenomenological models currently do not provide systematic uncertainties for the adjustable parameters they employ. A summary of the evolution of the single pion production (SPP) models used by the neutrino community is as follows:

\begin{enumerate}
    \item {\it Adler model (1968)} describes weak SPP in the first resonance region. In Adler's model, the cross sections are calculated from the multipoles, which are deduced from the dispersion relations for the $\Delta$ resonance and the Born terms for non-resonant interaction. \cite{Adler:1968tw}

  \item {\it The Rein-Sehgal (RS) model (1981)} describes resonance excitation in a relativistic quark model based on the work of Feynman, Kislinger, and Ravndal (FKR model), which accounts for SPP in electron and photon interactions \cite{rein:1980wg}. The cross sections are calculated from the helicity amplitudes of 18 resonances up to $W=2~\text{GeV}$. However, only one dipole form factor is used to describe all nucleon-to-resonance transitions. This dipole form factor was modified in Ref. \cite{Graczyk:2007bc} by equating the helicity amplitudes for the $\Delta$ resonance in the RS model and the Lalakulich-Paschos model and partially solving the equations. In the original RS model, the leptons are assumed to be massless. This assumption has been improved in Ref. \cite{berger:2007rq, Graczyk:2007xk, Kuzmin:2003ji}. The RS model with the mass correction \cite{berger:2007rq} is the primary model for resonant interactions in the NEUT and GENIE event generators.

    \item {\it Rein model (1987)}: This model calculates multipoles from the RS model and coherently adds them to the Born terms multipoles from the Adler model. Calculating these multipoles is computationally expensive, making the model unsuitable for use in event generators \cite{Rein:1987cb}.

    \item {\it Lalakulich-Paschos (LP) model (2005)}: This model describes resonance production in the first and second resonance regions using the Rarita-Schwinger formalism and employs Breit-Wigner amplitudes for the resonance decay. It utilises MAID data to parameterise the vector form factors of the resonances. Several other models have also utilised these form factors. \cite{Hernandez:2007ej, gonzalez-jimenez:2016qqq}.
    \item {\it Hernandez {\it et al.} (HNV) model (2007)}: This model incorporates non-resonant mechanisms from nonlinear chiral Lagrangian in addition to the resonant ones described in Refs. \cite{Fogli:1979cz, Alevizos:1977xf}, and extends them to fulfill chiral symmetry constraints at the threshold~\cite{Hernandez:2007qq, Hernandez:2010jf, Hernandez:2013jka, Hernandez:2016yfb}. The model is partially unitarized by imposing the Watson theorem~\cite{Alvarez-Ruso:2015eva}. Several other models have extended the HNV model by including additional resonances \cite{RafiAlam:2015fcw, Sobczyk:2018ghy, SajjadAthar:2022pjt}.

    \item {\it Dynamic Couple Channel (DCC) model (2015)}: In the dynamical approach, the hadronic rescattering processes are taken into account by solving a coupled channel equation for the $\Delta_{33}(1232)$ and higher resonances~\cite{Sato:2003rq,Nakamura:2015rt}. In this approach, a unified treatment of all resonance production processes satisfying unitarity is provided. A combined fit to electron-, photon-, and pion-induced reactions is used to construct the resonant and non-resonant amplitudes, as well as their relative phases, allowing the model to describe both single- and double-pion production consistently. However, systematic uncertainties on the fitted parameters are not evaluated.

    \item {\it Hybrid model (2017)}: This model extends the validity of the non-resonant model by incorporating a Regge approach. It combines low-W (ChPT) and high-W (ReChi) models in a phenomenological manner, creating a hybrid model~\cite{gonzalez-jimenez:2016qqq}.

    \item {\it MK model (2025)}: This model computes the helicity amplitudes of all resonances in the first and second resonance regions, as well as the two dominant resonances in the third region, and combines them with a non-resonant contribution taken from the Hybrid model. This construction establishes a consistent phase relation between resonant and non-resonant amplitudes~\cite{kabirnezhad:2017jmf,kabirnezhad:2020wtp,kabirnezhad:2022znc,Kabirnezhad:2024cor}. The transition form factors are constrained through a Meson Dominance framework, consistent with QCD expectations for nucleon form factors across both perturbative and non-perturbative regimes [402], providing a reliable description of the "mix of SIS quarks and hadron" region (see Fig. \ref{PhaseSpace2}). All form factors, as well as the phases linking resonant and non-resonant helicity amplitudes, are determined from a global fit to electron, photon, pion, and neutrino scattering data. The resulting parameter uncertainties and their correlations are fully propagated, yielding systematic uncertainties from the joint analysis.
    
\end{enumerate}

\subsubsection{neutrino-nucleus interactions}

Most approaches to the lepton-induced pion production on the nucleus are based on the impulse approximation (IA), i.e., one considers that the lepton interacts with only one nucleon in the nucleus. It is exactly the same idea used to describe the quasielastic one nucleon knocked out reaction. IA allows us to break down the problem into three parts: the elementary vertex, the nuclear framework, and final state interactions (FSI).
Models and challenges for the primary vertex (or neutrino-nucleon interaction) were summarized in Sections~\ref{neutrino-nucleon-SPP} and \ref{over:RES}, in what follows we discuss some of the features corresponding to the other two parts. For more information, we refer to the recent review articles~\cite{katori:2016yel, Nakamura:2016cnn,Mosel:2019vhx,Amaro:2019zos}.

The simplest nuclear model is the global relativistic Fermi gas (RFG). The RFG is able to capture the general features of the nuclear response, but multi-nucleon correlation effects are not included. Despite it simplicity, the RFG is still used in Monte Carlo (MC) event generators~\cite{MBNCpi010, MBCCpionC11, MBCCpion011, T2KCC1pi17, MINERvACCpi15, MINERvACCpi16}.
The local relativistic (or non-relativistic) Fermi gas (lFG) model is also widely used \cite{Ahmad06,Szczerbinska07,Zhang12,Hernandez:2013jka,Sobczyk2013,Martini14}. To improve its accuracy, different groups add additional ingredients on top of the bare model, such as nuclear correlations, mesonic and RPA effects.
Another model is the spectral function approach (SFA), in which the cross section is given as the product of the elementary lepton-nucleon cross section (for off-shell initial nucleon) and a realistic spectral function that describes the one-hole nuclear state~\cite{Benhar:2005dj,Rocco2019}. This factorized framework makes the SFA appealing to be used in MC event generators.
Finally, shell models or mean-field approaches are able to capture most of the nuclear dynamics by using single-particle wave functions for the initial and final states. These wave functions are obtained by solving the wave equation with some microscopic or phenomenological nuclear potential. They offer fair solutions to the problem of incorporating relevant nuclear effects, such as Pauli blocking, binding energies and elastic final-state interactions, in a consistent and quantum mechanical way~\cite{Gonzalez-Jimenez2019,Nikolakopoulos2023}. There are relativistic and non-relativistic versions. The drawback of these mean-field based models is that, compared to the plane-wave based approaches (RFG, lFG and SFA), the computational cost is relatively high.

In summary, there are a variety of nuclear frameworks that can be used to describe neutrino-induced SPP in the nucleus. The simplest nuclear models are not accurate enough to address the needs of the new era of neutrino oscillation experiments, and the more sophisticated ones have not yet found their way into the MC event generators, although some progress is being made~\cite{Nikolakopoulos2022, Achilles2023}.
It is important to mention that given the classical nature of MC event generators, in which the cross section is constructed by a two step process, namely, `elementary vertex $\times$ hadron propagation via cascade', the inclusive cross section remains unaffected by the cascade process. Consequently, it becomes imperative that the primary model, i.e. the one governing the elementary vertex, offers a robust and accurate inclusive response.
\\

{\bf Nuclear effects in SPP:}
\begin{enumerate}

    \item {\it Fermi motion:} Nucleons in a nucleus are moving. Its effect in the cross section consists of a blurring of the distributions, which become wider and, generally, lower in magnitude with respect to those for a nucleon at rest. The larger the nucleus, the larger this effect. Different models include Fermi motion in different ways. For example, plane-wave based approaches do an average of the single-nucleon cross section for a given probability distribution of the momentum of the nucleons (a step function in the global RFG or more realistic momentum distributions in the spectral function approach); in the case of shell model approaches, the information about the probability is in the single-particle wave functions, so the `average' over nucleon momentum is performed at the amplitude level. 

    \item {\it Elastic FSI and Pauli blocking:}
    The effect of Pauli blocking and distortion (or elastic FSI) of the outgoing nucleon in the resonance region has been studied in~\cite{Gonzalez-Jimenez2019,Nikolakopoulos2023,Garcia-Marcos2023}. It is shown that these nuclear effects notably modify the inclusive cross section and, therefore, they should be incorporated in the model of the primary vertex. These nuclear effects are particularly relevant when the outgoing nucleon has small kinetic energy (typically lower than $100$ MeV), this was expected since the wavelength associated with the nucleon is large, comparable to the distance between nucleons or even to the nuclear size; it is therefore not surprising that the plane-wave approximation, which is useful when the wavelength is small, fails under these circumstances.
    Despite that, it is shown that these effects are considerably smaller for inclusive cross sections after averaging over the neutrino energy.

    Similar effects are found due to the elastic FSI of the pion. Most of the works have been devoted to the study pion production induced by photons or electrons (see~\cite{Nagl91,Xiaodong1993,Drechsel:1999vh,Uden1998} and refs. there in), and some recent works have studied this effect in neutrino-induced coherent pion production~\cite{Singh2006,Alvarez-Ruso2007,Alvarez-Ruso2007NC,Amaro2009}. Addressing this effect in incoherent neutrino-induced pion production remains unexplored.

    \item {\it In-medium modification of the resonance properties:}
    The characteristics of resonances, i.e., the mass, decay width, branching ratio, coupling constants, form factors, could be modified when placed in a nuclear medium.
    It has been studied for the Delta resonance ($\Delta$) due to its dominant role in SPP.
    The method outlined in~\cite{Oset87,Nieves93,Gil97} is widely used in the neutrino-interaction community.
    In this approach, in the nuclear medium, due to the presence of other nucleons, the width of the $\Delta$ is reduced due to Pauli blocking, but on the other hand, it is increased by the collisions inside the nucleus via the processes $\Delta N\rightarrow NN$ and $\Delta NN\rightarrow NNN$. The $\Delta$ may produce a secondary pion through the reaction $\Delta N\rightarrow \pi NN$, instead of its decay to $N\pi$. These processes contribute to the modifications of the $\Delta$ width inside the nuclear medium.
    Other studies on the modification of the Delta properties are~\cite{Cugnon88, Arve94, Lee96, Kim97, Ghosh17}.

    The pionless decay channels, which in the framework of~\cite{Oset87,Nieves93,Gil97} correspond to the channels $\Delta N\rightarrow NN$ and $\Delta NN\rightarrow NNN$, is of pivotal importance for neutrino oscillation experiments since it contributes to the set of `QE-like' or $0\pi$ samples. Describing this process is not a trivial task. In MC event generators, within the classical picture of the cascade models, it is modelled as a real pion, produced in the primary vertex, for example due to a $\Delta$ decay, which is later absorbed in the nucleus due to (cascade) FSI. Given its relevance, this topic should be revisited and studied in greater depth.

    \item {\it Inelastic final-state interactions:} The pion produced in the primary vertex, apart from elastic rescattering, may undergo various interactions, including charge exchange, absorption, nucleon knockout, and even create new pions if the energy is large enough. In fact, the pion that is detected may not necessarily be the same one produced at the primary vertex but a result of secondary interactions. In MC event generators inelastic interactions are handled by the cascade models, which is the subject of Section~\ref{fsi}.

\end{enumerate}

\subsection{Status and prospects of measurements}
\subsubsection{nucleon (deuterium) target}

Experimental knowledge of the neutrino-nucleon cross section for single pion production in the GeV neutrino energy range, relevant for current and planned oscillation experiments, is limited and primarily comes from low-statistics bubble chamber experiments with hydrogen or deuterium targets. These key experiments are:
\begin{itemize}
    \item ANL: The data~\cite{Radecky:1981fn} were obtained from both hydrogen and deuterium fillings\footnote{The event rates are a combination of both hydrogen and deuterium fills of the detector} of the 12-ft bubble chamber. The wide-band neutrino beam peaks at 0.5 GeV and extends to 6 GeV.
    \item BNL: The data~\cite{Kitagaki:1986ct} were obtained from deuterium fillings of the 7-ft bubble chamber. The wide-band neutrino beam peaks at 1.2 GeV and extends to 15 GeV.
    \end{itemize}
The ANL and BNL cross-section data differ in normalization by approximately 30\%, a discrepancy that has been controversial for decades. Some studies suggest that this is due to differences in the flux predictions. However, there are additional reasons for this discrepancy:
\begin{enumerate}
    \item Energy Beam Differences: The ANL beam predominantly covers the hadron dominant region, while the BNL beam extends into the  non-perturbative quark-hadron transition region.
    \item Target Differences: The BNL event rate is based on deuterium fills, whereas the ANL event rate is a combination of both hydrogen and deuterium fills. Recent measurements with an electron beam~\cite{CLAS:2022kta} indicate that the hydrogen and deuterium cross sections differ by about 15\%.
\end{enumerate}
This data is crucial for constraining the axial form factor for pion production on free nucleons. The axial form factor at $Q^2>0.0~ (\text{GeV}/c)^2$
  cannot be constrained by any other data, leading to significant uncertainties in predictions for oscillation experiments. A new experiment is the most appropriate way to resolve this long-standing controversy.

\begin{itemize}
\item {\textbf{BEBC}
:} The single pion measurements were initially conducted on a hydrogen-filled bubble chamber with both neutrino and anti-neutrino wide-band beams peaking at 20~GeV. Subsequent data were collected using a deuterium fill. In total, 27,575 $\nu$ and 16,896 $\bar\nu$ events were recorded with the two beam settings. Although the BEBC energy range is not directly relevant for neutrino oscillation experiments, it is unique in having measured SPP cross-sections for both neutrinos and anti-neutrinos on hydrogen and deuterium targets. Consequently, the BEBC data on hydrogen targets at low $Q^2<1.0~ (\text{GeV}/c)^2$ \cite{WA21:1989gku} are particularly suitable for fitting the axial form factors in SPP models. The data show that the cross-sections for hydrogen and deuterium differ significantly in the first resonance region and non-perturbative regions.
\end{itemize}

\subsubsection{Recent and Prospective Measurements}
In only the last several years, a collection of analysis techniques have surfaced and matured which can better and more faithfully separate a wider array of model components and experimental variables. While pionless measurements have tended to drive the development and use these techniques, with the significant role that pion production will play in DUNE, it is important that increased effort now be devoted to exploring their application in resonance region.

In the era of modern high statistics nuclear target neutrino scattering experiments, more emphasis has been placed on pionless measurements, which have been more experimentally accessible and have a relatively cleaner event signature and theory. In general, any particular particle reconstruction such as individual pions poses additional challenges not faced in inclusive measurements. 
But even for experiments that have the capacity to reconstruct hadrons, nucleons reconstructed in pionless measurements are less sensitive to the numerous smearing and obfuscating final state effects, compared with the pion, which, as the lightest meson are strongly affected by them.

Experiments such as T2K and \minerva, the most mature experiments that have placed the most emphasis on cross-section measurements, have recently demonstrated increasing success and sophistication in reconstructing pions, but there is now much to do to explore the application of the full range of new and prospective measurement techniques to pions. Since the previous NuSTEC report, exciting progress has been made with T2K and \minerva each debuting new techniques that together probe each component of the neutrino interaction model: Transverse Kinematic Imbalance probes the nuclear initial state, T2K's Adler variables are sensitive to the nucleon-level interaction, and MINERvA's measurement of nuclear target cross section ratios are direct probes of nuclear effects.
\\

\paragraph{Recent Measurements and Analysis Techniques:}

\begin{itemize}
    \item \textbf{Transverse Kinematic Imbalance} - This technique, described above in \ref{sec:tki} provides a handle on the nuclear initial state, and was recently measured by T2K and \minerva for charged-current $\pi^+$ and $\pi^0$ channels \cite{T2K:2021naz, minerva:2020anu}. The measurements include the transverse kinematic imbalance (which was previously measured in CC QE-like production), as well as the double-transverse momentum imbalance ($\delta p_{TT}$). The results show that models have different predictions where the imbalance is small (around $\delta p_{TT}=0$) which suggests that the discrepancy comes from initial neutrino-nucleon interactions~\cite{lu:2015hea}.
    \item \textbf{Adler Variables} - Adler variables, defined in the hadronic $p\pi$ rest frame \cite{Adler:1968tw}, can provide some insight into the nucleon-level interaction and polarization of the $\Delta$, and they have been measured by \minerva in $\pi^0$ on CH \cite{MINERvA:2017okh} and more recently by T2K in $\pi^+$ on hydrocarbon in \cite{T2K:2019yqu}. Adler angles at the nucleon level are isotropic for each resonance, therefore, any asymmetry in their distribution is a result of the interference between resonances (resonances with the same spin and $N\pi$ orbital angular momentum interfere) and nonresonant interactions. This asymmetry has been shown in the old bubble chamber measurements with hydrogen target \cite{Aachen-Bonn-CERN-Munich-Oxford:1980iaz}. In the nuclear target, the Adler azimuthal angle ($\phi$-distribution) maintained its symmetry \cite{Sanchez:2015yvw}, therefore, the asymmetries in the $\phi_{Adler}$ measurements from \minerva and T2K experiments in the $\Delta$ region confirm the interference effect between the $\Delta$ resonance and the nonresonant interactions. 
    Compared with most of the generators, which include pion angles only from the $\Delta$ resonance (assume isotropy), data shows a preference for pion emission in the forward hemisphere. The T2K analysis, which is also summarized in \cite{Avanzini:2021qlx} measured an experimentally-accessible and somewhat less model dependent analogue to truth Adler variables and found overprediction from all generators. 
    
    \item \textbf{Nuclear Target Cross Section Ratios} - Most recently \minerva has published nuclear target ratios to scintillator of \numu~CC1$\pi^+$ \cite{MINERvA:2022djk} and shown preliminary results of nuclear target ratios to scintillator of \numu CC1$\pi^0$ \cite{minervanukepi0WC}. These results are the first of their kind, they directly probe nuclear effects, and they can be used to guide theoretical scaling from light to heavy nuclei such as argon. Cross section shape is found to be shared across nuclei for many distributions, for example $Q^2$ and $T_\pi$. GiBUU, NEUT, and GENIE3 do well at predicting the cross section ratios for heavy nuclei to scintillator, but no generator reproduces the correct absolute cross section normalization across nuclei. In particular, a strong observed suppression of heavier nuclei compared to lighter nuclei is not predicted. GiBUU, NEUT, and GENIE3, which use hN FSI, perform better at predicting this suppression compared with GENIE2 with hA-based FSI. Models prefer similar normalizations between $\pi^+$ and $\pi^0$ for each nuclei which are not observed in the data.
    \item \textbf{Model Independence Best Practices} - Much recent effort has been made to establish best practices for the reduction of model dependence in cross section measurements. 
    Observing these methods is important for theorists' and model builders' to more directly make use of accurate results. In pion production measurements, T2K has been the most explicit about analysis steps taken to model dependence. For example, in Ref. \cite{T2K:2021naz}, authors implement uncertainties from input neutrino interaction model predictions to cover variations in efficiency over the full phase space. The variations specially included resonant model factors ($M_A^{\text{RES}}$, $C_5^A$, and $I_{1/2}$), $\Delta^{++}$ decay width, DIS model parameters, several FSI parameters (modifying e.g. pion charge exchange), and ad hoc scale parameters which ``cover the efficiency's dependency on the initial state nuclear medium effects, which is not otherwise parameterized''.
\end{itemize}

\paragraph{Prospective Measurements and Analysis Techniques}
\begin{itemize}
    \item \textbf{Low Threshold, $T_\pi\sim0$, Measurement} – Measurements over the full range of pion phase space do not currently exist. In particular, pions at the lowest energy have not been measured. Indeed, data-model disagreement has been observed across experiments and pion production channels, with models overpredicting in the lowest bins $T_\pi$. This disagreement has been separately theorized to be a problem with the FSI model or a shift in the first resonance to lower invariant hadronic mass. To resolve this question, \minerva has developed a $T_\pi \sim 0$ reconstruction method that identifies low energy pions from its decay Michel electron signature alone. Alternatively, planned detectors such as the DUNE ND Phase II gas TPC and T2K's SuperFGD will have the capacity for a low-threshold measurement.
    \item \textbf{Neutron Detection} – A pion measurement with neutron detection would be helpful for separating neutrino and antineutrino interactions, and $\Delta^+$ vs $\Delta^{++}$ distinction would be very interesting for probing underlying physics. \minerva has shown some capacity for neutron detection, planned detectors, T2K's Gd-doped water cherenkov Titus detector and T2K's ND upgrade SuperFGD, will have increased capacity for neutron detection.
    \item \textbf{PRISM Measurements} – PRISM-style measurements will be valuable for measuring $E_{\nu}$ dependence, and prior to DUNE, TITUS and SBND will have the capacity for this measurement.
    \item \textbf{Theoretical Insights} – Theoretical developments play a central role in interpreting experimental data and motivating new measurements that can disentangle the complex reaction mechanisms contributing within the same kinematic region. In turn, new and increasingly differential measurements provide essential constraints for refining theoretical models and reducing their uncertainties. This synergy is particularly important in the broad resonance region, where resonant and non-resonant pion production, nuclear  effects, and the overlap between hadronic and partonic descriptions all contribute to the observed cross sections. Theoretical studies can also help explain apparent discrepancies between measurements by identifying the underlying physical mechanisms or differences in experimental phase-space coverage. For example, the longstanding discrepancy between the MiniBooNE and \minerva\ CC1$\pi^+$ cross sections \cite{sobczyk:2014xza} has been attributed, at least in part, to \minerva's broader $Q^2$ acceptance, which samples a wider kinematic region encompassing both hadronic and quark-level dynamics.
    \item \textbf{Correlated Measurements} – Much recent emphasis has been placed on measurements and model improvements of the $\nu_\mu CC 1\pi$ channel. \minerva, however with the \numubar~$\pi^-$ result in 2019, has completed a suite of CC$\pi$ measurements on scintillator of charged and neutral pions in neutrino and antineutrino modes. In general less emphasis has been placed on measurements and comparisons across chirality, flavor, pion channel. Section~\ref{pionless} has highlighted the value of cross-experiment, multi-channel, and cross-chirality analyses for identifying shared and distinct nuclear and interaction effects. There are strong future opportunities for such correlated measurements, particularly through joint analyses such as the ongoing T2K–NOvA effort.
\end{itemize}

\newpage
\section{Shallow and Deep Inelastic Scattering}
\label{sisdis}

At higher energy and momentum transfer beyond the resonance region, interactions become increasingly inelastic.  We currently define two primary categories of interaction in this region: the traditional ``deep" inelastic scattering, and a ``shallow'' inelastic scattering.  The separation of these regions is mainly phenomenological steered by momentum transfer $Q^2$ but the regions of phase space they cover pose different challenges to both theory and experiment.
At the DUNE experiment a significant fraction of events will be in these categories, making them critical to understand.

\subsection{Deep inelastic scattering}
When an (anti)neutrino interacts with a nucleon via the exchange of intermediate vector boson, at sufficiently large four momentum transfer squared~($Q^2$), the nucleon breaks up, completely loses its identity and produces a jet of hadrons, mainly mesons, in the final state along with the corresponding lepton. 
This process actually encompasses two different kinematic regions, but with no hard boundary between them.

These higher-$Q^2$ interactions are used in determining the parton distribution functions within the nucleon.  DIS cross sections are expressed in terms of the nucleon structure functions, which are in turn described in terms of the parton distribution functions~(PDFs). These nucleon
structure functions are written as a perturbative expansion of the strong coupling constant incorporating the leading order as well as higher order terms.
The expression for the differential cross section for the $\nu_l/{\bar\nu}_l - N$ scattering is given by:
\begin{eqnarray}\label{xsecsf}
 \frac{d^2\sigma_N}{dxdy}&=&\frac{G_F^2M_NE_\nu}{\pi(1+\frac{Q^2}{M_W^2})^2}
 \Big\{\Big[y^2x+\frac{m_l^2 y}{2E_\nu M_N}\Big]F_{1N}(x,Q^2)+
 \Big[\Big(1-\frac{m_l^2}{4E_\nu^2}\Big)-\Big(1+\frac{M_Nx}{2E_\nu}\Big)y\Big]F_{2N}(x,Q^2)\nonumber\\
 &\pm& \Big[xy\Big(1-\frac{y}{2}\Big)-
 \frac{m_l^2 y}{4E_\nu M_N}\Big]F_{3N}(x,Q^2)
 +\frac{m_l^2(m_l^2+Q^2)}{4E_\nu^2M_N^2 x}F_{4N}(x,Q^2)-\frac{m_l^2}{E_\nu M_N}F_{5N}(x,Q^2)\Big\}.\;\;\;
\end{eqnarray}
  where $G_F$ is the Fermi coupling constant, $F_{iN}(x,Q^2);~~i=1-5$ are the nucleon structure functions, $M_N$ is the target nucleon mass, $m_l$ is the charged lepton mass and $\nu$ is the energy transfer.
 The scaling variables $x\big(=\frac{Q^2}{2 p \cdot q}\big)$ and $y(=\frac{\nu}{E_\nu}=\frac{q_0}{E_\nu})$ lie in the range:
\begin{equation}\label{xydef}
 \frac{m_l^2}{2M_N (E_\nu - m_l)} \le x \le 1~~~~~~~~~\text{and}~~~~~~~~~~~
 a-b \le y\le a+b,
\end{equation}
where
\begin{eqnarray}
 a=\frac{1-m_l^2\Big(\frac{1}{2M_NE_\nu x}+\frac{1}{2E_\nu^2} \Big)}{2\Big(1+\frac{M_N x}{2E_\nu}\Big)}~~~~~~~~\text{and}~~~~~~~~
 b=\frac{\sqrt{\left(1-\frac{m_l^2}{2 M_N E_\nu x}\right)^2-\frac{m_l^2}{E_\nu^2}}}{2\Big(1+\frac{M_N x}{2E_\nu}\Big)}.
\end{eqnarray}
In the literature, different parameterizations of PDFs are
available to determine the nucleon structure functions such as CTEQ, MSTW, GJR, MMHT, etc., and are continuously being updated with the latest experimental data.

To roughly define this rigorous DIS region quantitatively, two requirements have to be met. First, to have sufficient
 $Q^2$ to scatter off a \st{single} quark within the nucleon without the necessity of including non-perturbative effects such as multi-quark scattering.
 \st{rather than interacting with multiple quarks or even the whole nucleon.}
 The second is reasonable probability of finding a quark off which to scatter at the kinematic point in question.
Note that although this definition of DIS has involved only $Q^2$, the hadronic mass of the final state hadronic system W is also significant.
The inclusion of W dependence in the definition of DIS enters qualitatively to avoid the region dominated by the resonance excitations and quantitatively through the second DIS requirement, suggesting the probability of finding a quark on which to scatter with the chosen $Q^2$ restriction.  Using the combined $Q^2$ and W requirements of the region in question, restrictions on $x_{Bj}$ and the corresponding distribution of quarks within the nucleon directly suggests the probability of finding a quark with which to interact.

The global QCD fitting community takes roughly $Q^2 \geq 4$~GeV$^2$ as the perturbative QCD region in lepton-nucleon/nucleus scattering involving scattering off a single quark enabling extraction of the PDFs. To understand if these high $Q^2$ interactions can take place at low W in the resonance region, we define the dominant resonance region as  $W\leq$ 1.5 GeV.  With these $Q^2$ and W
requirements, the resulting restriction on the $x_{Bj}$ distribution suggest that the probability of finding a quark in this  region is less than 1\% limiting the perturbative QCD DIS contribution in this dominant resonance W region.

\subsection{Shallow Inelastic Scattering}
The most direct way to describe Shallow Inelatic Scattering may be by defining what it is not.  It is not resonant meson production and it is not meson production by higher $Q^2$  perturbative QCD quark interactions.

Shallow inelastic scattering~(SIS) is then two different regions. The first region is the predominantly lower-$Q^2$ non-resonant meson production region. The second region begins at the somewhat higher $Q^2$ threshold of interactions within the nucleon ($Q^2 \approx 1$ GeV$^2$) and proceeds via increasing $Q^2$ through the non-perturbative QCD SIS interactions such as multi-quark interactions.   When $Q^2$ becomes large enough, we enter the quark perturbative QCD DIS region, which is around $Q^2 \approx $ 4 $GeV^2$ from PDF global fitting efforts. 
Within this "multi-quark" scattering SIS region
the nonperturbative QCD corrections such as Target Mass Correction (TMC) and dynamic higher twist effects become important.

As discussed in the previous (Resonance) section, the community suggested a cut of $W < 2$ GeV that isolates a region dominated by resonance production, which starts with the main contribution of the $\Delta$ resonance (W=1.232 GeV) and multiple smaller higher-W resonances.
The nonresonant meson production and non-perturbative multi-quark meson production from the SIS region intermix with the resonant meson production and there is no possible way
to separate, experimentally, individual events that produce mesons 
by these processes.  Therefore,  in practice, it is difficult to experimentally have a well-defined SIS region, clearly separate from the resonance region and the \st{true} perturbative DIS region.  The best way to define a region rich in SIS events is to remove reconstructed resonance events in the region and confirm that the $Q^2$ cut is efficient in separating non-perturbative and perturbative events.
The physics of this complex region has to be better understood both
theoretically and experimentally.

Except for neutrino beams from the Forward Physics Facility (FPF) at CERN,  most of the present day neutrino experiments are being performed in the few GeV energy region, quite similar to the energies of early neutrino experiments.
The current neutrino experimental community then often takes $Q^2$ $\geq$ 1 GeV$^2$ and W $\geq$ 2 GeV to define the "DIS" region. This lower $Q^2$ definition is a limit from very early (1970's) attempts to use the then defined ``scaling limit" at $Q^2$ $>$ $M_N^2$ (generalized to 1 GeV$^2$) as the start of DIS in early SLAC and neutrino bubble chamber experiments~\cite{Eichten:1973cs}. This is well below the perturbative QCD defined DIS interaction region that is a more contemporary physics-based definition. This lower restriction on $Q^2$ $\geq$ 1 GeV$^2$  implies the
inclusion of non-perturbative (SIS) events in their "DIS" sample suggesting that this lower-$Q^2$ cut sample be referred to as "experimental DIS" to avoid confusion with the higher-$Q^2$ perturbativve DIS region . This inclusion of SIS events in the experimental DIS region is demonstrated by the nCTEQ global analysis ~\cite{Segarra:2020gtj} to lower-$Q^2$/lower-W  data.
This study showed that to lower the $Q^2$ cutoff from 4 GeV$^2$ to even 1.7 GeV$^2$ in electroproduction, still above the 1.0 GeV$^2$ cut used in the current definition
of "experimental DIS", a reasonable global fit could only be obtained by including non-perturbative target mass corrections (TMC) and higher-twist effects. These nonperturbative QCD
corrections will be discussed in the following sections.

\subsubsection{Non-perturbative QCD correction in the SIS region}
\label{sec:DIS:HT}

In going from the higher-$Q^2$ DIS perturbative-QCD region to the lower-$Q^2$ SIS region, non-perturbative
phenomena become important for the precise modeling of cross sections.
Structure functions can be expressed in terms of powers of $1/Q^2$ (power corrections):
\begin{eqnarray}
F_{i}(x,Q^2) = F_{i}^{\tau = 2}(x,Q^2)
+ Q^{-2} H_{i}^{\tau = 4}(x)
+ Q^{-4} H_{i}^{\tau = 6}(x) + .....   \;\;\; i=1,2,3,
\label{eqn:ht}
\end{eqnarray}

\noindent where the first term ($\tau=2$) is known as the twist-two or leading twist (LT) term,
and corresponds to the scattering off a single free quark.
The dynamical higher twist (HT) terms with $\tau = 4,6$,\ldots
reflect the strength of multi-parton correlations ($qq$ and $qg$).
In addition to the dynamical HT terms, we also have
kinematic HT contributions associated with the finite mass of the target nucleon $M_N$,
which are mostly relevant when $Q^2$ approaches $M_N^2$. The corresponding target mass
corrections (TMC) involving powers of $1/Q^2$ are usually incorporated into the leading twist (LT)
term following the prescription of Refs.~\cite{Nachtmann:1973mr,Georgi:1976ve,Steffens:2012jx}. The inclusion of TMC does act as a form of higher twist addition as well.

\begin{figure}[!h]
	\centering
	\includegraphics[height=6 cm, width=8 cm]{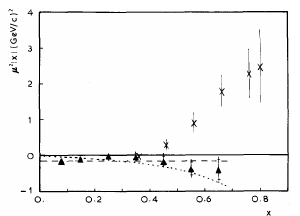}
	\caption{
Results of a fit to the x dependence of a general higher twist term of the form HT(x and $Q^2$) = $\mu^2 / Q^2$. Neutrino results are the solid triangles while electroproduction are the crosses.From ~\cite{Morfin:2021ujm}.}
	\label{fig:NuHighTwist}
\end{figure}

Although it has been shown in electroproduction that once the target mass
corrections are made, the effects of additional dynamical higher-twist terms are small, this is not surprising if indeed the TM correction is already acting as a form of higher twist correction.  It is interesting to note that there are
experimental results~\cite{Morfin:2021ujm} indicating the HT contributions themselves for $\nu$-A scattering are indeed different than the HT terms for charged lepton scattering as shown in figure~\ref{fig:NuHighTwist}. For electroproduction the nonperturbative
kinematic and dynamic QCD corrections are found significant in the region of high Bjorken $x$, where the dominant contribution comes
from the valence quarks and the nucleon/nuclear PDFs fall steeply. In this region
of $x$, due to experimental constraints as well as the lack in the understanding of high $x$ and low $Q^2$ region, PDFs are not yet very well determined in electroproduction and even more poorly constrained in $\nu$ - nucleus scattering.  The observed difference in HT between electro- and neutrino- production needs more experimental and theoretical study.

\subsubsection{Quark-Hadron Duality}
\label{sec:DIS:QHduality}

Since non-perturbative QCD is so poorly understood both theoretically and experimentally, the concept of QH duality has been considered as an alternative way to join DIS to SIS and Resonance regions.
QH duality is a concept first introduced by Bloom and Gilman~\cite{Bloom:1970xb, Bloom:1971ye} to describe observations in electron-proton scattering.  The concept notes that the description of scattering in the DIS region uses the language of quarks and gluons while in the resonance region the language is pions and nucleons. However the inclusive cross section unites the two kinematic regions so that the two descriptions must meet.  It is then interpreted that the resonance structure functions
in the low $Q^2$ region, suitably averaged over an energy interval, provides the same
result as the corresponding DIS structure functions at high $Q^2$, in the same energy interval.  This phenomenon then provides a connection between
quark-gluon description of the DIS formalism at high $Q^2$, and the pion-nucleon description
in the region of resonance excitation at lower $Q^2$. QH duality seems to hold fairly well for charged-lepton scattering but there are many difficulties and unanswered challenges in applying it to neutrino scattering such as QH duality not holding independently for neutrino-proton and neutrino-neutron interactions ~\cite{SajjadAthar:2020nvy}.
Until the application of QH duality to neutrino interactions is significantly better understood it is questionable to apply duality to neutrino interactions in event generators.
\subsection{Nuclear medium effects}
When the scattering takes place with a nucleon within the nuclear environment various nuclear medium effects come into the picture such as binding energy, Fermi motion, nucleon
correlations, shadowing, mesonic cloud contribution, etc. The DIS nuclear medium effects are found to be $x$ dependent and according to their dependence they are broadly divided into the relevant regions of $x$. For example,
shadowing effects which are considered to arise due to the destructive interference of amplitudes arising from the multiple scattering of the intermediate vector boson hadronic fluctuations off nucleons inside the nucleus, contributes in the region of
low $x$, i.e., $x\le 0.1$.

Since nucleons within the nucleus may interact with each other through strong interaction via the exchange of virtual mesons like $\pi,~ \rho, ~\omega,$ etc.,
there exists a possibility that the intermediate vector boson may have interaction with these virtual mesons instead of nucleons. This is known as
mesonic clouds contribution which is found to be significant in the intermediate region of $0.2\le x\le 0.6$. Beyond $x>0.7$, there is the Fermi motion effect. Furthermore, the nucleons within the nuclear environment are
not necessarily on-shell i.e. $p^2 \ne M^2$, therefore the structure functions will not
only depend upon the variables $x$ and $Q^2$ but also on the off shell virtuality. In the case of heavy nuclear target, where number of neutrons
 is larger than the number of protons, neutron excess corrections should be taken into account.
 This has recently been discussed in Refs.~\cite{SajjadAthar:2022pjt, SajjadAthar:2020nvy}.

The sum effect of these nuclear medium effects is that the PDFs of a nucleon in this complicated nuclear medium are not the same as the PDFs of a free nucleon.  These nPDFs have been calculated by several collaborations including the nCTEQ (\cite{Muzakka:2022wey}), EPS(\cite{Paukkunen:2013grz}) and HKN(\cite{Nakamura:2016cnn}) groups.
 These analyses examined the question of whether the medium effects in the weak sector would be the same as in the electromagnetic sector since the axial vector response of the hadronic current is important in the weak sector.  Both the nCTEQ and HKN analyses found that indeed there are statistically significant differences in the nPDFs of neutrino nucleus interactions compared to electromagnetic nucleus interactions while the EPS analysis using different statistical approaches including ignoring NuTeV correlated errors found only mild disagreements with electromagnetic results.

\begin{figure}[!h]
	\centering
	\includegraphics[scale=0.8]{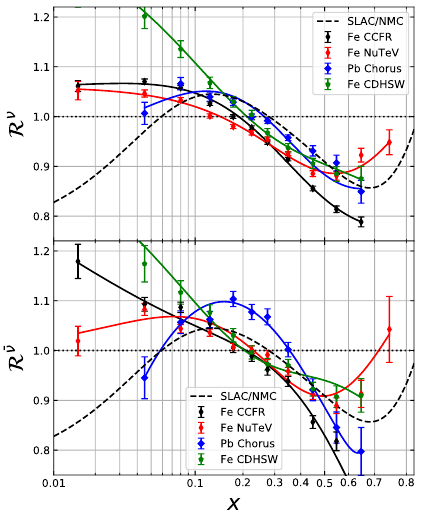}
	\caption{
The weighted average of the cross section for $Q^2 > 4   ~GeV^2$ and $W^2 > 12.25   ~GeV^2$  from Fe and Pb data compared to the free nucleon cross section computed using nCTEQ15 proton baseline. The black dashed line shows a prediction for purely vector interactions~\cite{Muzakka:2022wey}.}
	\label{fig:nCTEQR}
\end{figure}

 Using the nCTEQ results as an example, figure~\ref{fig:nCTEQR} shows the ratio of $\nu-A$  to $\nu-N$ for Fe and Pb.   Also shown is the SLAC/NMC electromagnetic curve.  It is obvious that the three Fe results differ significantly from the SLAC/NMC curve in the low-x shadowing region and, less significantly, in the EMC-Fermi motion transition region.  This nCTEQ analysis identified a subset of neutrino data which has no tension with the electromagnetic data and can be safely included in a combined global analysis.  Including just the scattering data from CHORUS Pb and the di-muon data from NuTeV and CCFR yields a much improved description of the strange quark PDF. Unfortunately, the majority of neutrino DIS cross-section data have been left unused in the process. The fact that the bulk of the DIS neutrino cross-section data is incompatible with the electromagnetic DIS data is established.
 That the CHORUS Pb results are in general agreement with SLAC/NMC results is quite surprisingly and needs understanding.  
 A new nCTEQ global analysis over an extended $Q^2$ and W region currently underway~\cite{Risse:2023wxr} will include both neutrino, namely the CHORUS Pb and di-lepton results, as well as electromagnetic results.

 Recently a group~\cite{nCTEQ:2023cpo} performed a first-ever global
QCD analysis of nuclear PDFs using a framework based on concepts from SRC nuclear models thereby testing modification of quark-gluon distributions in nuclei in correlated nucleon pairs.  The analysis
determined a universal distribution of partons in SRC nucleon
pairs as wll as the fractions of such SRC pairs for different A nuclei. It leads to similar, or better, data description as compared to the traditional parameterization, and enables a meaningful physical interpretation of the fit. This analysis represents a direct extraction of nuclear structure information from experimental observables directly probing quark-gluon nuclei dynamics. The fact that the obtained fractions of SRC pairs agree with previous extractions from the low-energy nuclear data establishes a direct link between high-
energy partonic properties and lower-energy nuclear physics. It thus presents a significant advance in our quest to understand atomic nuclei in terms of QCD.

\begin{figure}
\begin{center}
\includegraphics[width=0.65\textwidth]{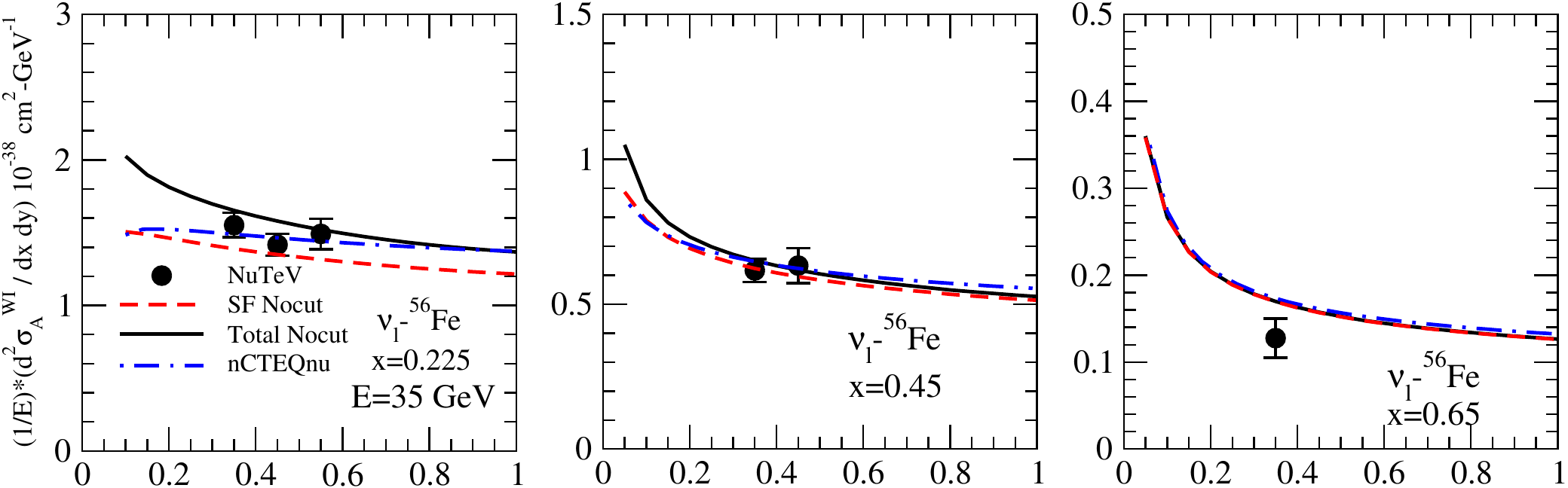}\\
\includegraphics[width=0.65\textwidth]{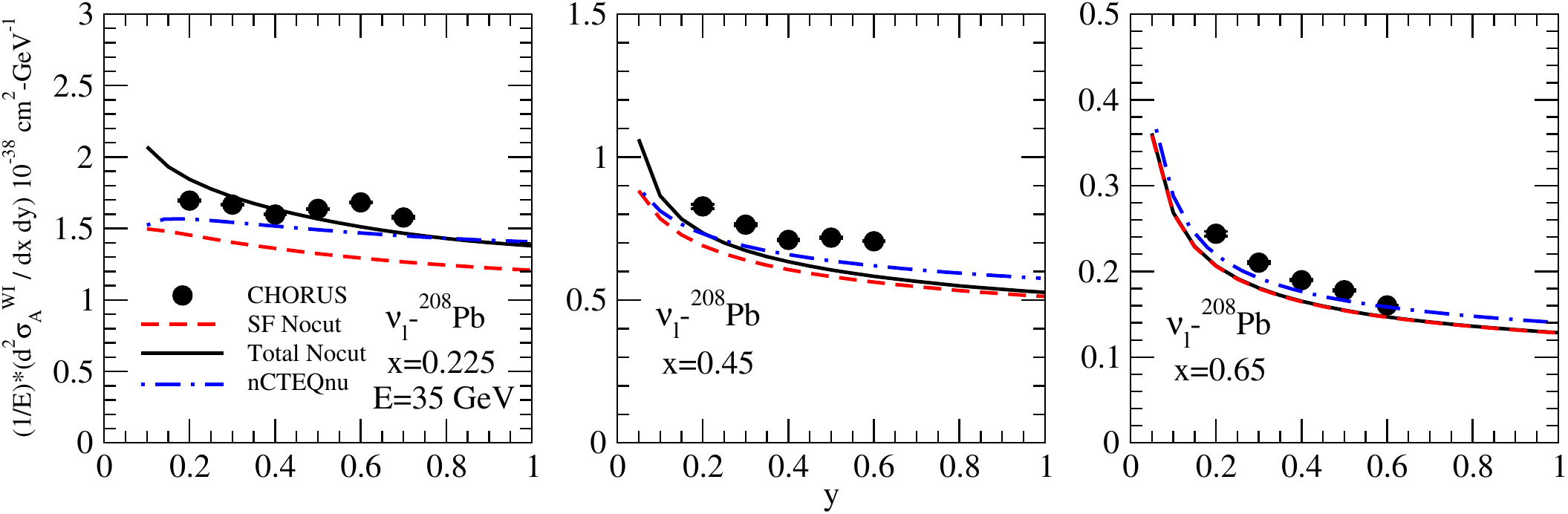}
\caption{Differential cross section vs y for $\nu_l-^{56}Fe$ (top panel)
and $\nu_l-^{208}Pb$ DIS (bottom panel) at the different values of x~\cite{SajjadAthar:2020nvy}.
Theoretical predictions are shown with the spectral function only (dashed line) and with the full model (solid line) at NNLO.
The blue dash-dotted line in the top row is the result from nCTEQnu nPDFs with $Q^2 \ge 1.0 ~GeV^2$. Solid circles with error bars
are the limited experimental data points of NuTeV~\cite{NuTeV:2005wsg} at this lower energy.}
\label{fig:35GeVFe-nu&nub}
\end{center}
\end{figure}

\begin{figure}
\begin{center}
  \includegraphics[height= 5 cm , width= 0.45\textwidth]{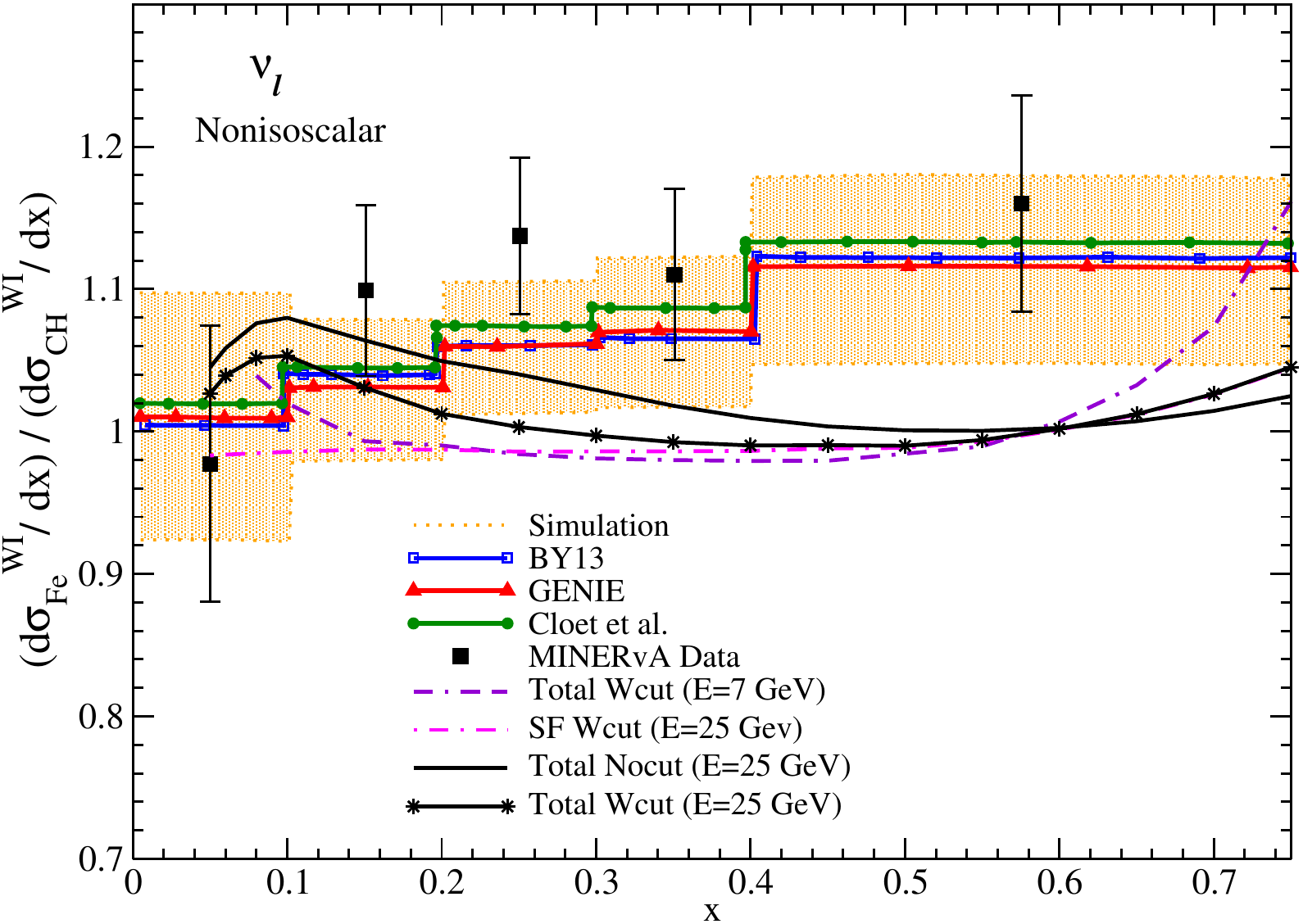}
  \includegraphics[height= 5 cm , width= 0.45\textwidth]{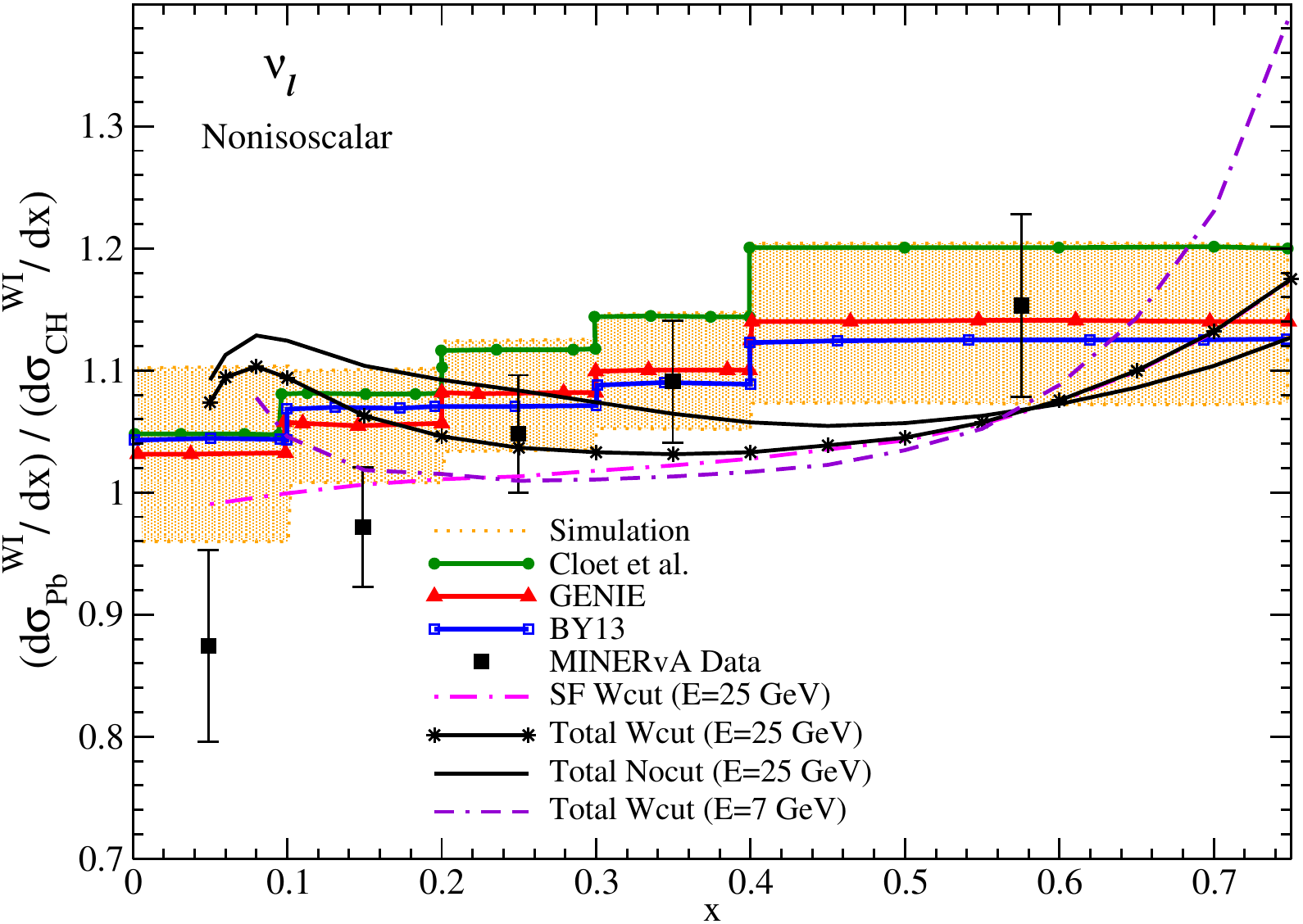}
\end{center}
 \caption{$\frac{d\sigma^{WI}_{A}/dx}{d\sigma^{WI}_{CH}/dx}~(A=$ $^{56}$Fe, $^{208}$Pb) vs $x$ for incoming neutrino beam of energies $E=7$ GeV
 and $25$ GeV off iron (left panel) and lead (right panel) target nuclei~\cite{Zaidi:2019asc}. The numerical results are
 obtained with the spectral function only (dash-dotted line: $E=25$ GeV) as well as with the full model (solid line: $E=25$ GeV, solid line with star: $E=25$ GeV
 and double dash-dotted line: $E=7$ GeV) at NNLO and are compared with the phenomenological results of Cloet et al.~\cite{Cloet:2006bq}, Bodek-Yang~\cite{Bodek:2010km},
 GENIE Monte Carlo~\cite{Andreopoulos:2009rq} and with the simulated results~\cite{MINERvA:2016oql}. The solid squares are the experimental points of MINERvA~\cite{MINERvA:2016oql}.
 The results are shown for the nonisoscalar nuclear targets.}
\label{fig6av}
\end{figure}

 In the theoretical sector, the nuclear medium effects for the (anti)neutrino-nucleus DIS process have only been studied by the Aligarh-Valencia group~\cite{SajjadAthar:2020nvy}, \cite{SajjadAthar:2022pjt},
 \cite{Zaidi:2019asc}, \cite{Zaidi:2021iam}, \cite{AtharSajjad:2022ipr} and the group of Kulagin and Petti~\cite{Kulagin:2007ju}. Aligarh-Valencia group has also studied the nuclear medium effects in charged lepton-nucleus DIS process, and found that the nuclear medium effects in electromagnetic and weak sectors are different especially in the region of low Bjorken $x$. This group has used a microscopic field theoretical model in which the effects of Fermi
motion, binding energy and nucleon correlations are incorporated through a relativistic nucleon spectral function. The spectral function that describes the energy and momentum distribution of the nucleons in nuclei is obtained by using the Lehmann's representation for the relativistic nucleon propagator and nuclear many body theory is used to calculate it for an interacting Fermi sea in the nuclear matter~\cite{FernandezdeCordoba:1991wf}. A local density approximation is then applied to translate these
 results to a finite nucleus. Furthermore, the contributions of the pion and rho meson clouds in a many body
field theoretical approach have also been considered. For incorporating the coherent nuclear effects Aligarh-Valencia group followed the works of Kulagin and Petti~\cite{Kulagin:2004ie,Kulagin:2007ju}.
 In Figs.~\ref{fig:35GeVFe-nu&nub} and \ref{fig6av}, theoretical results reported by the Aligarh-Valencia group for the charged current neutrino-nucleus deep inelastic differential scattering cross sections are presented at the different beam energies.

The theoretical results shown in Fig.~\ref{fig:35GeVFe-nu&nub}, for the double differential scattering cross section of $\nu_\mu-^{56}Fe$ (top panel) and $\nu_\mu-^{208}$Pb (bottom panel)
are obtained at E$_\nu$ = 35 GeV using only the spectral function (dashed lines) as well as using the full model (solid lines). On comparing these curves, it may be observed that the mesonic contributions play important role in the region of $x \le 0.5$, however, shadowing effect is negligible in the present kinematic region. These results are also compared with the predictions using the phenomenological approach of nCTEQnu nuclear PDFs~\cite{Morfin:private}.

In Fig.~\ref{fig6av}, theoretical results of Aligarh-Valencia group
for the ratio ($\frac{d\sigma^{WI}_{A}/dx}{d\sigma^{WI}_{CH}/dx}$) vs $x$ are shown for $\nu_\mu-A$ DIS scattering off $^{56}Fe$ (left panel) and $^{208}Pb$ (right panel) target nuclei. It is found that the deviation of the ratio from unity
is comparatively smaller in $ \frac{d\sigma^{WI}_{Fe}/dx}{d\sigma^{WI}_{CH}/dx}$ than in $\frac{d\sigma^{WI}_{Pb}/dx}{d\sigma^{WI}_{CH}/dx}$ which reflects the fact that nuclear medium effects are $A$ dependent. On comparing the results obtained using the spectral function only (labeled as ``SF Wcut'') and using the full model (labeled as ``Total Wcut'') with a cut of $W>2$ GeV, it may be realised that the effect of
mesonic cloud contributions become more pronounced for the heavier nuclear target at the low and intermediate $x$. The nuclear mass dependence as well as the isoscalarity corrections are found to be significant even at high values of neutrino beam energies. Furthermore, these results are relevant to understand the energy dependence and the impact of center of mass energy cut on the scattering cross section.
These numerical results are compared with  MINERvA's early low-energy experimental data, using the experimental DIS cuts for $Q^2$ and $W$, which have large error bars due to statistical uncertainties, simulated results (a wide band of $\approx 20\%$ uncertainty due to systematics) as well as with the phenomenological results of  Cloet et al.~\cite{Cloet:2006bq} (solid line with circle), Bodek et al.~\cite{Bodek:2010km} (solid line with square) and the GENIE MC generator~\cite{Andreopoulos:2009rq} (solid line with triangle). It is noticeable that neither the theoretical predictions nor the phenomenological results are able to satisfactorily describe the observed ratios of differential cross sections in the entire region of Bjorken $x$.

\subsubsection{Low $Q^2$ axial vector shadowing}
\label{sec:DIS:AXSHA}

\begin{figure}[!h]
	\centering
	\includegraphics[height=6 cm, width=8 cm]{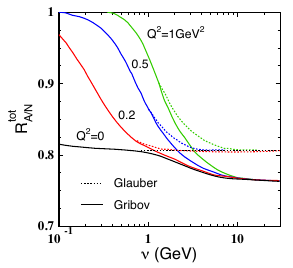}
	\caption{
The neon to nucleon ratio of total neutrino cross sections as a function of the hadron energy $\nu$ for various $Q^2$. Dashed and solid curves correspond to the
Glauber and Gribov corrected calculations (figure from ~\cite{kopeliovich:2012kw}).}
	\label{fig:axshad}
\end{figure}

Although this is below the $Q^2$ of even multiquark SIS, it is included here since axial shadowing could be quite important for the analysis of current and near future low-energy neutrino oscillation experiments. Due to considerations of the origins of PCAC, Kopeliovich ~\cite{Kopeliovich:2004px} demonstrates that shadowing can start at much lower energies for the axial current in neutrino production.   
The coherence time, the significant quantity for shadowing, for both the axial and vector current is proportional to the hadron energy transfer $\nu$, however it is inversley proportional to a term involving the pion mass squared for the axial current rather than the much larger mass of around 1 $GeV^2$ for the vector current.  The coherence time for the axial current then can be very long even at low energies of few hundred MeV. 
As an example, Kopeliovich calculated  expected shadowing using the Glauber approximation for neon depicted in Fig.~\ref{fig:axshad} by
solid curves as function of energy for different $Q^2$. 
As was anticipated, a rather strong shadowing occurs at small $Q^2$ in the low energy
region of a few hundreds of MeV. This remarkable behavior represents a striking and distinctive feature of the axial current. 

\subsection{Hadronization}

A DIS neutrino interaction double-differential cross-section yields both the outgoing lepton kinematics and defines the energy-momentum transfer vector to the hadronic system. Hadronization is the process of generating hadrons from this 4-momentum transfer vector, and for simplicity, it is decoupled from the rest of the formulation. Although hadronization is an important part of other particle physics sub-fields such as collider physics, it has been somewhat neglected in modern neutrino physics.

The importance of the neutrino hadronization process has been emphasized recently mainly because of the advent of new experiments focusing on the SIS-DIS energy region. These experiments include DUNE~\cite{dune:tdrvol2} and atmospheric neutrino experiments targeting the MSW resonance ($\sim$3-10~GeV, see Fig.~\ref{fig:oscillograms_abs}) including Hyper-Kamiokande~\cite{Hyper-Kamiokande:2025fci}, KM3NeT ORCA~\cite{KM3NeT:2021ozk}, and IceCube-Upgrade~\cite{IceCubeCollaborationP:2025rpl}. The visible hadron energy deposit is often used as an effective energy transfer. Precise measurement of it allows reconstruction of an effective inelasticity, which in turn allows a statistical $\nu-\bar\nu$ separation to enhance neutrino mass ordering sensitivity~\cite{Ribordy:2013xea,Olavarrieta:2024eaq}. This approach requires a prediction of final state hadron multiplicities and kinematics because they change energy distributions in the detectors where the detectors often sample a small phase space fraction of the hadron energy. For this, it is necessary to validate the hadronization model so that we can simulate the visible hadron energy deposit. 

Hadronization is described by phenomenological models. At low invariant mass region, data driven models are often used in generators, such as the KNO scaling model~\cite{Koba:1972ng} in GENIE~\cite{Yang:2009zx} and the multi-pi model in NEUT~\cite{Hayato:2021heg}.  At higher invariant mass region, PYTHIA~\cite{Sjostrand:2006za,Sjostrand:2007gs}, based on the Lund string fragmentation model, is widely accepted by all neutrino interaction generators. It has been pointed out~\cite{Kuzmin:2013tza} that all neutrino interaction generators underestimate averaged charged hadron multiplicity data from bubble chamber experiments. This is because default PYTHIA is tuned to higher energy experiments, and it requires tuning to adjust the model to lower energy experiments~\cite{katori:2014fxa}. A precise tuning eases the tension between hydrogen and deuterium bubble chamber data~\cite{Kuzmin:2013tza,GENIE:2021wox}. 

There are several tensions in bubble chamber data. The first one is the aforementioned hydrogen vs. deuterium data~\cite{Kuzmin:2013tza,GENIE:2021wox}. Although the mass number is small, deuterium is a nuclear target and comparison with hydrogen data requires some care. The second one is the tension between Fermilab 15' bubble chamber data and CERN-based data including Gargamelle and BEBC~\cite{Yang:2009zx}. The reason is unknown, however, old bubble chamber data tension between labs is not surprising in our community (cf., ANL-BNL data tension~\cite{Wilkinson:2014yfa}). The third one is the tension between charged hadron multiplicity and neutral pion multiplicity~\cite{katori:2014fxa}. The averaged neutral pion multiplicity follows the isospin symmetry ($\pi^++\pi^-:\pi^0=2:1$), and it is fixed with the averaged charged hadron multiplicity. Thus, it is difficult to simultaneously improve charged hadron and neutral pion multiplicities by changing PYTHIA parameters. Note, neutral pion data include heavier targets to minimize the radiation length where the nuclear effect may play some role.

\begin{figure} [tbp]
\centering
\includegraphics[width=0.49\textwidth]{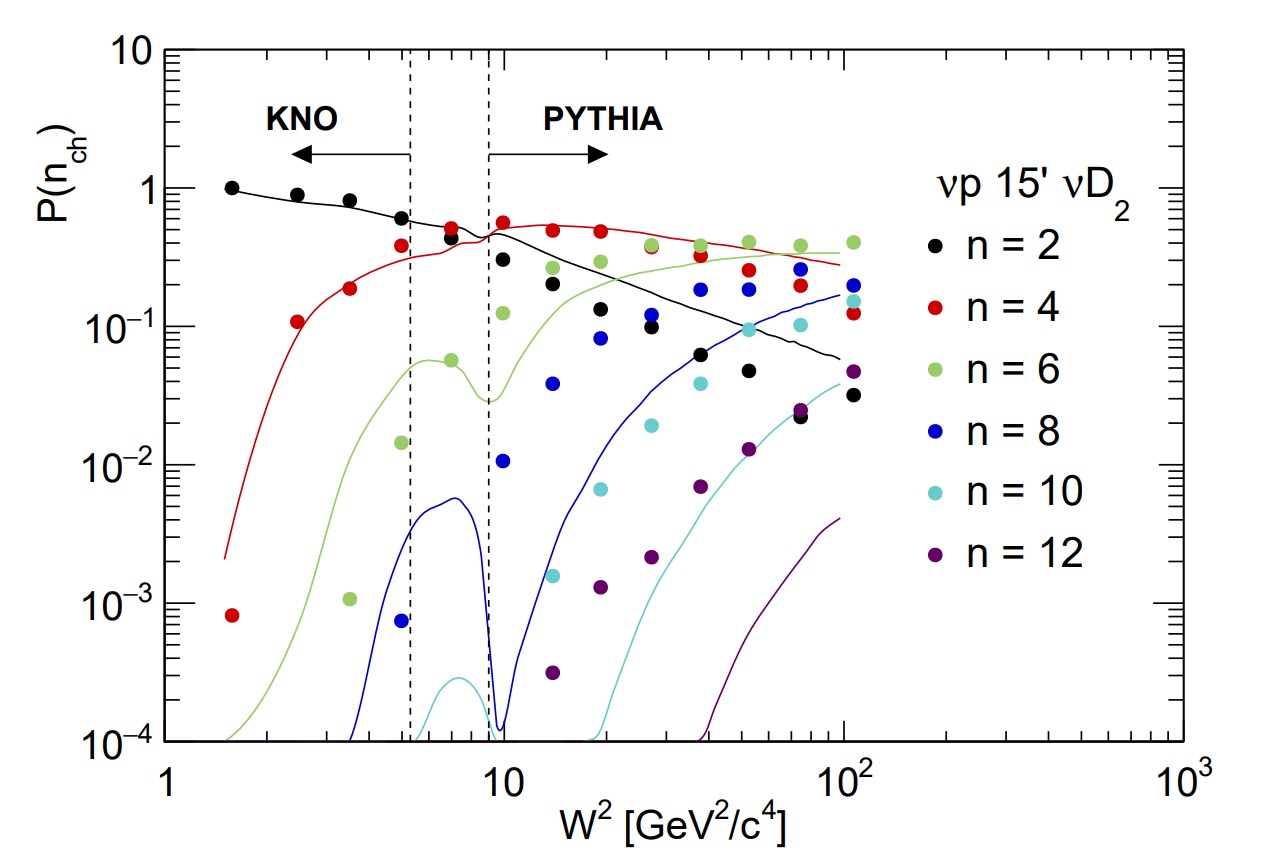}
\includegraphics[width=0.49\textwidth]{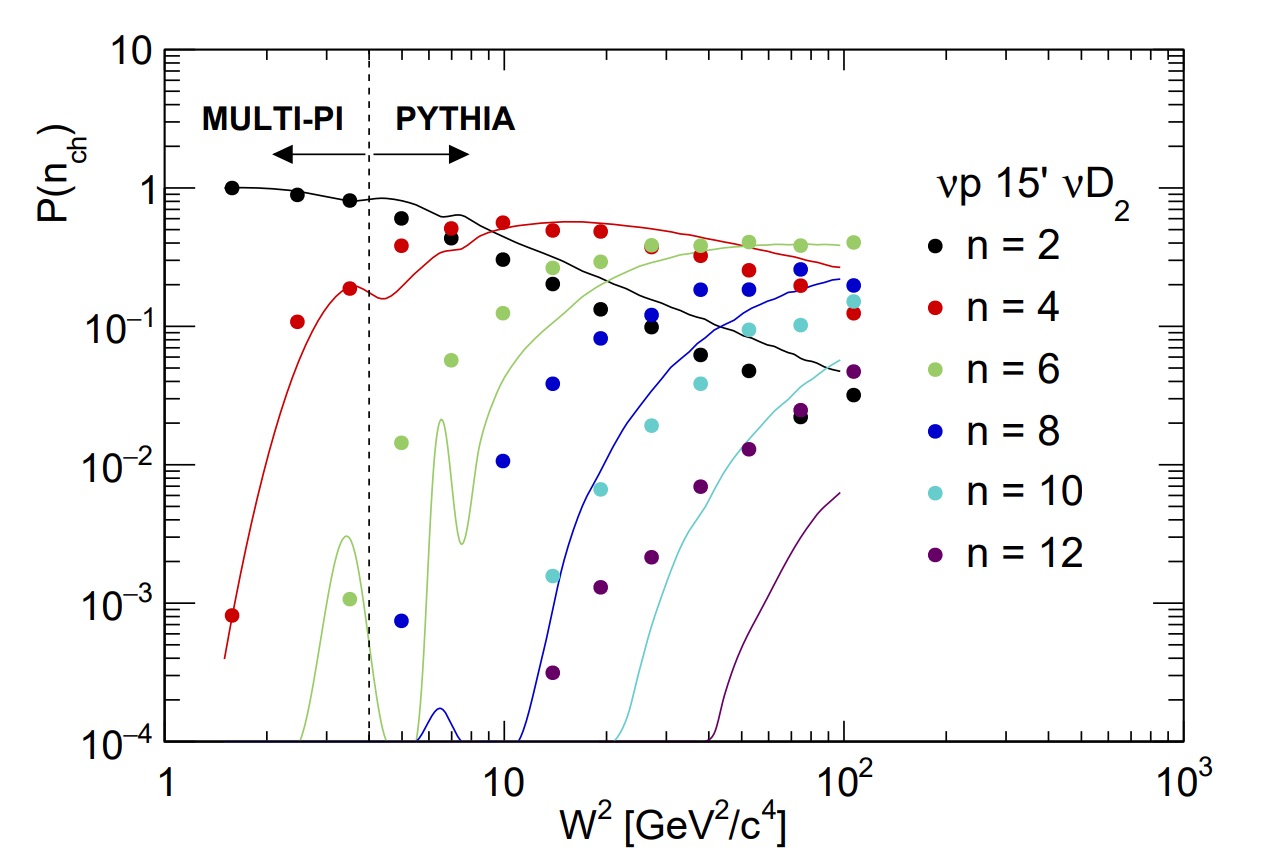}
\caption{\label{fig:SISTopo}
Data-simulation comparisons for the normalized topological cross-section data.  The Fermilab 15' bubble chamber data~\cite{Zieminska:1983bs} are compared with GENIE~\cite{GENIE:2021wox} (left, version 3 tune \texttt{G18\_02a\_02\_11a})) and NEUT~\cite{Bronner:2016gmz} (v5.4.0.1). both generators use custom methods to predict hadron multiplicity at very low invariant mass regions indicated by straight lines. Figures are taken from~\cite{Campbell:2022qmc}.}
\end{figure}

These data tensions focus on averaged hadron multiplicity. To perform event-by-event hadronization simulation, we need not only the average, but also the dispersion of multiplicity, {\it i.e.,} event-by-event probability of how many hadrons are generated with given invariant mass. PYTHIA cannot reproduce the neutrino hadron multiplicity dispersion data. This is demonstrated in Fig.~\ref{fig:SISTopo}, the topological cross-section prediction from GENIE and NEUT. This is because the hadron multiplicity dispersion predicted by PYTHIA is narrower than the neutrino hadron production data, and the simulation tend to produce lesser high-multiplicity events. This means, visible hadron energy deposit is likely to distribute more locally. The impact of such hadronization modeling on neutrino experiments depend on the detector design and performance. 

Understanding of neutrino hadronization in this energy region is an urgent topic. Without the correct prediction of hadron multiplicity including dispersion, the correct prediction of hadron energy deposit is impossible. This, hadronization problem, is just the first step of predicting the effective inelasticity in neutrino physics. In neutrino nucleus interactions, hadrons generated by primary interactions can undergo final state interactions within the nucleus, and then the hadrons may be re-scattered in the detector media before detection. We also see the difficulty of pursuing further by relying on old bubble chamber data. Modern hydrogen or deuteron bubble chamber data would solve all tensions~\cite{Alvarez-Ruso:2022exy}. Hadron multiplicity data from liquid argon TPC experiments including MicroBooNE~\cite{MicroBooNE:2018xad}, SBND~\cite{SBND:2020scp}, and ICARUS~\cite{ICARUS:2023gpo} can directly used to understand this problem for DUNE. Modern emulsion experiments, such as CHORUS~\cite{CHORUS:2007nlc}, 
DONuT~\cite{DONuT:2007bsg}, 
OPERA~\cite{OPERA:2017tas}, and NINJA~\cite{NINJA:2022zbi} can offer the highest precision hadron multiplicity data although with limited kinematics. 
Finally, high-energy neutrino experiments at CERN, including collider neutrino experiments such as FASER$\nu$~\cite{FASER:2023zcr,FASER:2024hoe,FASER:2024ref} and SND\@LHC~\cite{SNDLHC:2023pun}, 
and future SHiP~\cite{SHiP:2021nfo} and FPF~\cite{Feng:2022inv}, offer to study forward production of hadrons at high energy. 
These data have not been investigated yet, but they can also contribute to shed new light on our understanding of neutrino-induced hardonization. 

\subsection{Experimental Measurements}

Experimental measurements in the SIS and especially the DIS kinematic region require (anti)neutrino beams
of relatively high energies. Although historically the study of the DIS
region was one of the primary goals of early experiments, the focus on the measure of
 neutrino-oscillation parameters in modern experiments tends to emphasize
lower neutrino energies. For this reason, the opportunities to explore the SIS and DIS region in current experiments such as T2K and MicroBooNE are very limited. On the other hand, the current NOvA experiment - off-axis in the NuMI beam - and especially the MINERvA experiment - on-axis in the NuMI beam - do have the kinematic range to test the SIS region.  
Early  results from MINERvA~\cite{Norrick:2017tsh} suggested that simulation programs might reconsider their treatment of the SIS. A more recent study from MINERvA~\cite{MINERvA:2025hzq} explicitly of the SIS region demonstrates measured cross sections show significant deviations from current community simulation programs and suggests there is a considerable contribution from non-perturbative SIS interactions in the sample analyzed. To further the quantitative analysis of SIS interactions,  there is need to improve the scant information on resonance production in the higher W regions.  In addition, there has been no attempt to measure the hadron shower of SIS/DIS events to consider multiplicity and x, y and  $Q^2$ of the SIS multi-quark or DIS events.  
As for the future, it is worth noting that the understanding of the SIS and DIS regions is important for upcoming long-baseline and atmospheric oscillation experiments.
For instance, in the future DUNE experiment~\cite{Acciarri:2015uup} roughly 50\% of events will be in the W region beyond the Delta resonance.

Measurements from heavy nuclear targets are more abundant but are often limited by either the accumulated statistics, the
experimental detector granularity/resolution or both. Some of the existing higher statistics measurements
also provide somewhat conflicting results.
The first high statistics measurements (${\cal{O}}(10^7)$ events)
were performed by relatively coarse detectors like CDHS (iron)~\cite{Berge:1987zw,Berge:1989hr} and BCDMS~\cite{BCDMS:1989qop} as well as CHARM/CHARM II (marble/glass)~\cite{CHARM-II:1989nic}  based upon large
passive nuclear targets. The CCFR~\cite{Seligman:1997fe,CCRFNuTeV:2000wal} and
NuTeV~\cite{NuTeV:2005wsg,NuTeV:2001dfo}
experiments (iron) as well as the CHORUS experiment~\cite{KayisTopaksu:2011mx} using a lead calorimeter as a target~\cite{CHORUS:2005cpn} can be considered
as more modern experiments.
More recently the NOMAD experiment provided high resolution measurements from carbon and iron targets~\cite{Wu:2007ab,Samoylov:2013xoa} and the MINOS experiment performed cross section measurements in iron~\cite{Adamson:2009ju}, albeit with
somewhat limited experimental resolution.

The MINER$\nu$A experiment using the low-energy NuMI beam at Fermilab measured W $>$ 2 GeV and $Q^2$ $>$ 1 GeV$^2$ CC induced $\nu$-A cross sections on polystyrene, graphite,
iron and lead targets\cite{MINERvA:2016ing,MINERvA:2016oql}.  Currently, the MINER$\nu$A experiment using the much increased statistics and energy range of the NuMI medium-energy beam is producing a variety of both 1-D x-dependent and 2-D x,y dependent W $>$ 2 GeV and $Q^2$ $>$ 1 GeV$^2$ measurements on its full range of nuclear targets polystyrene, graphite, iron, and lead.  

The results of an earlier Shallow Inelastic Scattering  study~\cite{MINERvA:2025hzq} using the NuMI medium-energy beam and the higher statistics from the polystyrene target produced the first direct measurement of a predominantly SIS (called "multi-quark") region  with  a $Q^2$ cut of 1 GeV$^2$ to enhance the non-perturbative SIS contribution.  This study yielded the measurements of one-dimensional cross sections on hydrocarbon of $Q^2$, Bjorken x and muon momentum variables.  These measurements were compared with modified predictions from the GENIE 2 neutrino generator as well as predictions of other neutrino simulators GiBUU, NEUT, NuWro and an alternative GENIE 3 version. Significant discrepancies both in shape and magnitude between measurements and neutrino simulator predictions of all variables were observed.  It was also noted that in the SIS multi-quark region for all generators considered, the predicted resonance contribution is a relatively reduced fraction of the total cross section. This could suggest that besides any still unmodeled associated resonance production, there most likely is a significant contribution of SIS multi-quark interactions.

MINERvA is currently continuing the analysis of the SIS  1.5 GeV $< W <$ 2.0 GeV region with a study producing a variety of 2-D cross section measurements of $Q^2$, Bjorken x and muon momentum variables.  This 2D analysis will allow even deeper comparisons with current simulation efforts. 

\subsection{Comparisons between Models and Measurements}

Earlier, the total cross section $\sigma(E)$ was measured with good accuracy by
CDHS~\cite{Berge:1987zw}, CCFR~\cite{Seligman:1997fe}, and NuTeV~\cite{NuTeV:2005wsg} at high energies, resulting in a combined normalization
uncertainty of 2.1\% on $\sigma(E)/E$ for $E>40$ GeV safely in the DIS region. The  measurements by
NOMAD~\cite{Wu:2007ab} achieved good precision in the $3<E<40$ GeV region and MINOS~\cite{Adamson:2009ju},
MINER$\nu$A~\cite{MINERvA:2016ing}, and NOvA  achieved good precision down to $E\sim 1-2 $ GeV.

The measurements of the double differential cross sections $d \sigma / dxdy$ from CCFR (Fe)~\cite{CCRFNuTeV:2000wal}, NuTeV (Fe)~\cite{NuTeV:2005wsg} and CHORUS (Pb)~\cite{CHORUS:2005cpn} suggest
tensions among different data sets due to the quite a different nuclear target of Pb compared to Fe. The recent analysis by the nCTEQ collaboration~\cite{Muzakka:2022wey} confirms the earlier results \cite{Kovarik:2010uv} by the group showing that the existing $\nu A$ and $l^\pm A$ DIS data prefer different nuclear correction factors predominantly in the low-x region.
In particular, as summarized above, the data sets from the CCFR (Fe)~\cite{CCRFNuTeV:2000wal} and NuTeV (Fe)~\cite{NuTeV:2005wsg}  experiments consistently suggest that in
the small $x<0.08$ region (anti)neutrino cross sections are significantly different than
predictions obtained by a simple re-scaling of the charged lepton cross sections.  This excess
observed at small $x$ in the differential cross sections measured by NuTeV~\cite{NuTeV:2005wsg} and CCFR
may indicate a somewhat
reduced shadowing correction with respect to charged leptons.
In order to clarify these existing discrepancies, higher precision measurements and more thorough theoretical investigations are needed. 

The available measurements of the $F_2$ and $xF_3$ structure functions from
CCFR, NuTeV, and CHORUS are characterized by the same issues observed in the
differential cross sections mostly attributed to the different nuclei involved. However, since only cross-sections are directly observable
experimentally, the extraction of structure functions from measurements require some model-dependent assumptions.

\begin{figure}[!h]
	\centering
	\includegraphics[scale=0.8]{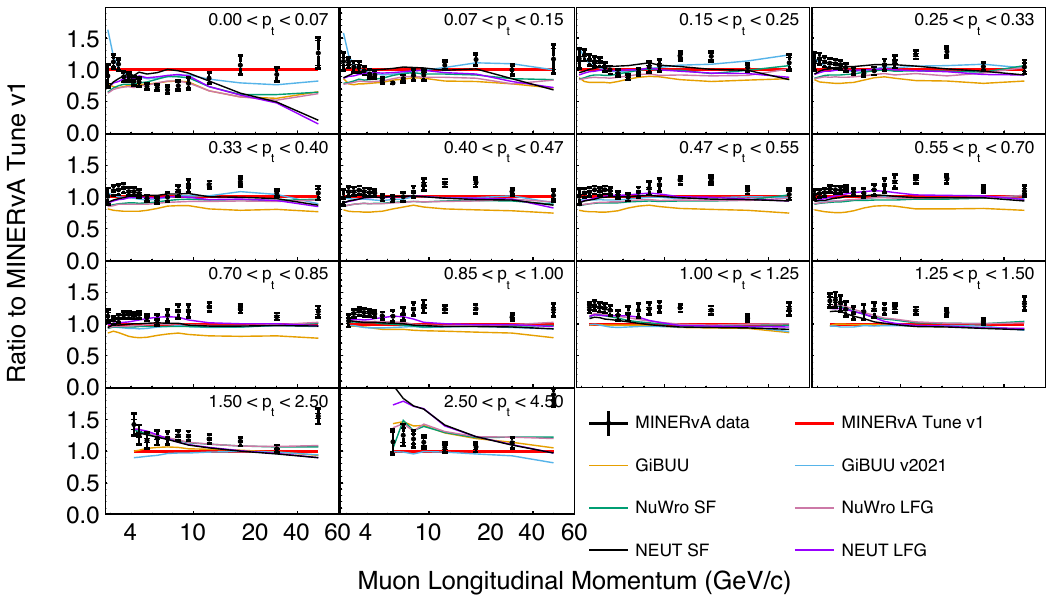}
	\caption{
The extracted cross section and external generator predictions displayed as a ratio to MINERvA Tune v1 from~\cite{MINERvA:2021owq}.}
	\label{fig:InclusiveGen}
\end{figure}

In general, inelastic cross sections are much better understood at high $Q^2$
than at relatively low $Q^2$ and $W$. This latter region is characterized by an interplay
between HT and nuclear corrections. The current experimental measurements of SIS and experimental DIS inelastic cross sections in this oscillation energy range are somewhat limited. However, the preliminary results of the \minerva experiment in both the SIS and experimental DIS region indicate inconsistencies in measurements compared to simulation programs.
For example, although not strictly an SIS/DIS measurement, \minerva has recently published the inclusive cross section of neutrino polystyrene interactions~\cite{MINERvA:2021owq}.  Figure~\ref{fig:InclusiveGen} demonstrates the comparison between \minerva measurements and all current generators.
In the kinematic region $p_L > 8~GeV$ and $p_t > 0.85~GeV$  dominated by experimental DIS ($W > 2~GeV$ and $Q^2 >1~GeV^2$) none of the current generators, including \minerva's tune of GENIE 2.12.6 labeled Tune V1, is consistent with the data. Since current and future neutrino oscillation experiments are predominately in this
low $Q^2$ and $W$ region, continued experimental and more specifically directed theoretical studies of this region are needed.

\subsection{Challenges}

\subsubsection{Modeling Issues}

\begin{itemize}
\item Optimize the description of the transition from the perturbative DIS region through the non-perturbative SIS region to  the resonance production region.

\item Understand the application of QH duality to (anti)neutrino-nucleus interactions with both $F_2$ and $xF_3$ structure functions.  Is $xF_3$ dual with resonance production while $F_2$ includes non-resonant production as well?
In particular how do we apply QH duality to nuclear targets with large neutron excess since Q-H Duality does not seem to hold for $\nu$-p or $\nu$-n individually? 

\item Study the impact of radiative corrections and their applicability in the transition region close to the kinematic limits of the parton treatment.

\item Study the impact of the structure functions $F_L$ and $R=F_L/F_T$ on (anti)neutrino cross-sections and violations of the Callan-Gross relation. Particularly the A dependence of R is important for (anti)neutrino - nucleus interactions. 

\item In both the theoretical and experimental study of SIS quantify the HT contributions, including nuclear HT, to the different structure functions $F_2, xF_3, F_T, F_L$ and introduce
comparisons with the corresponding HT terms in electromagnetic interactions. Existing extractions of HT in $F_2$ from neutrinos suggest a  different HT x-dependence. 

\item Improve hadronization models in modern generators including dispersion of multiplicities and momenta in order to describe exclusive hadron production at all W values.

 \item Study the interplay of the various nuclear effects (Fermi motion and nuclear binding, meson exchange currents, nuclear shadowing, off-shell effect, etc.) in different regions of Bjorken $x$ and $Q^2$ for neutrino and  antineutrino interactions off  protons and neutrons in the nuclear environment.

 \item Update SIS/DIS description in generators following the more recent parton distributions and models available. Systematic errors in the SIS/DIS region should be implemented in event generators including errors around the transition region. 
\item Hadronization models need to be tested with the old bubble chamber and modern neutrino hadron production data to evaluate their errors. \st{JGM: Can you suggest actual experimental results to be used in the testing and evaluation?  It may be that we have to weait for DUNE to cover a reasonable $Q^2$ - $\nu$ range.}

\end{itemize}

 \subsubsection{Experimental Issues}

\begin{itemize}

 \item Measurements of (anti-)neutrino cross sections in the shallow- and deep-inelastic regions on free nucleon targets only exist with limited statistics, limiting the ability to determine nuclear effects.  New data on free proton and deuteron targets would be invaluable.

 \item Need new more statistically precise measurements of both neutrino and anti-neutrino differential and total
 cross-sections off various nuclear targets. These measurements should have a wide  $x$ and $Q^2$ coverage, like the ones performed at Jefferson Laboratory using
 charged lepton beams, in order to compare the structure functions $F_2$ and $xF_3$, as well as the weak $F_2^{\nu(\bar \nu)}$ and the electromagnetic $F_2^{l^\pm}$. 

 \item The role of HT contributions, and potential applicability of QH duality, is limited by the lack of measurements in the transition region from DIS through SIS to resonance production.  In particular, total resonance production in the ($1.5~\le~W~\le~2$GeV) region are sorely needed.

 \item Need to perform detailed exclusive measurements of hadron production, multiplicities, angular and momentum distributions in order to constrain hadronization and FSI models. 

\item Need future theoretical work to clarify the inconsistent results from existing measurements
(BEBC, MINER$\nu$A, NuTeV, CHORUS) of nuclear effects at small $x$ values and in particular differences between (anti)neutrinos and charged leptons for the nuclear shadowing effect. 

\item Measure $\nu_e$ CC vs. $\nu_\mu$ CC vs. $\nu_\tau$ CC and test of lepton universality. 

\end{itemize}

Since precise DIS measurements typically require medium to high energy (anti)neutrino beams, it is worth noting that the only opportunity to study experimentally this region has been the Fermilab neutrino program using the Main Injector. The MINER$\nu$A experiment is in the process of analyzing these measurements in the DIS and SIS region off various nuclear targets  to address several topics listed above. The planned DUNE Near Detector complex can potentially cover most of the required measurements with unprecedented precision~\cite{Acciarri:2015uup}. In particular, measurements of neutrino and anti-neutrino scattering off free protons would be most welcome.

\newpage
\section{High (Higher) Energy Neutrinos from Natural Sources and at Colliders}
\label{sec:HighEnergy}
While the near future oscillation program will be mainly focused on neutrinos in the 100-MeV to few-GeV energy range, there are experiments studying energies significantly beyond this - above 100~GeV with a variety of physics goals.
The first detector performing such measurements was Gargamelle, a bubble chamber experiment, when operating at CERN's SPS in the 1970s~\cite{GargamelleSPS:1981hpd}. This program was continued in the 1980s and 90s with the CDHS~\cite{Berge:1987zw}, CHARM~\cite{CHARM:1982ods}, CHORUS~\cite{CHORUS:2005cpn} and NOMAD~\cite{NOMAD:2007krq} experiments at the SPS at CERN as well as the CCFR~\cite{Seligman:1997fe}, NuTeV~\cite{NuTeV:2005wsg} and DONuT~\cite{DONuT:2007bsg} detectors at the Tevatron at Fermilab.  
Together, these experiment  made major contributions to the field  by constraining the proton structure via neutrino DIS;  performing electroweak precision measurements, such as that of the weak mixing angle; observing the tau neutrino for the first time; and performing a variety of searches for new physics, all within the nuclear environment.

Over the past 20 years, two new sources of neutrinos for studying neutrino interactions have emerged.   Neutrino telescopes like IceCube have grown large enough to collect large samples of neutrinos, with up to a million events, across an energy range extending from 100 GeV up to above 100 PeV, far exceeding the energies accessible at particle accelerators.
Additionally, experiments have started to exploit the neutrinos produced at CERN's Large Hadron Collider (LHC).  Placed downstream from one of the LHC interaction points, experiments like FASER observe the neutrinos from the decays of forward pion, kaons and charmed particles. These experiments enable the study of neutrino interactions at energies reaching several TeV, extending the accelerator based neutrino program toward higher energies. Meeting the challenges of neutrino nucleus scattering at these energies has not only led to the introduction of specific higher energy interaction models but has also suggested concepts that would actually be beneficial over the entire DIS energy range.

\begin{figure}
\centering
\includegraphics[width=0.8\linewidth]{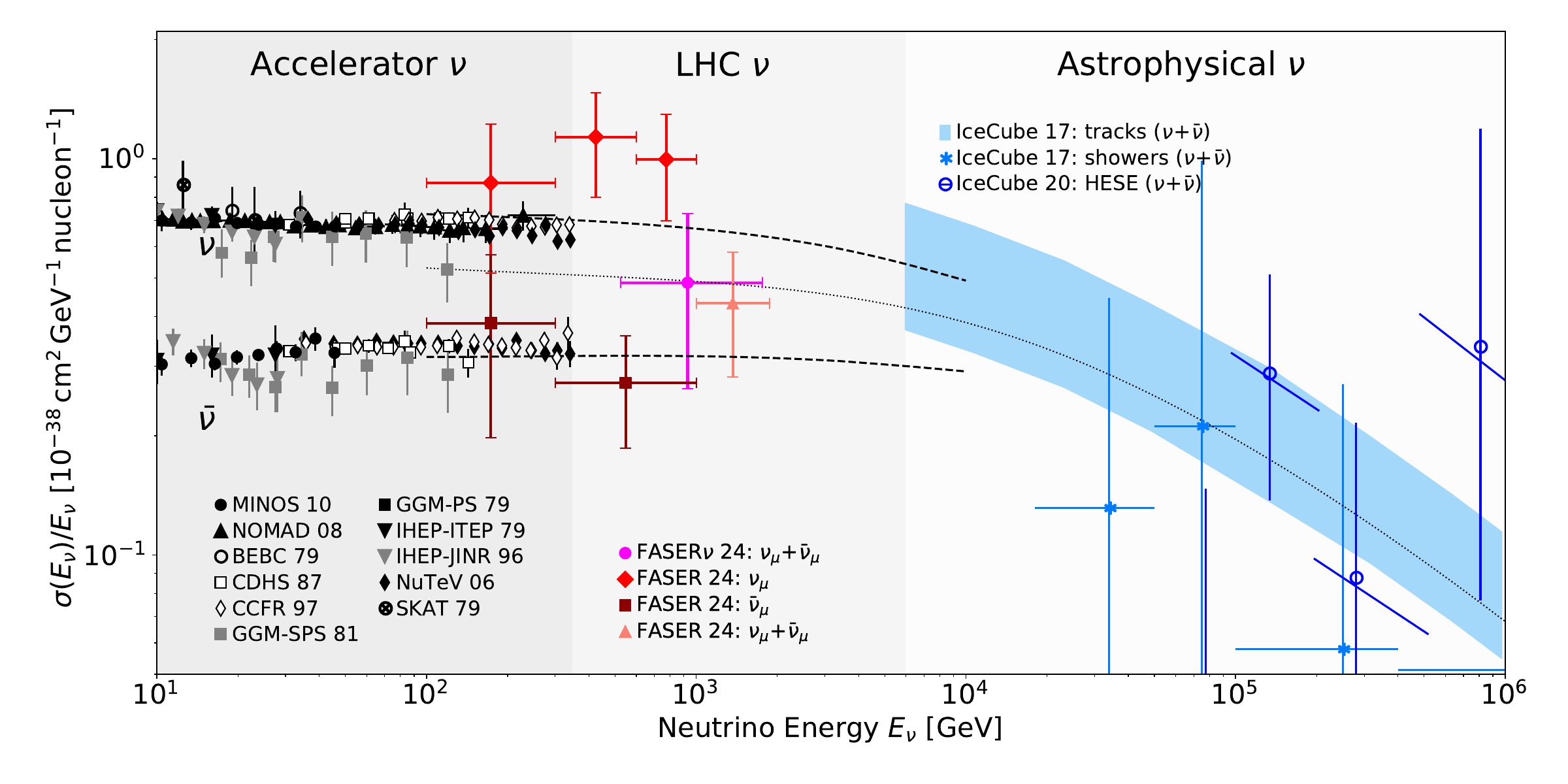}
\caption{Measurements of the muon charged-current neutrino-nucleon cross section as a function of the neutrino energy by  BEBC~\cite{deGroot:1978feq, Colley:1979rt}, CCFR~\cite{Seligman:1997fe}, CDHS~\cite{Berge:1987zw}, Gargamelle~\cite{GargamelleNeutrinoPropane:1979kqo, GargamelleSPS:1981hpd}, IHEP~\cite{Mukhin:1979bd, Anikeev:1995dj}, MINOS~\cite{MINOS:2009ugl}, NOMAD~\cite{NOMAD:2007krq}, NuTeV~\cite{NuTeV:2005wsg}, and SKAT~\cite{Baranov:1979255}, at accelerators, by FASER at the LHC~\cite{FASER:2024hoe, FASER:2024ref}, and by IceCube using astrophysical neutrinos~\cite{IceCube:2017roe, Bustamante:2017xuy, IceCube:2020rnc}. The dashed lines show the expected cross section using the Bodek-Yang model. }
\label{fig:highEnergy_XS}
\end{figure}

Figure \ref{fig:highEnergy_XS} shows the impact of these newer measurements of neutrino cross sections, along with fixed-target experiments.  The charged-current cross section have been measured over five orders of magnitude in energy, from 1 GeV to 1 PeV ($10^6$ GeV), with LHC neutrinos and natural neutrinos contributing respectively, 1 and 2 1/2 orders of magnitude to the spectrum.   

\subsection{Natural neutrinos at neutrino telescopes}

High-energy natural neutrinos fall into two categories \cite{Ackermann:2022rqc}: atmospheric neutrinos, produced by cosmic-ray interactions in the atmosphere (and discussed in Sec. \ref{subsec:atmospheric}), and astrophysical neutrinos, which come from astrophysical sources, likely including from active galactic nuclei \cite{IceCube:2018cha,IceCube:2022der} - massive galaxies with supermassive black holes at their center, which emit a relativistic jet along their axis.  They may also come from interactions between cosmic rays and photons, either from the cosmic microwave background radiation or from early stars. 

\subsubsection{Neutrino Telescopes}

The first detection of high-energy neutrinos at a neutrino telescope was by the AMANDA collaboration \cite{Andres:2001ty}, published in 2001.  the AMANDA Collaboration lowered photomultiplier tubes (PMTs) in pressure vessels into holes drilled in Antarctic ice.  The PMTs observed the Cherenkov light from charged particles produced in neutrino interactions. AMANDA proved the basic techniques of neutrino astronomy, but, with its diversity of technology, could not effectively scale to a larger instrument \cite{Halzen:2010yj}. 

\begin{figure}[btp]
\begin{center}
	\includegraphics[width=0.55\textwidth]{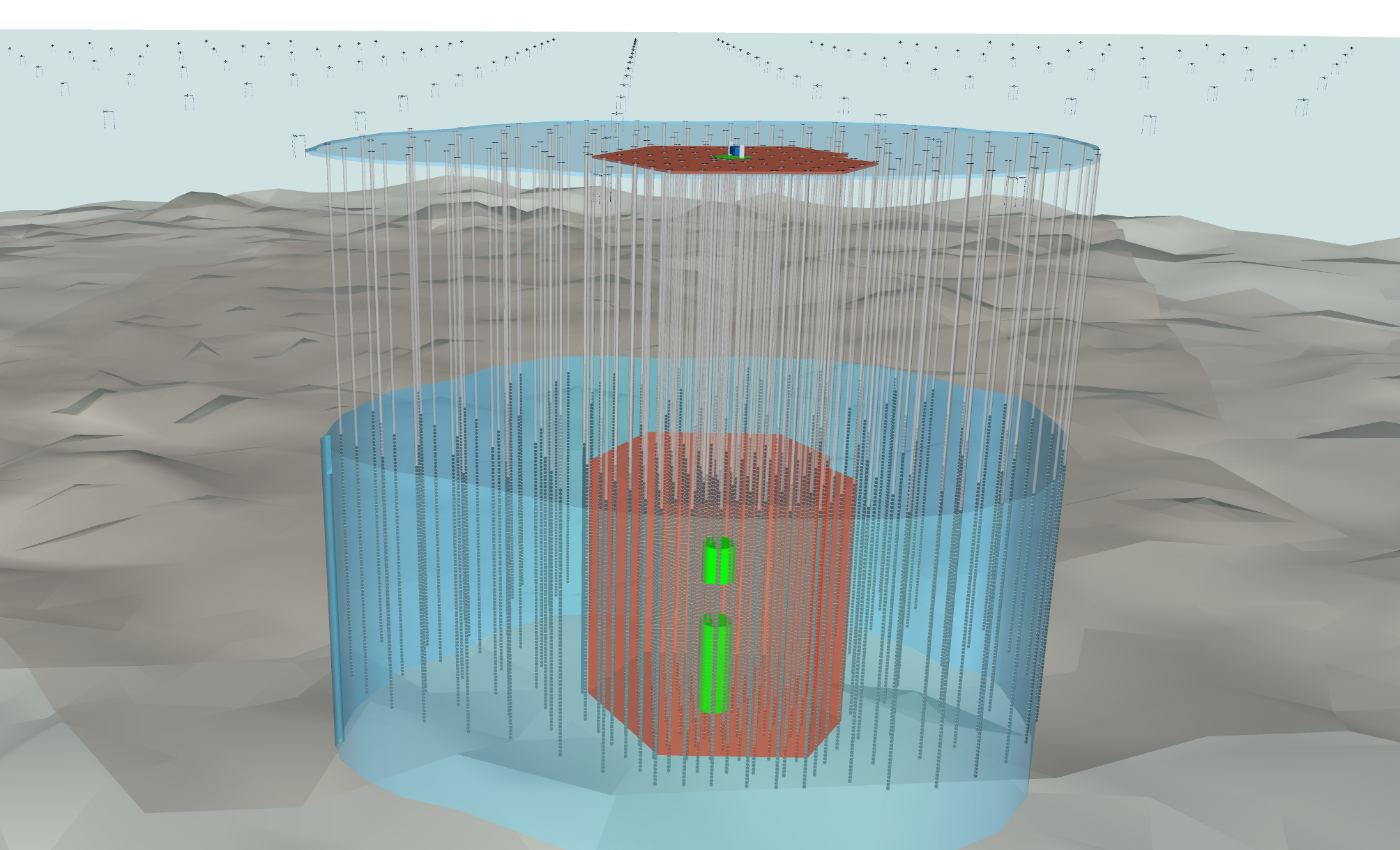}
    \caption{\color{black} A diagram of IceCube~\cite{IceCube-Gen2:2020qha}, showing the existing main detector (red shaded region) and DeepCore (green shaded region), along with the Upgrade (green shaded region) and the proposed Gen2 detector (blue shaded region).}
    \label{fig:highenergy:IceCube}
\end{center}
\end{figure}

The IceCube Neutrino Observatory instrumented 1 km$^3$ of ice with 5,160 10-inch PMTs, large enough, with enough optical collecting power to observe a large sample of atmospheric and astrophysical neutrinos.  As Fig. \ref{fig:highenergy:IceCube} shows, IceCube has multiple components.  The 78-string (vertical lines instrumented with 80 optical sensors) main array has an energy threshold around 100 GeV.  The denser Deep Core infill array lowered the threshold to about 10 GeV, over a small fraction of the array \cite{IceCube:2011ucd}, while the IceCube Upgrade, scheduled for deployment in late 2025/early 2026 will add 7 strings with more densely packed instrumentation with a threshold of a few GeV, albeit over a smaller volume encompassing about 2 megatons \cite{Ishihara:2019aao}.

The IceTop surface array, an array of 162 ice-filled tanks, spread over 1 km$^2$ above IceCube, detects Cherenkov radiation from air showers \cite{IceCube:2012nn}. It is used to measure the cosmic ray energy spectrum and, in conjunction with IceCube, the cosmic-ray composition (including searching for PeV gamma-rays \cite{IceCube:2019scr}), and to veto events containing air showers in IceCube. 

IceCube studies a wide variety of physics topics, including the observation of astrophysical neutrinos \cite{IceCube:2014stg}, with a key goal to find their sources \cite{IceCube:2018cha}, atmospheric neutrinos, searches for dark matter and other BSM physics, and also studies a number of neutrino properties.  IceCube is well matched to studies of neutrino mixing, since, for chords through the center of the Earth, maximal mixing occurs for a neutrino energy of 30 GeV, well placed in DeepCore's sensitivity.  The IceCube upgrade will extend IceCube capabilities to study neutrino oscillations, including with greatly improved sensitivity to $\nu_\tau$ appearance \cite{IceCube:2023ins}.

\subsubsection{Neutrino interaction physics at Neutrino Telescopes}
IceCube is also able to study the interactions of higher energy neutrinos using two different techniques \cite{Klein:2019nbu}.
First, it can measure the neutrino flux as a function of zenith angle, and thereby observe neutrino absorption in the Earth.  From this, the neutrino-nucleon cross-section can be inferred at energies from about 5 TeV up to about 1 PeV \cite{IceCube:2017roe,Bustamante:2017xuy,IceCube:2020rnc}. IceCube measures the ratio of the measured cross section to a reference, to account for the energy dependence of the cross section. These analyses require some assumptions, depending on the analysis.  Since they measure absorption,
one needs to account for both NC and CC interactions in the Earth; the cross-section ratio (to the standard model) is assumed to be the same for both types of interactions.  Assumptions are also needed for the $\overline\nu:\nu$ ratio.  For atmospheric neutrinos, this ratio is drawn from data-driven calculations, while for astrophysical neutrinos, analyses usually assume the ratio is $1:1$.  There is also an assumption that the astrophysical flux is isotropic.  

Currently, IceCube uses the Cooper Sarkar-Mersch-Sarkar (CSMS) calculation as a reference \cite{Cooper-Sarkar:2011jtt}.  CSMS is a NLO calculation that uses DGLAP evolution at low Bjorken$-x$.
Current measurements have about 30\% (or worse) precision, but near-future measurements should reach the 10\% level. Unfortunately, between the uncertainties in the atmospheric and astrophysical fluxes and uncertainties about the optical property of the ice in IceCube, improving precision beyond this would be very difficult \cite{Klein:2019nbu,IceCube:2021keu}.
Still, there are no beam-based experiments able to measure cross sections in this energy range.

IceCube is also able to make measurements of very-high energy neutrinos, with sensitivity to some basic properties of the interaction.  For example, in the reaction $\nu_\mu + N \rightarrow \mu + X$, it is possible to measure the cascade (hadronic shower) and muon energies separately, and thereby determine the inelasticity of the interactions at energies up to about 10 PeV .  At lower energies, the inelasticity measurements are in good agreement with data from accelerators \cite{IceCubeCollaborationSS:2025zgz}, while at higher energies, they are consistent with standard model predictions, assuming a $1:1$ $\overline\nu:\nu$ ratio \cite{IceCube:2018pgc}. Future measurements will be sensitive to the $\overline\nu:\nu$ ratio  \cite{Skrzypek:2025tmg}. 

For neutrino telescopes, inelasticity is a double-edged sword.  It can be used for important measurements, including one of the only ways to determine the $\overline\nu:\nu$ ratio for both astrophysical and atmospheric neutrinos \cite{Skrzypek:2025tmg}.  However, it is also an important element in using the measured muon energies from through-going events to determine the neutrino energy spectrum~\cite{IceCube:2017cyo,IceCube:2025ary}.  For this, it is important to know the inelasticity distribution, including corrections due to final-state radiation \cite{Plestid:2024bva} and nuclear effects \cite{Klein:2020nuk}.   If the $\overline\nu:\nu$ ratio is significantly different from $1:1$, then the inelasticity distribution will be altered, and neutrino energy spectra based on partially reconstructed events ({\it e. g.} $\mu$ from $\nu$ interactions) should be altered to account for the actual inelasticity distribution.

\begin{figure}
\centering
\includegraphics[width=0.85\linewidth]{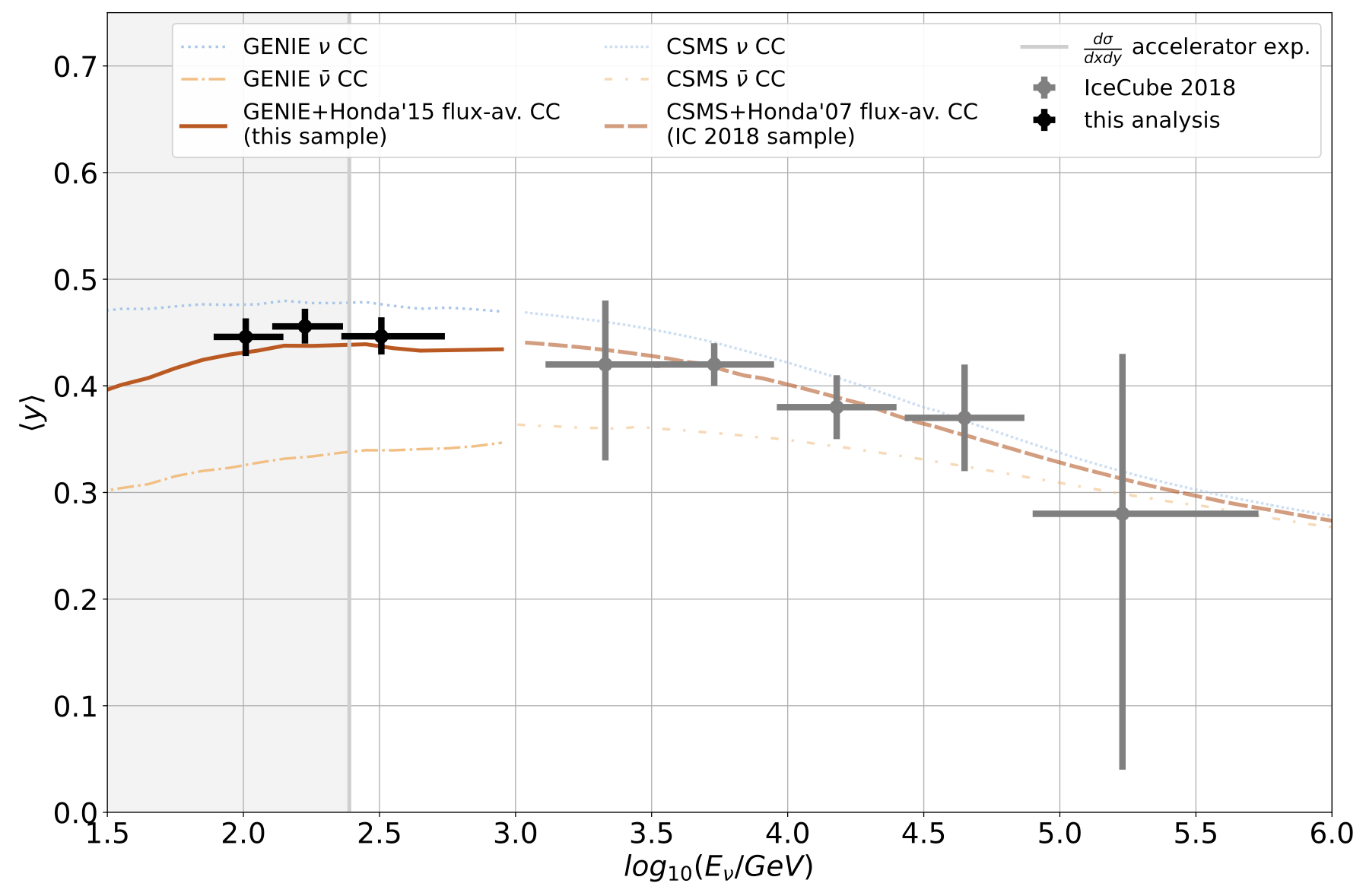}
\caption{An overview of the mean inelasticity measurements by IceCube, from 80 GeV to above 1 PeV.  At energies above 1 TeV, the data is compared with the CSMS calculation, while at lower energies GENIE is used.  Predictions are shown for $\nu$ and $\overline\nu$, along with the expected mixture (assumed to be $\overline\nu:\nu = 1:1$ for astrophysical neutrinos).  For $\overline\nu$, the GENIE and CSMS predictions do not completely agree at the 1 TeV cross-over.  From Ref. \cite{IceCubeCollaborationSS:2025zgz}.
}
\label{fig:IceCubeinelasticity}
\end{figure}

Figure \ref{fig:IceCubeinelasticity} shows the mean inelasticity measured by IceCube in H$_2$O, along with predictions from GENIE (for $E<1$ TeV) and the CSMS calculation \cite{Cooper-Sarkar:2011jtt}.  The calculations are different for $\overline\nu$ and $\nu$, so data vs. calculation comparisons depend on the assumed $\overline\nu:\nu$ ratio; IceCube uses $1:1$ for astrophysical neutrinos, and a model for atmospheric ones.  With those assumptions, the calculations are in good agreement with the data.  The CSMS and GENIE  $\overline\nu$ calculations do not agree perfectly, possibly due to GENIE's more sophisticated treatment of nuclei.

In the near future, the KM3NeT detector, in the Mediterranean Sea, should complement IceCube \cite{KM3Net:2016zxf}.  Compared to Antarctic ice, seawater has advantages - a longer optical scattering length and more uniform optical properties and the ability to deploy at deeper depths - but also additional challenges -  for example background light from $^{40}K$ and bioluminescence.  However, the biggest difference may be in the location. IceCube is located at the South Pole, giving it an unparalleled view of the northern sky, while KM3NeT can see most of the sky well.  Already, KM3NeT has observed an apparent neutrino with an energy of several hundred PeV, far higher than any events seen by IceCube \cite{KM3NeT:2025npi}.  Although the flux implied by this event is in tension with the IceCube data \cite{IceCubeCollaborationSS:2025jbi}, it has attracted much attention from theorists, and some possible explanations involve new BSM neutrino interactions.   The Lake Baikal GVD detector can also view much of the sky \cite{Baikal-GVD:2025rhg}.

Looking ahead, the proposed IceCube Gen2 detector will have an active volume for optical Cherenkov radiation several times bigger than IceCube \cite{IceCube-Gen2:2020qha}.  It will also have an even larger instrumented volume which will search for coherent radio Cherenkov radiation from neutrinos with energies above about $10^{17}$ eV.  At these energies, particle showers contain about 20\% more electrons than positrons; for frequencies below about 1 GHz, Cherenkov radiation is emitted coherently from the shower as a whole, leading to very large signals \cite{ParticleDataGroup:2022pth}.   Already the RNO-G detector is being deployed in Greenland \cite{RNO-G:2024esr};  building on previous smaller prototype arrays, RNO-G will serve as a proving ground for radio-detection of neutrinos, with a good chance of making the first positive observations.  

Radiodetection experiments probe very high energies, so are sensitive to both standard model parton densities down to very low Bjorken$-x$ values and also to many types of BSM physics \cite{Klein:2013xoa,Klein:2019nbu}.  A $10^{19}$ eV neutrino interaction, for example, probes quarks with $x\approx 10^{-6}$ at $Q^2\approx M_{W,Z}^2$.
Examples of BSM physics that are probed include leptoquarks \cite{Becirevic:2018uab}, sphalerons, and additional, rolled-up spatial dimensions \cite{Alvarez-Muniz:2001efi}.   For sphalerons, existing data from IceCube already provides tighter constraints than are available at the LHC \cite{Ellis:2016dgb}.   IceCube is also sensitive to neutrino interactions leading to some types of supersymmetry or Kaluza-Klein particles which could be observed as pairs of upward-going particles in the detector \cite{Albuquerque:2009vk,Kopper:2015rrp}. 
It is important to emphasize that any BSM (or standard model process) that impacts the neutrino cross section will also impact the inelasticity distribution.  Thus, cross sections and inelasticity measurements are complementary probes.  

\subsection{Neutrinos from the LHC}

\subsubsection{Experimental Program}

The LHC is the most powerful and highest-energy particle collider constructed to date, colliding protons at a center-of-mass energy of 13.6~TeV. Its primary aim is to study heavy SM particles, such as the Higgs boson, and to search for new heavy particles predicted by theories beyond the Standard Model, such as models of supersymmetry. The general-purpose detectors ATLAS and CMS were specifically designed to detect such signals. In addition, the LHC also produces the most energetic neutrinos created by humankind: proton-proton collisions generate an intense and strongly-collimated beam of highly-energetic neutrinos and antineutrinos of all three flavors in the forward direction along the collision axis. The possibility of detecting these LHC neutrinos was first pointed out in 1984~\cite{DeRujula:1984ns}, and revisited in several conceptual studies over the following decades~\cite{Winter:1990ry, DeRujula:1992sn, Vannucci:1993ud, Park:2011gh, Buontempo:2018gta, Beni:2019gxv}. However, it was not until the late 2010s that concrete steps toward an experimental program were initiated.

\begin{figure}
\centering
\includegraphics[width=0.95\linewidth]{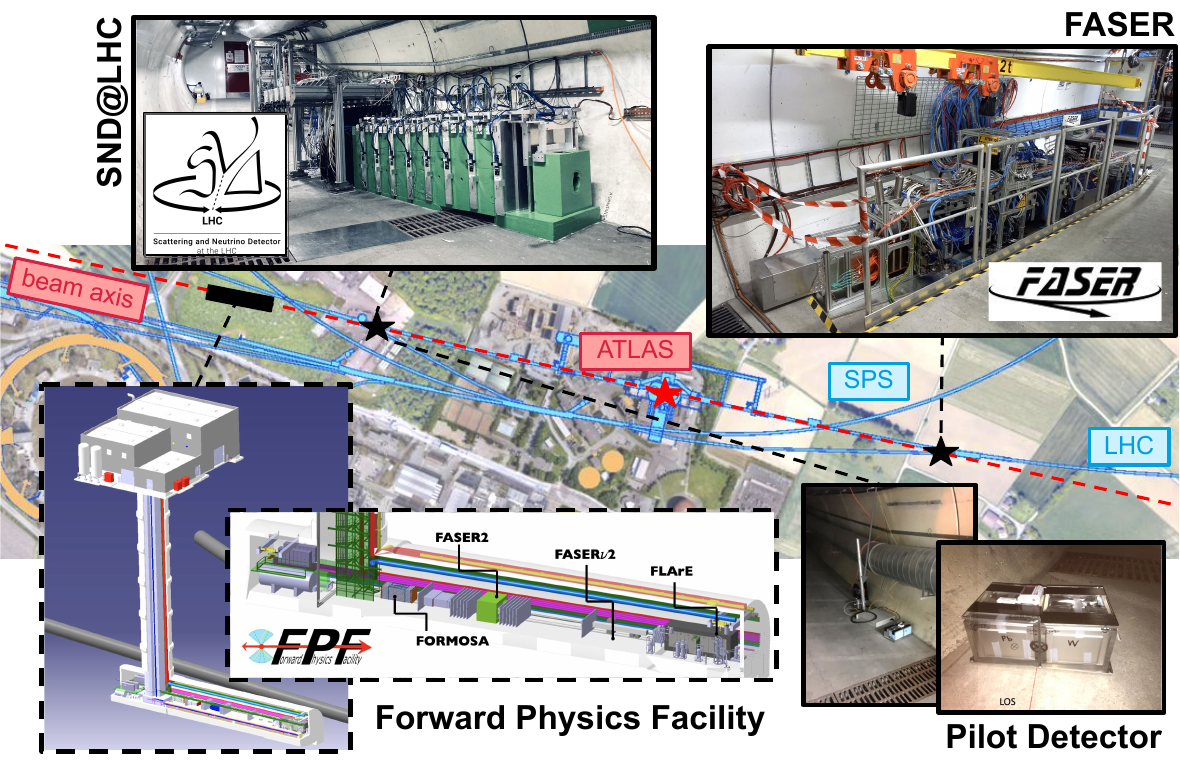}
\caption{\textbf{Location of the LHC Neutrino Detectors.} Proton-proton collisions at the ATLAS interaction point create a strongly collimated neutrino beam along the beam axis (red dashed line). This beam axis is accessible in two tunnels, connecting the SPS and LHC rings, about 480~m downstream of ATLAS (black stars). Pictures of the pilot emulsion detector, the FASER experiment as well as the SND@LHC experiment in these locations are shown. The proposed Forward Physics Facility would be located 620~m west of ATLAS.}
\label{fig:location}
\end{figure}

While the LHC tunnel gradually curves away, the neutrinos will continue to travel straight along the beam axis. As illustrated in Fig.~\ref{fig:location}, roughly 480 m downstream from the ATLAS IP, this beam axis intersects with the TI12/TI18 tunnels, which used to host the former injectors of the Large Electron-Positron Collider (LEP) but are not needed for LHC operations. This location provides a rare opportunity to access the beam collision axis and make use of the LHC neutrino beam. This opportunity was first recognized by the FASER collaboration, which in 2018 proposed installing a detector in this location to search for light, long-lived particles~\cite{Feng:2017uoz, FASER:2018ceo, FASER:2018bac, FASER:2018eoc, FASER:2019aik}. That summer, a small emulsion-based pilot detector with a target mass of 11~kg was deployed in the TI18 tunnel for just 4 weeks --- a photo is included in Fig.~\ref{fig:location}. Using the collected data, the FASER collaboration reported the first detection of collider neutrino candidates in 2021~\cite{FASER:2021mtu}, demonstrating that even small detectors at this location can complement the large multi-purpose detectors at the LHC~\cite{Foldenauer:2021gkm}.
Importantly, as is shown in Fig. \ref{fig:highEnergy_XS}, these neutrinos are in an energy regime below that accessible to most measurements of astrophysical neutrinos but exceeding the highest energies of previous accelerator-based measurements, making it a unique dataset.

To exploit this opportunity, two new LHC experiments were installed during Long Shutdown~2 (2019 - 2022) to take advantage of this opportunity: FASER~\cite{FASER:2022hcn} and SND@LHC~\cite{SNDLHC:2022ihg}. Pictures of both detectors are also shown in Fig.~\ref{fig:location}. Both experiments have been successfully taking data since the beginning of LHC Run~3 in the summer of 2022. In March 2023, the FASER collaboration reported the first observation of neutrino interactions at the LHC~\cite{FASER:2023zcr}. This observation was confirmed by the SND@LHC collaboration in May 2023~\cite{SNDLHC:2023pun}.

The FASER detector is aligned with the beam collision axis and covers pseudorapidities $\eta \gtrsim 8.5$. Positioned at its front is the FASER$\nu$ neutrino detector, which comprises a tungsten target with a target mass of approximately 1.2 tons, interleaved with emulsion films~\cite{FASER:2019dxq, FASER:2020gpr}. This detector provides high-resolution imaging of the charged particle tracks produced in neutrino interactions and enables both neutrino flavor identification and energy measurement~\cite{FASER:2025qaf}. The first observation of electron and muon neutrinos, along with a measurement of their interaction cross sections in the TeV range based on a small fraction of FASER$\nu$ data, has been reported ~\cite{FASER:2024hoe}. A veto system is located upstream of FASER$\nu$ to identify incoming particles, while downstream are FASER’s spectrometer and calorimeter system. Although the electronic detector is optimized for searches for long-lived particle decays, it also functions as a spectrometer for muons originating from neutrino interactions. The FASER collaboration has employed these electronic detector components to observe muon neutrinos through a muon appearance signature and to measure the muon neutrino and anti-neutrino flux as a function of neutrino energy~\cite{FASER:2023zcr, FASER:2024ref}.

The SND@LHC detector is located on the opposite side of ATLAS at an off-axis location, covering the pseudorapidity range $7.2 < \eta < 8.4$. Owing to its off-axis position and smaller overlap with the neutrino beam, it records approximately ten times fewer neutrino events than FASER~\cite{Ariga:2025qup}. The detector's target region consists of a hybrid system built around an 830~kg tungsten target, interleaved with emulsion and electronic trackers, which also serve as an electromagnetic calorimeter. A veto scintillator is placed upstream, and a hadronic calorimeter, which also provides muon identification, is located downstream.

Both FASER and SND@LHC will continue operating through the end of LHC Run~3 in 2026. Upgrades are planned for Run~4, beginning in 2030, with similarly sized detectors placed at the same locations~\cite{FASER:2025myb, Abbaneo:2926288}. Additionally, a continuation and expansion of this experimental program during the HL-LHC era with significantly larger detectors has been proposed as part of the Forward Physics Facility (FPF)~\cite{Anchordoqui:2021ghd, Feng:2022inv, Adhikary:2024nlv, FPFWorkingGroups:2025rsc,FPF:2025bor}. The FPF proposal includes a dedicated cavern to be constructed 620 m downstream of ATLAS, designed to host a suite of experiments, as illustrated in Fig.~\ref{fig:location}. This includes two dedicated neutrino experiments, FLArE and FASER$\nu$2, with FLArE employing a liquid argon time projection chamber with a 10~ton target mass followed by a calorimeter, and FASER$\nu$2 using a 20~ton tungsten target interleaved with emulsion films. Located at the back of the FPF is FASER2, which serves as a muon spectrometer for the two neutrino experiments. More recently, additional neutrino detectors have also been considered for placement at the surface exit points of the neutrino beam~\cite{Ariga:2020tnq, Kamp:2025phs}.

Looking further ahead, neutrino experiments have also been considered at future colliders. This includes neutrino detectors at the FCC-hh, whose neutrino beam would reach energies up to several tens of TeV. Due to its intensity, it may be feasible to also study neutrinos originating from heavy ion collisions or to observe neutrino scattering on smaller polarized targets~\cite{MammenAbraham:2024gun}. An even larger neutrino flux is expected at the proposed future multi-TeV muon colliders: the in-flight decay of beam muons provide a very intense neutrino beam source with a well-defined composition and energy spectrum~\cite{InternationalMuonCollider:2024jyv, InternationalMuonCollider:2025sys, Burk:2026fox, Choi:2026yzw}. As pointed out early on, these high-energy neutrinos could provide insight not only into neutrino cross sections but also allow for the determination of CKM matrix elements~\cite{Marzocca:2025inb}, electroweak precsion measurements~\cite{deGouvea:2025zfq}, refinement of parton distribution functions, non-standard interactions~\cite{Kling:2025zsb}, neutrino tridents~\cite{Kling:2026ghd, Francener:2026ukt}, and the potential discovery of new physics~\cite{Adhikary:2024tvl}. See also Refs.~\cite{King:1997dx, Bigi:2001xb, Mangano:2001mj, Bogacz:2022xsj} for historical reviews.

\subsubsection{Recent and Prospective Measurements}

The first observation of neutrinos at the LHC in 2023 marks the dawn of collider neutrino physics~\cite{Worcester:2023njy}. Since then, a first analysis using a small fraction of the data collected with the FASER$\nu$ emulsion detector in 2022 has reported measurement of both the electron and muon neutrino interaction cross section at TeV energies. An additional analysis using the muon momentum spectrum in the FASER spectrometer collected in 2022 and 2023 provided a measurement of the muon neutrino and anti-neutrino cross section in multiple bins of neutrino energy. The results are presented in Fig.~\ref{fig:highEnergy_XS}, alongside results from previously operating accelerator neutrino experiments as well as using astrophysical neutrinos at IceCube.  
While only a few hundred neutrinos have been analyzed so far, the existing experiments are projected to record around ten thousand neutrinos by the end of LHC Run~3 in mid-2026. Looking further into the future, about a hundred thousand neutrinos would interact in considered upgrades of the existing detectors during the HL-LHC era. Larger scale detectors proposed for the FPF could observe more than a million collider neutrinos. 

The energy of the LHC neutrino beam surpasses that of all other artificial neutrino sources and is exceeded only by atmospheric and astrophysical neutrinos. This enables LHC neutrino experiments to measure neutrino interaction cross sections at TeV energies that are otherwise inaccessible, thereby testing and validating neutrino cross section models~\cite{Candido:2023utz, Jeong:2023hwe}. These measurements can be carried out for all three neutrino flavors, including tau neutrinos. While only around ten tau neutrino interactions have been previously identified on an event-by-event basis by each of the DONuT and OPERA experiments~\cite{DONuT:2007bsg, OPERA:2018nar}, collider neutrino experiments have the potential to collect significantly larger event samples. This opens up a unique opportunity to perform precision studies of tau neutrino properties~\cite{MammenAbraham:2022xoc, MammenAbraham:2023psg}, to test lepton universality in neutrino scattering by comparing interaction rates across all three flavors~\cite{Falkowski:2021bkq, Kling:2023tgr}, and to measure the $F_4$ and $F_5$ structure functions~\cite{Francener:2024euo}. If equipped with charge identification capabilities, collider neutrino experiments could also, for the first time, separately observe tau neutrinos and tau antineutrinos via their muonic decays.

Beyond total cross sections, collider neutrino experiments will also constrain various exclusive channels. One example is charm-associated neutrino scattering, $\nu q \to \ell c$, where the secondary charm decay vertex can be identified using the emulsion detector. This process accounts for roughly 15\% of all neutrino interactions, corresponding to over a thousand such events in FASER. Additionally, collider neutrino experiments will be able to observe this process for the first time with electron and tau neutrinos and may even achieve the first observation of beauty-associated interactions, $\nu q \to \ell b$.

Neutrino interactions at the LHC mainly fall into the to DIS regime, with $W>2$~GeV and $Q > 1.65$~GeV, although the non-DIS processes can contribute up to 10\% at $E_\nu=100$~GeV.  A high-statistics sample of neutrinos allows to measure their corresponding interaction cross sections differentially in the DIS variables $x$, $y$ and $Q$. Notably, the neutrino energies at the LHC significantly exceed those of any other previous experiments studying DIS neutrino interactions, such as CDHS or NuTeV, and therefore provide an opportunity to extend their physics programs into a broader kinematic regime. Similar to the EIC and NuTeV programs, collider neutrino data can be used to constrain nuclear parton distribution functions 
and, through global analyses, extend sensitivity also to proton parton distribution functions. 
The corresponding sensitivity was investigated in Ref.~\cite{Cruz-Martinez:2023sdv}, and it was found that high-statistics collider neutrino scattering measurement could significantly reduce PDF uncertainties. In particular, high resolution emulsion experiments with charm hadron identification capabilities provide an opportunity to measure the target's strangeness content 
via the charm associated neutrino interactions $\nu_\mu s \to \mu c$. These measurements would then help to reduce PDF uncertainties for key measurements at the LHC central detectors~\cite{Cruz-Martinez:2023sdv} and allow to break degeneracies between PDFs and possible new physics when interpreting LHC data~\cite{Hammou:2024xuj}. 

In particular, neutrino DIS measurements on heavy target nuclei, currently tungsten, will offer complementary insights into nuclear PDFs and nuclear modifications, including shadowing at low $x$, anti-shadowing at intermediate $x$, and the EMC effect at high $x$. In addition, complementary muon DIS data can be obtained simultaneously by exploiting the large flux of high-energy muons passing through the neutrino detector~\cite{Francener:2025pnr}. Past observations revealed discrepancies between charged-lepton and neutrino scattering data, with the latter showing significantly weaker shadowing compared to electron and muon scattering~\cite{Schienbein:2009kk, Kovarik:2010uv, Muzakka:2022wey, SajjadAthar:2020nvy}. These differences in the shadowing region could possibly be expected with a difference in coherence length between V and A currents and the difference in mass of the IVBs involved.  Moreover, while models predict that shadowing vanishes in the large $Q^2$ limit, there is still no consensus on the magnitude of shadowing at intermediate values $Q^2 \sim 10^2$–$10^3$ GeV$^2$, which are relevant for LHC neutrino experiments. Additional complexity arises from differences in heavy quark mass effects between neutrino DIS and charged-lepton DIS structure functions~\cite{Risse:2025smp}. Neutrino DIS measurements at the LHC will provide new data across a broad kinematic range and potentially multiple nuclear targets to help clarify these issues.

Besides studies of neutrino interactions, additional insights are provided by measurements of the neutrino flux: as first such measurement was already performed by FASER in Ref.~\cite{FASER:2024ref, FASER:2025sge}.  LHC neutrinos mainly originate from the decay of charged pions, kaons, and charm mesons. The forward production of these particles at LHC energies has not been measured before. Neutrino flux measurements thus provide a novel way to probe and constrain forward hadron production, offering unique insights into the strong interaction in previously inaccessible kinematic regions. Forward pion and kaon production falls outside the scope of perturbative QCD and are instead described by phenomenological hadronic interaction model  designed to match the available data sets~\cite{Pierog:2013ria, Ostapchenko:2010vb, Riehn:2019jet, Fieg:2023kld}. LHC neutrino measurements will add complementary \jgm{to} these datasets. Improving forward particle production models is essential for astroparticle physics, where they are used to describe the production of high-energy particles in extreme astrophysical environments and the interactions of cosmic rays with Earth’s atmosphere. In particular, such measurements may help to resolve the muon puzzle~\cite{EAS-MSU:2019kmv, Soldin:2021wyv}, which refers to a long-standing tension between the number of muons observed in high-energy cosmic ray air showers and the number predicted by hadronic interaction models~\cite{Albrecht:2021cxw, Anchordoqui:2022fpn}. In contrast to light hadrons, forward charm production can be modeled using perturbative QCD methods~\cite{Bai:2020ukz, Buonocore:2023kna}. Charm quarks are predominantly produced via gluon fusion ($gg \to c\bar{c}$), where one gluon carries a large momentum fraction ($x \sim E_\nu/E_\text{proton} \sim 1$), and the other a very small fraction ($x \sim 4 m_c^2 / s \sim 10^{-7}$). 
The high-$x$ region is sensitive to intrinsic charm~\cite{Maciula:2022lzk} and constrain high-$x$ PDFs, while low-$x$ region is sensitive to the gluon PDF in regions far beyond the reach of current experiments, enabling tests of QCD phenomena such as BFKL dynamics and gluon saturation~\cite{Bhattacharya:2023zei}. Improved understanding of forward charm production also benefits astroparticle physics, particularly by reducing uncertainties in the prompt atmospheric neutrino flux~\cite{Bai:2022xad}. These neutrino flux measurements require the total neutrino cross section to be well-understood, and as such are limited by the several-percent theoretical uncertainties on those cross sections.  However they remain a powerful constraint as the only direct measurement.

Finally, the high energy of the neutrino beam and the potentially large number of interactions make collider neutrino experiments a promising laboratory for exploring new physics in the neutrino sector. A range of signatures and scenarios has been studied in the literature. These include searches for sterile neutrinos with multi-eV masses, which could lead to observable oscillation patterns~\cite{FASER:2019dxq, Anchordoqui:2021ghd, Bai:2020ukz, Anchordoqui:2023qxv}; searches for new neutrino-philic mediators that alter the expected tau-neutrino flux~\cite{Kling:2020iar, Batell:2021snh}; searches for sterile neutrinos  transition dipole moments~\cite{Ismail:2021dyp}; searches for non-standard neutrino interactions affecting neutrino production or scattering~\cite{Ismail:2020yqc, Falkowski:2021bkq, Kling:2023tgr}; searches for neutrino portal dark matter~\cite{Kelly:2021mcd, Berryman:2022hds}; and searches for neutrino trident production~\cite{Francener:2024wul, Altmannshofer:2024hqd}.

\subsection{Challenges with Neutrino Scattering at Higher Energies}

Neutrinos at the LHC and neutrino observatories have energies far exceeding those of other accelerator-produced neutrinos, and their interactions are therefore dominated by deep-inelastic scattering. This contrasts with other operating accelerator neutrino experiments, where DIS typically represents a subleading channel. 
As discussed in Sec.~\ref{sisdis}, the corresponding cross section is expressed in terms of the nucleon structure functions $F_{i}(x,Q^2)$. Various models for these structure functions have been proposed in the literature, some of which allow the formalism to be extended into the non-DIS regime.

A widely used phenomenological framework for describing inelastic neutrino–nucleon scattering cross sections in the multi-GeV energy range is the Bodek–Yang model~\cite{Bodek:2002vp, Bodek:2004pc, Bodek:2010km}. It is implemented in the GENIE~\cite{Andreopoulos:2015wxa} neutrino MC event generator, currently used at both FASER and SND@LHC, as well as in other widely used tools such as NEUT~\cite{Hayato:2021heg} and NuWro~\cite{Juszczak:2005zs}. The Bodek–Yang model builds on structure functions derived from effective leading-order GRV98 PDFs~\cite{Gluck:1998xa}. To account for mass effects, higher-order QCD corrections, nuclear effects, and to extend its applicability to the non-perturbative, low-$Q^2$ regime, various phenomenological corrections are applied. For LHC neutrino experiments, however, the Bodek–Yang model has certain limitations. It relies on outdated PDFs that do not include the constraints on proton and nuclear structure obtained over the past 25 years. It only provides PDFs for up, down, and strange quarks. It does not incorporate available higher-order QCD calculations. Finally, it was designed mainly for the multi-GeV energy range rather than for TeV-scale neutrinos. Refinements are therefore needed. To simulate hadronization, \texttt{GENIE} uses a hybrid model that depends on the value of $W$ in the interaction. At high $W$, parton showers are ignored and hadronization is modeled using \texttt{Pythia~6} with dedicated tunes, the first presented in Ref.~\cite{katori:2014fxa} and a recent update in Ref.~\cite{GENIE:2021wox}. GENIE also includes a formation zone and models final-state interactions for hadrons produced inside the nucleus using the intra-nuclear cascade approach.

Several other cross-section calculations based on NLO structure functions and modern PDF sets have been developed in the last two decades, mainly with applications in astroparticle physics in mind.  These calculations must extend to high energies, above $10^{20}$ eV, so need to model parton distributions at very low$-x$, down to $10^{-6}$ at large $Q^2$.   These calculations include the CSMS~\cite{Cooper-Sarkar:2011jtt} and BGR18~\cite{Bertone:2018dse, Garcia:2020jwr} models, as well as calculations using the CT18~\cite{Xie:2023suk} PDFs. These results are available for simulations in the high-energy DIS (HEDIS) module of GENIE~\cite{Garcia:2020jwr, GENIE:2021npt}. HEDIS, unlike the default DIS module of GENIE, does not include non-DIS contributions, formation zones, or final-state interactions, and currently relies on the latest version of Pythia 6 without tuning, with a transition to Pythia 8 planned for the future.

Even more recently, new neutrino event generators incorporating higher-order QCD corrections and matched to modern parton showers have been presented in Refs.~\cite{vanBeekveld:2024ziz, FerrarioRavasio:2024kem, Buonocore:2024pdv}. These employ POWHEG and Pythia8, widely used in the collider physics community, but currently describe only the DIS component and do not include non-DIS contributions or final-state interactions.  Efforts are needed to combine these tools for DIS with others for non-DIS processes, ensuring a consistent treatment of the transition between the non-perturbative low-$Q$ and perturbative high-$Q$ regimes of the neutrino DIS structure function, while avoiding double counting across regions.

A recent study~\cite{Hernandez-Sainz:2026rxc} systematically compare the predictions of NLO general-purpose Monte Carlo event generators for neutrino DIS at the LHC (including Sherpa~\cite{Sherpa:2024mfk}, Herwig7~\cite{Bewick:2023tfi}, and POWHEG) against analytical structure-function calculations provided by YADISM~\cite{Candido:2024rkr} and dedicated neutrino simulation packages (GENIE, NuWro, GiBUU). Modern event generators were found to agree within sub-percent level for inclusive and differential DIS observables once common settings are adopted. In contrast, GENIE and the other dedicated neutrino simulators was found to agree with these NLO calculations at the inclusive level only through an accidental cancellation while its differential distributions with respect to the lepton kinematics deviate from the NLO calculations by up to about 10\%, implying serious limitations of GENIE. Differences between generators remain for observables sensitive to the hadronic final state.

One important question concerns nuclear effects, such as shadowing. Shadowing in neutrino interactions has been observed by earlier bubble chamber experiments~\cite{BEBCWA59:1989nmi} and more recently by the MINERvA experiment~\cite{MINERvA:2016oql} down to $x \approx 0.05$, which lies at the beginning of the shadowing region for MINERvA's low $Q^2$. Shadowing can be calculated using perturbative QCD and electroweak theory~\cite{kopeliovich:2012kw} by comparing cross-section predictions obtained with different parameterizations of the parton distributions, with and without nuclear shadowing. Some calculations of the cross section and inelasticity have found that shadowing for higher-energy neutrinos is relatively small, at the few-percent level~\cite{Klein:2020nuk}. These results were based on the EPS09 and EPPS16 parameterizations, which are fits to data showing little shadowing at high $Q^2$. However, a recent calculation of neutrino cross sections using the newer EPPS21 parameterizations found a significantly larger shadowing effect, up to 10\% for gold/lead and about 5\% for iron~\cite{Weigel:2024gzh}. This calculation also reported 10–20\% deviations from the CSMS model~\cite{Cooper-Sarkar:2011jtt}, exceeding the estimated uncertainties in either work. A more recent study investigating the influence of shadowing on the astrophysical neutrino flux reported that shadowing can reduce neutrino fluxes from heavy nuclei by a factor of 2–3 compared to proton-only scenarios, indicating its importance~\cite{Liu:2025kkq}. Experimentally, the difference arises from the choice of data used in the parameterizations. This suggesting that future data, especially from the collider neutrino experiments, will be important. In addition, it would also be valuable to have a theoretical uncertainty estimate for the expected smallness of shadowing at $Q^2 \approx M_W^2$. 

Another relevant aspect concerns hadronic and nuclear final-state effects. The LHC neutrino detectors, particularly the FASER$\nu$ emulsion detector with its high spatial resolution, will be able to resolve the kinematics of the final state, such as the multiplicity and energies of charged particles. This is sensitive to the modeling of the parton shower, hadronization, and the interaction of both partons and hadrons as they traverse the dense matter of the target nucleus.  Many considerations of meson formation time/zone suggests that at LHC neutrino energies, unless at very small $y$, hadronization of mesons could be outside of the nucleus suggesting minimal  final-state interactions. However, the quarks and gluons that eventually form hadrons must still propagate through the nucleus, and as colored particles their interactions are not negligible. As discussed in Ref.~\cite{Mosel:2022tqc}, several open questions remain, including on i) the modeling of parton shower and hadronization inside cold nuclear matter; ii) the description of formation time and formation zones; iii) the validity of color transparency; and iv) the role of final-state interactions and the baryon avalanche effect. These same physics questions are also been explored in the context of electron-ion collisions at the EIC~\cite{Accardi:2012qut}. 

At high energies, additional processes beyond DIS begin to contribute. At PeV energies, $\overline\nu_e$ can interact with atomic electrons in the target, producing real $W^-$.  This process has a resonance, known as the Glashow resonance, at an energy around $E \sim m_W^2/(2 m_e)\sim 6.3$~PeV. IceCube has observed one event consistent with this process, indicating that at least some astrophysical neutrinos are antineutrinos. Similar resonances occur at TeV energies, corresponding to the charged vector mesons. In another new type of process, neutrinos can fluctuate to a charge lepton plus a $W^\pm$, and then interact electromagnetically with the target.  These cross sections of these interactions is at most a few percent of the DIS cross sections \cite{Zhou:2019frk}.  However, they have rather different event characteristics than DIS, and so it is important to identify and correctly classify these events.

Finally, uncertainties in the modeling of neutrino interactions must be well controlled for both precision SM and BSM measurements. Ideally, LHC and astrophysical neutrino measurements would use multiple event generators and various implementations of the underlying physics assumptions described above to systematically assess their impact on measurements, quantify systematic uncertainties, and, when possible, compare predictions to data. In this regard, the data collected by FASER and SND@LHC during LHC Run 3 will be valuable for constraining the interaction physics in generators. Additionally, these generators should share a common interface and output format, preferably an established standard like HEPMC~\cite{Buckley:2019xhk} or LHEF~\cite{Alwall:2006yp}, to facilitate their integration into experimental analysis frameworks.

\subsection{Challenges and Open Question}

FASER$\nu$ and the neutrino telescopes probe much higher neutrino energies than was previously possible, making neutrino interactions sensitive for the first time to quarks at low Bjorken-$x$ and high $Q^2$. These studies have raised a number of theoretical questions:
\begin{itemize}
\item It would be valuable to have a theoretical uncertainty estimate for the expected size of shadowing at $Q^2 \approx M_W^2$.
\item A theoretical model describing parton showers evolve inside dense nuclei and how partons interact with cold nuclear matter is currently lacking.
\end{itemize}

Neutrino interaction Monte Carlo event generators are vital for analyzing data from both the LHC neutrino experiments and neutrino observatories. These generators need to be updated such that they can all offer:
\begin{itemize}
\item Inclusion of higher order corrections, mass effects, and modern PDFs (including nuclear effects) in DIS models.
\item Ability to generate events over the entire kinematic range, including non DIS contributions, and accounting for a smooth transition between the different regions 
\item Modelling of parton shower and hadronization, considering interplay with the cold nuclear matter, formation zones, final state interactions, and dissociation of the nucleus remnant
\item Uncertainties for all the above-mentioned effects, for example via a variety of tunes or models
\item Flexibility to introduce model variations and tunes 
\item It would be beneficial to use common file formats and interfaces to other tools like GEANT4 that allow easy implementation into the analysis frameworks and tuning tools like Rivet. 
\end{itemize}

Additionally, there remain a number of experimental challenges to these programs: 

\begin{itemize}
\item High-energy neutrino telescopes have to be extremely massive, limiting where and how they can be built (usually using naturally occurring target material such as ice or water).  It becomes financially challenging to instrument these detectors finely enough to see details of e.g. hadron showers, limiting the information that can be learned.
\item There exists only one source of very high-energy neutrinos from colliders, limiting the space for experiments to the existing tunnels near the LHC.  The FPF will significantly increase the flexibility of this program.
\item A detailed understanding of the neutrino fluxes from the LHC is challenging due to the fact that these neutrinos are primarily produced by decays of particles the LHC detectors cannot measure.
\item The current FASER$\nu$ design, utilising emulsion, provides exquisite detail around the interaction vertex to enable the measurement of tau neutrinos and charmed hadron production.  Unfortunately this technology requires development and scanning of these emulsions which can take a considerable time.  The development of new detector technologies that are able to identify tau neutrinos without the need for emulsion scans should be encouraged.
\item The possibility to use several target  materials in the same detector, ranging from low-$Z$ to high-$Z$, should be considered for the collider neutrino experiments and the FPF in order to study nuclear effects such as shadowing. 
\end{itemize}

\newpage
\section{Strange Particle Production}
\label{strange}
The production of strange particles (mesons and baryons) in neutrino interactions was investigated in the initial period of investigations of the neutrino interactions, with a number of cross section results released by bubble chamber experiments~\cite{Erriquez:1978pg,Erriquez:1977tr,Eichten:1972bb,SKAT:1989nel,Fanourakis:1980si,Ammosov:1986jn}. More recently, there have been measurements of the polarisation of $\Lambda$ baryons produced in neutrino interactions by the NOMAD experiment~\cite{NOMAD:2009qmu}, and a measurement of single $\Lambda$ production in the MicroBooNE experiment~\cite{MicroBooNE:2022cls}. 

While these processes are relatively rare, they provide a unique complementary probe of a variety of effects.
Additionally, an important search for proton decay is the $p \rightarrow K^+ + \bar{\nu}$ channel, which will contend with an uncertain rate of atmospheric neutrino-incuded charged kaons.

\subsection{Cabbibo suppressed quasi-elastic processes}
\begin{figure}
    \centering
    \begin{subfigure}{0.47\linewidth}
        \includegraphics[width=\linewidth,trim={8cm 6cm 0 0},clip]{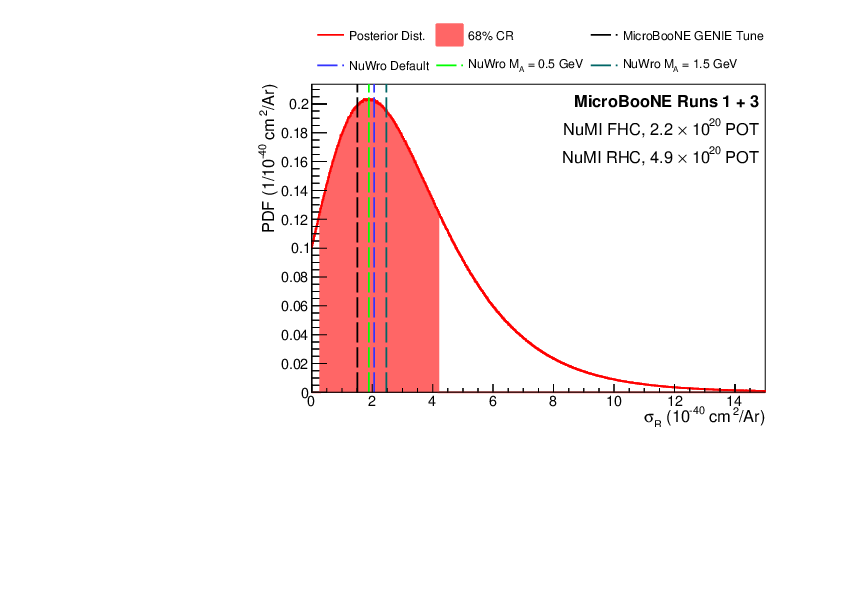}
        \caption{Extracted cross section posterior.}
    \end{subfigure}%
    \begin{subfigure}{0.47\linewidth}
        \includegraphics[width=\linewidth,trim={8cm 6cm 0 0},clip]{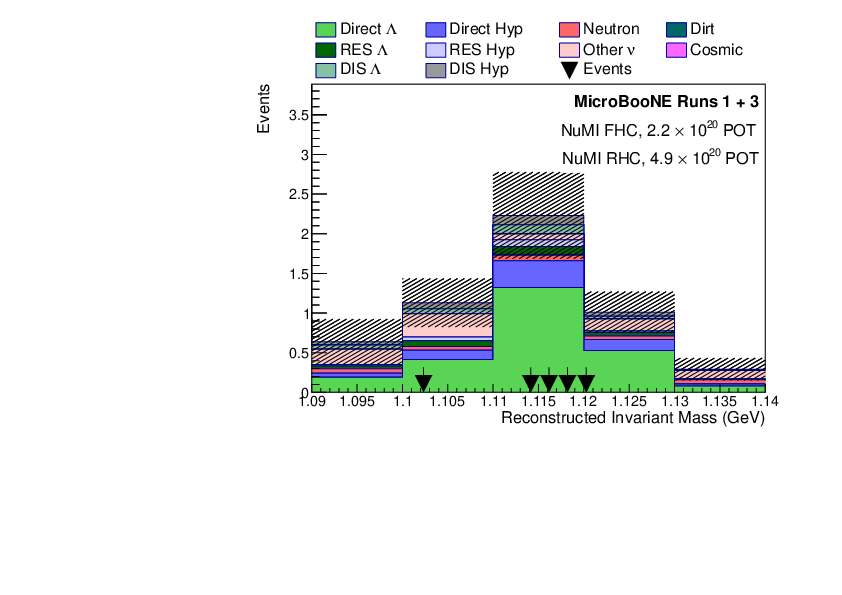}
        \caption{Reconstructed invariant masses of the 5 $\Lambda$ candidates compared with MC prediction.}
    \end{subfigure}
    \caption{Results from the MicroBooNE experiment for the quasi-elastic production of $\Lambda$\cite{MicroBooNE:2022cls}.}
    \label{fig:MicroBooNELambda}
\end{figure}

\begin{figure}
    \centering
    \begin{subfigure}{0.47\linewidth}
        \includegraphics[width=\linewidth]{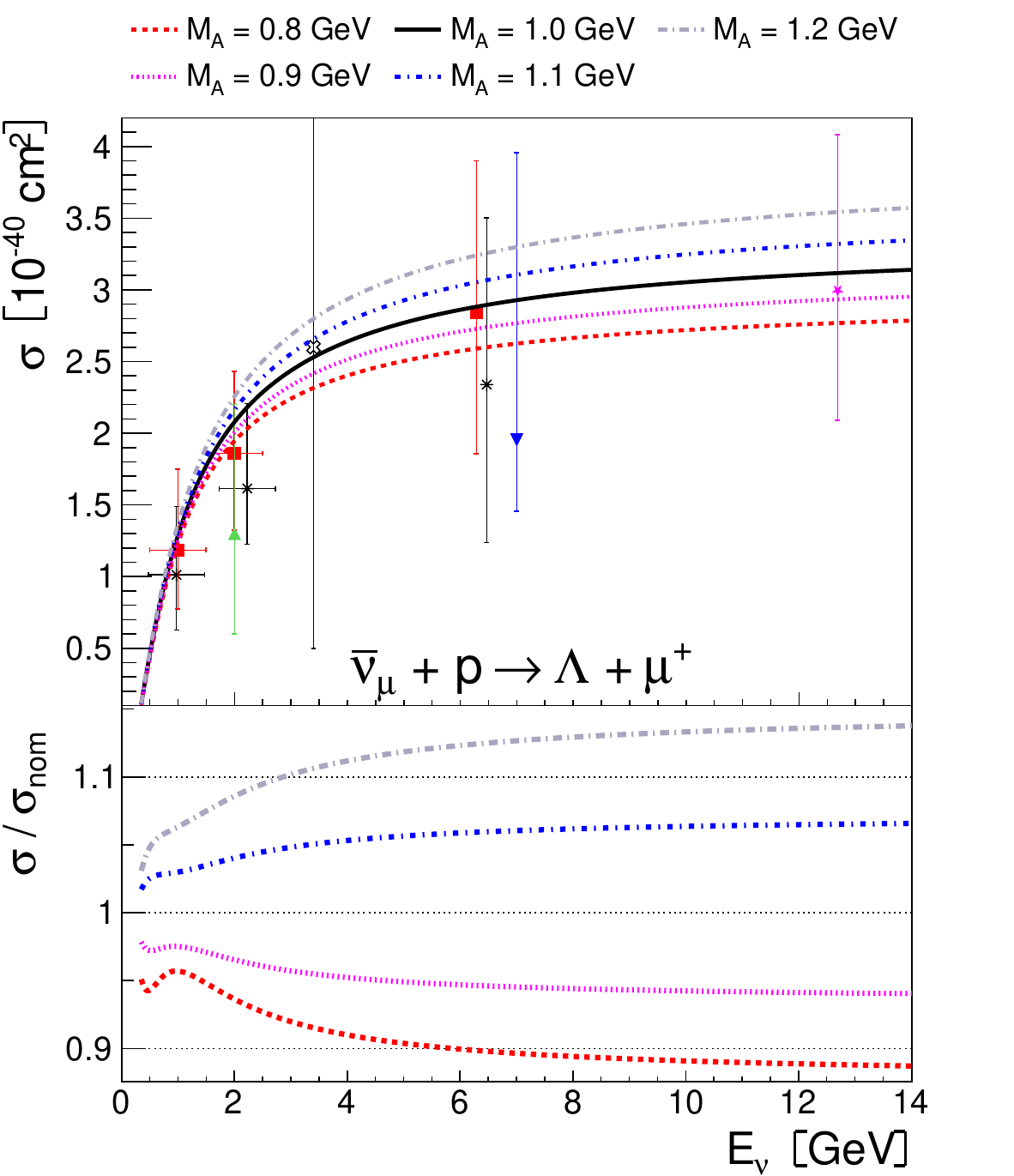}
        \caption{Predictions for the cross section for $\Lambda$ baryon production versus data}
    \end{subfigure}%
    \hspace{0.5cm}%
    \begin{subfigure}{0.47\linewidth}
        \includegraphics[width=\linewidth]{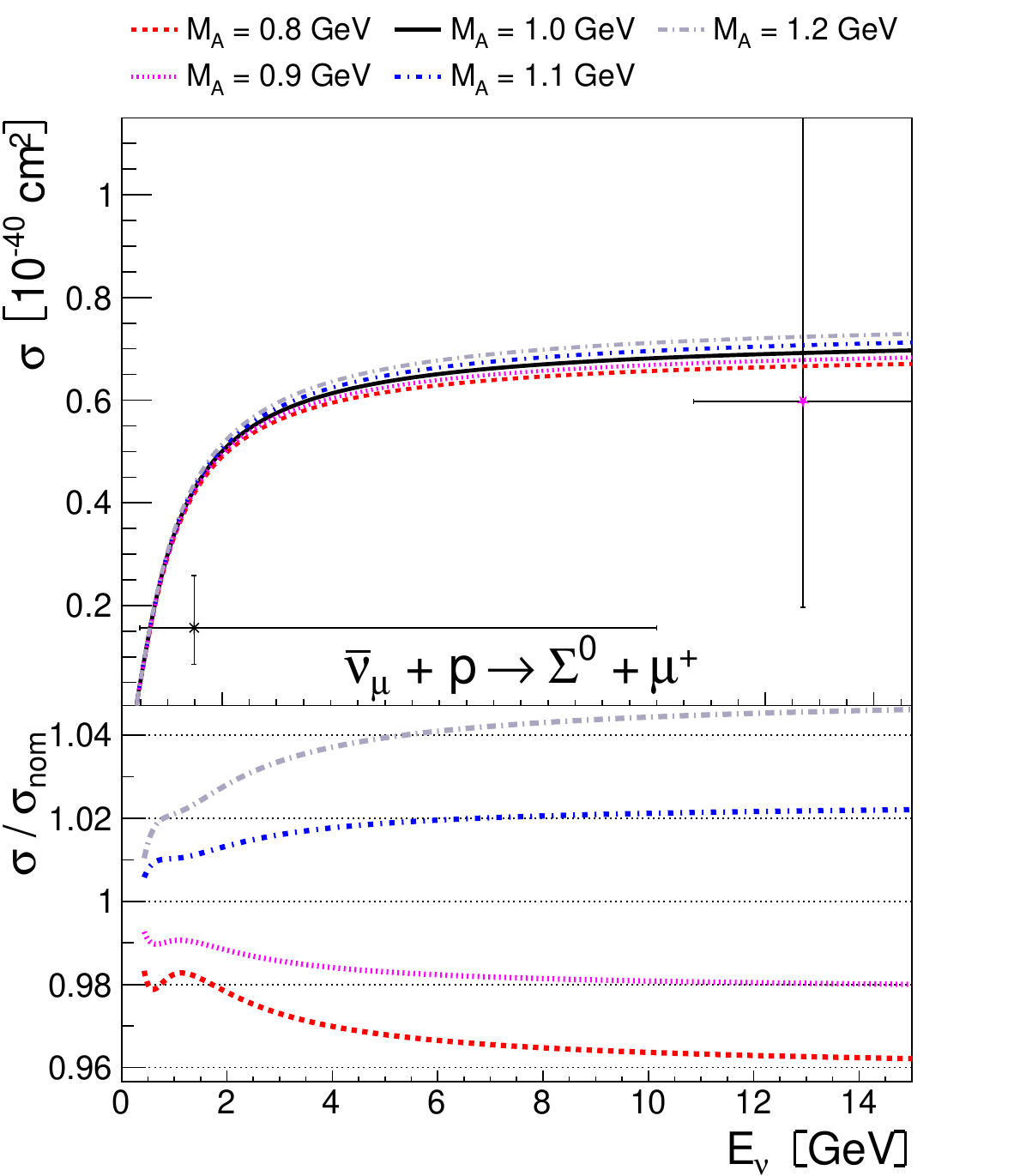}
        \caption{Predictions for the cross section for $\Sigma^0$ baryon production versus data.}
    \end{subfigure}
    \caption{Comparisons of cross sections calculated with the NuWro neutrino event generator to bubble chamber measurements of Cabibbo suppressed hyperon production. Data is taken from Ammosov\cite{Ammosov:1986jn} (FNAL Bubble Chamber, pink star), Brunner\cite{SKAT:1989nel} (SKAT, triangle down), Erriquez\cite{Erriquez:1978pg,Erriquez:1977tr} (Gargamelle, red square and black x respectively), Eichten\cite{Eichten:1972bb} (Gargamelle, triange up) and Fanourakis\cite{Fanourakis:1980si} (BNL Bubble Chamber, white cross).}
    \label{fig:CCQEFreeNucleon}
\end{figure}

The process with the lowest neutrino energy threshold leading to the production of hyperons (baryons with some valence quark content from the second and third generations) is ``CCQE-like'' hyperon production, for which there exist three strangeness producing channels involving free nucleons:
\begin{equation}
\begin{array}{l}\bar{\nu}_{l}+p \rightarrow l^{+}+\Lambda, \\ \bar{\nu}_{l}+p \rightarrow l^{+}+\Sigma^{0}, \\ \bar{\nu}_{l}+n \rightarrow l^{+}+\Sigma^{-}.\end{array}
\end{equation}
The $\Delta S = \Delta Q$ selection rule means these interactions are only available for anti-neutrinos. The existing sets of measurements of these interactions are shown in Figures~\ref{fig:CCQEFreeNucleon} and~\ref{fig:MicroBooNELambda}, compared with predictions from MC simulations. The majority of this data was produced by bubble chamber experiments conducted during the 1960's and 1970's, with the most recent measurement for free nucleons being published by SKAT in 1990. Recently a first measurement of $\Lambda$ production from Ar nuclei was published by the MicroBooNE collaboration~\cite{MicroBooNE:2022cls}. All existing measurements are limited in sensitivity by low statistics.

The cross section model typically employed to simulate these interactions is a modified form of the Llewellyn Smith model~\cite{LlewellynSmith:1971uhs}, with form factors obtained by exploiting the SU(3) quark flavour symmetry~\cite{Pais:1971er,Sobczyk:2019uej,Fatima:2018tzs,Thorpe:2020tym}. These models typically assume the same form factor parameterisation as the $\Delta S=0$ CCQE interactions, though as Figure~\ref{fig:CCQEFreeNucleon} shows, existing measurements are insufficient to test this assumption.

\subsection{Associated production}
\begin{figure}[h]
  \centering
  \includegraphics[scale=0.6]{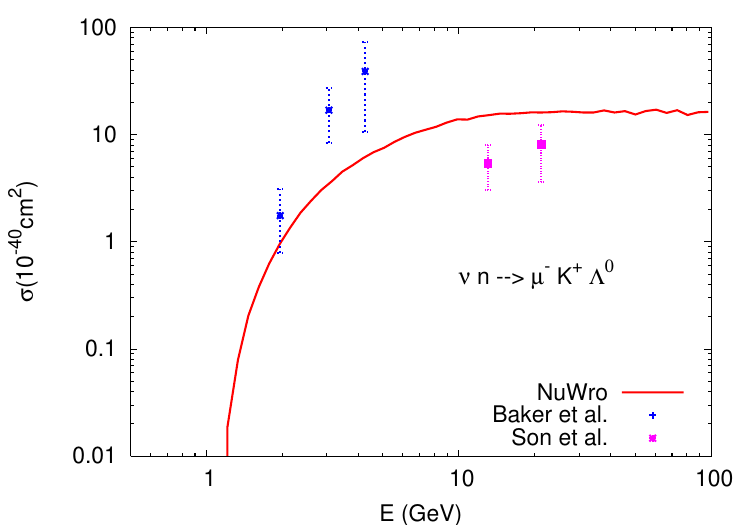} \\
  \caption{Cross sections for channel $\nu n \to \mu^- K^+ \Lambda$. Data points are taken from \cite{Baker:1981tx,Son:1983xh}}. The  figure taken from \cite{Nowak:2006xv}
  \label{strange1}
\end{figure}

Once a neutrino energy is above the $s\bar{s}$ production threshold ($1.095$~GeV), so-called associated production of strange particles will occur.
In Fig. \ref{strange1}, the generator's predictions for the cross section for $\nu n \to \mu^- K^+ \Lambda$ reaction are shown. The data was taken from \cite{Baker:1981tx,Son:1983xh}, and is not consistent (notice the log scale). The MC results are in general agreement with the data.

\subsection{Single kaon production}
The MINERvA collaboration measured charged kaon production in the charged current, neutral current, and coherent production modes.
The measurement is a combination of cabibbo-suppressed single-kaon production, and associated production.

\begin{figure}[h]
  \centering
  \includegraphics[scale=0.5]{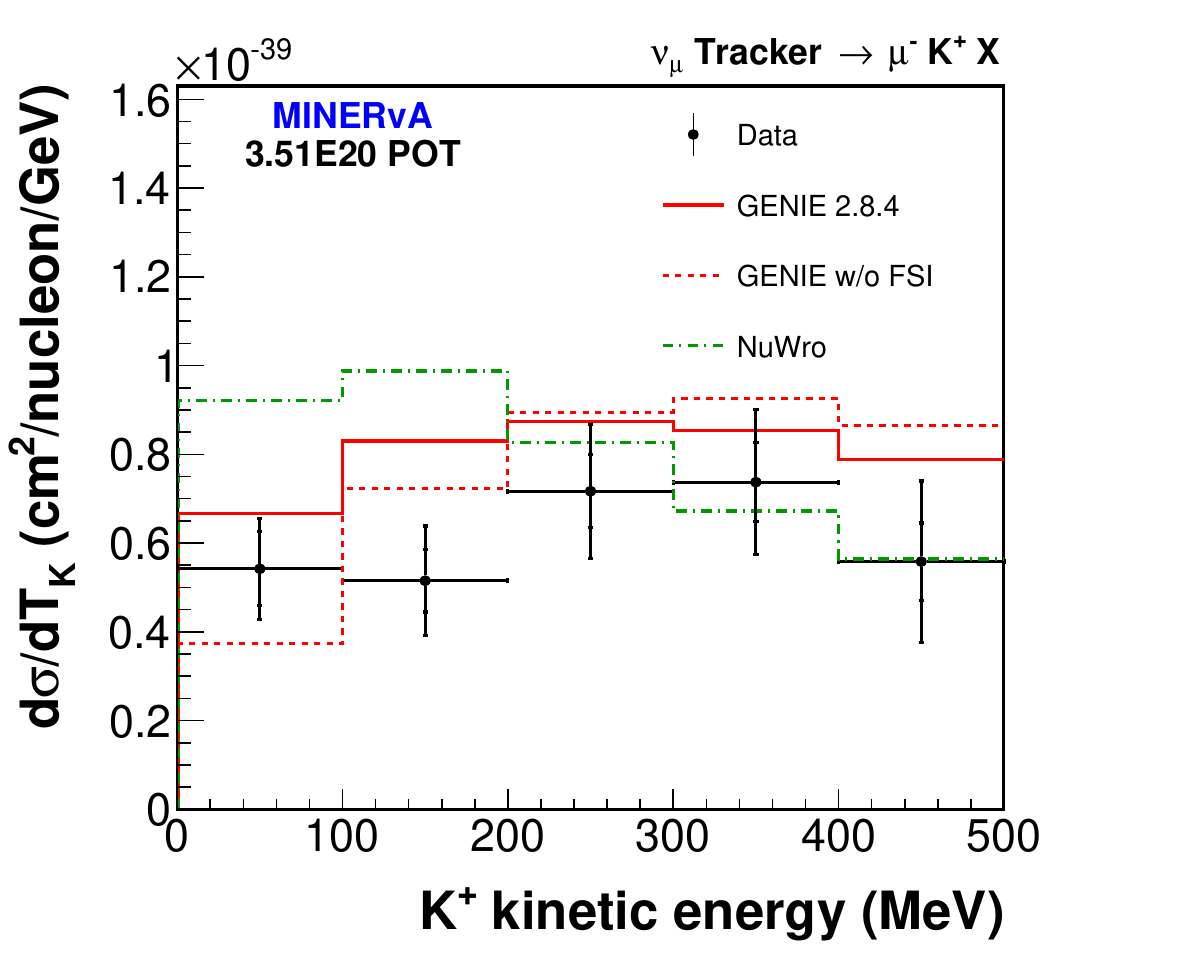}
  \caption{A measurement of charged kaon production in the MINERvA detector. Taken from \cite{MINERvA:2016zyp}}
  \label{Mnv_kaons}
\end{figure}

Fig. \ref{Mnv_kaons} shows the charged current measurement, compared to generator predictions.
While the uncertainties remain large, there is clearly tension with the generators.
Understanding this tension will require additional data, and measurements of the other properties of the events (such as the momentum and direction of the muon, the presence of other particles, etc).
Interestingly, a subset of events from this measurement provided the first evidence (at the level of 3$\sigma$) of coherent kaon production in neutrino interactions.

\subsection{SIS/DIS scattering on strange sea quark/antiquark}

\begin{figure}
\centering
\begin{tabular}{c c}
    \includegraphics[scale=0.5]{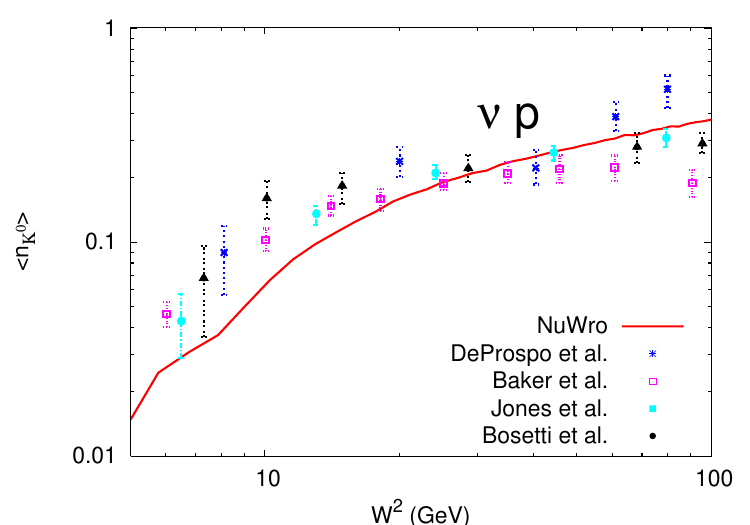}
 &     \includegraphics[scale=0.5]{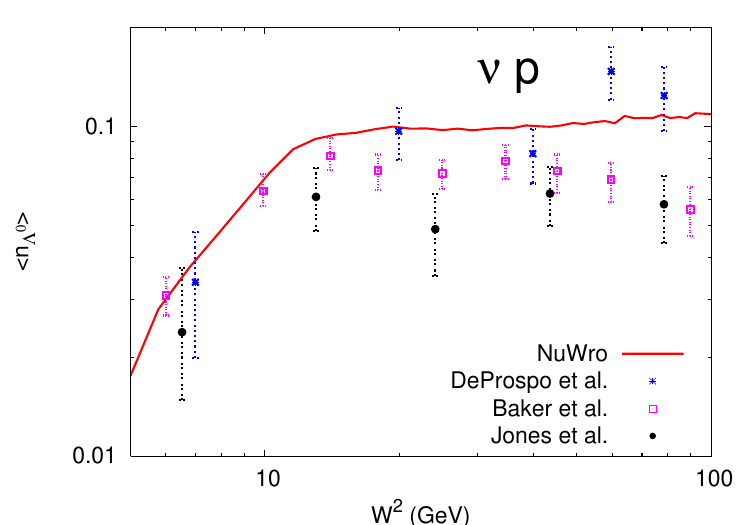} \\
\end{tabular}
  \caption{A comparison of NuWro prediction to the data of the average multiplicity for $K^0$ and $\Lambda^0$ particles in neutrino scattering on proton.The  figure taken from \cite{Nowak:2006xv}   }
  \label{strange2}
\end{figure}

At sufficiently high $Q^2$ and W, neutrinos scatter off individual quarks, allowing them to probe the contributions of strange sea quarks.
In the Fig. \ref{strange2} we show the cross section for neutral strange particles $K^0$
and $\Lambda^0$. Data points are taken from \cite{Baker:1986xx,Aachen-Bonn-CERN-Democritos-London-Oxford-Saclay:1982tho,E632:1994njp,Jones:1992bm}. We notice the
good agreement of the generator's predictions with the data.

\subsection{Nuclear effects}

Understanding the final state interactions of strange hadrons is important to interpreting the data from neutrino experiments.
Some data exists to constrain the re-interactions of charged kaons, but most other strange baryons have minimal data regarding their interactions with nuclei.
Theoretically, the addition of a strange quark to a baryon leads to it experiencing a different nuclear potential, and no Pauli blocking effects (unlike protons and neutrons).

Finally, strange particles can be produced in final state interactions of non-strange particles, although this is almost exclusively through $s\bar{s}$ production.

\subsection{Challenges}
The primary challenges in using the production of strange particles to learn about neutrino interactions are:

\begin{itemize}
  \item Lack of statistics.  All existing measurements are from a small number of events, leading to large statistical uncertainties.  New data from intense sources such as those collected with SBND would be critical.
  \item Understanding the different nuclear effects for strange baryons requires either collection of new data, theoretical development, or both
  \item Many channels are only accessible to antineutrinos, meaning running beam experiments in antineutrino-enhanced configurations is important
\end{itemize}

\newpage
\section{Coherent Processes}
\label{cohdiff}
\subsection{General Introduction}

Coherent scattering refers to the situation in which the target nucleus remains in its ground state after an interaction. It can be elastic or inelastic. Coherent Elastic Neutrino-Nucleus scattering (CEvNS) was first observed by the COHERENT collaboration in 2017 using neutrinos from stopped pions produced at the Oak Ridge spallation neutron source~\cite{COHERENT:2017ipa}. Several other experiments searching for CEvNS are underway (see Ref.~\cite{Abdullah:2022zue} for a compilation). 
CEvNS is only observable through measurements of very small nuclear recoil energies.
In the few GeV region, where recent and future accelerator-neutrino experiments operate, this process represents a tiny, likely undetectable fraction of the cross section. Thus, it has no direct impact on the detector response and oscillation analyses that represent the primary concern of the present review. However, CEvNS studies have an indirect impact on neutrino oscillations owing to their potential to constrain some non-standard neutrino interactions (NSI). These NSI can alter matter effects in neutrino propagation and potentially mimic a CP-violating signal~\cite{Farzan:2017xzy}.

In coherent inelastic scattering, other particles are created: mesons (pions, kaons, $\rho$ mesons), but also photons, dilepton pairs (trident scattering) and, in some Beyond-Standard Model (BSM) scenarios, heavy neutral leptons (HNL) or other new particles. Charge conservation implies that charged positive (negative) particles can only be emitted in CC scattering induced by neutrinos (antineutrinos); in the same way neutral particles arise from NC interactions. The absence of tree-level flavor-changing NC suppresses the coherent production of neutral (anti)kaons or any other strange particles below observable rates. 

NC photon and $\pi^0$ production are common background sources in 
$\overset{\scriptscriptstyle (-)}{\nu_e}$ appearance searches with Cherenkov detectors. In this context, the relatively large (in the forward direction) coherent pion production reaction has been extensively investigated (reviews can be found in Refs.~\cite{Alvarez-Ruso:2014bla,NuSTEC:2017hzk}). 
Coherent processes are also backgrounds for the observation of some BSM processes. In particular, coherent $\gamma$ emission could coexist with a similar signal from exotic candidates to explain the low energy excess observed by MiniBooNE (more details in \ref{subsubsec:Upsc} below). Furthermore, under certain conditions explored in Ref.~\cite{Jung:2025bqh}, coherent pion production could be a tool for constraining the neutrino flux.

In general the scattering amplitude of coherent single- particle production $\overset{\scriptscriptstyle (-)}{\nu_l}(k) \, A(p)  \rightarrow l(k') \, m(p_m) \, A(p')$ can be cast in terms of form factors which are themselves functions of the four-momentum transferred to the nucleus squared, $t=(p'-p)^2$. These form factors exhibit a diffractive pattern, typical of processes that probe the nucleus as a whole, with a maximum at $t=0$ and rapidly decreasing in $|t|$ (at least up to the first diffractive minimum). As discussed in Refs.~\cite{Alvarez-Ruso:2014bla,NuSTEC:2017hzk}, this feature defines the global kinematic properties to be expected in coherent observables: (negligibly) small nuclear recoils, forward peaked (with respect to the incoming neutrino direction) angular distributions and large cross section reduction as the mass of the emitted particle increases. The scaling with the nucleus mass is non-trivial and strongly depends on the process.

\subsection{Coherent meson production}
\label{subsecCohPiTh}
Among weak coherent processes, the production of pseudoscalar mesons, mainly pions but also kaons, is the most extensively scrutinized one, both experimentally and theoretically. 

The conservation of the axial current in the chiral limit (where quark masses are zero) implies that the coherent scattering amplitude at $q^2 =0$ can be written in terms of an unphysical meson-nucleus scattering amplitude $T_{m A} (W^2, t, q^2=0, m_m^2=0)$, where $W$ denotes the meson-nucleus invariant mass. 
Relating the coherent meson production differential cross section at $q^2 = (k - k')^2 =0$  to the meson-nucleus cross section, as done in the class of models coined as PCAC models in the literature~\cite{Rein:1982pf,Berger:2008xs,Paschos:2009ag} requires an additional approximation $T_{m A} (W^2, t, m_m^2=0, q^2=0) \approx T_{m A} (W^2, t, m_m^2, q^2=m_m^2)$. It is expected to be accurate within a few percent for pion production, given the smallness of the pion mass, but it is less precise for kaon production. The extrapolation to $q^2 \neq 0$ is usually accounted by an ad-hoc form factor\footnote{An early discussion on the physics of this form factor can be found in Ref.~\cite{Belkov:1986hn}}, assuming that other terms in the amplitude are negligible. This is known to lead to unrealistically broad pion angular distributions~\cite{Amaro:2008hd,Hernandez:2009vm}. Finally, this approach requires knowledge of the pion-nucleus elastic cross section for the particular nucleus. At present, a comprehensive analysis of pion-nucleus scattering in terms of phase shifts and inelasticities is only available for $^{12}$C~\cite{Dumbrajs:1983it,Schlaile:1997vw}. An assumption on the A-dependence of the coherent cross section is then required but, as shown in \ref{sec:CohMesonExp}, it cannot be merely reduced to a simple scaling.

The alternative approach is to model the coherent reaction directly, relying on the same microscopic input used to describe pion production on nucleons and performing the coherent sum over all nucleonic currents~\cite{Kelkar:1996iv,Singh:2006br,Alvarez-Ruso:2007rcg,Amaro:2008hd,Leitner:2009ph,Nakamura:2009iq,Zhang:2012xi}. 
There are benefits to this approach, namely consistency with other pion production mechanisms, correct treatment of interactions at $q^2 \ne 0$, and natural handling of different nuclei (especially in the case where there exists limited or no pion scattering data).
However, there are challenges with this approach too - as in the description of incoherent pion production, the microscopic approach has to handle many different resonances, non-resonant contributions, and their interferences. Furthermore, a consistent meson-nucleus optical potential is required to deal with the distortion of the outgoing pion wave function.  
Historically, this class of models has focused on coherent pion production in the $\Delta(1232)$ resonance region, limiting their applicability to pion energies below $\sim 500$~MeV.
This method has also been applied to coherent $K^\pm$ production~\cite{Alvarez-Ruso:2012kmi}.  

\subsubsection{Experimental status}
\label{sec:CohMesonExp}

\paragraph{\bf Pion production.} 

Coherent pion production was first observed in the neutral current channel early 1983 by the Aachen-Padova spark-chamber experiment~\cite{Faissner:1983ng} and confirmed by the Aachen Gargamelle group~\cite{Isiksal:1984vh}. 
Further NC coherent $\pi$ production experiments with \numu and \numubar  followed.  The SKAT experiment performed the first observation of CC coherent pion production \cite{Grabosch:1985mt, Nahnhauer:1986xh}. 
MINERvA was the first experiment to apply kinematical constraints coming directly from the dynamics of coherent pion production, namely that $|t|\lesssim 1/R^2$, where $R$ denotes the nuclear radius. 
Under the assumption that nuclear recoil is negligible, the neutrino energy, and the absolute value of its 3-momentum is given by $\mid \bm{k} \mid = E_\mu + E_{\pi}$ and the energy transfer, $q^0$, is equal to the pion energy, $E_{\pi}$. 
The absolute value of $t$,
\begin{equation}
    \mid t \mid \approx ( \bm{k} - \bm{k'} - \bm{p}_\pi )^2
\end{equation}
can then be determined experimentally in terms of the measured muon and pion energies, muon and pion angles $\theta_{\mu}$ and $\theta_{\pi}$  with respect to the neutrino beam direction and the opening angle between the muon and the pion track, $\theta_{\pi\mu}$.
Details of these and other earlier experiments can be found in Sec.~9 of the  NuSTEC White Paper~\cite{NuSTEC:2017hzk}. 
 
More recently, new measurements of Coh$\pi$ have been performed by NOvA~\cite{NOvA:2019bdw}, 
MINERvA~\cite{MINERvA:2022esg} and T2K~\cite{T2K:2023xlh}.
The NOvA experiment has performed a NC coherent $\pi^0$ measurement in a carbon-dominated target, with an effective target mass of 13.8, at the narrow-band off-axis NuMI beam with an average neutrino energy of 2.7 GeV~\cite{NOvA:2019bdw}. 
The reported flux averaged $\sigma=13.8 \pm 0.9 \mathrm{(stat)} \pm 2.3 \mathrm{(sys)} \times 10^{-40}$~cm$^2$ is the most precise NC coherent $\pi^0$ measurement in the few-GeV region. Preliminary results of a new precise CC coherent pion production measurement at an average A of 15.8 can be found in Ref.~\cite{Kuruppu:2023gtt}.

\begin{figure}[h!]
\centering
\includegraphics[width=0.44\textwidth]{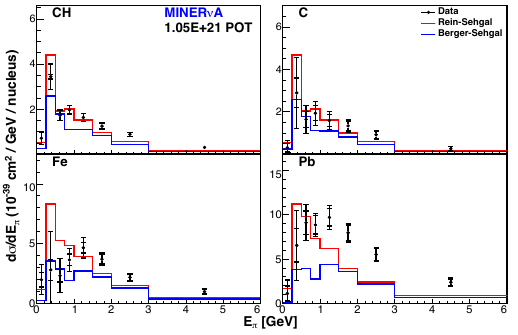} \hspace{0.1\textwidth}  \includegraphics[width=0.44\columnwidth]{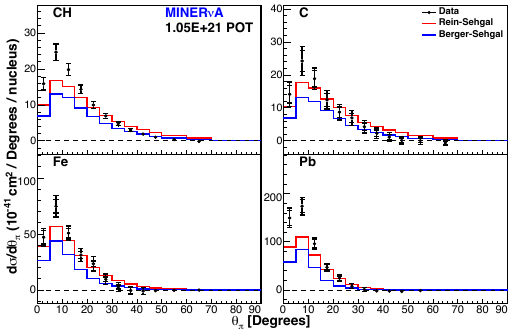} 
    \caption{Left panel: Differential cross sections as function of $E_\pi$ on CH, C, Fe, and Pb targets.  
    Right panel: Differential cross sections as a function of $\theta_\pi$: CH, C, Fe, and Pb. Both figures are taken from Ref.~\cite{MINERvA:2022esg} and compared to the RS (red) and BS (blue) models. } 
\label{fig:MINERvA1}
\end{figure}

\begin{figure}[h!]
\centering
\includegraphics[width=0.44\textwidth]{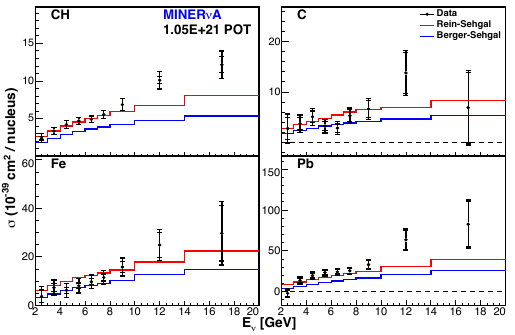} \hspace{0.1\textwidth}  \includegraphics[width=0.44\columnwidth]{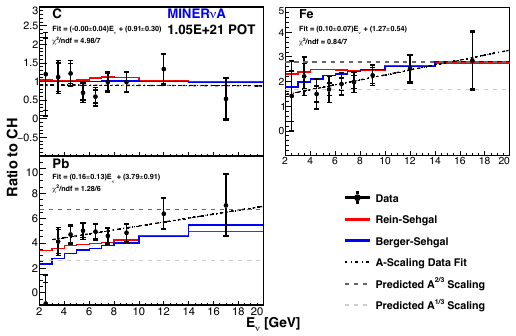}
    \caption{Left panel: Total cross sections as a function of $E_\nu$ on CH, C, Fe, and Pb,  compared to the RS (red) and BS (blue) models.   
    Right panel: Cross section ratios of A/CH as a function of $E_\nu$ for C, Fe and Pb. The upper (lower) dashed line is the ratio predicted by an $A^{2/3}$ ($A^{1/3}$) scaling. The slope is the best A-scaling fit. Both from Ref.~\cite{MINERvA:2022esg}.}
\label{fig:MINERvA2}
\end{figure}

The latest quite high-statistics coherent study from MINERvA using neutrinos from the NuMI medium-energy (ME) beam ($E_{\nu} \approx$ 2 to 20 GeV) concentrated on measurement of the neutrino CC  coherent $\pi^+$ cross section simultaneously in hydrocarbon (CH), graphite (C), iron (Fe) and lead (Pb) targets, a range in A from 12 to 207. As shown in Fig.~\ref{fig:MINERvA1} and the left panel of Fig.~\ref{fig:MINERvA2}, these measurements exceed the predictions of both the Rein-Sehgal (RS)~\cite{Rein:1982pf} and Berger-Sehgal (BS)~\cite{Berger:2008xs} PCAC based models at produced $\pi^+$ energies ($E_\pi > 1$ GeV) energies, $\pi^+$ angles $\theta_\pi < 10$ degrees and at higher neutrino energies (mainly $E_\nu > 8$ GeV ).  

While the disagreement is fairly minor for carbon, the disagreements are much larger for iron and lead.
These MINERvA data show an interesting feature when considering the cross section ratio of Fe or Pb relative to CH:
as shown in the right panel of Fig.~\ref{fig:MINERvA2} the A dependence actually seems to increase from an approximate $A^{1/3}$ scaling at few GeV to an $A^{2/3}$ scaling for $E_\nu > 10$ GeV.  This change in A-dependence with $E_\nu$ is consistent with expectations of the model in reference~\cite{Belkov:1986hn}, which relies on the eikonal approximation and the Glauber formalism for multiple scattering to model the elastic pion-nucleus cross section and the $q^2$ dependent nuclear form factor.

\begin{figure}[h!]
	\centering
	\includegraphics[scale=1]{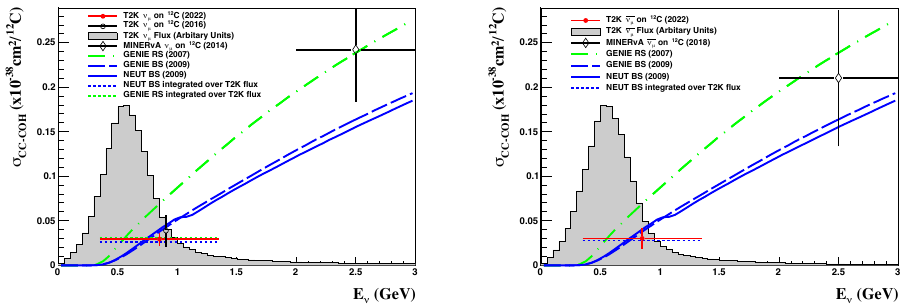}
	\caption{The T2K neutrino (left) and antineutrino (right) CC coherent $\pi$ production cross section measurement on $^{12}$C (scaling with $A^{1/3}$)~\cite{T2K:2023xlh}.}
	\label{fig:T2K}
\end{figure}

The recent T2K study updates the earlier measurement of neutrino induced CC coherent $\pi^+$ production. It also performs the first measurement of antineutrino CC coherent $\pi^-$ production at a mean antineutrino energy less than 1~GeV ($< E_\nu > = 0.85$~GeV). As can be seen in Fig.~\ref{fig:T2K}, the results are reported on $^{12}$C, the main detector component, assuming a scaling of A$^{1/3}$ for other targets. The $\nu_\mu$ and $\bar{\nu}_\mu$ cross section are found to be largely equivalent as expected from the dominance of the axial current.
The data are consistent with both the NEUT BS and the GENIE RS model predictions within the measurement uncertainties (see Fig.~\ref{fig:T2K}). 

The combination of the recent T2K and MINERvA results suggests we now have an experimental picture of Coh$\pi^+$ production over $E_\nu$ from less that 1 GeV to 20 GeV, covering the range of most future experiments although  not a theoretical model which describes the whole range of energies and target masses.
\\

\paragraph{\bf Coherent Production of heavier mesons.} 
As the probed $|t|$ values increase with the mass of the emitted meson, the cross sections for coherent production of mesons heavier than pions are small and very difficult to measure.
Following an earlier measurement of coherent $\rho^+$ production~\cite{Tian:2013ogd}, 
the NOMAD collaboration has studied the $\rho^0$ channel. The signature for coherent $\rho^0$ production is a pair of charged pions from the $\rho^0$ decay and nothing else in the detector. The relative colinearity of the outgoing vector meson with respect to the incoming neutrino direction is a distinguishing feature of coherence. 
A strong signal of $\rho^0$ has been observed, with a rate of (3.89 $\pm$ 0.9) × $10^{-4}$ with respect to $\nu_\mu$ CC events ~\cite{Kullenberg:2020wbt}.
\\

In comparison to CC coherent $\pi$, neutrino-induced CC coherent kaon production, $\nu_{\mu}A\rightarrow\mu^{-}K^{+}A$ has an even lower rate due to Cabibbo suppression ($V_{us}=0.22 < V_{ud} = 0.97$) and due to the kaon mass being significantly larger than the pion mass.
The MINERvA experiment isolated this channel~\cite{MINERvA:2016cun} with a $3.0\, \sigma$ significance.  No new results have been reported since then.
\\

\paragraph{\bf Diffractive contribution to meson production.}

Diffractive is the name given to scattering off a nucleon (as opposed to a nucleus in coherent scattering) when small momentum is transferred to the nucleon in the collision and the wavelength is large enough to see the nucleon as a whole.  Unlike the Coh$\pi$ case, the relatively small nucleon mass makes the outgoing nucleon (in principle) experimentally detectable, as it has been the case in MINERvA (see the NuSTEC white paper~\cite{NuSTEC:2017hzk} for a summary).

A study of this process by the MINERvA collaboration resulted in a estimation of the diffractive cross section for both \numu and \numubar  ~\cite{MINERvA:2017ipy} and~\cite{MINERvA:2016uck}. 
The measured cross section for $E_{\pi} \geq 3$~GeV,
is comparable to that for NC coherent $\pi^{0}$ production from carbon.
This process can be important for the background studies of oscillation experiments, but can only occur in a detector with free hydrogen. 

\subsection{Coherent photon emission}
\label{subsec:cohgamma}

While taking place on strongly interacting nucleons and nuclei, single-photon emission in NC (anti)neutrino scattering encompasses electromagnetic and weak vertices, to which this process owes its small cross section. Nevertheless, NC photon emission events can constitute a background in $\overset{\scriptscriptstyle (-)}{\nu_e}$ appearance measurements when the photons are misidentified as electrons or positrons from CC interactions. In particular, at MiniBooNE, NC photon production constitutes the second largest background in the kinematic region corresponding to the yet-unexplained excess of electron-like events \cite{MiniBooNE:2018esg}\footnote{MicroBooNE data have been found to be inconsistent with an interpretation of the excess based on $\nu_e$ interactions \cite{MicroBooNECollaboration:2024cpi}}. 

Coherent photon emission, 
$\overset{\scriptscriptstyle (-)}{\nu} \, A \rightarrow \overset{\scriptscriptstyle (-)}{\nu} \, A \, \gamma$, accounts for a small but by no means negligible fraction of the total NC photon yield. Actually, the relative relevance of the coherent channel with respect to the incoherent one is comparable, if not greater than in pion production reactions. For example, according to Refs. \cite{Zhang:2012xn,Wang:2014nat}, the coherent mechanism represents 10-18\% (13-27\%) of the photon background to the MiniBooNE excess of events in $\nu$ ($\bar\nu$) mode. Moreover, the relative importance of the coherent process increases significantly in the forward direction, particularly for the $\bar\nu$ mode, as shown in Fig. 4 of Ref. \cite{Wang:2015ivq}, and Fig. 7 of Ref. \cite{Wang:2014nat} for T2K and MiniBooNE, respectively. In addition, coherent photon emission can be a background for the study of $\nu_{\mu}e \rightarrow\nu_{\mu}e$, which is a "standard candle" to constrain neutrino fluxes and a process of interest by itself.  

\begin{figure}[h!]
    \centering
    \includegraphics[width=0.5\linewidth]{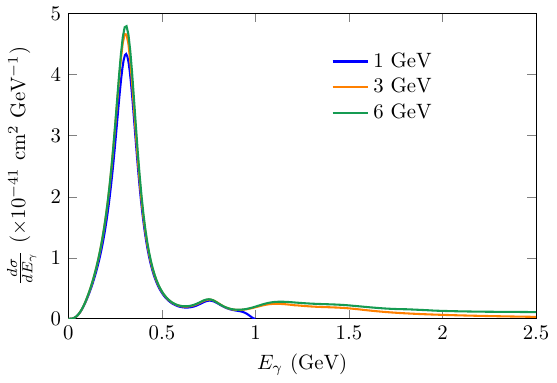}
    \caption{Photon energy distributions for coherent photon emission on $^{12}$C for 1, 3 and 6 GeV neutrinos.}
    \label{fig:cohgamma}
\end{figure}

Because of the photon polarization, the structure of the current is more involved than in the case of pseudoscalar meson production. It is therefore not possible to obtain a simple expression that relates coherent photon production to other processes although connections to Compton scattering and coherent pion photoproduction were explored \cite{rein:1981ys}. The MiniBooNE anomaly inspired theoretical work on NC photon emission and several microscopic calculations of the coherent channel were performed \cite{Hill:2009ek,Zhang:2012xi,wang:2013wva}. These studies are termed microscopic because they rely on the $Z \, N \rightarrow N \, \gamma$ amplitudes on nucleons, which are added coherently. Calculations show that the excitation of the $\Delta(1232)$ is the most important ingredient, even more than in the incoherent case because of a partial cancellation of the non-resonant incoherent contributions. It is also worth noting that at the most favorable kinematics for the nuclear form factor ($q^2=0$), the amplitude is strongly suppressed. Therefore, the optimal kinematics is the result of a balance which leads to a nontrivial angular dependence (see Fig. 12 of Ref. \cite{wang:2013wva}). Baryon resonances heavier than the $\Delta(1232)$, specifically $N(1520)$ but also $\Delta(1700)$ and $\Delta(1959)$, have a small contribution in the high-energy tails of the photon distributions (Fig. \ref{fig:cohgamma}, adapted from Ref. \cite{Alvarez-Ruso:2021dna}). 
This process also receives a contribution from the axial anomaly of QCD through a $Z\gamma\omega$ vertex. While this contribution is much smaller than the $\Delta(1232)$ one \cite{Hill:2009ek,Zhang:2012xi,rosner:2015fwa} and by no means can explain the MiniBooNE excess, its actual strength and energy dependence are uncertain and remain an open question.

\paragraph{Experimental update.} 
Photon emission from neutrino-nucleus scattering has not yet been observed. At high $\langle E_\nu \rangle \sim 25$~GeV energy and in the forward direction, the NOMAD experiment established an upper limit~\cite{NOMAD:2011gyy}. In the $\langle E_\nu \rangle \sim 0.6$~GeV region, T2K has obtained an upper limit with the ND280 detector~\cite{t2k:2019odo}, while MicroBooNE has conducted searches ~\cite{MicroBooNE:2021zai,MicroBooNE:2025ovj} at $\langle E_\nu \rangle \sim 0.8$~GeV with different hadronic final states. MicroBooNE has also performed the first dedicated search for the coherent channel~\cite{MicroBooNE:2025rsd}. This study has led to the first upper limit on argon nuclei and neutrino energies below 1 GeV, of $\sigma  < 1.49 \times 10^{-41}$~cm$^2$ at 90\% C.L., corresponding to $\sim 20$ times the prediction of Ref.~\cite{wang:2013wva}. This analysis was performed with nearly half of the POT collected by the experiment. Tighter constraints can therefore be expected from MicroBooNE but also from the high-statistics data from SBND. To help in these analyses, the dominant contribution to the coherent photon emission cross section according to the calculation of Ref.~\cite{wang:2013wva} has been implemented in GENIE~\cite{GENIE:2021npt}.

\subsection{Beyond-the-Standard Model coherent processes}

\subsubsection{Neutrino trident scattering}
Neutrino trident scattering refers to the neutrino-induced production of dilepton pairs in the Coulomb field of the nucleus~\cite{Czyz:1964zz,Lovseth:1971vv,Fujikawa:1971nx,Brown:1971qr,Koike:1971tu,Brown:1973ih}, namely $\nu_\alpha A \to \nu_\alpha \ell^+_\beta \ell^-_\gamma A$ 
where $\alpha, \beta, \gamma \in \{e,\mu,\tau\}$ represent the lepton flavor. 
At typical accelerator neutrino energies, this is predominantly a coherent process, with the exception of tau production, which is much rarer.
Whenever the two charged leptons are of different flavors, $\beta \neq \gamma$, the process can only proceed via charged-current reactions, while both neutral and charged currents contribute when $\alpha = \beta = \gamma$.
While not a BSM process in itself, it has attracted a lot of attention as it provides a sensitive probe of a variety of new-physics models~\cite{Altmannshofer:2014pba,Magill:2016hgc,Ballett:2019xoj}.
The rate is also relatively well understood as it depends primarily on the electromagnetic nuclear form factor and the SM electroweak parameters. 
Indeed, the virtual leptons exchange a photon with the electric charge in the nucleus, correlating this process with the nuclear form factors measured with electron beams. The photon exchange brings along a coherent enhancement.

At high energies, such as those relevant for neutrino telescopes and atmospheric neutrino experiments, trident production is complemented by the production of real $W^\pm$ bosons in the Coulomb field of the nucleus.
The latter becomes an important contribution to the total neutrino cross section, reaching several percent of the total neutrino cross section above $100$~TeV~\cite{Zhou:2019frk,Zhou:2019vxt}.
This is particularly relevant for neutrino telescopes like IceCube~\cite{IceCube:2013low,IceCube:2017roe} and KM3NET~\cite{KM3Net:2016zxf} (see section \ref{sec:HighEnergy} for more details of high-energy neutrinos).

The first measurement of this rare process was performed by CHARM II~\cite{Geiregat:1990gz} in the CERN wide-band neutrino beam. 
Using their own quoted SM prediction and summing the quoted statistical and systematic uncertainties in quadrature, their result is
\begin{equation}
    \frac{\langle \sigma_{\rm CHARM-II} \rangle}{\langle \sigma_{\rm SM} \rangle} = 1.58 \pm 0.64,
\end{equation}
where $\langle \dots \rangle$ indicates that the cross section has been averaged over the neutrino flux.
Soon after, the CCFR collaboration reported a more precise determination of the cross section using the Fermilab Tevatron neutrino beam,
\begin{equation}
    \frac{\langle \sigma_{\rm CCFR} \rangle}{\langle \sigma_{\rm SM} \rangle} = 0.82 \pm 0.28.
\end{equation}
Note that the significance of these measurements is between $2.5\sigma$ and $3\sigma$ and that they are both statistically limited.
These experiments operated in traditional accelerator neutrino beams composed predominantly of muon neutrinos and antineutrinos, and so were only sensitive to the $\nu_\mu$ and $\overline{\nu}_\mu$ initiated channels.
To the best of our knowledge, no other search for any of the remaining trident channels has been performed to date.

Neutrino trident production can provide a unique test of the SM electroweak sector~\cite{Brown:1973ih}.
This is especially true in channels like $\nu_\mu A \to \nu_\mu \mu^+ \mu^- A$, where the neutral-current and charged-current contributions interfere, leading to a cancellation of about $40\%$ in the total rate.
In addition, the trident rate can be sensitive to new physics that can alter the interactions of neutrinos with the charged leptons produced~\cite{Mishra:1991bv,Gaidaenko:2000hg,Altmannshofer:2014pba,Kaneta:2016uyt,Ge:2017poy,Magill:2017mps,Falkowski:2018dmy,Ballett:2019xoj}.
In many cases, such new physics effects are very effectively constrained by elastic neutrino-electron scattering measurements~\cite{Lindner:2018kjo,Bilmis:2015lja}, unless the new physics is not flavor universal or simply does not couple to the first generation of leptons.
This is the case for a new force carrier, $Z^\prime$, associated with the muon or tau flavors, such as when the conserved current $L_\mu - L_\tau$ is promoted to a new gauge symmetry~\cite{Altmannshofer:2014pba,Kaneta:2016uyt,Ballett:2019xoj}.
Precision measurement of dimuon tridents, in this case, can constrain the $Z^\prime$ interaction strength to be considerably below that of weak interactions.
Sensitivity studies have been performed for DUNE~\cite{Altmannshofer:2019zhy}. 

\subsubsection{Neutrino upscattering and other new-physics channels}
\label{subsubsec:Upsc}

\paragraph{Neutrino upscattering}
Through coherent scattering with nuclei, neutrinos can produce new light particles.  This process can proceed via the exchange of the SM Z boson, new light mediators, or through effective operators like a transition magnetic moment.
In what follows, we discuss a few scenarios considered in the literature and their synergies with the existing physics program to study coherent interactions.

A large neutrino magnetic moment, schematically given by $\mathcal{L} \supset \mu_\nu \overline{\nu_L} \sigma_{\mu \nu} F^{\mu \nu} \nu_R$ in the case of Dirac neutrinos, can be best constrained in CEvNS or neutrino-electron low-energy measurements.  
However, if the magnetic moment is a \emph{transition} magnetic moment involving a new heavy particle, such as a heavy neutral lepton $N$, the interaction term would have the form $\mathcal{L} \supset \mu_{\rm tr} \overline{\nu_L} \sigma_{\mu \nu} F^{\mu \nu} N_R$, where $N$ can have a Majorana or Dirac mass~\cite{Magill:2018jla}. In this case, there is a production threshold 
and accelerator experiments play a crucial role in the search for the new particle, especially if $N$ can produce subsequent visible signatures. 
This scenario is often referred to as neutrino upscattering and proceeds primarily through coherent interactions with nuclei.
In the simplest models, the transition magnetic moment is responsible for both the production and decay of the $N$ particle, $\nu A \to (N \to \nu \gamma) A$. 
The signature constitutes a coherent scattering production point, either upstream of the detector in the dirt and surrounding material or inside the detector, followed by the production of a single photon inside the fiducial volume of the experiment.
The signal is the same as in coherent single-photon emission discussed in Sec.~\ref{subsec:cohgamma}. 
Part of the motivation behind such scenarios is the MiniBooNE low-energy excess of electron-like events~\cite{MiniBooNE:2018esg,MiniBooNE:2020pnu}.
Indeed, the low-energy nature of the excess can be effectively explained by single photons produced in the decay of new particles~\cite{Gninenko:2009ks,Gninenko:2010pr,Gninenko:2012rw,Masip:2012ke,Radionov:2013mca,Magill:2018jla,Schwetz:2020xra,Vergani:2021tgc,Alvarez-Ruso:2021dna,Kamp:2022bpt}.
While successful at explaining the energy spectrum, these models are less successful at reproducing its angular spectrum.
These scenarios can also be constrained by experiments like MINERvA and the Short-Baseline Neutrino Program but, so far, no dedicated searches have been performed.

Another similar scenario arises in models where HNLs can interact with matter via new light mediators. 
In that case, the new mediator can enhance upscattering of neutrinos to HNLs, and lead to the decay of the HNL to dilepton final states~\cite{Bertuzzo:2018itn,Ballett:2018ynz,Ballett:2019pyw,Datta:2020auq,Dutta:2020scq,Abdullahi:2020nyr,Abdallah:2020biq,Abdallah:2020vgg,Hammad:2021mpl}.
In the most minimal models, the overall signature is $\nu A \to (N \to \nu \ell^+\ell^-) A$, where $\ell \in \{e, \mu\}$.
Other exotic final states such as $\gamma \gamma$ or $\pi^+\pi^-$ can also be created depending on the branching ratios of the produced HNL.
When the dominant decay channel of the HNL is into $e^+e^-$ final states, the signature resembles the single photon signature closely.
The overlap between $e^+e^-$ and single photon signals will depend on the ability of each detector to distinguish between these two final states.
A recent search by MicroBooNE sets the first limits on the coherent upscattering to HNLs with a dedicated search for $e^+e^-$ final states~\cite{MicroBooNE:2025khi}.

\paragraph{Neutrino polarizability} We also comment on another possibility where neutrinos can coherently interact with nuclei to produce single photons.
This scenario was discussed in Refs.~\cite{Bansal:2022zpi,Gehrlein:2025tko} and referred to as neutrino polarizability due to the structure of the two-photon operator, e.g., $\overline{\nu}_L \nu_R F_{\mu \nu} F^{\mu\nu}$ for Dirac neutrinos.
One UV completion of the operator includes a scalar mediator $\phi$ with that interacts with neutrinos as well as two photons via a $\phi \gamma \gamma$ vertex.
In that case, the virtual exchange of this scalar by a neutrino on the Coulomb field of the nucleus induces a single photon.
This is akin to the inverse-Primakoff reaction, except it is induced by a virtual scalar particle.
The signature is a single photon emitted preferentially in the forward direction.
These photons also tend to have slightly higher energies than other coherent channels thanks to the high dimensionality of the operator.

\paragraph{Decay in flight signatures}
The production of new particles at the target can also be followed by their decay inside neutrino detectors.
Neutrino experiments have been extremely successful at covering the parameter space of dark sectors below the GeV scale thanks to the high intensity and large detector fiducial volumes.
A few examples include light scalar particles~\cite{MicroBooNE:2019izn} and heavy neutral leptons~\cite{MicroBooNE:2021usw,ArgoNeuT:2021clc}. 
Although these are not neutrino-scattering events, their signatures mimic coherent scattering due to the lack of a hadronic vertex (especially the lack of protons or neutrons).
Final states include $e^+e^-$, $\mu^+\mu^-$, $\pi^+ \pi^-$, single $\gamma$, as well as charged-current-like channels like $e^\pm\pi^\mp$ and $\mu^\pm\pi^\mp$.
Clearly, to search for such signatures experimentally requires an improved understanding of the coherent neutrino interactions that exist as backgrounds.

\subsection{Open questions}
\label{subsec:compare}

Coherent and diffractive meson production, with mainly pions but also kaons and $\rho$-mesons in the final state, have been measured by several experiments. Theoretical studies using different assumptions and techniques have complemented these measurements. Furthermore, other coherent inelastic processes leading to photons or dileptons in the final state have been investigated within the SM and in BSM scenarios. Progress in both theory and experiment has been significant, but outstanding questions remain open.   

\begin{itemize}
\item From the MINERvA experiment, operating with the medium energy configuration of the NuMI beam, we now have not only detailed information about the energy and angular distribution of coherently produced pions but also about the target-mass dependence of the cross section. The revealed discrepancies, particularly in the A-dependence are a challenge for model builders and generator developers: 

\begin{itemize}
    \item In spite of the limitations spelled out in Sec.~\ref{subsecCohPiTh} the simplicity of PCAC models makes them valuable, but the A-dependence should be better understood. For this purpose, new pion-nucleus elastic scattering data, particularly on argon, would be valuable. 
    \item In generators, the presence of multiple and inconsistent implementations of a given model is harmful and should be avoided. 
    \item Microscopic models have the potential to explain the target-mass dependence and describe angular distributions more accurately, but they need to be extended to higher energies to cover the kinematic range probed by MINERvA and DUNE in the future. 
    \item Although the microscopic model of Ref.~\cite{Alvarez-Ruso:2007rcg} is available in GENIE, the implementation is  computationally burdensome. More efficient implementations of this and other microscopic models are therefore needed.
\end{itemize}

\item It would be ideal to have an alternative experiment to check the available results over a comparable neutrino energy range. Unfortunately, there is currently no second experiment to check the latest MINERvA measurements, but MicroBooNE and SBND can complement the T2K measurement with argon data at lower energies.

\item Accomplishing the first measurement of photon emission in coherent (anti)neutrino-nucleus NC scattering is a paramount experimental challenge. Nevertheless, the first upper limit obtained by MicroBooNE is a solid step in this direction and should be followed by more stringent constraints from MicroBooNE and SBND. Such limits may already become useful in constraining some BSM scenarios. On the theoretical side, a more precise  assessment of the anomalous $Z\gamma\omega$ vertex and its contribution to the coherent photon emission cross section, particularly its energy dependence, is still needed.

\item There is currently a significant effort in the theoretical physics community to explore beyond-the-SM scenarios at neutrino experiments. Although experiments have begun to respond to these efforts by developing new dedicated searches, much ground remains to be covered. One major challenge in developing new analyses to search for new physics is the need for appropriate Monte-Carlo (MC) tools that can simulate the underlying signatures. There have been several efforts to extend existing Monte-Carlo generators like \textsc{GENIE}~\cite{Andreopoulos:2009rq,GENIE:2021zuu} to include new physics processes, as well as the creation of new custom tools like \textsc{Achilles}~\cite{Isaacson:2021xty,Isaacson:2022cwh}, \textsc{DarkNews}~\cite{Abdullahi:2022cdw}, \textsc{LeptonInjector}~\cite{IceCube:2020tcq}, and \textsc{MadDump}~\cite{Buonocore:2018xjk}. While this represents good progress, there are still significant gaps in new physics coverage and, in many cases, a lack of \emph{integration} between new MC and experimental analysis tools.

\end{itemize}


\newpage
\section*{Acknowledgements}

We thank D. Schmitz, C. Wilkinson and Z. Vallari for valuable comments on sections of this manuscript.
A. P. F was supported by the U.S. Department of Energy, Office of Science, under award DE-SC0012069.
The work of S. R. K. was supported in part the U.S. Department of Energy, Office of Science, Office of Nuclear Physics, under contract number DE-AC-76SF00098 and the National Science Foundation under grant number PHY-2209445
R.G.-J. was supported by project RYC2022-035203-I funded by MCIN/AEI/10.13039/501100011033/FEDER and FSE+, UE; and by “Ayudas para Atracción de Investigadores con Alto Potencial-modalidad A” funded by VII PPIT-US.
G.D.M. acknowledges support from the European Union’s Horizon research and innovation programme under the Marie Sk$\l$odowska-Curie Actions (MSCA) HORIZON-MSCA-2023-SE-01-01 - MSCA Staff Exchanges 2023, Grant Agreement ID: 101183137 (JENNIFER3 project), from the Spanish Ministerio de Ciencia, Innovaci\'on y Universidades and ERDF (European Regional Development Fund) under contract PID2023-146401NB-I00, by the Junta de Andalucia (grant No. FQM160), and from the University of Tokyo ICRR’s Inter-University Research Program FY2025 and FY2026 (Refs. 2025i-J-001 and 2026i-J-001).
N.J.~ acknowledges support from Ghent University Special Research Fund and the Fund for Scientific Research Flanders (FWO-Flanders). 
A.A was supported by ERC grant (NeutrinoNuclei, 101078772). 
A.B.B. was supported in part by the U.S. National Science Foundation Award PHY-2411495.
R.J.H. was supported by the U.S. Department of Energy, Office of Science, Office of High Energy Physics, under Award DE-SC0019095.
F.K. was funded by the Deutsche Forschungsgemeinschaft (DFG, German Research Foundation) under the grants SCHO~1527/13-1 and KL~3880/1-1.
Support for participating scientists was provided by
UKRI Science and Technology Facilities Council.
C.W. is supported by The Royal Society under award URF\textbackslash R1\textbackslash 241892 and the Science and Technology Facilities Council. 
C.W. acknowledges discussions during ECT* workshops, supported by EURO-LABS,
which received funding from the European Union’s Horizon Europe Research and Innovation programme under grant agreement No 101057511, the NuXTract
workshop at CERN, Pitt-PACC workshops hosted by the University of Pittsburgh, and event generator workshops at Fermilab.
M.K. acknowledges support from the Royal Commission for the Exhibition of 1851 through a Research Fellowship and from the Royal Society through a University Research Fellowship.
N.J. is supported by the Science and Technology Facilities Council.
JTS was partially  supported by the National Science
Centre under grant UMO-2021/41/B/ST2/02778.
This manuscript has been authored by FermiForward Discovery Group, LLC under Contract No. 89243024CSC000002 with the U.S. Department of Energy, Office of Science, Office of High Energy Physics.
The United States Government retains and the publisher, by accepting the article for publication, acknowledges that the United States Government retains a non-exclusive, paid-up, irrevocable, world-wide license to publish or reproduce the published form of this manuscript, or allow others to do so, for United States Government purposes.

\bibliography{all}
\end{document}